\documentclass[longauth]{aa}  
\usepackage{graphicx}
\usepackage{txfonts}
\usepackage{hyperref}
\usepackage{enumitem}
\usepackage{ulem}

\usepackage{pgfplotstable,tabularx,booktabs}
\usepackage{colortbl,xcolor,array}
\usepackage{makecell}
\pgfplotsset{compat=1.17}
\usepackage{txfonts}
\usepackage{xr}
\usepackage[export]{adjustbox}
\usepackage{pdflscape}
\pgfkeys{/pgf/number format/.cd,1000 sep={}}
\usepackage{orcidlink}
\graphicspath{ {./Plots/} }

\begin{document}

\title{ALMAGAL III. Compact source catalog: Fragmentation statistics and physical evolution of the core population}

   
\titlerunning{ALMAGAL III. Compact source catalog and physical analysis}
\authorrunning{A. Coletta et al.}

\author{A. Coletta\inst{\ref{iaps},\ref{sapienza}}\fnmsep\thanks{\email{alessandro.coletta@inaf.it}}\orcidlink{0000-0001-8239-8304}, S. Molinari\inst{\ref{iaps}}\orcidlink{0000-0002-9826-7525}, E. Schisano\inst{\ref{iaps}}\orcidlink{0000-0003-1560-3958}, A. Traficante\inst{\ref{iaps}}\orcidlink{0000-0003-1665-6402}, D. Elia\inst{\ref{iaps}}\orcidlink{0000-0002-9120-5890}, M. Benedettini\inst{\ref{iaps}}\orcidlink{0000-0002-3597-7263}, C. Mininni\inst{\ref{iaps}}\orcidlink{0000-0002-2974-4703}, J.~D.~Soler\inst{\ref{iaps}}\orcidlink{0000-0002-0294-4465}, Á. Sánchez-Monge\inst{\ref{csic},\ref{ieec}}\orcidlink{0000-0002-3078-9482}, P. Schilke\inst{\ref{koln}}\orcidlink{0000-0003-2141-5689}, C. Battersby\inst{\ref{uconn}}\orcidlink{0000-0002-6073-9320}, G.~A. Fuller\inst{\ref{uman},\ref{koln}}\orcidlink{0000-0001-8509-1818}, H. Beuther\inst{\ref{mph}}\orcidlink{0000-0002-1700-090X}, Q. Zhang\inst{\ref{cfa}}\orcidlink{0000-0003-2384-6589}, M.\ T.\ Beltr\'an\inst{\ref{arcetri}}\orcidlink{0000-0003-3315-5626}, B. Jones\inst{\ref{koln}}\orcidlink{0000-0002-0675-0078}, R.~S.~Klessen \inst{\ref{ITA},\ref{IWR}}\orcidlink{0000-0002-0560-3172}, S. Walch\inst{\ref{koln},\ref{datacologne}}\orcidlink{0000-0001-6941-7638}, F. Fontani\inst{\ref{arcetri},\ref{lerma},\ref{mpe}}\orcidlink{0000-0003-0348-3418}, A. Avison\inst{\ref{ska},\ref{uman},\ref{almauk}}\orcidlink{0000-0002-2562-8609}, C.~L. Brogan\inst{\ref{nraova}}\orcidlink{0000-0002-6558-7653}, S.~D.~Clarke\inst{\ref{asiaa}}\orcidlink{0000-0001-9751-4603}, P. Hatchfield\inst{\ref{jpl}}\orcidlink{0000-0003-0946-4365}, P. Hennebelle\inst{\ref{saclay}}\orcidlink{0000-0002-0472-7202}, P.~T.~P. Ho\inst{\ref{asiaa},\ref{hawaii}}\orcidlink{0000-0002-3412-4306}, T.~R. Hunter\inst{\ref{nraova}}\orcidlink{0000-0001-6492-0090}, K.~G. Johnston\inst{\ref{linc}}\orcidlink{0000-0003-4509-1180}, P.~D. Klaassen\inst{\ref{edin}}\orcidlink{0000-0001-9443-0463}, P.~M. Koch\inst{\ref{asiaa}}\orcidlink{0000-0003-2777-5861}, R. Kuiper\inst{\ref{duis}}\orcidlink{0000-0003-2309-8963}, D.~C. Lis\inst{\ref{jpl}}\orcidlink{0000-0002-0500-4700}, T. Liu\inst{\ref{shanghai}}\orcidlink{0000-0002-5286-2564}, S.~L. Lumsden\inst{\ref{leeds}}\orcidlink{0000-0001-5748-5166}, Y. Maruccia\inst{\ref{napoli}}\orcidlink{0000-0003-1975-6310}, T. Möller\inst{\ref{koln}}\orcidlink{0000-0002-9277-8025}, L. Moscadelli\inst{\ref{arcetri}}\orcidlink{0000-0002-8517-8881}, A. Nucara\inst{\ref{torver},\ref{iaps}}\orcidlink{0009-0005-9192-5491}, A.~J.~Rigby\inst{\ref{leeds},\ref{cardiff}}\orcidlink{0000-0002-3351-2200}, K.~L.~J. Rygl\inst{\ref{inafbo}}\orcidlink{0000-0003-4146-9043}, P. Sanhueza\inst{\ref{naoj},\ref{meguro}}\orcidlink{0000-0002-7125-7685}, F. van der Tak\inst{\ref{gron},\ref{sron}}\orcidlink{0000-0002-8942-1594}, M.~R.~A. Wells\inst{\ref{mph}}\orcidlink{0000-0002-3643-5554}, F. Wyrowski\inst{\ref{bonn}}\orcidlink{0000-0003-4516-3981}, F. De Angelis\inst{\ref{iaps}}\orcidlink{0009-0002-6765-7413}, S. Liu\inst{\ref{iaps}}\orcidlink{0000-0001-7680-2139}, A. Ahmadi\inst{\ref{leiden}}\orcidlink{0000-0003-4037-5248}, L. Bronfman\inst{\ref{unichile}}\orcidlink{0000-0002-9574-8454}, S.-Y.~Liu\inst{\ref{asiaa}}\orcidlink{0000-0003-4603-7119}, Y.-N.~Su\inst{\ref{asiaa}}, Y. Tang\inst{\ref{asiaa}}\orcidlink{0000-0002-0675-276X}, L. Testi\inst{\ref{ubo}}\orcidlink{0000-0003-1859-3070}, and H. Zinnecker\inst{\ref{uniautochile}}\orcidlink{0000-0003-0504-3539}
}

\institute{INAF - Istituto di Astrofisica e Planetologia Spaziali (IAPS), Via Fosso del Cavaliere 100, I-00133 Roma, Italy \label{iaps}
\and Dipartimento di Fisica, Sapienza Universita’ di Roma, Piazzale Aldo Moro 2, I-00185, Rome, Italy \label{sapienza}
\and Institute of Space Sciences (ICE-CSIC)
Carrer de Can Magrans s/n, E-08193, Barcelona, Spain \label{csic}
\and Institut d'Estudis Espacials de Catalunya (IEEC), E-08860, Castelldefels (Barcelona), Spain \label{ieec}
\and Physikalisches Institut der Universit\"at zu K\"oln, Z\"ulpicher Str. 77, D-50937 K\"oln, Germany \label{koln}
\and University of Connecticut, Department of Physics, 2152 Hillside Road, Unit 3046 Storrs, CT 06269, USA \label{uconn}
\and Jodrell Bank Centre for Astrophysics, Oxford Road, The University of Manchester, Manchester M13 9PL, UK \label{uman}
\and Max Planck Institute for Astronomy, K\"onigstuhl 17, 69117 Heidelberg, Germany \label{mph}
\and Center for Astrophysics, Harvard \& Smithsonian, 60 Garden Street, Cambridge, MA 02138, USA \label{cfa}
\and INAF-Osservatorio Astrofisico di Arcetri, Largo E. Fermi 5, 50125, Firenze, Italy \label{arcetri}
\and Universit\"{a}t Heidelberg, Zentrum f\"{u}r Astronomie, Institut f\"{u}r Theoretische Astrophysik, Albert-Ueberle-Str. 2, 69120 Heidelberg, Germany \label{ITA}
\and 
Universit\"{a}t Heidelberg, Interdisziplin\"{a}res Zentrum f\"{u}r Wissenschaftliches Rechnen, Im Neuenheimer Feld 205, 69120 Heidelberg, Germany \label{IWR}
\and Center for Data and Simulation Science, University of Cologne, Germany \label{datacologne}
\and Laboratoire d’\'Etudes du Rayonnement et de la Mati\`ere en Astrophysique et Atmosph\`eres (LERMA), Observatoire de Paris, Meudon, France \label{lerma}
\and Max-Planck-Institute for Extraterrestrial Physics (MPE), Garching bei M\"unchen, Germany \label{mpe}
\and SKA Observatory, Jodrell Bank, Lower Withington, Macclesfield SK11 9FT, UK \label{ska}
\and UK ALMA Regional Centre Node, M13 9PL, UK \label{almauk}
\and National Radio Astronomy Observatory, 520 Edgemont Road, Charlottesville VA 22903, USA \label{nraova}
\and Institute of Astronomy and Astrophysics, Academia Sinica, 11F of ASMAB, AS/NTU No.\ 1, Sec.\ 4, Roosevelt Road, Taipei 10617, Taiwan \label{asiaa}
\and Jet Propulsion Laboratory, California Institute of Technology, 4800 Oak Grove Drive, Pasadena, CA 91109, USA \label{jpl}
\and Universit\'e Paris-Saclay, Universit\'e Paris-Cit\'e, CEA, CNRS, AIM, 91191 Gif-sur-Yvette, France \label{saclay}
\and East Asian Observatory, 660 N.\ A'ohoku, Hilo, Hawaii, HI 96720, USA \label{hawaii}
\and School of Engineering and Physical Sciences, Isaac Newton Building, University of Lincoln, Brayford Pool, Lincoln, LN6 7TS, United Kingdom \label{linc}
\and UK Astronomy Technology Centre, Royal Observatory Edinburgh, Blackford Hill, Edinburgh, EH9 3HJ, UK \label{edin}
\and Faculty of Physics, University of Duisburg-Essen, Lotharstraße 1, D-47057 Duisburg, Germany \label{duis}
\and Shanghai Astronomical Observatory, Chinese Academy of Sciences, 80 Nandan Road, Shanghai 200030, China \label{shanghai}
\and School of Physics and Astronomy, The University of Leeds, Woodhouse Lane, Leeds LS2 9JT, UK \label{leeds}
\and INAF - Astronomical Observatory of Capodimonte, Via Moiariello 16, I-80131 Napoli, Italy \label{napoli}
\and Dipartimento di Fisica, Università di Roma Tor Vergata, Via della Ricerca Scientifica 1, I-00133 Roma, Italy \label{torver}
\and Cardiff Hub for Astrophysics Research \& Technology, School of Physics \& Astronomy, Cardiff University, Queen's Buildings, The Parade, Cardiff CF24 3AA, UK \label{cardiff}
\and INAF-Istituto di Radioastronomia, Via P. Gobetti 101, I-40129 Bologna, Italy \label{inafbo}
\and National Astronomical Observatory of Japan, National Institutes of Natural Sciences, 2-21-1 Osawa, Mitaka, Tokyo 181-8588, Japan \label{naoj}
\and Department of Earth and Planetary Sciences, Institute of Science Tokyo, Meguro, Tokyo, 152-8551, Japan \label{meguro}
\and Kapteyn Astronomical Institute, University of Groningen, 9700, AV Groningen, The Netherlands \label{gron}
\and SRON Netherlands Institute for Space Research, Landleven 12, 9747, AD Groningen, The Netherlands \label{sron}
\and Max-Planck-Institut f\"ur Radioastronomie, Auf dem H\"ugel 69, D-53121 Bonn, Germany \label{bonn}
\and Leiden Observatory, Leiden University, PO Box 9513, 2300 RA Leiden, The Netherlands \label{leiden}
\and Departamento de Astronomía, Universidad de Chile, Casilla 36-D, Santiago, Chile \label{unichile}
\and Dipartimento di Fisica e Astronomia, Alma Mater Studiorum - Universit\`a di Bologna \label{ubo}
\and Universidad Autonoma de Chile, Avda Pedro de Valdivia 425, Santiago de Chile \label{uniautochile}
   }

   \date{Received xxx; accepted xxx}


\abstract
{The physical mechanisms behind the fragmentation of high-mass dense clumps into compact star-forming cores and the properties of these cores are fundamental topics that are heavily investigated in current astrophysical research. The ALMAGAL survey provides the opportunity to study this process at an unprecedented level of detail and statistical significance, featuring high-angular resolution $1.38$ mm ALMA observations of $1013$ massive dense clumps at various Galactic locations. These clumps cover a wide range of distances ($\sim2-8$\,kpc), masses ($\sim10^{2}-10^{4}\,\mathrm{M_{\odot}}$), surface densities ($0.1-10$ g $\mathrm{cm^{-2}}$), and evolutionary stages (luminosity over mass ratio indicator of $\sim0.05<L/M<450\,\mathrm{L_{\odot}/M_{\odot}}$). 
Here, we present the catalog of compact sources obtained with the \textit{CuTEx} algorithm from continuum images of the full ALMAGAL clump sample combining ACA-$7$m and $12$m ALMA arrays, reaching a uniform high median spatial resolution of $\sim1400$ au (down to $\sim800$\,au). We characterize and discuss the revealed fragmentation properties and the photometric and estimated physical parameters of the core population. 
The ALMAGAL compact source catalog   includes $6348$ cores detected in $844$ clumps ($83\%$ of the total), with a number of cores per clump between $1$ and $49$ (median of $5$). The estimated core diameters are mostly within $\sim800-3000$ au (median of $1700$ au). We assigned core temperatures based on the $L/M$ of the hosting clump, and obtained core masses from $0.002$ to $345\,\mathrm{M_{\odot}}$ (complete above $0.23\,\mathrm{M_{\odot}}$), exhibiting a good correlation with the core radii ($M\propto R^{2.6}$). 
We evaluated the variation in the core mass function (CMF) with evolution as traced by the clump $L/M$, finding a clear, robust shift and change in slope among CMFs within subsamples at different stages. This finding suggests that the CMF shape is not constant throughout the star formation process, but rather it builds (and flattens) with evolution, with higher core masses reached at later stages. 
We found that all cores within a clump grow in mass on average with evolution, while a population of possibly newly formed lower-mass cores is present throughout. The number of cores increases with the core masses, at least until the most massive core reaches $\sim10\,\mathrm{M_{\odot}}$. 
More generally, our results favor a clump-fed scenario for high-mass star formation, in which cores form as low-mass seeds, and then gain mass while further fragmentation occurs in the clump.
}

\keywords{ISM: structure -- Stars: formation -- Submillimeter: ISM -- Methods: observational -- Techniques: interferometric -- Surveys
        }

\maketitle
   
   
%

\section{Introduction}

The majority of stars, including the Sun (\citealt{Adams10, Gounelle+12, Pfalzner+15}), form within crowded clusters hosting at least one high-mass star ($M_\star\geq8\,\mathrm{M_\odot}$, see, e.g., \citealt{Carpenter00, Lada&Lada03}). 
Therefore, studies of the physical properties of high-mass star-forming regions (HMSFRs) and, in particular, the mechanisms regulating their evolution, provide crucial information on the birth processes of stars and, ultimately, of planetary systems. 

Massive stars influence their surrounding environment in multiple ways (see \citealt{Krumholz+14,Rosen+20}), for instance, through gravitational, mechanical (winds, outflows), and radiative (radiation pressure) interaction, and eventually through supernova explosions that enrich the interstellar medium (ISM) with heavy elements. 
Stellar feedback processes turn out to be relevant not only locally, but also on Galactic scales (\citealt{Bolatto+13,Girichidis2016,Rathjen+21}), and their effects vary as the physical and chemical properties of the involved regions change with evolution (e.g., \citealt{Caselli05, Molinari+08, Klessen+16, Molinari+19, Coletta+20}). 

The formation of high-mass stars (see, e.g., reviews by \citealt{Stahler+00,Beuther+07,Zinn&York07,Motte+18}) takes place in cold and dense regions within interstellar molecular clouds (MCs) called clumps, which typically have sizes of the order of $0.1-2$ pc, temperatures of $\sim10-60$ K, masses of $\sim10^2-10^4\,\rm{M_\odot}$, and average densities of $\sim10^3-10^6\,\mathrm{cm^{-3}}$ (e.g., \citealt{Elia+17,Elia+21}). 
In detail, massive stars form within the densest, relatively hot compact substructures called cores ($n\geq10^6$ cm$^{-3}$, $R\leq0.05$ pc, e.g., \citealt{Garay&Lizano99,Beuther07,Zhang+09,Sanchez-Monge+13b,Yamamoto17,Sanhueza+19,Sadaghiani+20,Pouteau+22}), which can be heated by the forming protostar(s) to temperatures up to $\sim100$ K or even more (e.g., \citealt{Beuther+05,Kurtz+00,Cesaroni+07,Beltran+11,Silva+17}). 

The formation of a (proto)cluster involves the breaking up of the molecular clump into cores via dynamical fragmentation during its gravitational collapse (\citealt{Stahler&Palla04}). 
The outcome of an individual collapse (e.g., size, mass, and spatial distribution of the formed compact fragments, i.e., cores) depends on the initial conditions of the parent clump, for example, in terms of morphology, mass, and density. Observationally, clumps present a variety of shapes and emission patterns (ranging from extended and diffuse, to filamentary-like, to compact structures) and different degrees of fragmentation into substructures (e.g., \citealt{Sanhueza+19,Svoboda+19}). 
Besides a pure gravitationally driven scenario, other effects have been also brought into play to explain the fragmentation mechanisms leading to the formation of high-mass stars, such as turbulent support (\citealt{Mckee+03,Zhang+09}), interstellar magnetic fields (\citealt{Commercon+11,Hennebelle+11,Zhang+14b,Tang+19,Palau+21}), radiative feedback \citep{Krumholz+09,Peters2010,Hennebelle+20}, and mass flow along filaments (\citealt{Peretto+13,Smith2014, Smith2016,Lu+18,Wells+24}). 

The competitive accretion scenario (or clump-fed, e.g., \citealt{Zinnecker82,Klessen+00,Klessen2001,Bonnell+01,Bonnell+04,Bonnell+06,Smith+09,Vazquez+19,Padoan+20,Traficante+23,Morii+24}) depicts a dynamic, multi-scale hierarchical framework, where the material that accretes onto the cores (and ultimately onto one or more protostars) is infalling from the larger scale intra-clump medium. Multiple small, low-mass seeds resulting from the fragmentation of a massive clump compete to access and accrete the available reservoir. This implies that not all the fragments will manage to produce high-mass stars. Moreover, objects that succeed in consistently accreting mass will grow more efficiently with time (e.g., \citealt{Bonnell+01,Wang+10,Battersby+17}), thanks to their increasing gravitational attraction on larger portions of the surrounding material. 
This scenario is nowadays generally favored over the core (monolithic) accretion theory (or core-fed, e.g., \citealt{Mckee+03,Tan+03,Krumholz+05,Tan+14}), where the reservoir for accretion is entirely initially available within a massive, isolated core. 
While in fact substantial proofs of existence of massive prestellar cores are missing, despite dedicated efforts (e.g., \citealt{Zhang+09,Wang+14,Zhang+15,Sanhueza+17,Sanhueza+19,Svoboda+19,Morii+23,Mai+24}), multiple evidence in support of clump-fed mechanisms have been produced, both theoretically (e.g., \citealt{Bonnell+06,Bonnell+07,Smith+09,Grudic+22,Hennebelle+22}) and observationally (e.g., \citealt{Sanhueza+19,Anderson+21,Liu+23,Traficante+23,Morii+24,Wells+24,Xu+24}). These and other works revealed clump-to-core infall motions and core mass growth with evolution. 

Clumps and cores are observed at millimeter and submillimeter (mm and submm) wavelengths via the continuum emission of dust grains. 
The introduction of modern radio mm/submm interferometers, such as the Submillimeter Array (SMA, \citealt{Ho+04}), the NOrthern Extended Millimeter Array (NOEMA, \citealt{Chenu+16}), and, especially, the Atacama Large Millimeter/submillimeter Array (ALMA, \citealt{Wootten+09}), has recently made it possible to investigate statistically significant samples of high-mass star-forming regions at an unprecedented level of detail. High-resolution ($<1''$) observations are required to locate and resolve the compact substructures (cores) within massive clumps, which are typically located at large distances ($>1$ kpc). 
Various surveys have recently been conducted to study the fragmentation properties of candidate star-forming clumps at different evolutionary stages. 

For example, InfraRed Dark Clouds (IRDCs) were targeted with the SMA (e.g., \citealt{Zhang+09,Zhang+11,Wang+11,Wang+14,Sanhueza+17}). The CORE survey (\citealt{Beuther+18}) observed $20$ protostellar regions at $\lesssim1000$ au spatial resolution with NOEMA. 
Among the surveys employing ALMA, ASHES (\citealt{Sanhueza+19,Morii+23,Morii+24}) performed mosaic observations of $39$ $70\,\mu$m-dark regions at $\sim2000$ au resolution, ATOMS (\citealt{Liu+20,Liu+22a,Liu+22b}) studied $146$ hyper- and ultra-compact HII regions (HC/UCHIIs) at $\sim2000$ au resolution, while SQUALO (\citealt{Traficante+23}) inspected a varied sample of $13$ regions at different evolutionary stages (from $70\,\mu$m-dark to HII) at a $\sim2000$ au resolution. The TEMPO survey (\citealt{Avison+23}) observed $38$ massive star-forming regions with a wide range of evolutionary stages at a $\sim2000$ au resolution. Most recently, ASSEMBLE (\citealt{Xu+24}) targeted $11$ evolved massive clumps at a $2200$ au maximum resolution, while the DIHCA survey (\citealt{Ishihara+24}) observed $30$ clumps at a $\sim900$ au resolution. 
These and other studies (e.g., \citealt{Palau+14,Svoboda+19}) have revealed a variety of clump fragmentation degrees, going from fields with only $1$ compact fragment to more crowded ones (with up to $40$ fragments). Wide ranges of core sizes (from $\sim600$ to a few thousands of au) and masses (from $\sim0.1$ to $\sim300\,\mathrm{M_\odot}$) were also estimated. 
The core mass function (CMF) was recently examined in some detail by, for example, \citet{Sanhueza+19}, \citet{Lu+20}, \citet{Sadaghiani+20}, and \citet{Pezzuto+23}. In particular, the ALMA-IMF survey (\citealt{Motte+22}) inspected $15$ massive protoclusters, discussing their inner and overall CMF and its variation with evolution (\citealt{Pouteau+22,Pouteau+23,Louvet+24}). 

It must be noted that such analyses are always sensitive to the achieved spatial resolution (see, e.g., \citealt{Sadaghiani+20,Louvet+21}), mass sensitivity, and employed observational technique (e.g., single pointing or mosaic). Moreover, importantly, different algorithms are usually used to extract compact sources from continuum maps; for instance, \textit{CuTEx} (Curvature Thresholding Extractor, \citealt{Molinari+16b,Elia+17,Elia+21}), \textit{astrodendro} (e.g., \citealt{Sanhueza+19,Svoboda+19,Anderson+21,Morii+23}), \textit{getsf} (e.g., \citealt{Pouteau+22,Xu+24}), and \textit{hyper} (\citealt{Traficante+23}). 
Each approach inherently defines the properties of objects to be revealed (e.g., in terms of size and shape). This aspect must be considered when comparing results based on different assumptions. 

However, a comprehensive understanding of the clump fragmentation process and the mechanisms of massive star formation is still lacking. 
Large, representative samples observed with high resolution and sensitivity are needed to investigate the physical properties of the cores, their relation with hosting clump properties, and how they evolve over time (e.g., \citealt{Csengeri+17,Fontani+18b,Sanhueza+19,Anderson+21,Traficante+23,Xu+24}). 

The ALMAGAL survey (see \citealt{Molinari+25}, \citealt{Sanchez+25}), an ALMA Cycle $7$ Large Program (2019.1.00195.L, PIs: Sergio Molinari, Cara Battersby, Paul Ho, Peter Schilke), is now providing the opportunity to study all the different aspects of the star formation process in our Galaxy with an unprecedented level of statistical relevance, robustness, and detail. This is thanks to the size and variety of its target sample, featuring $1013$ massive dense clumps spread across the Galactic plane and spanning the full evolutionary sequence from IRDCs to HII regions, and its high-resolution and high-sensitivity $1.38$ mm observations, making it possible to assess fragmentation down to the typical compact cores scales ($\lesssim1000$ au). 

In this paper, we investigate the clump fragmentation process and its outcome by taking full advantage of the advanced capabilities provided by the ALMAGAL survey. 
The paper is structured as follows. 
In Sect. \ref{obs}, we present the full ALMAGAL target sample and the observations carried out with ALMA, also characterizing the main properties of the dust continuum images. In Sect. \ref{SE}, we describe the compact source extraction procedure, which was developed using the \textit{CuTEx} algorithm (\citealt{Molinari+11}) and applied to the maps to reveal the dense substructures (cores) within the whole ALMAGAL clump sample. In Sect. \ref{photo_cat}, the obtained ALMAGAL catalog of compact sources is presented, along with an analysis of the revealed fragmentation properties and measured photometric parameters of the sources (e.g., fluxes and angular sizes). We derive and discuss the main physical properties of the cores (e.g., physical sizes, masses, and densities) in Sect. \ref{core_phys_props}. The evolution of the CMF is investigated in Sect. \ref{CMF_evol}. Further detailed analysis on the evolution of core masses and its relation with fragmentation is performed in Sect. \ref{correlations}. 
Lastly, in Sect. \ref{sum_concl}, we provide a summary of the analyses and the main results of the paper and draw suitable conclusions. 


\section{Observations}
\label{obs}

\subsection{The ALMAGAL clump sample}\label{sample}

The ALMAGAL target sample (see \citealt{Molinari+25} for full description) includes $1013$ dense clumps with declination $\delta\leq0^\circ$, which have been extracted as candidate hosts of high-mass star formation from the large infrared surveys of young massive protostellar regions: Herschel Infrared Galactic Plane Survey (Hi-GAL, \citealt{Molinari+10,Molinari+16b,Elia+17,Elia+21}, $915$ targets) and Red MSX Source survey (RMS, \citealt{Hoare+05,Urquhart+07,Lumsden+13}, $98$ targets). 

The sample covers a broad range of heliocentric distances ($\sim2-8\,\mathrm{kpc}$), masses ($\sim10^2-10^4\,\mathrm{M_\sun}$), and surface densities ($\sim0.1-10\,\mathrm{g\,cm^{-2}}$, compatible with high-mass star formation, \citealt{Kauffmann+10, Butler+12, Krumholz+14, Tan+14}). The ALMAGAL clumps probe different Galactic environments, as they are distributed across the disc of the Milky Way, from the central bar to the outskirts of the outer spiral arm, spread over $3-14$ kpc in Galactocentric distance. Moreover, target clumps cover the full evolutionary path, representing different evolutionary stages across the star formation process, from IRDCs (e.g., \citealt{Kauffmann+10,Sanhueza+12,Barnes+21}) to HII Regions (e.g., \citealt{Hoare+05}); namely, from prestellar to protostellar phases (see \citealt{Elia+17,Elia+21}). This feature is affirmed by the wide range of luminosity over mass ratio covered ($\sim0.05<L/M<450\,\mathrm{L_{\odot}/M_{\odot}}$), that can be adopted as an evolutionary proxy (e.g., \citealt{Molinari+08,Elia+10,Molinari+16,Elia+17,Elia+21,Traficante+23}). 

The extensive and uniform coverage of all these physical parameters, which are key aspects of high-mass star formation environments, ensures our sample and results have a very high level of statistical significance and robustness that is certainly unprecedented in this field of study.

\subsection{ALMA observations}
\label{alma_obs}

Full details of the ALMAGAL observations can be found in \citet{Molinari+25} and \citet{Sanchez+25}. We provide an overview in the following. Single-pointing observations of the $1013$ ALMAGAL targets were conducted from October 2019 to March 2020, and from March 2021 to July 2022 with the ALMA interferometer (\citealt{Wootten+09}) in the spectral Band $6$ at around $1.38$\,mm ($\sim217$\,GHz, \citealt{Kerr+14}), using both ACA-7m (Atacama Compact Array, \citealt{Iguchi+09,Kamazaki+12}) and $12$m (\citealt{Escoffier+07}) arrays. 
The observing time for each target ranged from $1$ to $4$ minutes. 
Two distinct observing setups consisting of different configurations of the ALMA arrays were employed: i) near sample targets ($d\lesssim4.7\,\mathrm{kpc}$, $535$ fields) were observed using the $7$m array of ACA (also labeled as 7M throughout the paper) plus the C-2 and C-5 configurations of the $12$m array; ii) far sample targets ($d\gtrsim4.7\,\mathrm{kpc}$, $478$ fields) were observed using the 7M array plus the C-3 and C-6 configurations of the $12$m array. 
The C-2 and C-3 (hereafter also referred to as TM2) are the compact configurations of the $12$m array (baselines up to $500$ m). Instead, C-5 and C-6 (hereafter also referred to as TM1) represent extended configurations of the $12$m array (i.e., featuring the longest baselines, up to $2.5$ km). 
The maximum (i.e., highest) angular resolution achieved in Band $6$ with the two setups, defined by C-5 and C-6 configurations, is $\sim0.3''$ and $\sim0.15''$, respectively. 
The maximum recoverable scale (MRS) is set by the minimum baseline length ($9$ m, given by the 7M array) and is equal to $\sim29''$, corresponding to average physical scales of $\sim0.5$ pc in the near sample and $\sim0.8$ pc in the far sample fields. 
By combining such observational configurations, we can recover the range of spatial scales of interest for the clump fragmentation process, with the (sub)parsec clump-scale structured and diffuse emission revealed by the $7$m array and more compact configurations, and the smaller scale structure (down to the core scales of $\sim1000$ au) traced by more extended arrays (TM2 and especially TM1 configurations). 
Most importantly, the two setups allow us to obtain a nearly uniform resulting spatial resolution of $\sim800-2000$ au (median of $\sim1400$ au) for all clumps (see Sect. \ref{contmapprop}), despite their wide spread in distance. This has made it possible to perform a consistent and detailed analysis of fragmentation statistics and physical properties over the whole sample. The main specifications of the observing setups used are summarized in Table \ref{obs_summary}. \\

\begin{table*}[ht!]
    \caption{Observations of the ALMAGAL target clumps performed with ALMA, for each of the two observing setups used.}
    \centering
    \begin{tabular}{c c c c c c}
    \hline\\[-9pt]
    Subsample & N. of targets & Heliocentric distance & ALMA array configurations & Max. angular resolution & Max. recoverable scale \\ [2pt]
    & & range (kpc) & & ($''$) & ($''$) \\ [2pt]
    \hline\hline\\[-8pt]
    Near & 535 & $\sim2-4.7$ & ACA-7m + C-2 + C-5 & 0.3 & $29$ \\
    Far & 478 & $\sim4.7-8$ & ACA-7m + C-3 + C-6 & 0.15 & $29$ \\
    \hline\\[1pt]
    \end{tabular}
    \label{obs_summary}
\end{table*}

\subsection{Continuum maps properties}
\label{contmapprop}

\begin{figure*}[t]
    \centering
    \includegraphics[width=1\columnwidth]{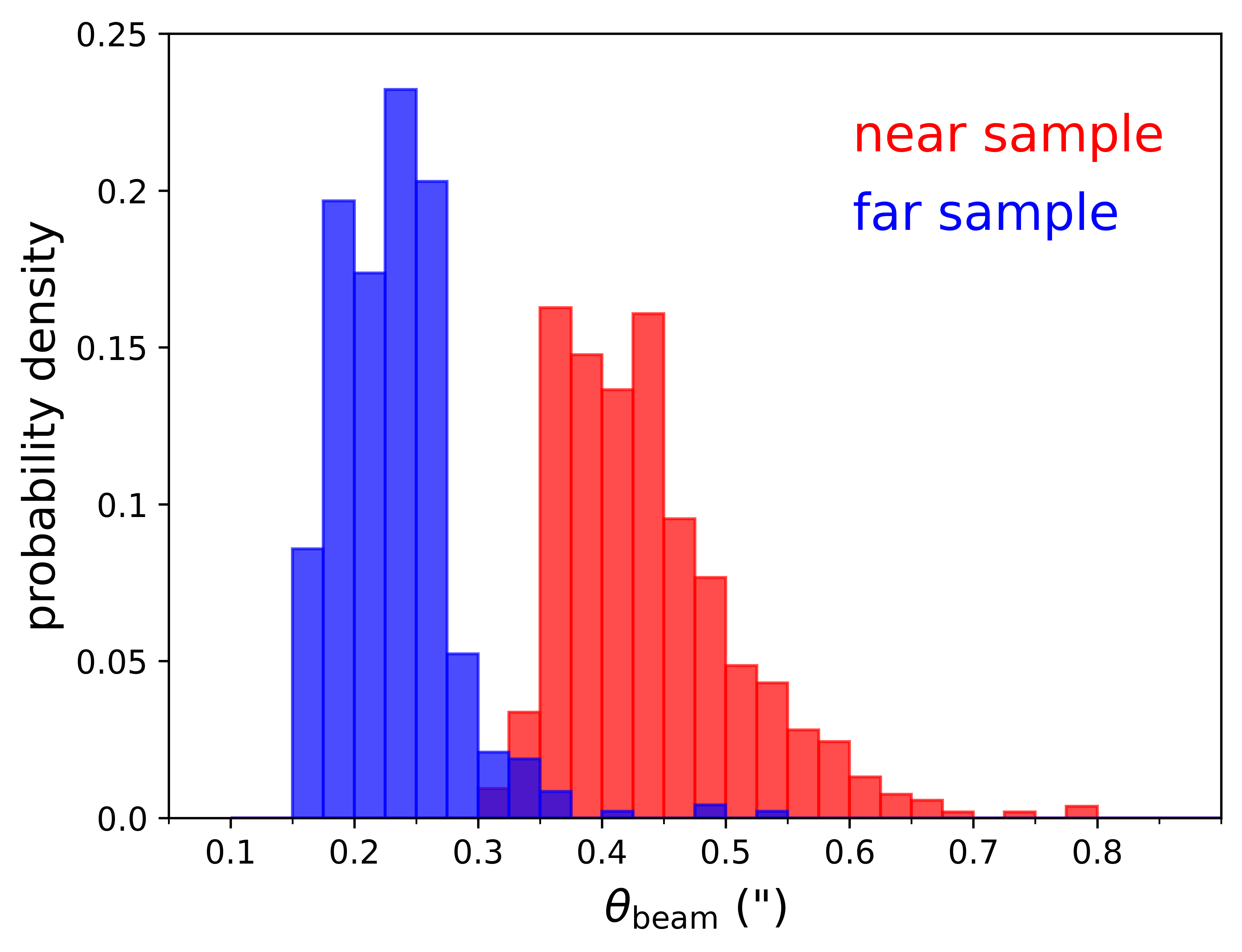}
    \includegraphics[width=1\columnwidth]{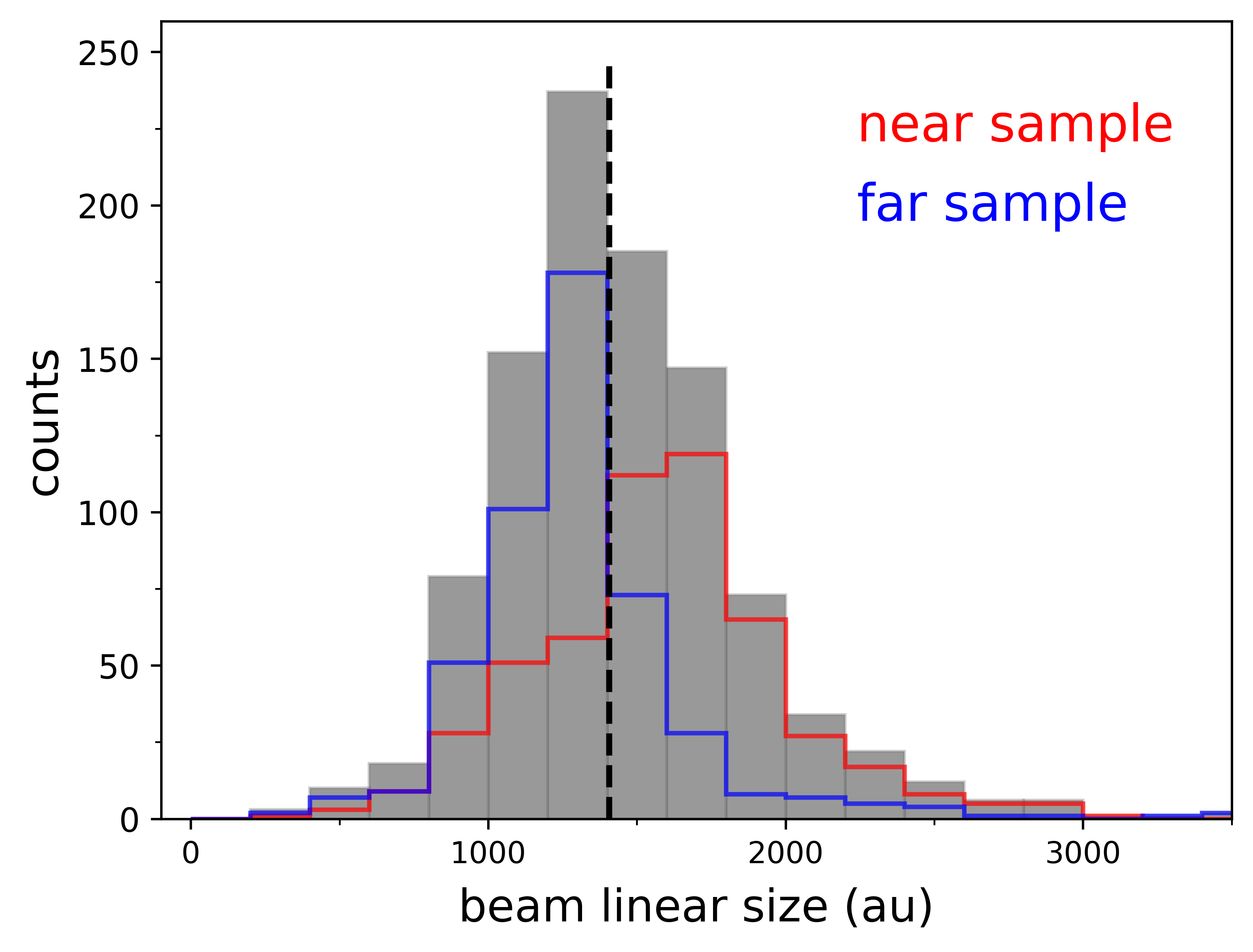}

    \caption{Beam properties of the ALMAGAL continuum maps. \textit{Left panel}: Probability density distribution of the circularized angular sizes of the map beam, separately for near (red) and far (blue) observational samples, respectively (see \citealt{Sanchez+25} for more details). 
    \textit{Right panel}: Distribution of the corresponding linear sizes (i.e., achieved spatial resolution). The grey histogram represents the overall distribution, with the vertical dashed black line marking its median value ($1400$ au). Near and far subsamples are overplotted with the same color coding of the left panel. 
    It can be noted that, due to the distance variation across our sample, the bimodal distribution in the left panel converts into a fairly uniform beam linear size overall distribution (see text for details).}
    \label{beam_circ_props_plots}
\end{figure*}

Data from the different observational configurations employed were combined to produce continuum maps (called 7M+TM2+TM1 images) using the Common Astronomy Software Applications (\textit{CASA}, \citealt{McMullin+07,CASATeam22}) package. 
A fully detailed technical description of the ALMAGAL data acquisition and processing pipeline, and characterization of the obtained continuum maps can be found in \citet{Sanchez+25}. Here, we discuss the properties that are most relevant to the interpretation of the analysis and results of this paper. 

The left panel of Fig. \ref{beam_circ_props_plots} shows the distribution of the angular size (i.e., circularized full width at half maximum, FWHM) for the synthesized beam of the maps ($\theta_\mathrm{beam}$). 
As anticipated in Sect. \ref{obs}, due to the different observing setups used, the distribution is clearly bimodal, peaking at $\sim0.2''$ for far sample fields and at $\sim0.4''$ for near sample ones. 
Overall, the maximum angular resolution, achieved through the addition of the most extended baselines configurations (TM1) to the combined data, is $\sim0.15''$, which leads in our sample to a corresponding maximum spatial resolution $\lesssim1000$ au, more than sufficient to resolve the typical core scales ($<0.05$ pc or $\sim10000$ au, e.g., \citealt{Sanchez-Monge+13b,Beuther+18,Sanhueza+19,Sadaghiani+20,Pouteau+22}). 
We refer here to circularized sizes, derived as the geometric mean of the major and minor axes of the elliptical map beam (see \citealt{Sanchez+25}). Such form is more meaningful for this work since circularized angular sizes have been used to estimate the physical sizes of the extracted compact sources (see Sect. \ref{Dcore}). 

Using those two different setups is justified when converting beam angular sizes into corresponding linear sizes, shown in the right panel of Fig. \ref{beam_circ_props_plots}. 
Although the near and far subsamples do not perfectly overlap, the median values of their distributions (about $1600$ au and $1300$ au, respectively) are comparable within $\sim20\%$, and combine for a fairly homogeneous overall distribution, thus ensuring consistent comparison among targets at considerably different distances. 
The resulting range of resolved physical scales is within $\sim800-2000$\,au in most cases ($\sim90\%$), with a median of $1400$ au. 
Fields showing very high ($<500$ au) or low ($>3000$ au) linear resolutions are outliers ($<5\%$) mainly produced by the refinement of clump distances performed between the time of the proposal preparation and the present work (see \citealt{Molinari+25} and Benedettini et al. in prep. for more details). That caused $81$ targets ($\sim8\%$ of the sample) which were observed with the near sample setup to assume a new distance compatible with the far sample (thus degrading the corresponding linear resolution) or vice versa (leading to linear sizes well below $1000$ au, see Sect. \ref{contmapprop}). 
As a consequence, in the latter case, the survey resolved even smaller structures than what was originally planned. 
However, such variations do not affect the statistical significance of the analysis performed in this work. 

\begin{figure}[t!]
    \centering
    \includegraphics[width=1\columnwidth]{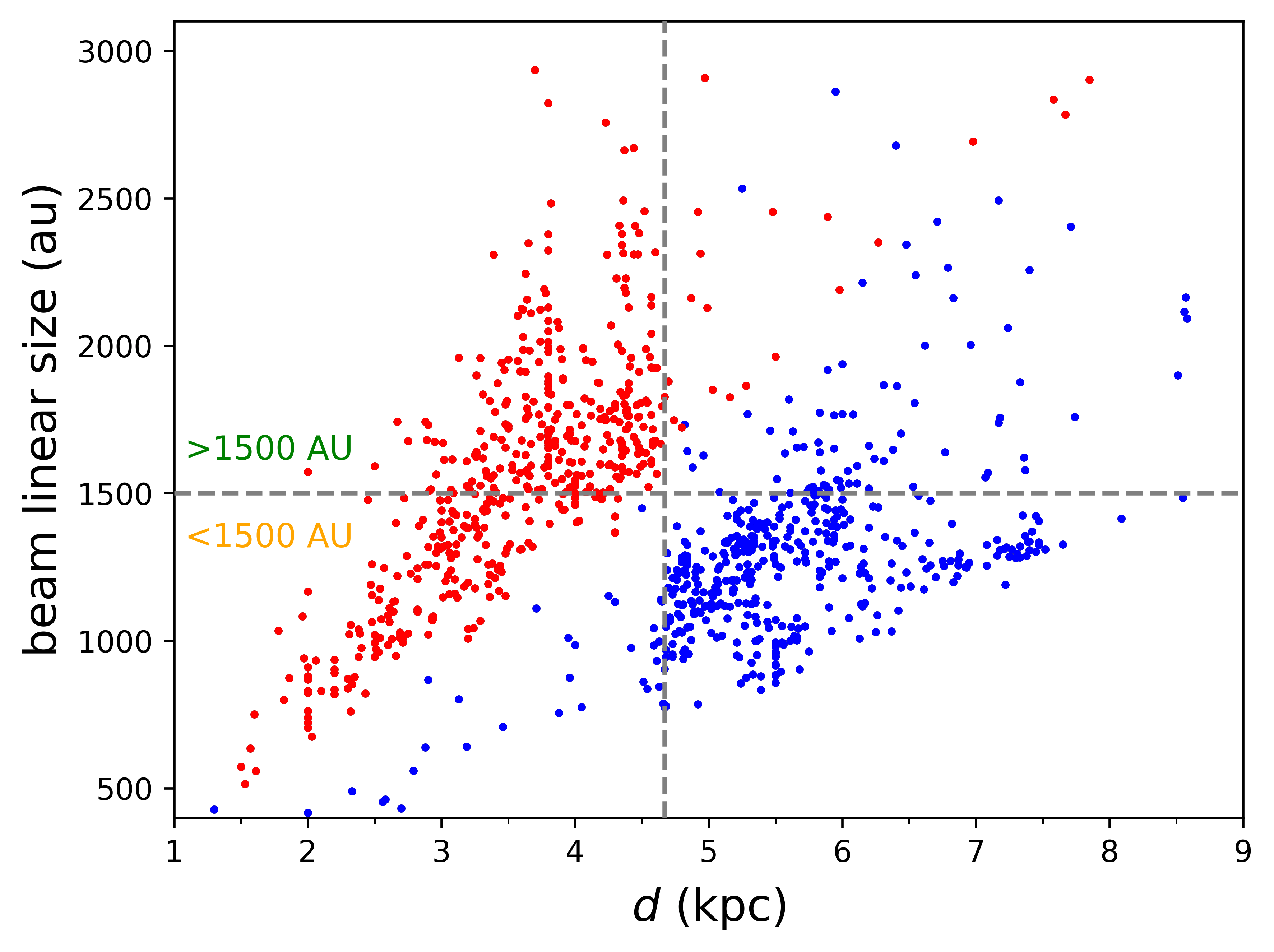}
        \caption{Linear size of the map beam as a function of the heliocentric distance of the ALMAGAL clumps. Targets observed with near or far sample configurations appear as red or blue points, respectively. The vertical dashed grey line marks the nominal $\sim4.7$ kpc distance limit separating the near and far subsamples (with the few exceptions described in Sect. \ref{contmapprop}). The horizontal dashed grey line separates higher spatial resolution fields (i.e., beam linear size $<1500$ au) from lower resolution ones ($>1500$ au). Note: for better visualization purposes, distance and/or resolution outliers are not shown in this plot ($\sim30$ targets, $\lesssim3\%$).}
    \label{beam_size_dist_plot}
\end{figure}

Figure \ref{beam_size_dist_plot} reports the beam linear size (i.e., circularized diameter) as a function of the target distance. The two increasing trends within near and far subsamples are due to the effect of distance, while the spread is due to the beam size distribution (Fig. \ref{beam_circ_props_plots}, left panel). 
In addition, for visualization and future analysis purposes, we split the sample into two groups according to the achieved spatial resolution, using as discriminator $1500$ au, which roughly corresponds to the mode of the beam size distribution (right panel of Fig. \ref{beam_circ_props_plots}). Moreover, we distinguish between targets having distances below or above $\sim4.7$ kpc (see Sect. \ref{alma_obs}). Due to the distance refinements described above, these two ranges do not correspond to the near/far sample observational setups for all ALMAGAL fields. 

\begin{figure}[h!]
        \includegraphics[width=\columnwidth]{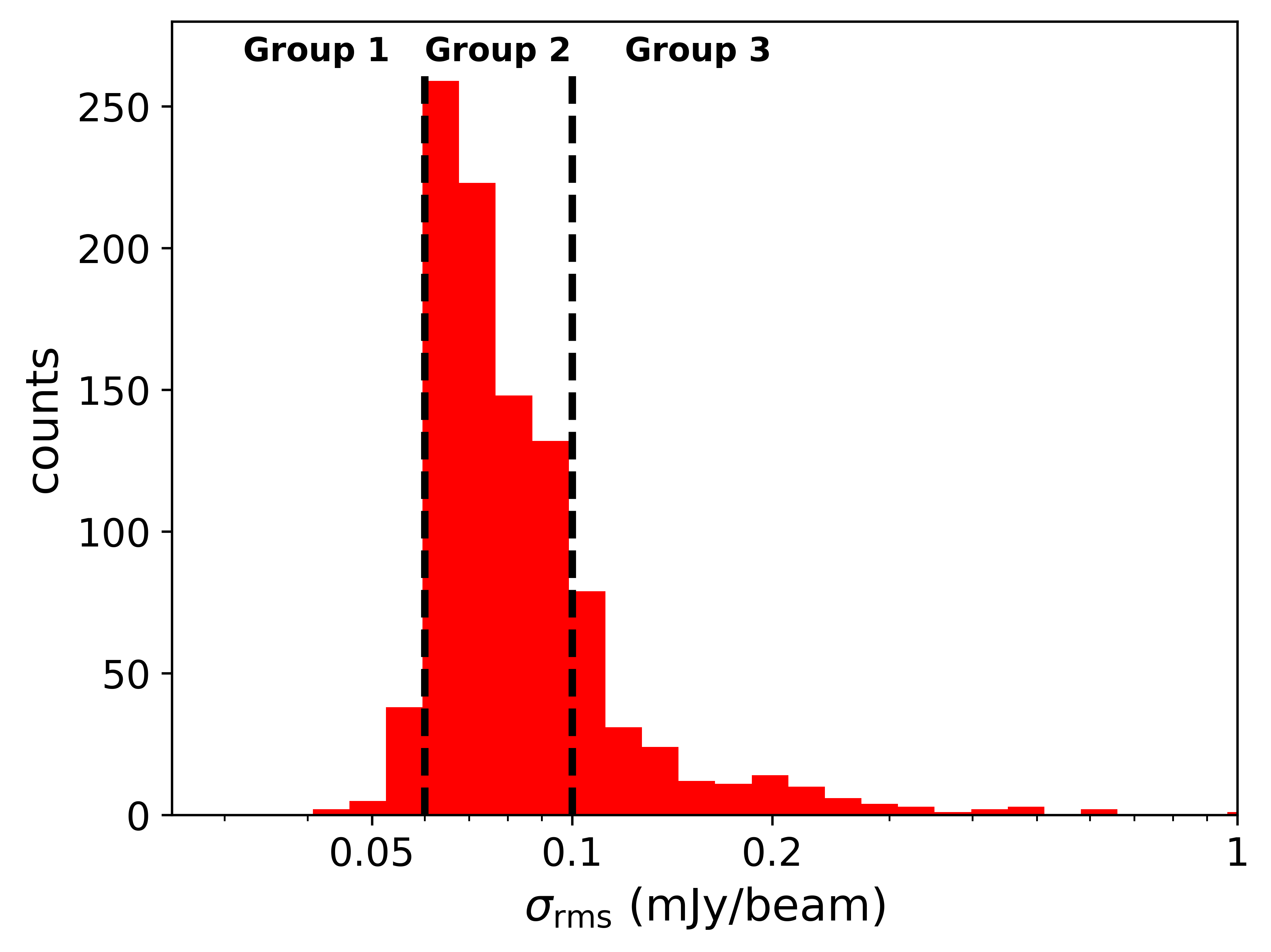}
        \caption{Distribution of the map rms noise level $\sigma_\mathrm{rms}$ across the ALMAGAL maps sample (see \citealt{Sanchez+25} for more details). The vertical dashed lines mark the subdivision into the three reported groups of fields that are employed in Sect. \ref{FcomplPhotacc}.}
    \label{map_RMS_hist}
\end{figure}

The root mean square (rms) noise level of the maps ($\sigma_{\rm{rms}}$) was estimated as the standard deviation of the residual image once masked to exclude regions with significantly bright emission in the final intensity image, and thus corresponds to the AGSTDREM keyword stored in the header of the FITS images of ALMAGAL targets (see \citealt{Sanchez+25}). 
These estimates have been used for flux thresholding during the compact source extraction procedure (see Sect. \ref{SE}). The distribution of rms noise levels across our sample is reported in Fig. \ref{map_RMS_hist}. Values mostly range between $\sim0.05$ and $\sim0.2$ mJy/beam. The subdivision shown in the plot, identifying three different groups of targets based on their rms level, is introduced and used in Sect. \ref{FcomplPhotacc}. 
Flux sensitivity, intended as the minimum peak flux detected at $1\,\sigma$ level of the map rms noise, is then $\sim$\,0.05\,mJy/beam, satisfying (and actually improving) the requested sensitivity of $\sim0.1$ mJy/beam (see \citealt{Sanchez+25}).

\section{Compact source extraction}\label{SE}

\subsection{Source extraction strategy}
\label{SE_intro}

The ALMAGAL survey poses a considerable challenge in terms of adopting the best technique to perform automatic extraction and flux estimation of compact sources across an extensive sample of star-forming clumps in the Galactic plane. 
The continuum images of the ALMAGAL clumps are characterized by various morphologies, spatial emission distributions, and background properties, the latter being particularly relevant due to the thermal emission of cold dust dominating these environments at mm wavelengths. 

Moreover, choosing a particular algorithm inherently selects the nature and properties of objects that will be extracted, for example: size, shape, and compactness. 
For instance, the widely used code \textit{astrodendro} (\citealt{Rosolowsky+08}; see, e.g., \citealt{Sanhueza+19}; \citealt{Svoboda+19,Anderson+21,Morii+23,Ishihara+24}) does not make any assumption on the shape of the sources, usually resulting in more irregular structures in which also more extended, diffuse emission might be included. These features make it more suitable for studying the hierarchical structuring of star-forming regions (see Wallace et al. in prep.), rather than performing a physical characterization of the population of compact, coherent cores. Furthermore, \textit{astrodendro} performs a plain sum of the emission within the structure contour starting from the zero-level flux and does not include by default any background emission estimation and subtraction (\citealt{Rosolowsky+08}); thus, source fluxes tend to be frequently overestimated. 
Instead, codes such as \textit{getsf} (\citealt{Men'shchikov+21}; see, e.g., \citealt{Pouteau+22,Xu+24}), \textit{hyper} (\citealt{Traficante+15}; see \citealt{Traficante+23}), and the here used \textit{CuTEx} (\citealt{Molinari+11,Molinari+17}; see, e.g., \citealt{Molinari+16b,Elia+17,Elia+21}, and below), although differing in the flux estimation method, focus on more compact, roundish (elliptical) sources. This latter approach is particularly suited for ALMA continuum images, since they are obtained through a cleaning procedure in which the model image is convolved with an elliptical Gaussian beam (see, e.g., \citealt{Hogbom74,Cornwell08}), so that ultimately small-scale substructures tend to resemble, to some extent, the size and shape of the beam. 
Such different approaches in extracting substructures make it essential to specify which kind of "compact" sources the detection is aimed at and on what physical scales (given the resolution available in the observations). Namely, it is essential to first clarify what is intended as a core. These aspects have to be properly taken into account when comparing the properties of objects detected under different assumptions. 
Here, we define as cores the dense, compact sources of roundish (or slightly elliptical) shape with a centrally peaked brightness profile, detected above a certain flux threshold at $\sim1400$ au average scales (see Fig. \ref{beam_circ_props_plots}). We fit the sources with a 2D Gaussian, and, to properly recover the true source flux, we simultaneously perform an estimation and subtraction of the background emission (see Sect. \ref{cutex} for full description). 

Moreover, to test two different approaches, we injected synthetic sources into selected ALMAGAL maps and conducted source extraction runs with \textit{CuTEx} and \textit{astrodendro}. After comparing their performance, \textit{CuTEx} proved to be more efficient in detecting compact sources and accurately measuring their flux. Additionally, we performed analogous tests with \textit{hyper}, which is based on aperture photometry. Full details of these tests are given in Appendix \ref{AppSE_codes_tests}.

\subsection{Source detection and photometry pipeline}
\label{cutex}

\begin{figure}[t]
    \center
    \includegraphics[width=1\columnwidth]{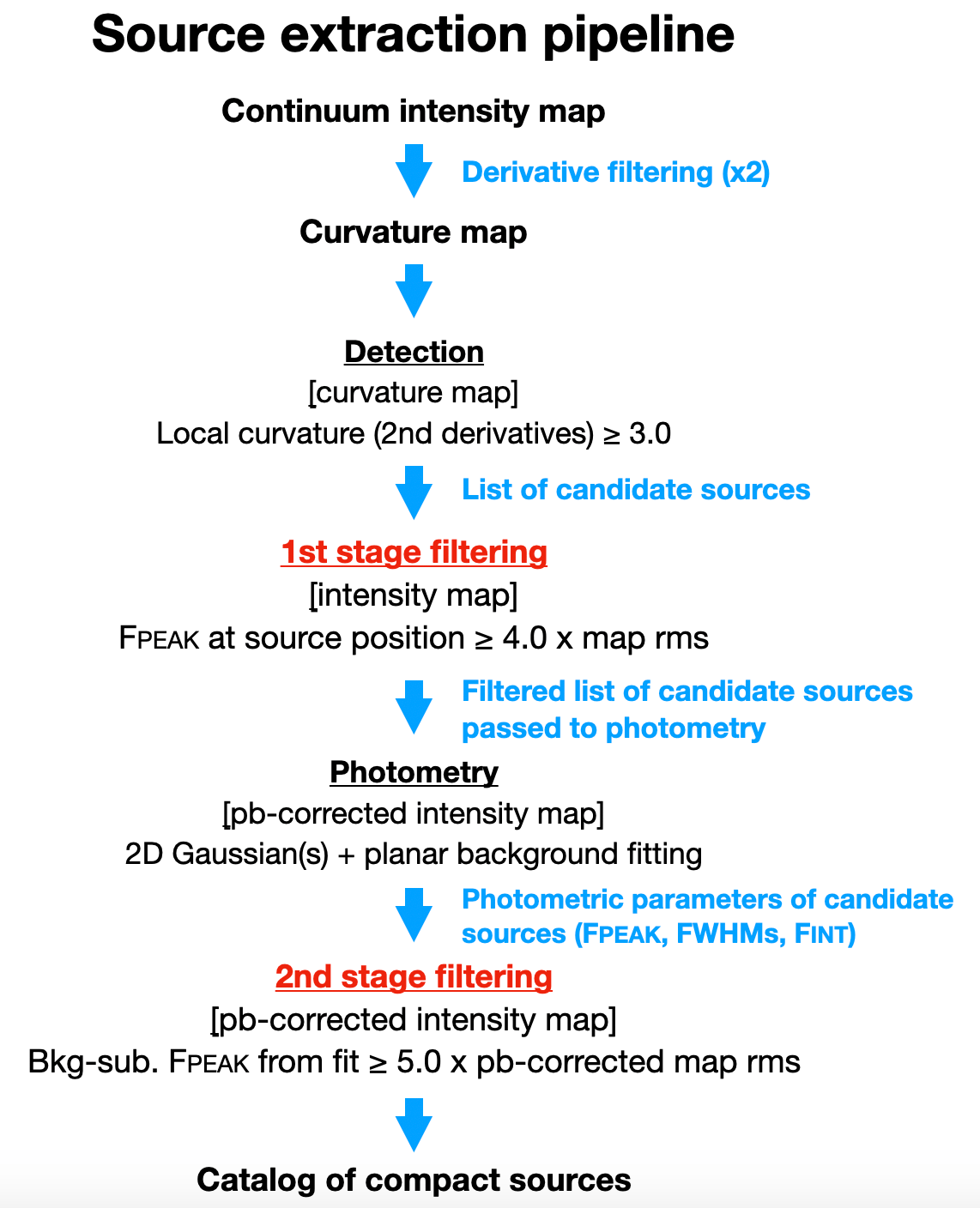}
    \caption{Diagram of the main phases of the source detection and photometry pipeline developed and used in this work. Underlined black steps were already present in the original \textit{CuTEx} version (but were still further implemented in this work), while those in red have been newly introduced in the context of the present work.}
    \label{cutex_workflow}
\end{figure}

\begin{figure*}[t!]
    \center
    \includegraphics[width=0.8\columnwidth]{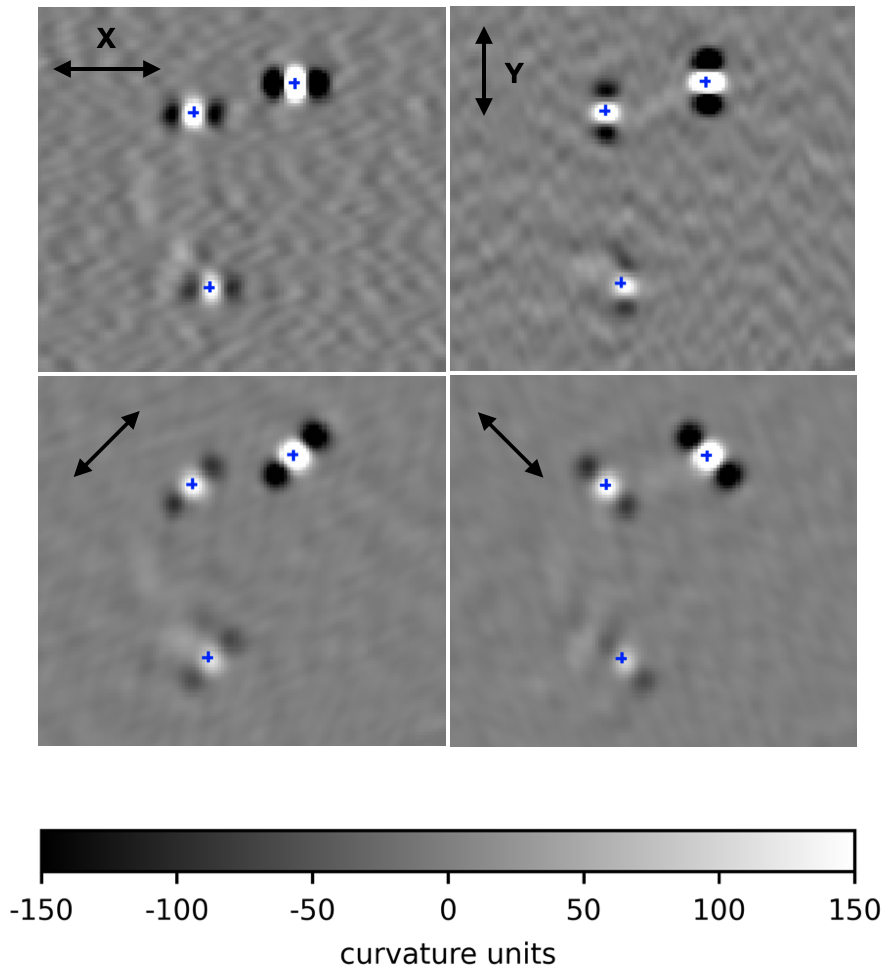}
    \includegraphics[width=0.795\columnwidth]{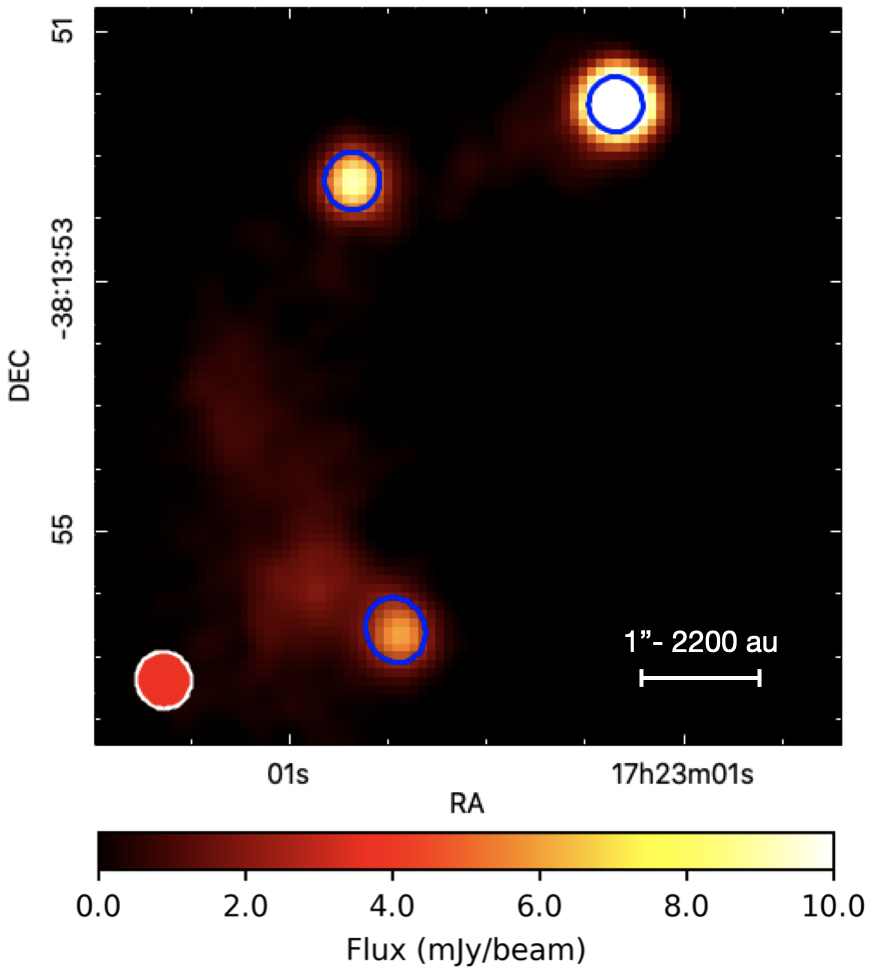}
    \caption{Example of the source extraction procedure performed with \textit{CuTEx} (see Fig. \ref{cutex_workflow}) for the ALMAGAL field AG349.6440-1.0948. \textit{Left panel}: Detection phase performed on curvature images along the reported four different directions; the blue crosses mark the candidate compact sources. \textit{Right panel}: Outcome of the photometry procedure performed on the corresponding intensity image; the blue ellipses represent the estimated shape of the compact sources (in terms of FWHMs and position angle, $\lambda_{\rm{PA}}$) as a result of the 2D Gaussian fitting procedure (see Sect. \ref{cutex} for details). Note: the color scale of the image was adjusted to highlight all the three sources at the same time, so their visual appearance in terms of emission extent might not correspond to their actual size (ellipses). The map beam is shown in red in the bottom-left corner of the plot. The angular and physical scales are also reported.}
    \label{cutex_example}
\end{figure*}

\begin{figure*}
    \center
    \includegraphics[width=1\textwidth]{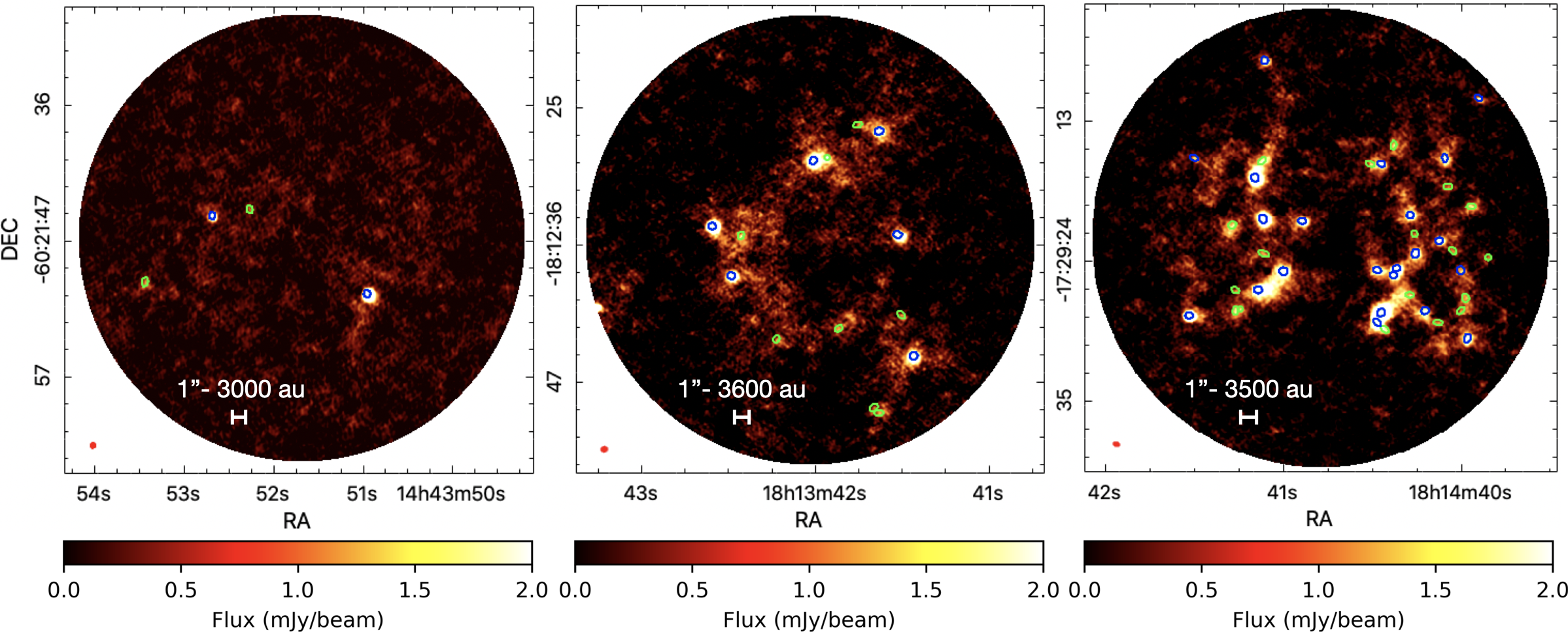}
    \caption{Outcome of the source extraction procedure performed with \textit{CuTEx} on selected ALMAGAL clumps (from left to right, fields AG316.4045-0.4657, AG012.4977-0.2233, AG013.2428-0.0854). As in Fig. \ref{cutex_example}, the blue ellipses mark the compact sources detected at $5\,\sigma_\mathrm{rms}$. The three clumps showcase the variety of fragmentation degree, morphologies of structures and patterns of emission revealed across the ALMAGAL sample. The \textit{CuTEx} algorithm was optimized and tested in order to efficiently perform in all these different conditions. The green ellipses mark the sources that are recovered by allowing a signal-to-noise ratio (S/N) cut of $3\,\sigma_\mathrm{rms}$. The map beam is reported in red in the bottom-left corner of each panel.}
    \label{cutex_outcome}
\end{figure*}

The identification and measurement of compact sources were performed on the processed continuum images with the \textit{CuTEx} algorithm \citep{Molinari+11,Molinari+16b,Molinari+17}. 
A detailed characterization and operational description of the original version of \textit{CuTEx} is reported in \citet{Molinari+11}. The code has been specifically further developed and improved during this work to best adapt to the properties of the ALMAGAL maps, which partly differ from IR/submm images, like the Hi-GAL maps for which it was originally designed. The key difference is that the beam is sampled with $\sim6$ pixels in ALMAGAL maps (rather than $\sim3$ pixels in Hi-Gal maps). As a consequence, the derivative filter which is applied to the signal image to obtain the "curvature" image (see \citealt{Molinari+16b} for details) has been modified in this \textit{CuTEx} version to take into account the beam sampling while preserving the ability to make $\sim$beam-size compact sources stand out, by selectively filtering out the larger spatial scales. 

Figure \ref{cutex_workflow} reports the workflow of the final source extraction pipeline that was applied to the ALMAGAL 7M+TM2+TM1 continuum images. 
The detection procedure uses the non-primary beam-corrected intensity images, ensuring a homogeneous noise level across the entire field. 
In more detail, the detection (see Fig. \ref{cutex_example}, left panel) is carried out on the curvature images obtained with a double differentiation of the intensity image in four different directions ($x$, $y$, and the two diagonals). 
This approach, peculiar with respect to the most commonly used algorithms based on plain thresholding on direct intensity image, allows for an easier identification (using a curvature threshold) and size estimation of compact sources, which stand out in the curvature image, whereas slower varying diffuse and background emission is strongly dampened. 
Adjacent pixels exceeding a curvature threshold (in our case three times the rms of the curvature image, $\sigma_{\partial^2}$) are grouped and taken as candidate source(s). The threshold must be exceeded over all four differentiation directions to ensure that possible filamentary-like structures are not extracted as compact objects. 
The position of the compact source(s) contained within the cluster of pixels is placed at significant local maxima (curvature value at least $1$ $\sigma_{\partial^2}$ above one of all surrounding pixels) on the average of the four curvature images or, if none, at the average coordinates of all pixels in the cluster. 
The list of sources produced by the detection phase undergoes a first stage of filtering requiring that their peak flux ($F_\mathrm{PEAK}$) is higher than four times the intensity map rms $\sigma_\mathrm{rms}$. 
This step provides the advantage of running detection with a relatively relaxed curvature threshold, but then obtaining a more robust catalog with only significant sources, while excluding possible false positives. Moreover, the local value of the curvature is linked to both the source intensity and compactness (size), so that the group of pixels above the threshold could be either sufficiently bright and/or compact. Applying the first stage filtering allows us to keep, among the initially selected cluster of pixels, only the ones that are sufficiently bright to be cleaned during the imaging. In this way, spurious sources produced, for example, by correlated noise and small blobs can be excluded. 

Source photometry (see Fig. \ref{cutex_example}, right panel) is based on the assumption that the source brightness profile can be represented by a 2D elliptical Gaussian with variable parameters, including size and orientation. As anticipated in Sect. \ref{SE_intro}, this approach is suited to detect and measure compact sources in ALMA images, given that the cleaning procedure involves a Gaussian beam. 
A relatively accurate initial guess for source size parameters (i.e., source FWHM values and position angle, $\lambda_{\rm{PA}}$, of the Gaussian ellipse) can be obtained using information from the curvature image, namely, by locating the points where the second derivatives reach a minimum that corresponds to a change in the brightness profile curvature, indicating that the source is starting to merge with the underlying background. The eight points resulting from the four curvature images are used to fit an ellipse with variable semi-major and semi-minor axes and orientation, which provides the photometry with the initial guesses for the source size (in terms of the standard deviations $\sigma_{\rm{a/b}}$) and corresponding orientation, $\lambda_{\rm{PA}}$, of the Gaussian fit. Additionally, to account for the fact that detected sources may be lying on intense and spatially varying backgrounds, each source is simultaneously fit with a plane of variable inclination degree and direction, to ensure an accurate estimation of the background emission to be then subtracted from the measured brightness of the source. Furthermore, to account for possible partially blended or nearby sources that may contaminate one another, a detected source is simultaneously fit together with its neighbors (through a multiple Gaussians fit) within a given radius, set to $0.75$ times the guessed source size. 
The fit is performed over an image area that is three times the beam width around the source position, thus providing sufficient space to reliably estimate the local background as the median of the pixel fluxes outside the source(s) mask. 
In particular, the fit is executed on the primary beam-corrected image using a total of nine parameters: six needed to describe the assumed Gaussian profile of each source simultaneously fitted (peak flux $F_\mathrm{PEAK}$, central coordinates $x_0$ and $y_0$, standard deviations $\sigma_{\rm{a}}$ and $\sigma_{\rm{b}}$, and orientation $\lambda_{\rm{PA}}$) and three for the planar background plateau. 
To permit its refinement on the signal image, the source size estimated by the fit is left free to vary within $0.5-1.5$ times the initial guess, assuming a defined lower limit of $0.95$ times the map beam size (in terms of the minimum axis of the Gaussian ellipse, $\theta^{\rm{min}}_\mathrm{beam}$). 

An extensive and varied series of tests were performed on both observed and synthetic maps, generated by injecting Gaussian sources at different flux levels into ALMAGAL images to reproduce in a controlled way the conditions addressed in the "real" source extraction run (see Appendices \ref{AppSE_codes_tests} and \ref{AppFcomplPhotacc} for details). These testing and tuning runs were performed to maximize source detection efficiency and photometry accuracy by finding the most suitable set of extraction parameters, able to account for the diverse physical properties of the targets (e.g., source emission and background structure) and the technical features of the maps (e.g., noise and dynamic range properties). For each of the test runs, the outcome of the source extraction procedure across the whole clump sample was evaluated through a thorough visual inspection. The integrated flux $F_{\rm{INT}}$ of the detected source is computed as (\citealt{Molinari+11})
\begin{align}
\label{FINT_eq}
F_{\rm{INT}} = 2\pi\sigma_{\rm{a}}\sigma_{\rm{b}}\,F_{\rm{PEAK}}\,/\,\Omega_\mathrm{beam}\,,
\end{align}
where $F_{\rm{PEAK}}$ is the source peak flux subtracted by the background at the source location (expressed in Jy/beam), $\sigma_{\rm{a/b}}$ are the standard deviations of the fitted source profile, and $\Omega_\mathrm{beam}$ is the beam solid angle:
\begin{align}
\Omega_\mathrm{beam}=\frac{2\pi}{8\ln2}\,\theta^{2}_\mathrm{beam}\,,
\end{align}
where $\theta_\mathrm{beam}$ is the circularized FWHM of the beam, computed as the geometric mean of its major and minor axes: 
\begin{align}
\theta_\mathrm{beam}=\sqrt{\theta^{\rm{maj}}_\mathrm{beam}\cdot\theta^{\rm{min}}_\mathrm{beam}}\,.
\end{align}
Lastly, a second and final filtering stage, which we specifically introduced in this improved version of \textit{CuTEx}, is applied on the source catalog, requiring that the background-subtracted peak flux from photometry is higher than five times the map rms $\sigma_\mathrm{rms}$ (which is computed on the non-primary beam-corrected image), once $\sigma_\mathrm{rms}$ is also corrected for the local value of the primary beam. We imposed this rather conservative S/N threshold to obtain a more robust and reliable core catalog, which still provides very large statistics. Based on thorough visual inspection, lower S/N would lead to a significant number of spurious detections. 
The output of the procedure for the whole target sample, resulting in the catalog of ALMAGAL compact sources, is presented in detail in Sect. \ref{photo_cat}. 

Figure \ref{cutex_outcome} shows the outcome of the source extraction procedure for selected ALMAGAL fields, representing different fragmentation and emission patterns. 
We note that besides the sources that were detected, other relatively bright and centrally peaked structures might appear visually. While such objects seemingly did not satisfy the S/N and/or compactness requirements, they are in most cases recovered when allowing a lower S/N (e.g., $3\,\sigma_\mathrm{rms}$, as shown in Fig. \ref{cutex_outcome}).

\subsection{Flux completeness and photometric accuracy}
\label{FcomplPhotacc}

\begin{figure*}
    \includegraphics[width=0.725\columnwidth]{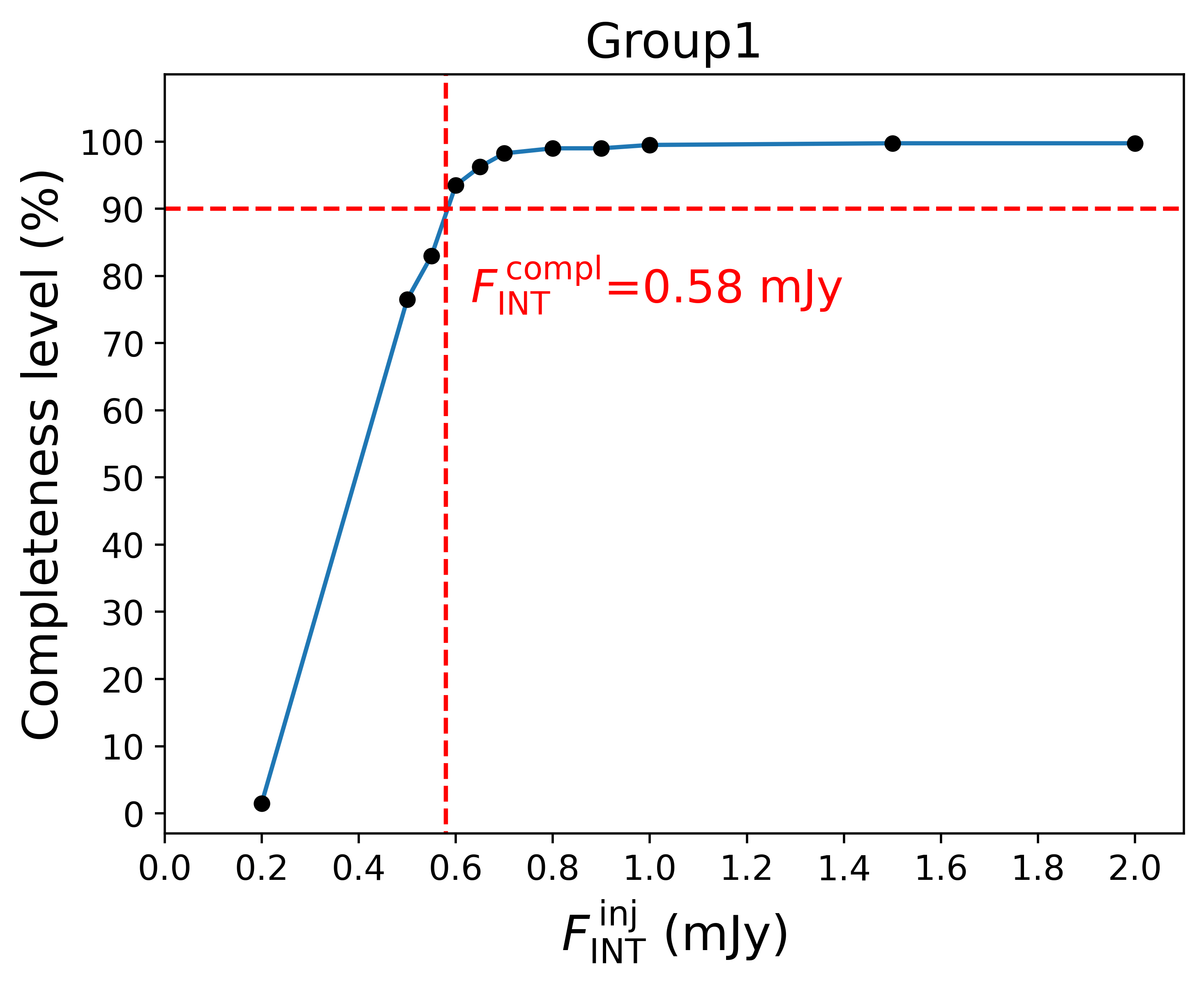}\hspace{-3pt}
    \includegraphics[width=0.69\columnwidth]{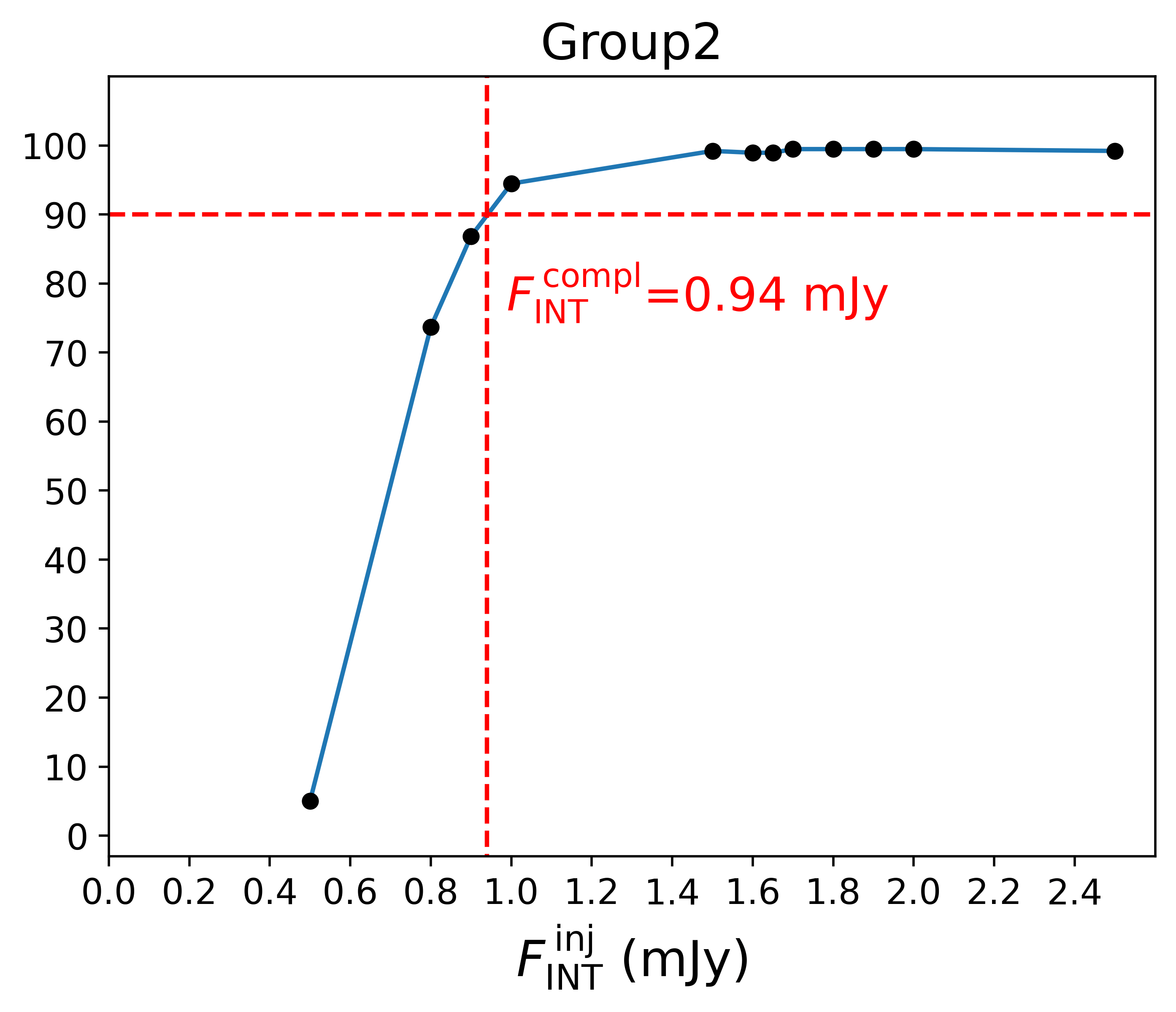}\hspace{-3pt}
    \includegraphics[width=0.69\columnwidth]{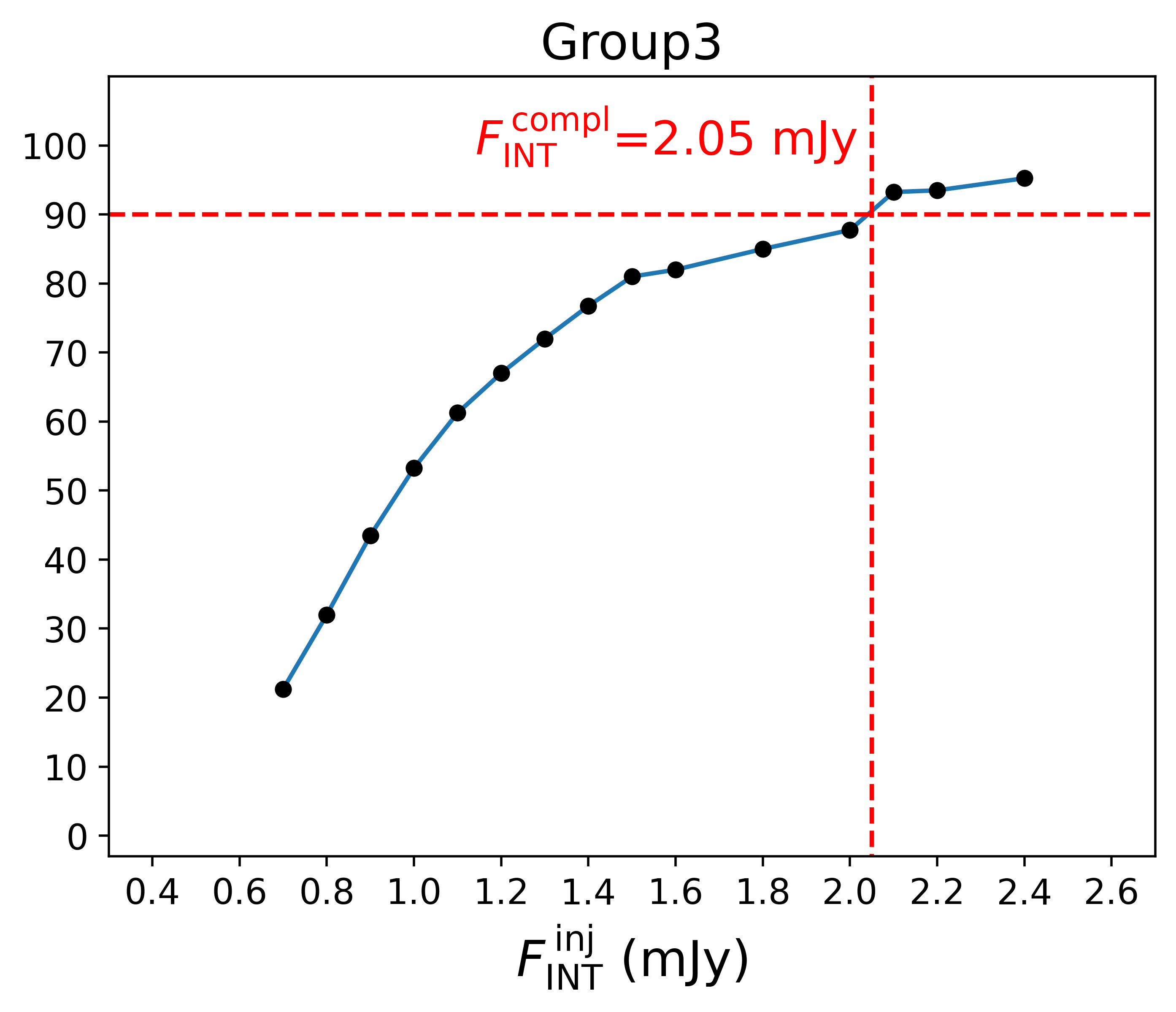}
        
    \caption{Completeness levels (i.e., percentage of detection) as a function of the "true" integrated flux of the synthetic sources injected into the maps ($F_{\rm{INT}}^{\,\rm{inj}}$) for the three different groups of targets selected based on their rms noise, as introduced in Sect. \ref{contmapprop}: Group 1 (left panel), 2 (middle panel), and 3 (right panel), respectively. The two dashed red lines in each panel identify the estimated completeness flux, $F_{\rm{INT}}^{\,\rm{compl}}$, which is reported in red.}
    \label{fl_compl_plots}
    \end{figure*}

Flux completeness levels for our compact source catalog were estimated through synthetic source extraction experiments. 
As a preliminary step, we visually divided the map sample in three groups according to their rms noise properties, as shown in Fig. \ref{map_RMS_hist}: i) low noise maps ($\sigma_{\rm{rms}}$\,$<$\,$0.06$\,mJy/beam, $53$ fields or $5\%$ of the sample, hereafter Group 1), ii) medium noise maps ($0.06\leq\sigma_{\rm{rms}}\leq0.1$ mJy/beam, including the peak of the distribution, $769$ fields or $76\%$, hereafter Group 2), and iii) high noise maps ($\sigma_{\rm{rms}}>0.1$ mJy/beam, $191$ fields or $19\%$, hereafter Group 3). This was done in view of performing a dedicated analysis for each of the groups to obtain more accurate flux completeness estimates. 
We then injected synthetic 2D elliptical Gaussian sources ($20$ per field) at multiple (from $10$ to $15$) integrated flux levels into $60$ selected ALMAGAL fields representing a variety of emission morphologies, which we divided into three groups of $20$ according to the rms noise classification presented above. In total, we generated $\sim4000$ synthetic sources for each group. 
We ran our extracting algorithm on the whole set of generated images using the same parameter setup applied on native fields (see Sect. \ref{cutex}) and compared the photometry output obtained for the simulated sources with their "truth tables", annotating the number of detections recorded for each generated flux level ($F_{\rm{INT}}^{\,\rm{inj}}$). 
Detailed description and motivations for this methodology are provided in Appendix \ref{AppFcomplPhotacc}. 
The results are shown in Fig. \ref{fl_compl_plots} for the three different groups of fields. 
The flux completeness level ($F_{\rm{INT}}^{\,\rm{compl}}$) is defined as the integrated flux value at which a $90\%$ rate of detection is reached. 
Completeness flux is $0.58$\,mJy for Group 1 (low noise fields, left panel), $0.94$\,mJy for Group 2 (medium noise fields, mid panel), and $2.05$\,mJy for Group 3 (high noise fields, right panel), respectively. 
A summary of the flux completeness analysis is given in Table \ref{fl_compl_stats}. Corresponding mass completeness is addressed in Sect. \ref{Msens-compl}. 

The flux completeness estimation procedure also enables us to characterize the accuracy of our source extraction method in terms of flux and size recovery. Indeed, for injected synthetic sources the estimated integrated flux is on average within $\sim20\%$ from the generated one, while the source size is within $\sim20\%$, both consistent with the true values given the adopted uncertainties. Dedicated plots are shown in Appendix \ref{AppFcomplPhotacc}. 

\begin{table}[h!]
    \caption{Flux completeness analysis.}
    \label{fl_compl_stats}
    \centering
    \begin{tabular}{c c c c c}
    \hline\\[-9pt]
    Group & N. of targets & $\sigma_{\mathrm{rms}}$ range & $F_{\rm{INT}}^{\,\rm{inj}}$ & $F_{\rm{INT}}^{\,\rm{compl}}$ \\[3pt]
    & & (mJy/beam) & (mJy) & (mJy) \\[2pt]
    \hline\hline\\[-9pt]
    $1$ & $53$ & $<0.06$ & $0.2-2$ & $0.58$ \\
    $2$ & $769$ & $0.06-0.1$ & $0.5-2.5$ & $0.94$ \\
    $3$ & $191$ & $>0.1$ & $0.7-2.4$ & $2.05$ \\
    \hline\\[1pt]
    \end{tabular}   
        \tablefoot{ 
        Summary of the flux completeness analysis for the three different groups of fields selected based on their rms noise level (Col. 3). Columns 4 and 5 report the range of integrated flux levels injected and the derived completeness limit, respectively. The integrated flux ranges were chosen in order to more accurately sample in each group the specific range where the $90\%$ detection threshold defining the flux completeness level lied. 
        }
\end{table}

\section{Compact source catalog analysis}\label{photo_cat}

\subsection{ALMAGAL catalog of compact sources and their photometric properties}
\label{cat}


Table 3 reports the catalog and the main photometric properties of the compact sources (cores) extracted from the ALMAGAL continuum images with the pipeline described in Sect. \ref{cutex}. The table is only available at the CDS in machine-readable form. 
The content of the table is the following. 
Column 1) hosting clump name; 
Col. 2) ordinal core identifier number within the clump; 
Cols. 3-4) right ascension and declination (FK5 equatorial coordinates); 
Cols. 5-6) Galactic longitude and latitude; 
Col. 7) background-subtracted peak flux ($F_{\rm{PEAK}}$); 
Col. 8) value of the fitted planar inclined background at flux peak position ($F_{\rm{BKG}}$); 
Cols. 9-10) full widths at half maximum of the fitted ellipse (non-beam-deconvolved, $FWHM_{\rm{a}}$ and $FWHM_{\rm{b}}$); 
Col. 11) counterclockwise position angle (i.e., orientation) of the fitted ellipse, starting from the x-axis ($\lambda_{\rm{PA}}$); 
Col. 12) integrated flux ($F_{\rm{INT}}$); 
Cols. 13-14) major and minor axes of the map beam, respectively ($\theta^{\rm{maj}}_\mathrm{beam}$ and $\theta^{\rm{min}}_\mathrm{beam}$); 
Col. 15) beam position angle ($\lambda^{\rm{beam}}_{\rm{PA}}$); 
Cols. 16-17) flags reporting the presence (Y/N) of strong cm radio continuum emission in correspondence of the core footprint (left flag) or the hosting clump (right flag), based on the analysis of CORNISH and CORNISH-South counterparts of ALMAGAL targets (see Sect. \ref{core_sample_sel} and Appendix \ref{Appff_analysis} for details). Clumps/cores for which no radio counterpart was available for inspection are flagged as N. 

The cores are progressively ordered within the hosting clump, thus indicating the number of detections reported in each target, which is discussed in Sect. \ref{NFRAG_stats}. All fluxes have been computed on the primary beam-corrected image. The background flux $F_{\rm{BKG}}$ can be intended as a tentative proxy of the local conditions where the compact source has been detected and measured. 
For example, a relatively high $F_{\rm{BKG}}$ may suggest the potential presence of a diffuse emission plateau or other nearby bright sources under or in close proximity to the detection location on the map, so that its flux estimate may be considered to be less robust. A few $F_{\rm{BKG}}$ values ($<3\%$) turn out to be negative. These values could, in principle, affect the goodness of the source flux estimate, but this is not an issue in our catalog since we verified that they are not significant as they stay within the local fluctuations on the map $\sigma_\mathrm{rms}$. The parameters $FWHM_{\rm{a}}$ and $FWHM_{\rm{b}}$ refer to the non-beam-deconvolved full sizes of the axes of the fitted Gaussian ellipse (rotated with respect to the $x$-$y$ coordinate system of the map by the position angle $\lambda_{\rm{PA}}$). Once converted to $\sigma_{\rm{a}}$ and $\sigma_{\rm{b}}$, respectively, they are used in Eq. \ref{FINT_eq} to compute the source integrated flux ($F_{\rm{INT}}$). 
The main properties of the corresponding map elliptical beam (major and minor axes extent, and orientation) are also reported to give an indication of the size of the source relative to the beam. The distribution of this ratio is shown and discussed in Sect. \ref{FWHMs}. As detailed in Sect. \ref{cutex}, detected sources were selected to have a minimum FWHM above $0.95$ times the beam minor axis, so that it may happen that in some cases $\theta^{\rm{min}}_\mathrm{beam}$ turns out to be greater than the source minimum extent between $FWHM_{\rm{a}}$ and $FWHM_{\rm{b}}$. All included core peak fluxes are above the $5\,\sigma_\mathrm{rms}$ cut applied in the last photometric filtering stage. After a thorough visual inspection of the detection maps, however, five fields (namely AG025.8252-0.1777, AG029.9558-0.0161, AG274.0659-1.1488, AG305.3676+0.2128, AG331.5117-0.1024) were found to be likely slightly affected by some artifacts coming from the continuum map cleaning process (see \citealt{Sanchez+25}). These fields then required a dedicated tuning of the extraction parameters in order to optimize the robustness of the compact source extraction. As a consequence, the cores detected within these clumps present fluxes that were selected to be above $6\,\sigma_\mathrm{rms}$ (for the four targets AG025.8252-0.1777, AG029.9558-0.0161, AG274.0659-1.1488, and AG331.5117-0.1024) or $8\,\sigma_\mathrm{rms}$ (for target AG305.3676+0.2128).

\subsection{Detection and fragmentation statistics}
\label{NFRAG_stats}

\setcounter{table}{3}   

\begin{table*}[t!]
    \caption{Detection statistics from the source extraction run performed on all ALMAGAL targets.}
    \label{table:det_stats}
    \centering
    \begin{tabular}{c c c c c c c c c c c}
    \hline\\[-8pt]
    $N_{\mathrm{clumps}}$ & $N_{\mathrm{clumps}}$ & $N_{\mathrm{frag}}$ & $N_{\mathrm{frag}}$ & $N_{\mathrm{frag}}$ & $N_{\mathrm{frag}}$ & $N_{\mathrm{frag}}$ & $N_{\mathrm{frag}}$ & $N_{\mathrm{frag}}$ & $N_{\mathrm{frag}}$ & $N_{\mathrm{frag}}$ \\ [2pt]
    & with detections (\%) & & range & mean & median & 1 & 2 & $3-9$ & $10-20$ & $>20$
    \\[3pt]
    \hline\hline\\[-7pt]
    1013 & 844 (83\%) & 6348 & $1-49$ & $8$ & $5$ & 130 (15\%) & 110 (13\%) & 378 (45\%) & 168 (20\%) & 58 (7\%)
    \\[1pt]
    \hline
    \vspace{1mm}
    \end{tabular}
    \end{table*}

    \begin{figure*}[t!]
    \includegraphics[width=\columnwidth,valign=t]{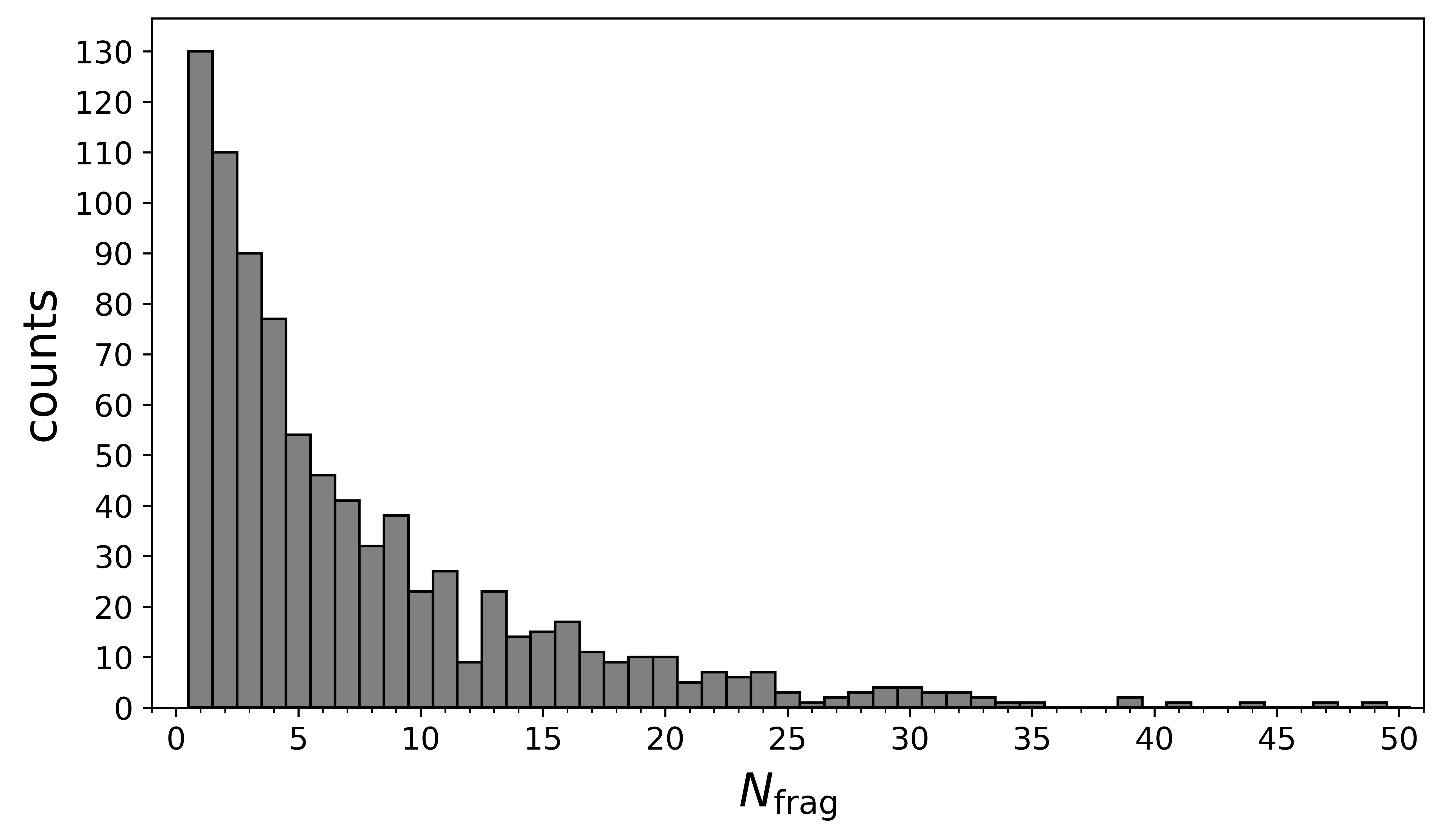}
    \includegraphics[width=1.015\columnwidth,valign=t]{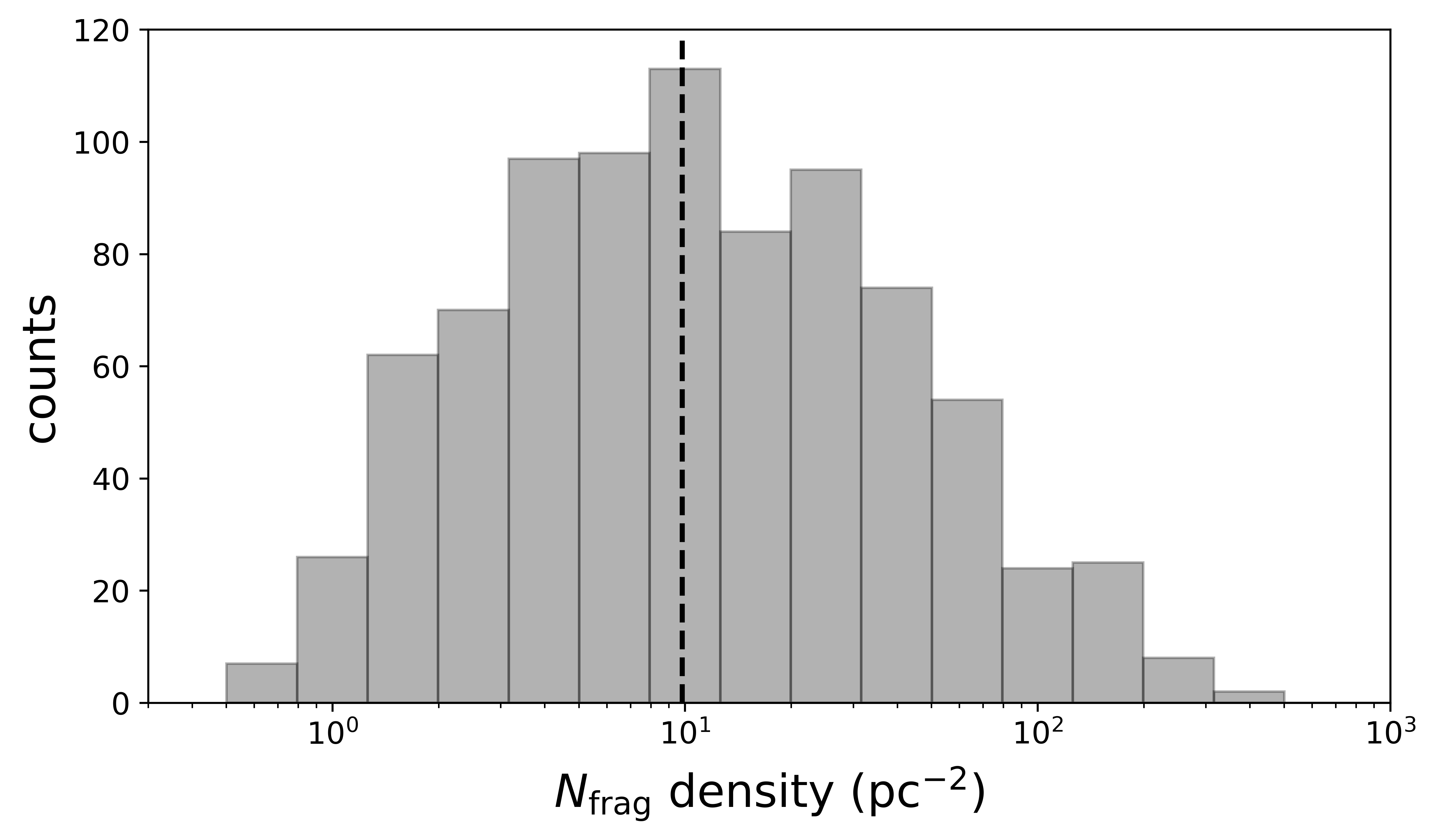}
    
    \caption{Fragmentation statistics deduced from the compact source catalog (Table 3). \textit{Left panel}: Overall distribution of the number of cores detected in each clump; detailed statistics are reported in Table \ref{table:det_stats}. 
    \textit{Right panel}: Overall distribution of the derived spatial density of the cores in each clump, computed within the entire FOV. The vertical dashed black line marks the median value of $\sim10$ pc$^{-2}$. 
    In both panels, only clumps revealing at least $1$ detection are included.}
    \label{NFRAG_stats_hists}
    \end{figure*}

The ALMAGAL compact source catalog features $6348$ cores detected in $844$ clumps from the total $1013$ inspected ($\sim83\%$). Therefore, $169$ fields ($\sim17\%$) reported no detections. 
Such evidence does not imply the lack of signal in the continuum maps, but rather that such emission did not satisfy the requirements of our source extraction procedure, for example, in terms of morphology, compactness, and contrast. 
This may be related to the physical properties of those targets (e.g., surface density, evolutionary stage, IR fluxes) as investigated in detail in other ALMAGAL works (\citealt{Molinari+25}, Elia et al. in prep.). 

Figure \ref{NFRAG_stats_hists} (left panel) shows the distribution of the number of cores ($N_\mathrm{frag}$) revealed across the inspected clumps. The overall $N_\mathrm{frag}$ range is $1-49$, while the average number of detections per clump is $8$ and the median is $5$, respectively. Moreover, $130$ clumps ($15\%$) showed only $1$ detection, while $58$ clumps ($7\%$) appear to be rather crowded, reporting more than $20$ detected cores. 
Table \ref{table:det_stats} summarizes the overall detection statistics. The right panel of Fig. \ref{NFRAG_stats_hists} shows the distribution of the spatial densities of cores within the hosting clumps (computed within the entire FOV), which ranges between $\sim0.5$ and $\sim500$ sources pc$^{-2}$, with a median value of $\sim10$\,pc$^{-2}$. 

The large statistics and broad range of physical parameters provided by ALMAGAL reveal the fragmentation properties of a wide variety of high-mass star-forming regions, which may be characterized by varied morphologies, physical properties and evolutionary stages. This most probably results in the large range of fragmentation degrees that we report in our catalog, much wider than that found by other recent ALMA studies, which feature much smaller samples. 
\citet{Sanhueza+19} (ASHES) detected $\sim300$ cores in $12$ massive $70\,\mu$m-dark prestellar clump candidates ($>500\,\mathrm{M_{\odot}}$) observed with mosaics at $\sim4800$ au average resolution, with a number of fragments per clump between $13-39$. Later, \citet{Morii+23} extended the sample to $39$ clumps, revealing $\sim840$ cores in total, with a range of $8-39$ fragments per clump (median of $20$). \citet{Svoboda+19}, instead, found $67$ cores in $12$ massive ($>400\,\mathrm{M_{\odot}}$) starless clump candidates with a fragmentation degree ranging from $1$ to $11$, observed at $\sim3000$ au spatial resolution. In a sample of $146$ late-stage clumps (HII regions), \citet{Liu+22b} (ATOMS) detected a total of $450$ cores (candidate hot cores/HC-UCHIIs). 
\citet{Traficante+23} (SQUALO) inspected a representative sample of $13$ massive clumps with masses above $170\,\mathrm{M_{\odot}}$ at various evolutionary stages ($0.1\leq L/M\leq100\,\rm{L_{\odot}/M_{\odot}}$, covering from prestellar to advanced phases) with a maximum resolution of $\sim2000$ au, within which they identified $55$ dense fragments (ranging from $1$ to $9$ fragments per clump). \citet{Avison+23} (TEMPO) found $287$ cores at $\sim2000$ au resolution in their $38$ HMSFRs spanning $\sim0.5\leq L/M\leq700\,\rm{L_{\odot}/M_{\odot}}$. Most recently, \citet{Xu+24} (ASSEMBLE) detected $248$ cores in their $11$ evolved massive clumps. 
Furthermore, \citet{Beuther+18} (CORE) observed $20$ clumps with distances below $5.5$ kpc and masses above $40\,\mathrm{M_{\odot}}$ at a similar $\sim1000$ au maximum resolution with NOEMA, revealing diverse fragmentation properties, going from fields with a single fragment to crowded fields with up to $20$ detections. 

Therefore, overall, in our clump sample we find a wider range of fragmentation degrees with respect to other similar recent high-resolution studies. 
However, the statistical and physical significance (in terms of number and range of physical properties of the observed clumps) and spatial resolution achieved by these and other interferometric works are considerably lower than the ones provided by the ALMAGAL sample. 
The capability to observe highly fragmented clumps is indeed expected, thanks to the high resolution and high sensitivity of the ALMAGAL observations. Nevertheless, a significant difference with previous studies is the high fraction of clumps revealing a small number of fragments ($28\%$ with $1$ or $2$ fragments). 
We underline, in this respect, that we focused our source extraction strategy on detecting only compact substructures. 
However, this might be also related to the properties of the clump, for instance, in terms of mass, surface density, and evolutionary stage. The relationships between the observed fragmentation and the physical properties of the clumps and their embedded cores are investigated in Sect. \ref{correlations}, while a more extensive analysis is given in Elia et al. (in prep.). 

Ultimately, in more general terms, it has to be taken into account that the capability of detecting dense fragments always depends to some extent on several aspects, such as modality (mosaic or single-pointing), sensitivity, and resolution of the observations, design of the source extraction algorithm, and uncertainties on source size and flux estimates (see, e.g., \citealt{Sadaghiani+20,Louvet+21}), in addition to the inherent physical properties of the targets (e.g., morphology, spatial emission distribution, evolutionary stage; see \citealt{Molinari+25}). In Sect. \ref{NFRAG_bias}, we analyze the potential impact on the revealed fragmentation properties of the different resolutions and distances featured in our data.

\subsubsection{Sensitivity and physical effects on the observed fragmentation statistics}
\label{NFRAG_bias}

    \begin{figure*}[t!]
    \includegraphics[width=\columnwidth,valign=t]{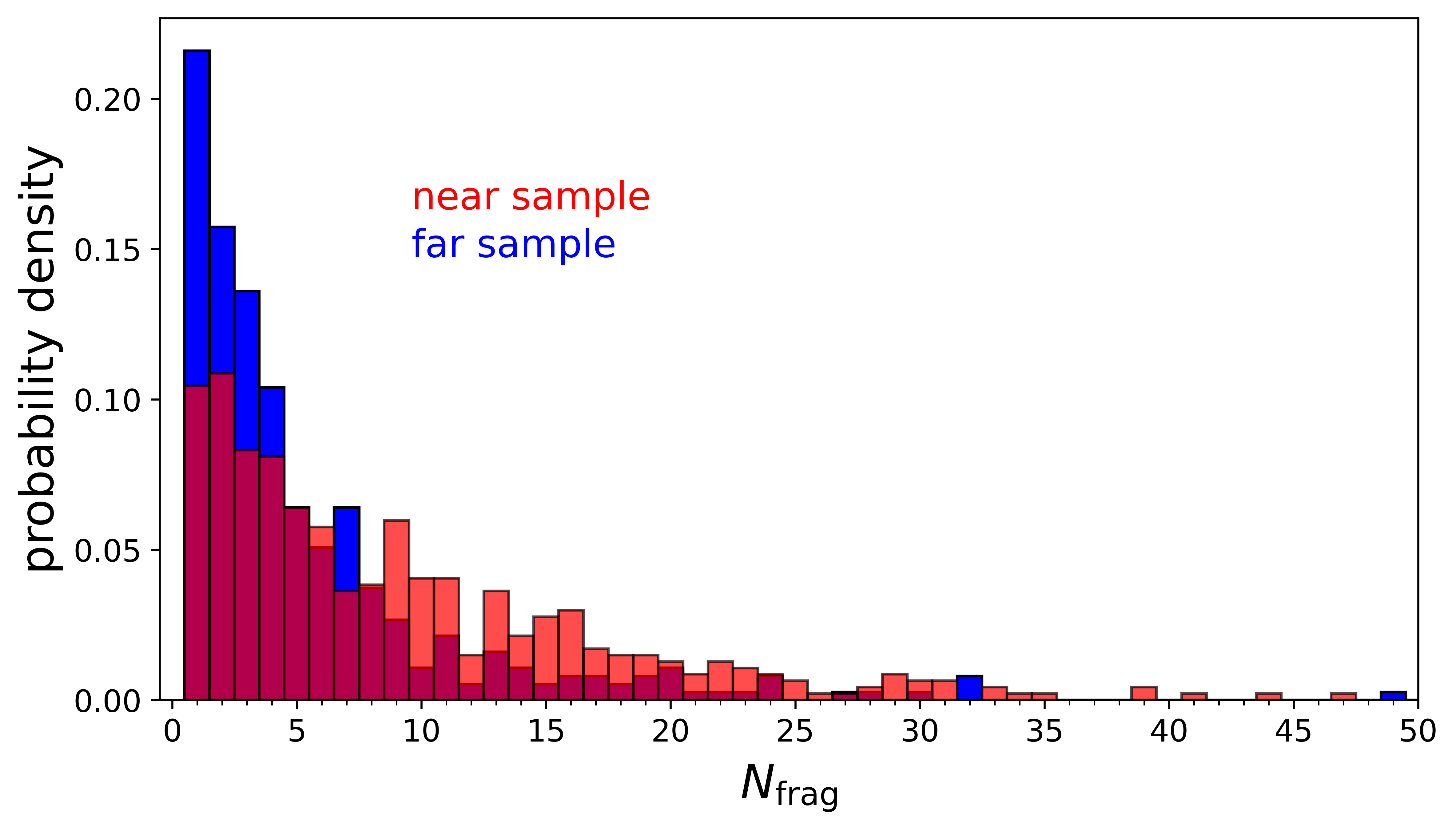}
    \includegraphics[width=\columnwidth,valign=t]{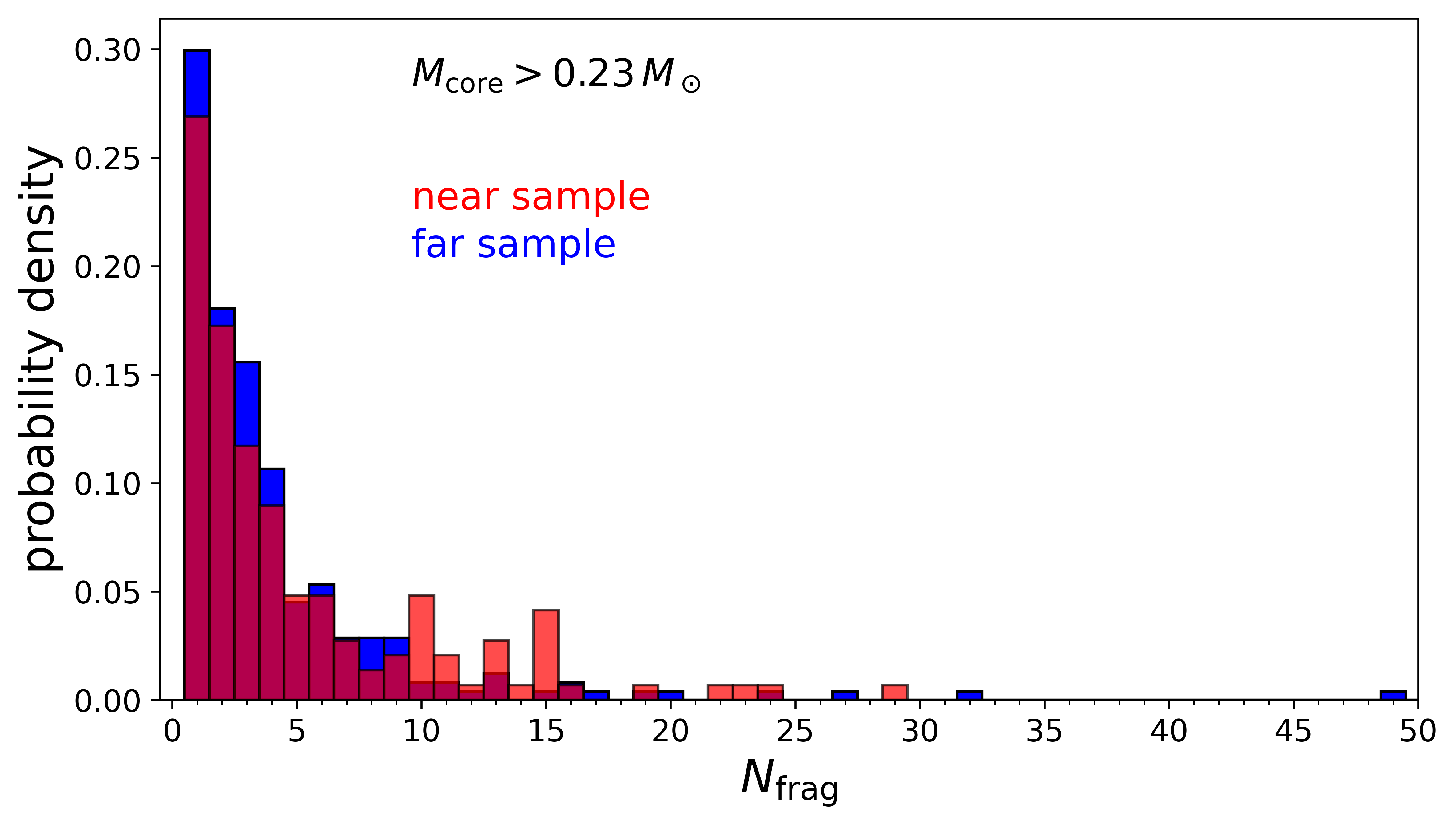}

    \caption{Fragmentation statistics evaluated for near sample and far sample sources separately. 
    \textit{Left panel}: Probability density distribution of the number of cores detected in each clump for near sample (red bars) and far sample (blue bars) sources, respectively. 
    \textit{Right panel}: Same as left panel, but only including clumps containing core masses above $0.23\,\mathrm{M_{\odot}}$ (see text for interpretation). Detailed statistics for the individual distributions are reported in Table \ref{table:frag_stats}.}
    \label{NFRAG_bias_hists_1}
    \end{figure*}

    \begin{table*}[h!]
    \caption{Detailed fragmentation statistics for Fig. \ref{NFRAG_bias_hists_1} distributions.}
    \label{table:frag_stats}
    \centering
    \begin{tabular}{l c c | c c}
    \hline\\[-7pt]
    & \multicolumn{2}{c|}{all cores} & \multicolumn{2}{c}{$M_\mathrm{core}>0.23\,\mathrm{M_{\odot}}$}\\ [2pt]
    & near sample & far sample & near sample & far sample
    \\[3pt]
    \hline\hline\\[-7pt]
    $N_{\mathrm{clumps}}$ & 469 & 375 & 145 & 244 \\
    $N_{\mathrm{frag}}$ & 4280 & 2068 & 744 & 1008 \\
    $N_{\mathrm{frag}}^{\rm{min}}$ & 1 & 1 & 1 & 1 \\
    $N_{\mathrm{frag}}^{\rm{max}}$ & 47 & 49 & 29 & 49 \\
    $N_{\mathrm{frag}}^{\rm{mean}}$ & 9 & 6 & 5 & 4 \\
    $N_{\mathrm{frag}}^{\rm{median}}$ & 7 & 3 & 3 & 3 \\
    $N_{\mathrm{frag}}=1$ & 49 (10\%) & 81 (22\%) & 39 (27\%) & 73 (30\%) \\
    $N_{\mathrm{frag}}=2$ & 51 (11\%) & 59 (16\%) & 25 (17\%) & 44 (18\%) \\
    $N_{\mathrm{frag}}=3-9$ & 197 (42\%) & 181 (48\%) & 53 (36\%) & 109 (44\%) \\
    $N_{\mathrm{frag}}=10-20$ & 127 (27\%) & 41 (11\%) & 24 (17\%) & 14 (6\%) \\
    $N_{\mathrm{frag}}>20$ & 45 (10\%) & 13 (3\%) & 4 (3\%) & 4 (2\%) 
    \\[2pt]
    \hline
    \end{tabular}
\tablefoot{
    The first two numeric columns refer to the left panel of Fig. \ref{NFRAG_bias_hists_1}, while the latter two to the right panel (see Fig. \ref{NFRAG_bias_hists_1} for further explanation). 
}       
    \end{table*}
    
    \begin{figure*}[h!]
    \includegraphics[width=0.98\columnwidth,valign=t]{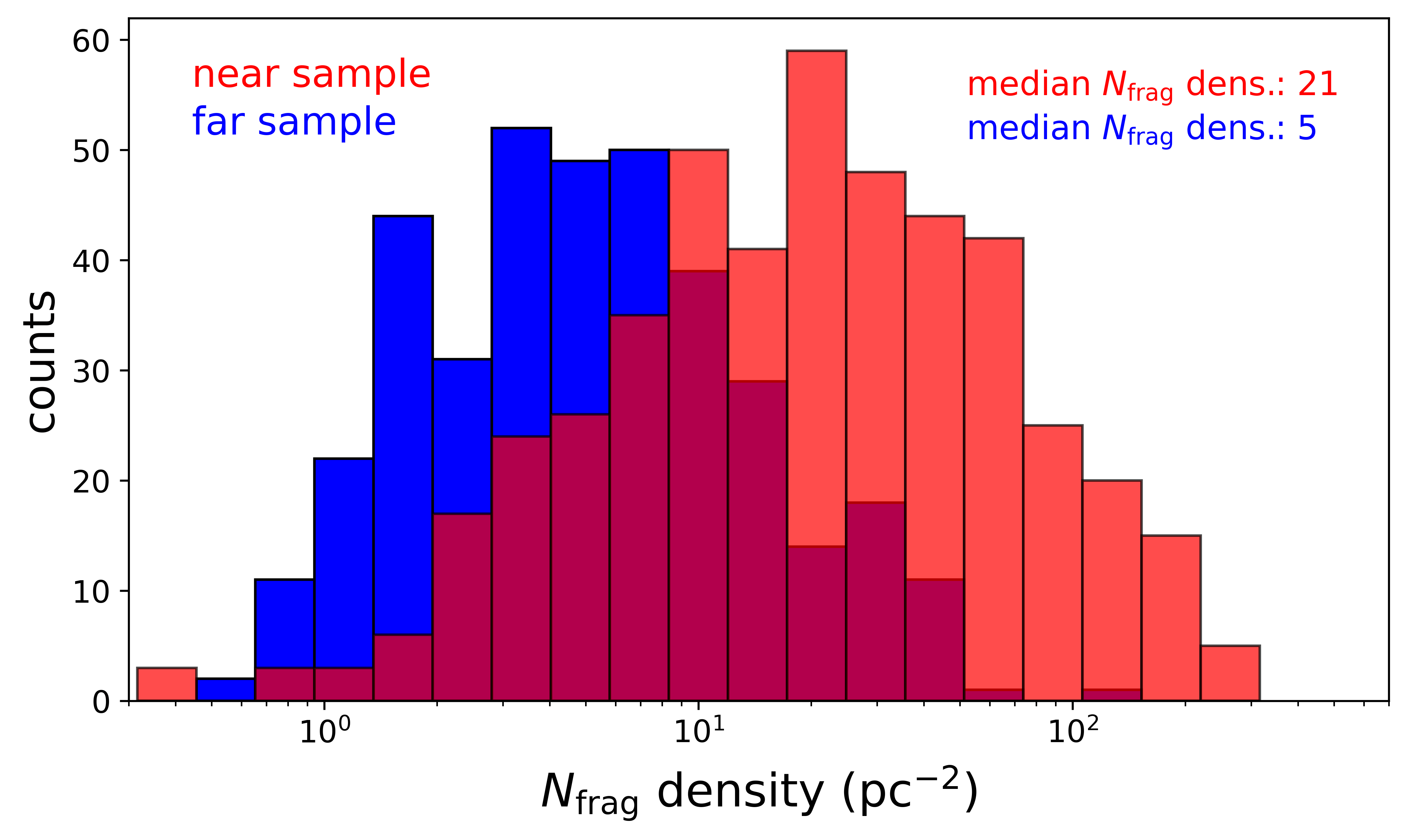}    
    \includegraphics[width=0.98\columnwidth,valign=t]{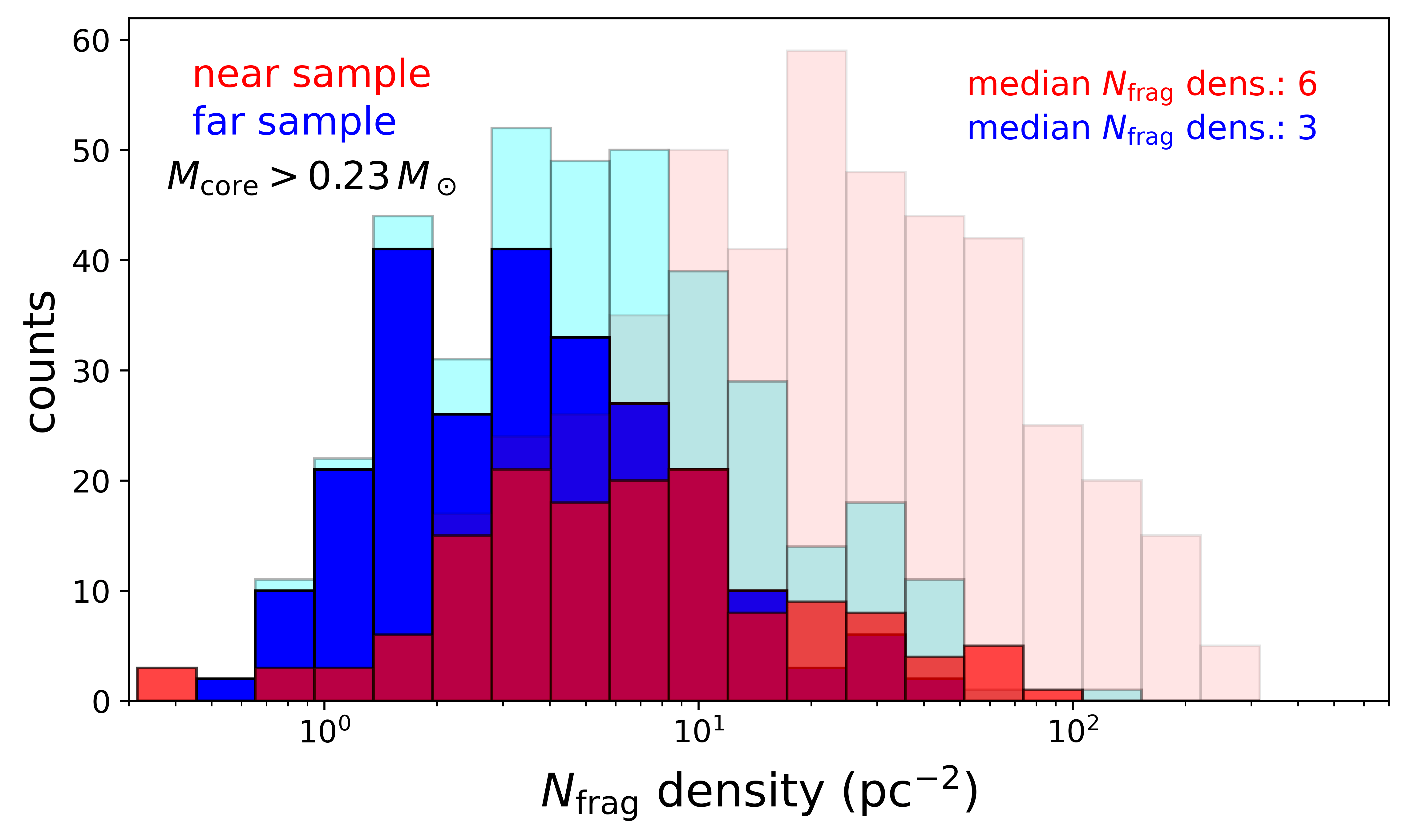}
    
    \caption{Analysis of the derived spatial density of the cores. \textit{Left panel}: Distribution of the spatial density of the cores within each clump for near sample (red bars) and far sample (blue bars) sources, respectively. 
    \textit{Right panel}: Same as left panel, but only including clumps containing core masses above $0.23\,\mathrm{M_{\odot}}$. For comparison, the overall distributions of the left panel appear here in transparency in the background as pink (near sample) and cyan (far sample) bars, respectively. In both panels, overlapping areas between the two distributions are marked by the magenta bars. 
    The median values of the near and far sample distributions are also reported in each panel.}
    \label{NFRAG_bias_hists_2}
    \end{figure*}

The angular resolution and target distance are critical parameters for the interpretation of our data, given the broad range of clump distances covered and the different specific observational setups used for near and far sample sources (see Sect. \ref{obs}). 

Figure \ref{NFRAG_bias_hists_1} (left panel) shows the distributions of the probability density of the number of fragments within each clump, for the two near and far subsamples separately (i.e., observed with the two different observational setups). The plot displays clear discrepancies between the two distributions. Detailed statistics for each distribution are reported in Table \ref{table:frag_stats}. The total number of fragments detected in near sample targets is more than twice that found in far sample ones, the gap being not merely justified by the different number of clumps included in each subsample. As for the probability density distribution, although the overall range of $N_\mathrm{frag}$ is nearly the same, the near sample distribution is clearly more shifted towards higher $N_\mathrm{frag}$ values with respect to the far sample one. The latter is indeed strongly peaked towards $N_\mathrm{frag}$ below $5$ and in particular, importantly, reports nearly twice the number of fields with only 1 detection. This trend is also confirmed by the different $N_\mathrm{frag}$ mean and median values of the two distributions (considerably higher for the near sample) and the percentage distribution of counts among different ranges of $N_\mathrm{frag}$. Since the angular FOV of the maps is the same ($\sim35''$) for both near and far sample targets, due to the different distances we are actually sampling a wider spatial scale in far sample fields (on average $\sim1$ pc compared to $\sim0.6$ pc of near sample fields). This could in principle result in a higher number of detections in far sample fields (assuming a uniform distribution of fragments in the sky). The fact that this expected trend is not seen by comparing the two distributions tells us that another effect is actually overpowering resolution in dominating the fragmentation properties that we observe as a function of distance, which we identify with the change in mass sensitivity across our sample. As it is directly shown in Sect. \ref{Msens-compl}, the constant sensitivity in flux of the sample translates into an increasing mass sensitivity threshold with distance, which gradually reduces the possibility to reveal lower-mass cores going from the nearest to the farthest clumps. From the fragmentation viewpoint, the wider range of masses available for detection naturally leads to a higher mean number of detections within near sample targets with respect to the far sample. Indeed, we verified (right panel of Fig. \ref{NFRAG_bias_hists_1} and Table \ref{table:frag_stats}) that near and far sample distributions of $N_\mathrm{frag}$ become almost identical when selecting only fields hosting derived core masses above the completeness limit ($\sim0.23\,\mathrm{M_{\odot}}$, see Sect. \ref{Mcore}), thus proving that mass sensitivity is indeed the predominant factor in the observed fragmentation statistics. 

The left panel of Fig. \ref{NFRAG_bias_hists_2} reports the two subsample distributions of spatial density of fragments. Again, a clear shift between the two profiles is found, largely inherited from the discrepancy in the number of fragments seen in Fig. \ref{NFRAG_bias_hists_1} (left panel), with near sample counts oriented towards higher densities than far sample ones, as testified by the corresponding median values ($21$ pc$^{-2}$ for near sample sources, $5$ pc$^{-2}$ for far sample sources). This behavior is for the most part ascribable to the same mass sensitivity effect discussed above, and the balance between the two subsamples is indeed nearly recovered when considering core masses above completeness (right panel of Fig. \ref{NFRAG_bias_hists_2}). Evidently, most of the counts of the near sample distribution that caused the considerable shift with the far sample in the overall distributions are now lost, proving that they mostly pertained to low-mass sources, which we are actually not able to reveal at larger distances. However, a slight shift between the two subsamples (quantified by the different median values) remains also when filtering the detections by mass. Therefore, the influence of an additional physical effect on this trend cannot be ruled out. If one assumes the fragments to be spatially distributed in a rather uniform fashion in the clump, one would expect to find constant spatial densities between near and far sample maps, even if they cover a different range of spatial scales. The fact that we observe a discrepancy, instead, could suggest that fragments are preferentially concentrated within the central, relatively spatially limited regions of the clumps, so that we find lower densities when expanding the sampled FOV. 

The separation and spatial distribution of the fragments revealed within the ALMAGAL clumps will be investigated in detail in a following work (Schisano et al. in prep.), to make further considerations on the possible physical mechanisms governing the fragmentation process (see, e.g., \citealt{Kainulainen+17,Sanhueza+19,Svoboda+19,Gieser+23b,Morii+23,Traficante+23,Ishihara+24}).

\subsection{Photometric properties}
\label{phot_props}

\subsubsection{Integrated fluxes}\label{Fint}

    \begin{figure*}[t!]
    \includegraphics[width=\columnwidth]{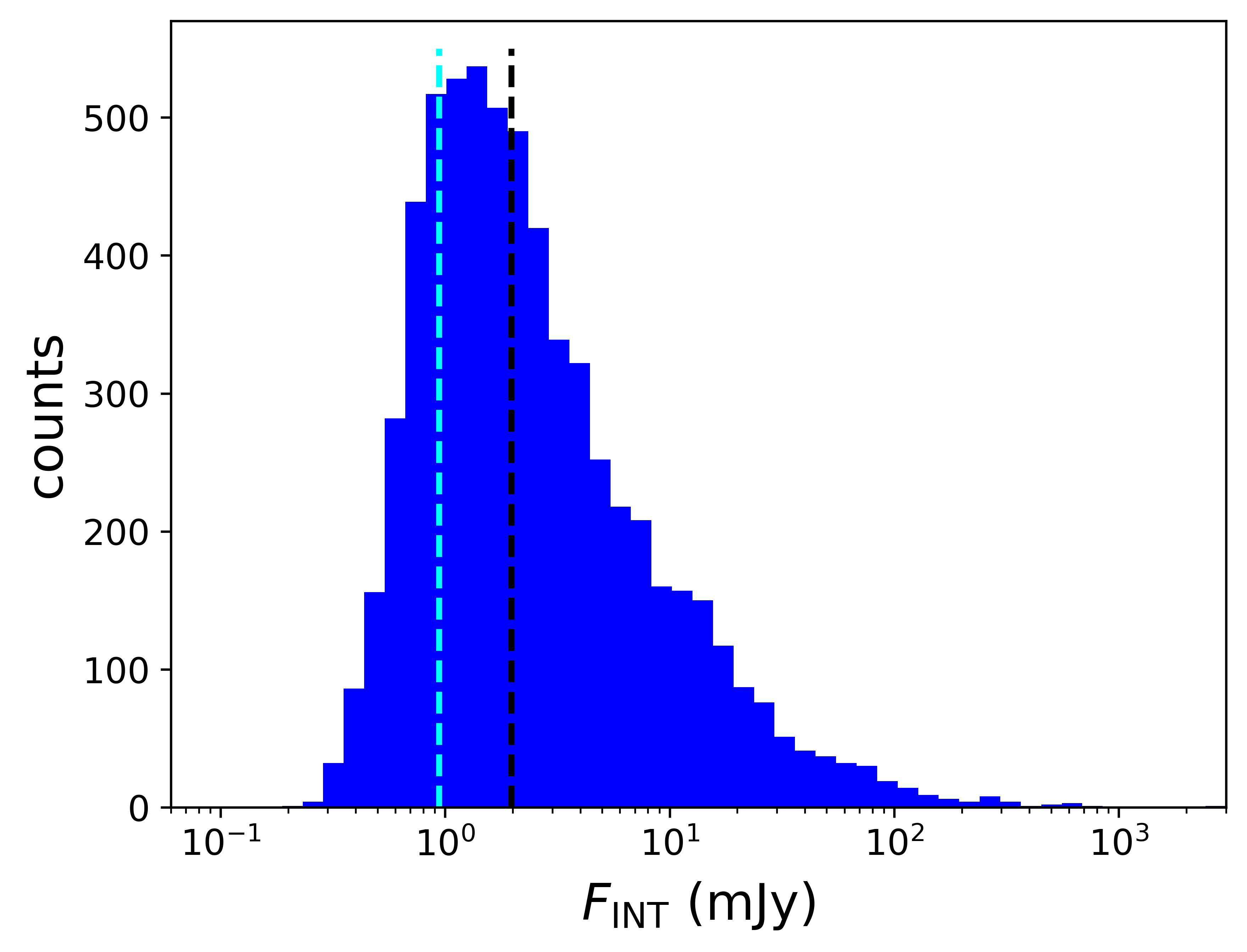}
    \includegraphics[width=1.025\columnwidth]{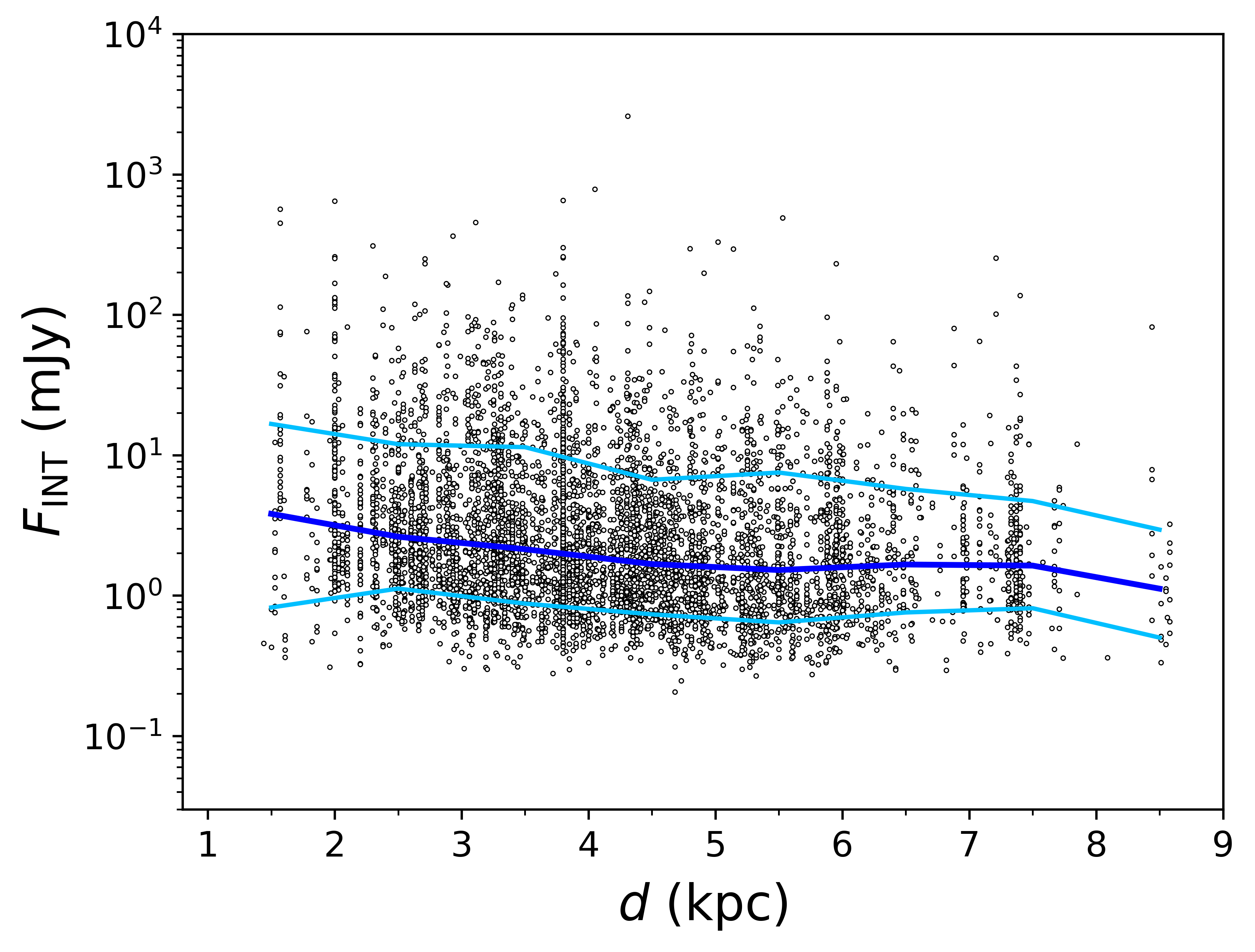}
    \caption{Integrated fluxes of the extracted compact sources as estimated with \textit{CuTEx}. \textit{Left panel}: Overall distribution of the integrated fluxes of the sources. The adopted flux completeness value of $0.94$ mJy (see Sect. \ref{FcomplPhotacc}) is marked by the vertical dashed cyan line. The vertical dashed black line marks the median value of the distribution ($2$ mJy). 
    \textit{Right panel}: Measured integrated fluxes of the sources as a function of the hosting clump distance. The solid blue line traces the median $\mathrm{F_{INT}}$ values within given bins of distance. Light blue lines correspond to the 15th (bottom) and 85th (top) percentiles of the flux distribution in the same bins of distance. Note: for improved visualization, distance outliers ($<3\%$) are not shown in this plot.}
    \label{Fint_plots}
    \end{figure*}

The integrated fluxes of the extracted compact sources, as estimated through the \textit{CuTEx} photometry procedure described in Sect. \ref{cutex} by assuming a Gaussian source profile, are reported in Col. $12$ of Table 3. The overall distribution of integrated fluxes is shown in Fig. \ref{Fint_plots} (left panel). The dashed cyan line indicates the flux completeness limit of $0.94$ mJy, estimated for our catalog as described in Sect. \ref{FcomplPhotacc}. We decided to adopt the completeness limit computed for Group 2 targets as overall reference, since such group represents the vast majority of the ALMAGAL maps ($76\%$, Fig. \ref{map_RMS_hist}), characterized by a "normal" level of noise. 
The measured integrated fluxes range from $\sim0.3$ to $\sim10^3$ mJy. The median of the integrated flux distribution is $2$ mJy. The faintest source detected is $\sim2$ times above the requested flux sensitivity of $\sim0.1$ mJy, corresponding to a S/N of $\sim5.5$. The brightest source has a S/N of $\sim1500$ (see \citealt{Sanchez+25}). 
The uncertainty of estimated integrated fluxes is $\sim20\%$, as deduced by comparing the fluxes of synthetic sources measured with \textit{CuTEx} with the true, generated ones at different flux levels (see Sect. \ref{FcomplPhotacc} and Appendix \ref{AppFcomplPhotacc} for details). 

The right panel of Fig. \ref{Fint_plots} reports the measured integrated fluxes of the sources as a function of the distance of their hosting clumps. The trend traced by the blue line shows that the median measured integrated flux is roughly constant (i.e., within a factor of $\sim2$) with the source distance. This was indeed designed for the ALMAGAL survey in order to ensure that consistent and comparable fluxes can be recovered across the whole clumps sample despite its wide spread in distance (see \citealt{Molinari+25}). A similar trend is found for 15th and 85th percentiles of the flux distribution within bins of distance, indicating that detected objects have roughly the same statistical distribution in flux independently from their distance. Moreover, the range of fluxes recovered at the different distances is roughly the same. The measured integrated fluxes have been used to estimate the mass of the extracted cores, as described in Sect. \ref{Mcore}.

\subsubsection{Angular sizes}
\label{FWHMs}

\begin{figure*}[ht!]
    \includegraphics[width=\columnwidth,valign=t]{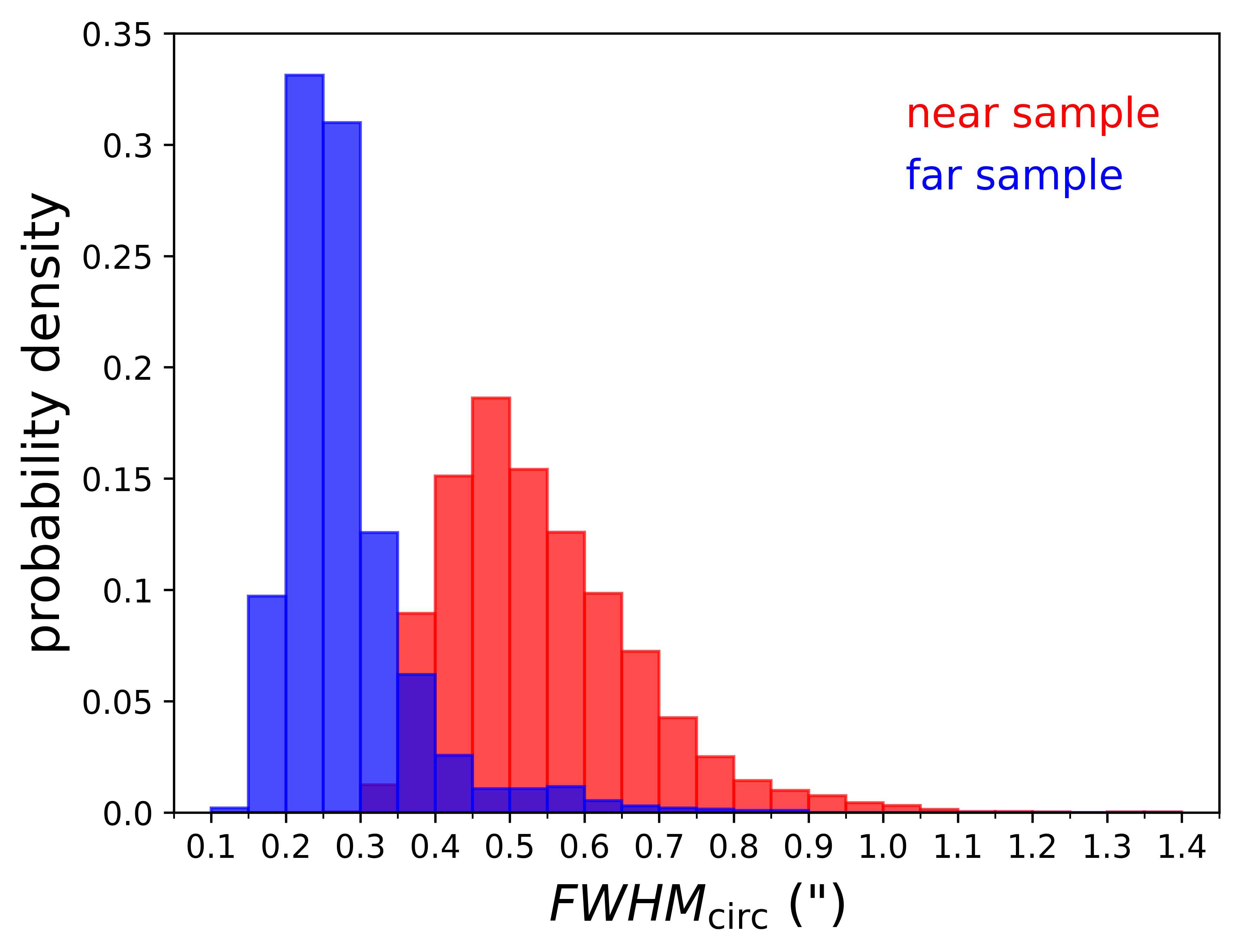}
    \includegraphics[width=1.016\columnwidth,valign=t]{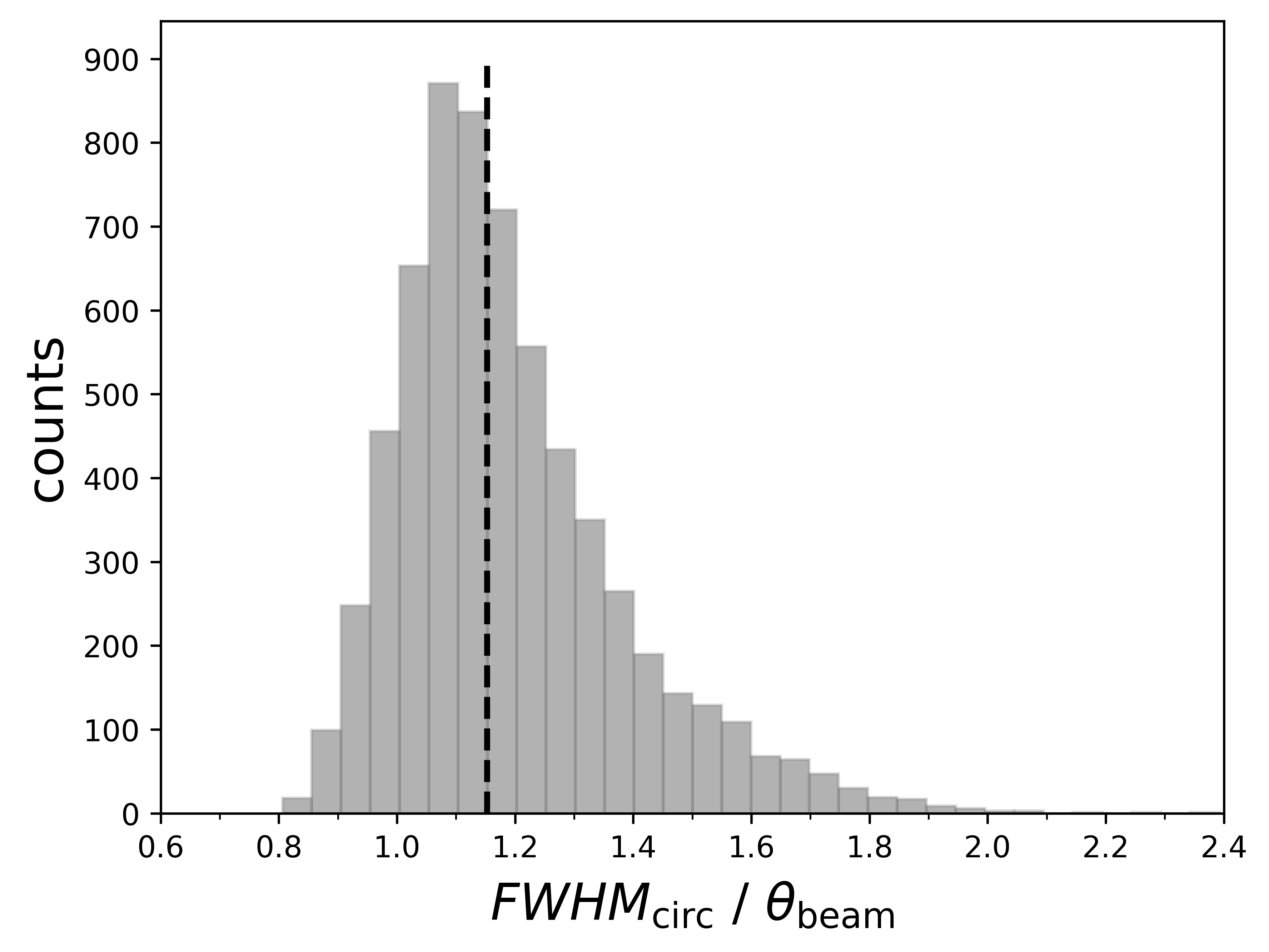}
    
    \caption{Angular sizes of the extracted sources. \textit{Left panel}: Probability density distribution of the circularized FWHM of the sources. Red and blue bars include sources detected within the near sample and the far sample targets, respectively (see Sect. \ref{obs}). 
    \textit{Right panel}: Distribution of the ratio between the estimated angular size of the source and the map beam size (both in circularized form), providing a resolution-independent estimate of the source sizes. The vertical dashed black line marks the median value of $1.15$ for the ratio.}
    \label{FWHM_plots}
    \end{figure*}

The circularized angular size of the extracted sources, $FWHM_{\rm{circ}}$, was computed as the geometric mean of the FWHM of the fitted source ellipse reported in Cols. $9-10$ of Table 3. The overall distribution of angular sizes ranges between $\sim0.15-1.4''$. The two different observing setups adopted in the ALMAGAL survey (described in Sects. \ref{obs} and \ref{contmapprop}) translate into two distinct distributions of angular sizes for detected sources belonging to each of the two samples. 
This can be noted from the histogram of Fig. \ref{FWHM_plots} (left panel), where the two distributions overlap only marginally. Focusing on each of the individual subsamples, far sample sources show angular sizes within $\sim0.15-0.9''$ (with approximate mean and median values of $\sim0.25''$), while near sample sources within $\sim0.3-1.4''$ (with mean and median values of $\sim0.5''$). 
Although the effective spatial resolution is rather uniform across our sample (right panel of Fig. \ref{beam_circ_props_plots}), the dependence of the physical size estimate from the target distance is not completely removed, as it is shown in Sect. \ref{Dcore}. 
A resolution-independent view of the source sizes can be obtained by normalizing the angular size by the beam size of the corresponding map, as reported in Fig. \ref{FWHM_plots} (right panel). In these terms, detected sources have sizes from $\sim0.8$ to $\sim2.4$ times the beam, with a median value of $1.15$ (i.e., $15\%$ discrepancy). These values essentially match the expected outcome of the algorithm, as such compact source sizes are the ones \textit{CuTEx}, by design, is more sensitive to (see \citealt{Molinari+16b} for details). 
The absence of a rigid lower limit at ratio $\sim1$ for the peak of the size distribution tells us that most of the detected compact sources are resolved. They turn out to mostly be slightly larger than the beam, implying that they are compact and centrally peaked objects. 
We note that for this ratio, even values below the imposed minimum source size threshold are allowed, since we are comparing with the circularized beam, while the lower boundary limit for the source size is set to $0.95$ times the beam minor axis (see Sect. \ref{cutex}). We also caution that, since we are expressing quantities in circularized form for consistency with the subsequent physical analysis, ratios above $1$ may in principle result, for some sources, from the combination of a non (or barely) resolved minimum FWHM, and a more extended maximum FWHM. 

\begin{figure}[ht!]
    \includegraphics[width=\columnwidth]{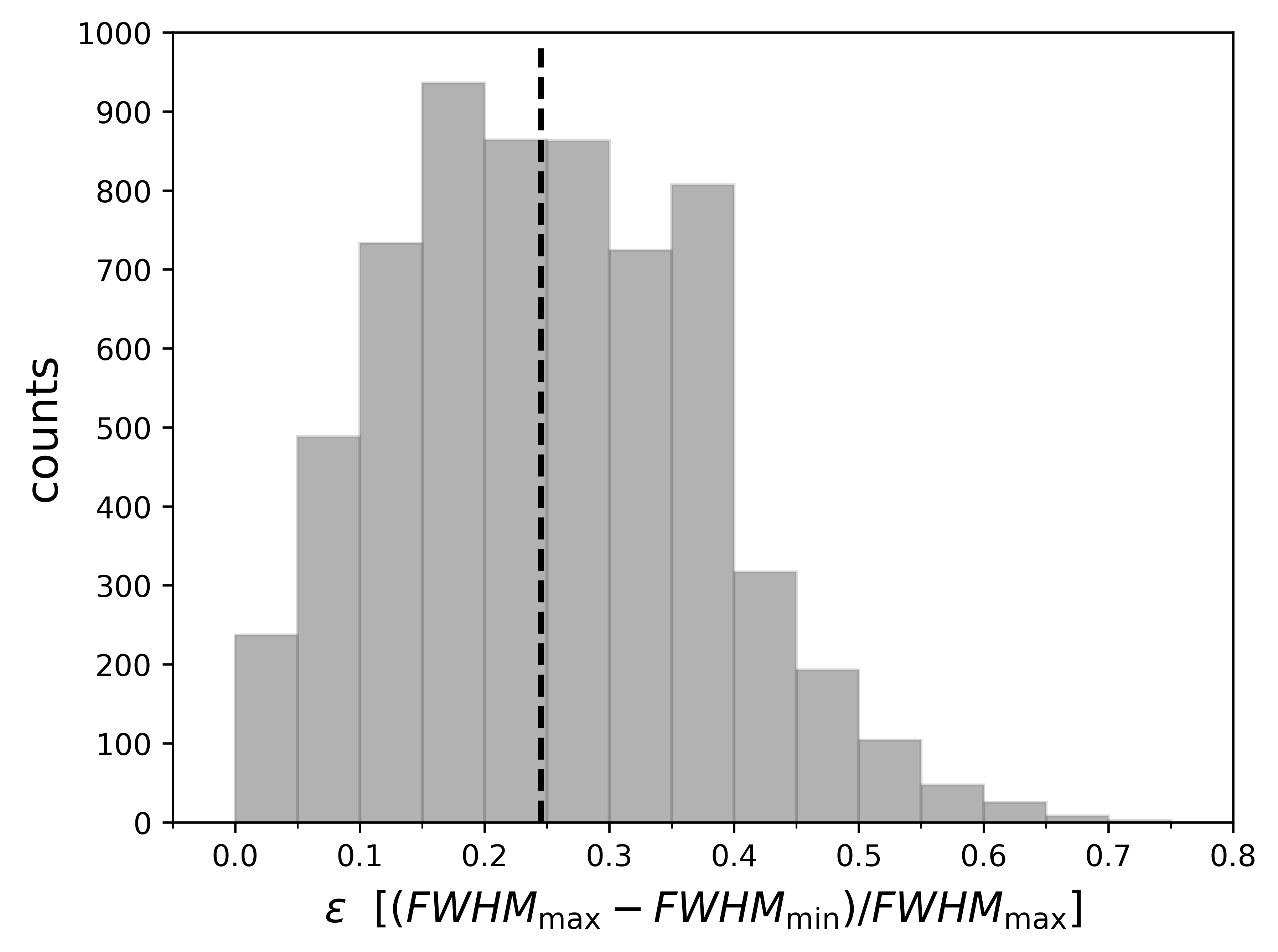}
    \caption{Histogram of the ellipticities of the sources ($\epsilon$), defined as the relative discrepancy between the maximum and the minimum axes (i.e., the two FWHMs) of the fitted ellipse. Ellipticity $\epsilon=0$ means perfectly circular source, whereas large $\epsilon$ indicate highly elongated ones. The vertical dashed black line marks the median value $\epsilon=1.25$.}
    \label{ell_hist}
    \end{figure}

The overall distribution of source ellipticities ($\epsilon$), defined as the relative discrepancies between the maximum and minimum axes of the fitted ellipse, i.e., ($FWHM_{\rm{max}}-FWHM_{\rm{min}}$)/$FWHM_{\rm{max}}$, is shown in Fig. \ref{ell_hist}. A null ellipticity corresponds to perfectly circular sources, while values close to $1$ indicate highly elongated sources. Observed $\epsilon$ range from just above $0$ to $\sim0.7$, meaning that the most elongated sources have a ratio of $\sim3$ between their major and minor extents. The median value of $\epsilon$ across the whole sample is $0.25$ (i.e., a $1.25$ ratio between the two axes). A physical interpretation can be made of these estimates, although some caveats related to the technical properties of the maps and the features of the source extraction method have to be considered. On the one hand, lower ellipticities could possibly be tracing mainly compact, centrally concentrated and isolated sources, while higher values could represent more diffuse ones, potentially residing in larger scale and/or filamentary-like structures. On the other hand, it must be considered that the map beam is always elliptical to some extent (mostly within $\sim1$ and $\sim1.4$ in terms of ratio between major and minor beam axes in our case, see \citealt{Sanchez+25}), and this has an effect on the photometry algorithm when searching for the best fitting of the source ellipse, so that we will never obtain perfectly circular sources. 

\section{Physical properties of the cores}
\label{core_phys_props}

In this section, parameters obtained from the source detection and photometry procedure, reported in the compact source catalog of Sect. \ref{cat}, are used, together with the properties of the hosting clumps, to estimate the main physical parameters of the extracted core population. In particular, the distributions of core physical sizes and masses are shown and discussed. 

\subsection{Core physical sizes}
\label{Dcore}

 \begin{figure*}[t!]
    \includegraphics[width=\columnwidth,valign=t]{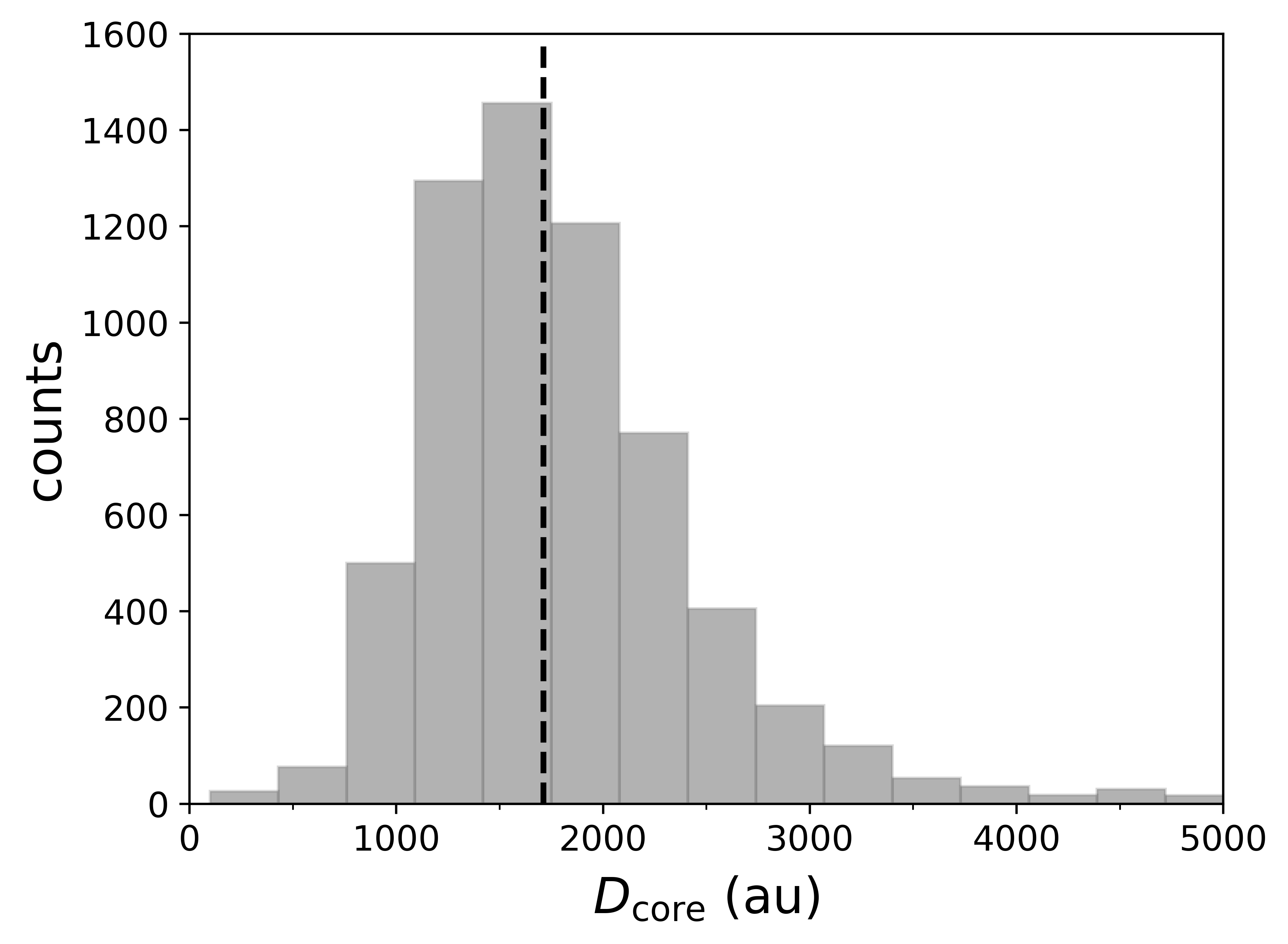}
    \includegraphics[width=\columnwidth,valign=t]{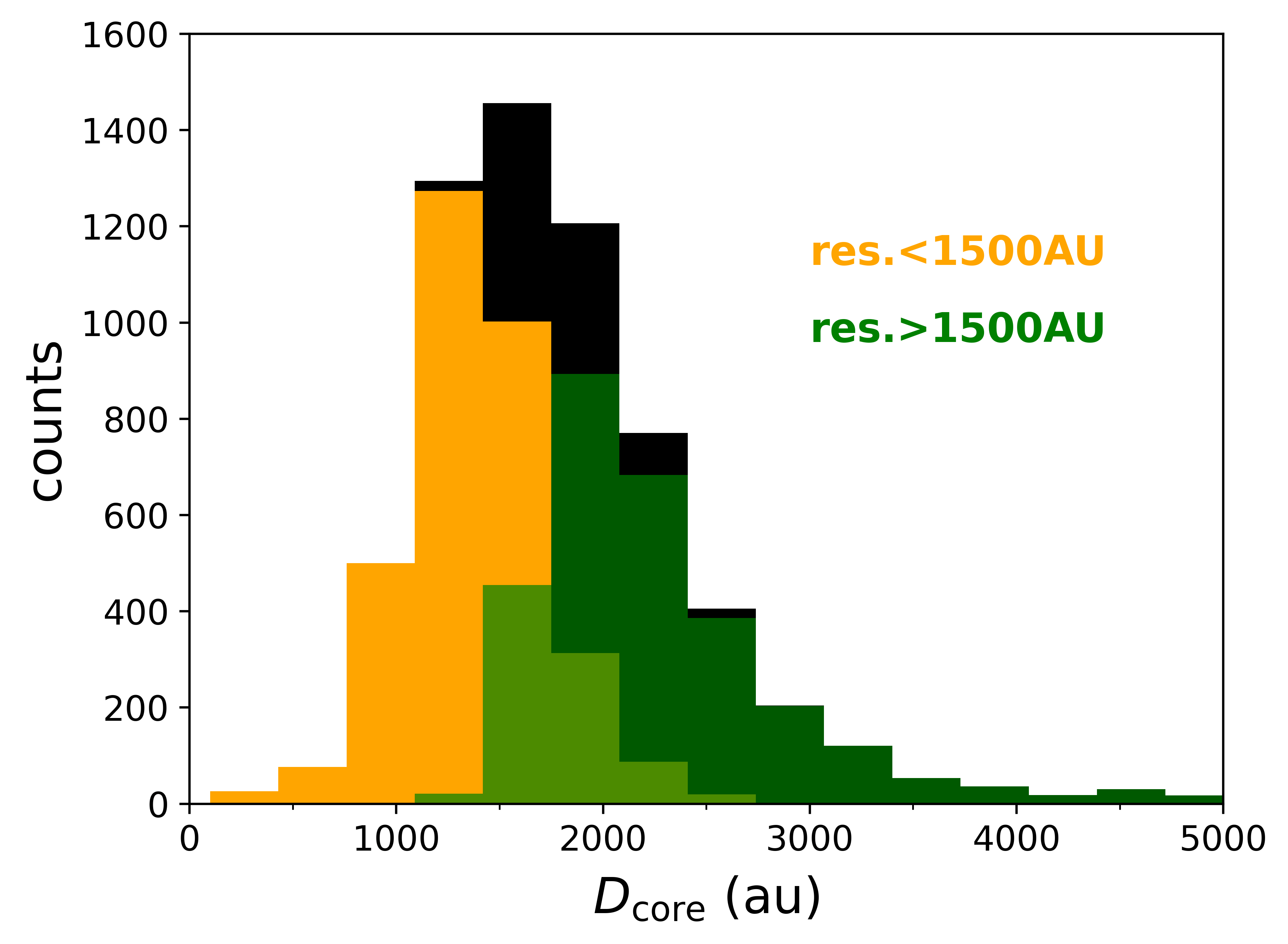}
    \caption{Distribution of the cores' estimated physical sizes (i.e., diameters). 
    \textit{Left panel}: Overall distribution of core sizes. The vertical dashed black line marks the median value of the distribution ($1700$ au). For better visualization purposes, the few outliers above $5000$ au ($\sim2\%$) are not shown. 
    \textit{Right panel}: Same distribution of left panel (black), but overplotted with the core size distributions resulting from better resolved fields (gold, below $1500$ au) and less resolved ones (dark green, above $1500$ au). The two histograms overlap in the lighter green area.}
    \label{Dcore_hists}
    \end{figure*}

From the measured circularized angular size of the detected source ($FWHM_{\rm{circ}}$, Sect. \ref{FWHMs}), its corresponding physical size (i.e., diameter) can be computed as
\begin{align}
&D_\mathrm{core}\,\mathrm{(au)}=FWHM_{\rm{circ}}\,\mathrm{(\rm{''})}\,\,d\,\mathrm{(pc)}\,,
\end{align}
where $d$ is the clump heliocentric distance. 

The estimated sizes of the cores are listed in Col. 3 of Table 7. 
Figure \ref{Dcore_hists} (left panel) shows the overall distribution of sizes estimated for the cores. 
Core sizes range within $\sim800-3000$ au in most cases ($\sim90\%$), with a mean value of $\sim1900$ au and a median value of $\sim1700$ au, roughly where the distribution peaks. A few sources ($\sim2\%$) show sizes even below $800$ au (down to $\sim250$ au), corresponding to the cases when the best possible resolution is achieved in the observations thanks to the combination of both good angular resolution and small target distance. Similarly, a small population of cores ($\sim3\%$) with estimated sizes above $\sim4000$ au is also present, corresponding to lower angular resolution and large distance cases. 
As explained in Sect. \ref{contmapprop}, however, some of these extreme cases could correspond to targets for which the distance estimates were recently corrected, thus switching between near and far samples. 
The uncertainty of estimated sizes coming from the photometry procedure is within $\sim20\%$, as deduced by comparing the angular sizes of a large population of synthetic sources measured by \textit{CuTEx} with their true simulated values (see Sect. \ref{FcomplPhotacc} and Appendix \ref{AppFcomplPhotacc}). 
The relative error on the clump heliocentric distance, originating from the $V_{\rm{LSR}}$ estimation at clump level (see \citealt{Molinari+25}), is on average $10\%$. 

Recent subparsec resolution studies of the fragmentation of high-mass star-forming clumps, mostly performed with the ALMA, NOEMA or Very Large Array (VLA) interferometers both through dust continuum and line observations, have measured fragment sizes over a wide range of spatial scales, from $\sim0.1$ pc ($\sim20000$ au, e.g., \citealt{Sanchez-Monge+13b,Anderson+21,Avison+23,Morii+23,Traficante+23}, all featuring maximum spatial resolutions within $2000-5000$ au) down to $\sim2000-3000$ au (e.g., ALMA studies \citealt{Sanhueza+19,Svoboda+19,Pouteau+22,Morii+23,Xu+24}), and even $\lesssim1000$ au (\citealt{Beuther+18}, observed with NOEMA). In more detail, \citet{Sanhueza+19} and later \citet{Morii+23} measured fragment sizes between $3500-10000$ and $2000-20000$ au, respectively, in early stage $70\,\rm{\mu}m$-dark clumps within $6$ kpc distance at a $>2000$ au resolution. \citet{Pouteau+22} found $\sim2500-10000$ au sizes in the single, large complex W43, hosting the young and more evolved protoclusters MM2 and MM3, observed at a $2500$ au resolution. In their sample of evolved ($L/M>20\,\rm{L_{\odot}/M_{\odot}}$) massive protostellar objects observed at high resolution ($\lesssim1000$ au), \citet{Beuther+18} revealed a wide range of sizes, from $\sim600$ to $\sim14000$ au. \citet{Traficante+23} reported fragments from $\sim4000$ to $20000$ au in size across their sample covering different evolutionary stages, from $70\,\rm{\mu}m$-dark to HII regions, observed at $\sim2000$ au maximum resolution. \citet{Xu+24} (ASSEMBLE) measured core sizes between $\sim2200-6200$ au. 
It must be noted that, since our source extraction method is sensitive to compact sources up to nearly three times the beam wide (Sect. \ref{SE}), we may be actually missing possible more extended structures which are instead revealed by other literature works. 

The two distinct observational setups chosen for the near and far samples allowed us to obtain a rather uniform distribution of spatial resolutions across the whole clump sample, which resulted in uniform estimated core physical sizes, without clear separation between sources belonging to different subsamples. 
However, the spatial resolution still has an effect, as it can be seen from Fig. \ref{Dcore_hists} (right panel), where cores of fields observed with a spatial resolution better than $1500$ au are plotted as gold bars, while cores coming from less resolved fields (spatial resolution above $1500$ au) appear as dark green bars. The two distributions partly overlap, but a slight shift is visible (a factor of $2$ discrepancy between the two medians), proving that indeed the smallest cores revealed are potentially a bi-product of the better resolution available in those targets, whereas less resolved ones are responsible for the high-size end of the global distribution. Therefore, in less resolved clumps, we generally extract larger scale substructures than in better resolved fields. 
In other words, at higher spatial resolutions the centrally condensed/contrasted sources tend to match the beam size (see also right panel of Fig. \ref{FWHM_plots}), so that it is likely that most of the emission actually comes from a region smaller than the beam itself. 
In any case, from a statistical point of view, the vast majority of targets have effective spatial resolutions consistent within a factor of $\sim2$ (or $3$ at most, see right panel of Fig. \ref{beam_circ_props_plots}), so that within such limits we can be confident to be tracing similar kinds of physical objects. Nevertheless, it is still possible that at least in a few extreme cases (particularly at lower spatial resolutions) we could be actually probing different types of objects from one field to another. Such structures, in reality, may represent an upper level in a hierarchical framework of fragmentation, as well as unresolved smaller scale source(s).

\subsection{Core masses}
\label{Mcore}

From the measured source integrated flux ($F_\mathrm{INT}$, Sect. \ref{Fint}) the core mass has been estimated assuming optically thin dust continuum emission through the formula (e.g., \citealt{Kainulainen+17})
\begin{equation}
\label{Mcore_eq}
M_{\rm{core}} = \frac{R_{\rm{gd}}\,F_{\rm{INT}}\,d^2}{B_{\nu}(T)\,\kappa_{1.3}}\,,
\end{equation}
where $R_{\mathrm{gd}}=100$ is the widely assumed gas-to-dust mass ratio (see Sect. \ref{Mcore_err}), and $\kappa_{1.3}=0.9$\,$\mathrm{cm^2\,g^{-1}}$ is the dust absorption coefficient at $1.3$ mm wavelength, describing the opacity of dust grains with thin ice mantles at the typical gas densities of these regions ($\gtrsim10^6\,\mathrm{cm^{-3}}$, \citealt{Ossenkopf&Henning94,Sanhueza+19}, see Sect. \ref{Mcore_err} for details). $B_{\nu}(T)$ is the Planck function at the core temperature $T$ (see Sect. \ref{Tcore_model}). 

The uncertainties on the core mass estimate brought by the assumptions and parameters involved in the mass calculation are discussed in Sect. \ref{Mcore_err}.

\subsubsection{Core temperature model}
\label{Tcore_model}

A key parameter for the computation of the core mass is the temperature of the dust envelope enclosing the core, which determines its observed emission through the Planck function $B_{\nu}(T)$ (Eq. \ref{Mcore_eq}). 
Ideally, direct measurements of the temperature for each source should be used, obtained from molecular line emission of high-density tracers such as $\mathrm{NH_3}$ (ammonia), $\mathrm{H_2CO}$ (formaldehyde), $\mathrm{CH_3OH}$ (methanol), or $\mathrm{CH_3CN}$ (methyl cyanide) (see, e.g., \citealt{Zhang+02,Sanchez-Monge+13b,Lu+14,Beuther+18,Cesaroni+19,Olguin+21,Brouillet+22,Li+22,Gieser+23,Taniguchi+23,Izumi+24}, and \citealt{Mininni+25}). However, these estimates are not yet available for the whole target sample and all extracted cores at the time of this work, as they will be presented in a subsequent ALMAGAL work (Jones et al. in prep.). 

Without such core-level estimates, educated assumptions on dust temperature for the cores must be made. Assigning a single, universal temperature to the whole core sample would represent a too coarse and inaccurate (although widely used) approximation, given the very wide range of different evolutionary stages and physical properties covered in our sample. Instead, we chose to base our assumptions on the properties of the respective hosting clumps, as it has been done in some other recent ALMA studies. In particular, several authors directly assigned to the cores the temperature estimated at larger scale for the hosting clumps. For example, \citet{Sanhueza+19} used $10-15$ K in their $70\,\mu$m-dark early stage high-mass clumps, whereas \citet{Svoboda+19} used $10-14$ K and \citet{Zhang+21} $10-20$ K in high-mass starless clump candidates. \citet{Csengeri+17} estimated dust temperature from the clump luminosity of their early stage massive clumps, then adopting a unified value of $25$ K. \citet{Traficante+23}, featuring a limited but varied sample of clumps in terms of evolutionary stages, split the sample in three groups according to the clump evolutionary indicator $L/M$, assigning increasing temperature levels ($20$, $30$ and $40$ K, respectively) to the cores belonging to more evolved clumps. 
For this work, we decided to adopt a similar but further developed approach, attributing to the cores of each clump a temperature according to its evolutionary stage, based on the calibration of the clump $L/M$ with the rotational temperature of the high-density tracer $\mathrm{CH_3C_2H}$ performed in \citet{Molinari+16} (see Appendix \ref{AppTcoremodel}). In our model we assign: i) $20$ K to cores belonging to less evolved targets (clump $L/M\leq1\,\mathrm{L_{\odot}/M_{\odot}}$); ii) $35$ K to core pertaining to intermediate sources ($1<L/M\leq10\,\mathrm{L_{\odot}/M_{\odot}}$); iii) a temperature following the relation (with $L/M$ expressed in $\mathrm{L_{\odot}/M_{\odot}}$) 
\begin{align}
\label{Tcore_3_eq}
T_{\rm{core}}=21.1\cdot\left(L/M\right)^{0.22}\,\mathrm{K,}
\end{align}
to cores in more evolved targets ($L/M>10\,\mathrm{L_{\odot}/M_{\odot}}$), thus proportional to a power law of the clump $L/M$, scaled with a proportionality factor ($21.1$) to best fit to the observed points in the \citet{Molinari+16} model. This corresponds to a temperature of $35$ K for clumps having $L/M=10\,\mathrm{L_{\odot}/M_{\odot}}$. 
The three groups of clump $L/M$ and the corresponding reference temperatures, $T_\mathrm{core}$, adopted for the cores are summarized in Cols. $1$ and $2$ of Table \ref{Tcore_tab}. The number of cores for each group is shown in Col. $5$. The resulting distribution is reported in Appendix \ref{AppTcoremodel}. 

With this approach, we are able to broadly account for the wide range of different evolutionary stages and physical conditions covered by our clump sample, which to some extent reflects on the properties of the internal population of cores. In the first group, i.e., with $L/M\leq1\,\,\mathrm{L_{\odot}/M_{\odot}}$, we do not expect the dust temperature to significantly vary from the $\sim20$ K that typically characterizes pristine clouds in prestellar phases (see, e.g., the above cited ALMA works). A slightly larger variance in temperature may be expected in the intermediate group, $1<L/M\leq10\,\,\mathrm{L_{\odot}/M_{\odot}}$, according to the number of protostellar sources already formed within the clump, but the fairly constant and scarcely scattered trend found in the \citet{Molinari+16} model ensures a good robustness to our approximation. A wider range of temperatures is indeed possible in most evolved clumps, with $L/M>10\,\,\mathrm{L_{\odot}/M_{\odot}}$, where formed protostars and young HII regions resulting from densest cores may be efficiently heating up the dust envelopes and clump molecular environment at large. By not setting a unique temperature for those sources, but rather a value proportional to the clump luminosity (and $L/M$), we intend to account for these effects. 

Our model results in an overall range of core temperatures between $20$ and $81$ K, with a median value of the distribution of $35$ K. Such values are consistent with those used in other recent ALMA surveys, such as \citet{Anderson+21} ($18-76$ K in IR-dark hub-filaments systems) and \citet{Pouteau+22} ($19-65$ K within W43-MM2\&MM3). It must be noted nonetheless that all those works feature lower spatial resolutions ($\geq2500$ au) with respect to our observations ($1400$ au on average). 

Ultimately, a caveat has to be raised about these temperature assumptions. Cores belonging to the same clump will have a distribution of individual evolutionary stages (varying from still quiescent regions to active star-forming sites) and related physical properties (including temperature), which will not necessarily coincide with the overall behavior of the hosting clump, whose estimated physical parameters are mostly dominated by the brightest embedded object (see \citealt{Elia+21}, Elia et al. in prep.). It also remains possible that some hot cores in much evolved clumps could reach temperatures even higher than $80$ K ($\sim100$ K or even more, e.g., \citealt{Beuther+05,Beltran+11,Silva+17}). 
Moreover, the presence of young HII regions may also have an influence, at least locally, on the properties of individual nearby cores, namely raising their effective temperature, potentially leading to misleading flux and mass estimates (see Sect. \ref{core_sample_sel} and Appendix \ref{Appff_analysis}). 

The uncertainties associated with our temperature assumptions, and their impact on estimated core masses, are discussed in Sect. \ref{Mcore_err} and Sect. \ref{CMF_evol}.

\subsubsection{Core mass distribution}

\begin{figure}[h!]
        \includegraphics[width=\columnwidth]{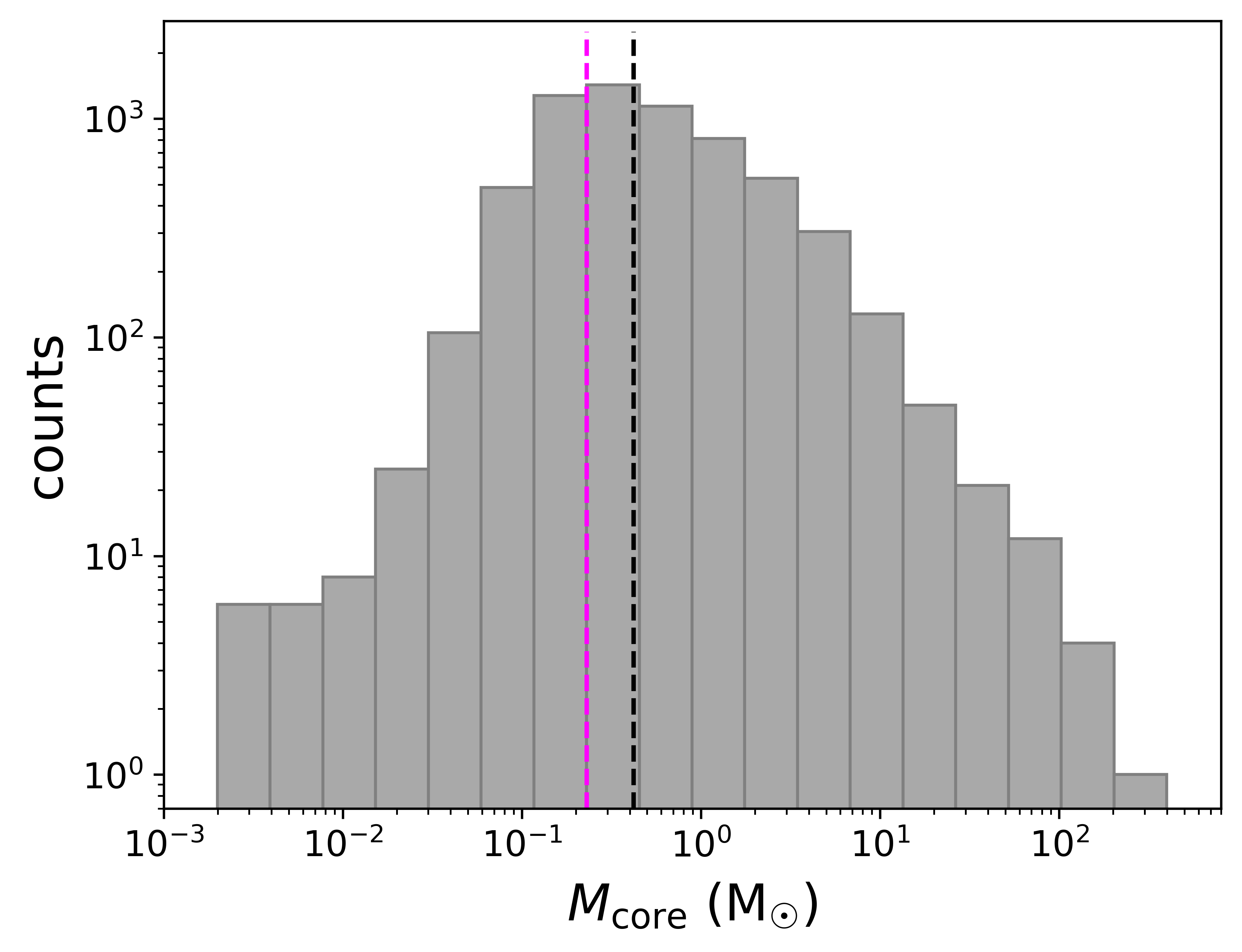}
        \caption{Overall distribution of the estimated core masses. The vertical dashed magenta line marks the average mass completeness limit of $0.23\,\mathrm{M_{\odot}}$ estimated for our sample (see Sect. \ref{Msens-compl}), while the vertical dashed black lines marks the median value of the distribution ($0.4\,\mathrm{M_{\odot}}$).}
    \label{Mcore_hist}
\end{figure}

The core masses, estimated with Eq. \ref{Mcore_eq} adopting the temperature assumptions described in Sect. \ref{Tcore_model}, are listed in Col. $6$ of Table 7. Their overall distribution (CMF) is shown in Fig. \ref{Mcore_hist}, and ranges $\sim0.002-345\,\mathrm{M_\sun}$, with a mean value of $1.5\,\mathrm{M_\sun}$ and a median of $0.4\,\mathrm{M_\sun}$. The CMF peaks at $\sim0.3\,\mathrm{M_\sun}$. Thanks to the high spatial resolution and high sensitivity of the ALMAGAL observational setup, we are thus able to reveal dense, compact fragments across nearly $5$ orders of magnitude in mass with large statistics. The sample is complete for core masses above $0.23\,\mathrm{M_{\odot}}$ (see Sect. \ref{FcomplPhotacc} and \ref{Msens-compl} for more details). 

The observed mass range is wider than that found by \citet{Traficante+23} at slightly lower spatial resolutions ($0.4-300\,\mathrm{M_\sun}$ on $>2000$ au scales). As discussed in Sect. \ref{NFRAG_stats}, clumps show a variety of degrees of fragmentation, and more crowded fields also report significantly wide internal distributions of core masses. We reveal massive cores ($>30\,\mathrm{M_\sun}$) only in $28$ cases (contained in $25$ clumps), corresponding to $0.05\%$ of the core catalog. The largest core masses we report are higher than those found by other recent ALMA and NOEMA studies with similar (or slightly lower) spatial resolutions, such as, \citet{Beuther+18} reporting up to $40\,\mathrm{M_\sun}$, and \citet{Pouteau+22} with $70\,\mathrm{M_\sun}$ at most. More generally, several interferometric studies have revealed core masses from $\sim0.1$ to some tens (up to $70$) of Solar masses (e.g., \citealt{Sanchez-Monge+13b,Beuther+18,Sanhueza+19,Svoboda+19,Pouteau+22,Morii+23,Xu+24}). Other authors measured masses of $100\,\mathrm{M_{\odot}}$ (e.g., \citealt{Sadaghiani+20} in NGC6334 molecular cloud) or above, in particular up to $\sim150\,\mathrm{M_{\odot}}$ (\citealt{Gieser+23}, in a sample of $11$ HMSFR from IRDC to HIIs), and $\sim300\,\mathrm{M_{\odot}}$ (\citealt{Traficante+23}). 

Detailed analysis of the uncertainties and observational biases influencing the measured core masses follows in Sect. \ref{Mcore_err}. The variation of core masses with evolution is analyzed in Sects. \ref{CMF_evol} and \ref{mcore_growth}. 

\subsubsection{Uncertainties and biases on core mass estimation}
\label{Mcore_err}

The mass estimate for the cores is potentially affected by different sources of uncertainties and observational biases (e.g., \citealt{Battersby+10}). 
The choice of the dust temperature which is assigned to the cores clearly plays a crucial role in this respect, representing in our case the largest source of uncertainty. 
To quantify its impact, similarly to what was done by \citet{Traficante+23}, we set three ranges of temperatures around the reference values used for the three groups of sources described above (see Sect. \ref{Tcore_model}). 
The three ranges are: I) $10-40$ K; II) $20-60$ K; III) $T_{\rm{core}}\pm30$ K (only allowing resulting values above $20$ K for this group, for consistency). The overall range of temperatures explored was therefore $10-111$ K. The three ranges were chosen in a rather conservative manner, being wide enough to allow for a proper exploration of the related mass variation while preserving reasonable fundamental assumptions on physical properties of objects given their evolutionary stage, that is, not allowing prestellar sources to go beyond $T=40$ K or evolved ones to drop below $T=20$ K (\citealt{Traficante+23}). 
Then, for each source, we computed the masses corresponding to the extreme temperature values of the range assigned to its group using the same Eq. \ref{Mcore_eq} ($M_{\mathrm{lim}}$), and evaluated the ratios between the reference core mass ($M_\mathrm{core}$, computed using the reference $T_\mathrm{core}$) and the two extreme mass values. We obtained mass discrepancies from a factor of $\sim1.8$ to $\sim3$, the latter cases occurring when the lowest temperature in the explored range is used, leading to higher masses than the reference ones (i.e., $M_{\mathrm{core}}/M_{\mathrm{lim}}<1$). 
Columns $3$ and $4$ of Table \ref{Tcore_tab} report, respectively, the range of temperatures explored to quantify the mass estimate variation and the corresponding range of obtained $M_{\mathrm{core}}/M_{\mathrm{lim}}$ ratios, for each of the three source groups based on clump $L/M$. 

Uncertainty associated with integrated flux measurement is $\sim20\%$ (see Sect. \ref{FcomplPhotacc} and Appendix \ref{AppFcomplPhotacc}). It is worth noting that, at least in more evolved regions, continuum flux contamination from nearby HII regions (mostly from free-free radio emission) might occur. This issue is discussed in Sect. \ref{core_sample_sel} and Appendix \ref{Appff_analysis}. 

\begin{figure}[h!]
    \includegraphics[width=\columnwidth]{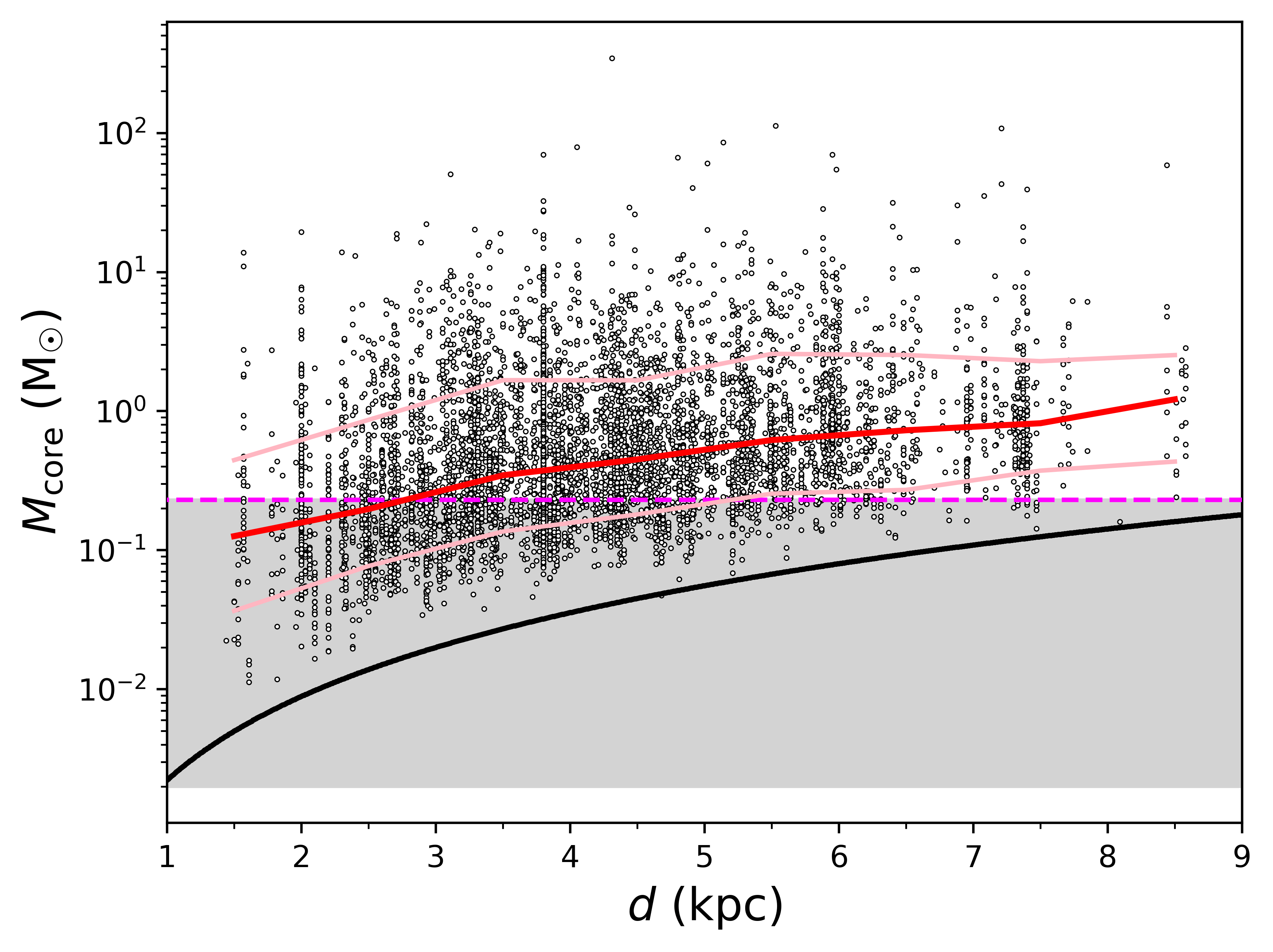}
    \caption{Estimated core masses as a function of the hosting clump heliocentric distance, highlighting the variation of mass sensitivity across distance. The red line draws the trend of the median value among core masses contained within $1$ kpc bins of distance, and the two pink lines mark the position of the $15$th (bottom) and $85$th (top) percentiles of the mass distribution within the same distance bins. 
    The horizontal dashed magenta line marks the mass completeness limit of $0.23\,\mathrm{M_{\odot}}$, while the solid black curve is drawn to approximately follow the trend of the minimum core mass revealed with target distance. The grey shaded area highlights the incomplete part of the sample, and, in particular, the region below the black curve represents the "forbidden" (i.e., not detectable) range of masses based on revealed objects at the different target distances. Note: for better visualization purposes, distance outliers ($<3\%$) are not shown in this plot.}
    \label{Mcore_bias}
\end{figure}

The error in the clump distance (also included in Eq. \ref{Mcore_eq}) is on average $10\%$ (Sect. \ref{Dcore}), which brings a $20\%$ potential uncertainty in the computed core mass. 
Nevertheless, distance has the effect to practically delimit the range of masses that we are able to reveal across the whole sample of targets. 
This can be easily noted in Fig. \ref{Mcore_bias}, where the estimated core masses are plotted against the distance of the respective clumps, so that the various cores of a given clump are aligned along the vertical direction. While the highest core masses are almost constant across the whole sample, the lower masses are inferiorly bounded, and this limit, due to the dependency relation between flux and distance, increases with the square of distance. 
This behavior translates into a variation of the mass sensitivity of about $1$ order of magnitude in total across the sample, implying that we are more sensitive to low masses (down to $\sim10^{-2}\,\mathrm{M_{\odot}}$) in closer targets with respect to farther ones, where only $M_{\mathrm{core}}\gtrsim0.1\,\mathrm{M_{\odot}}$ are accessible (see Sect. \ref{Msens-compl} for details). 
Such trends are easily noticeable from the lines representing the median value, and the $15$th and $85$th percentiles of the core mass distribution within given bins of distance (see Fig. \ref{Mcore_bias} for full description), which also give an idea of how objects are distributed over the covered mass range. In detail, the $15$th percentile value (i.e, in the lower-mass range) increases by about $1$ order of magnitude in total, whereas the $85$th percentile (i.e., higher masses) by a factor of $\sim3$. 
As a consequence, the core mass distribution within the far sample is more oriented towards higher masses, while the near sample will cover a broader mass range. 
Moreover, given the mass-distance trend of Fig. \ref{Mcore_bias} and the linear dependence between the beam size and the distance, lower spatial resolutions lead on average to higher core masses, with a discrepancy of up to a factor $\sim100$ between minimum masses obtained in most ($\lesssim1000$ au) and least ($>3500$ au) resolved fields. 

Another source of uncertainty in the mass estimate is the choice of dust opacity, which depends on the grain and gas density model adopted (see, e.g., \citealt{Mathis+77,Ossenkopf&Henning94,Draine11}). 
For our mass computation, we adopted the value $\kappa_{1.3}=0.9$\,$\mathrm{cm^2\,g^{-1}}$ for the dust absorption coefficient at $1.3$\,mm wavelength of a distribution of coagulated MRN grains (Mathis, Rumpl \& Nordsieck, \citealt{Mathis+77}, i.e., grains as homogeneous spheres of mixed composition) with thin ice mantles at gas densities $\gtrsim10^6\,\mathrm{cm^{-3}}$ 
(\citealt{Mathis+77,Ossenkopf&Henning94}). 
This value allows for a consistent comparison with the results of several other interferometric studies of fragmentation at core scales, such as \citet{Palau+13}, \citet{Beuther+18}, \citet{Sanhueza+17}, \citet{Sanhueza+19}, \citet{Svoboda+19}, and \citet{Morii+23}. 
However, gas particle densities in our core sample are mostly found in the range $\sim10^7-10^8\,\mathrm{cm^{-3}}$ (see Sect. \ref{MR_surfd_n_core}). Adopting a value of $\kappa_{1.3}\simeq1.0-1.1$\,$\mathrm{cm^2\,g^{-1}}$, as obtained from the estimates reported in \citet{Ossenkopf&Henning94} for those densities, would cause a systematic decrease of $\sim10-20\%$ of the estimated core masses. 
The uncertainty brought by opacity would be lower or at most similar to those due to temperature, flux, and distance. The correction factor would nevertheless change up to a factor of $\sim2$ (used by, e.g., \citealt{Palau+13} and \citealt{Pouteau+22}) if pre-coagulation or different grain composition models (such as naked grains or ice mantles with larger thickness, see \citealt{Ossenkopf&Henning94}) would be considered. 

We assumed a widely used gas-to-dust mass ratio $R_{gd}$\,$=$\,$100$ (e.g., \citealt{Kainulainen+17,Sanhueza+17,Sanhueza+19,Motte+22,Morii+23,Traficante+23}), which at best constitutes only a rough estimate. 
While such value is quite well established for the Solar surroundings, its extension to the entire Galaxy is still under debate, as it has actually been found to vary as a function of the Galactocentric radius and local metallicity (see \citealt{Giannetti+17}). Other values that can be found in literature range between $\sim110-150$ (e.g., \citealt{Beuther+18,Svoboda+19,Gieser+23}), and would imply an up to $\sim50\%$ upward correction for calculated masses. 

Ultimately, it is clear that, under our assumptions, mass uncertainty due to temperature is potentially significantly higher than the contributions of the other parameters involved. 
The true impact of temperature, but also of distance and spatial resolution, on the observed core mass function are further investigated in Sect. \ref{CMF_evol_trend} and Appendix \ref{AppCMFevoltests}.

\begin{table}[h!]
    \caption{Main parameters of the core temperature model used for mass estimation and related analysis (see text for explanation).}
    \label{Tcore_tab}
    \centering
    \begin{tabular}{c c c c c}
    \hline\\[-8pt]
    $(L/M)_{\rm{clump}}$ & $T_{\rm{core}}$ & $\Delta T$ & $M_{\rm{core}}/M_{\rm{lim}}$ & $N_{\rm{cores}}$ \\
    ($\rm{L_{\odot}/M_{\odot}}$) & (K) & (K) & & \\[2pt]
    \hline\hline\\[-8pt]
    $\leq1$ & $20$ & $10-40$ & $0.37-2.30$ & $1261$ \\
    $1<\,L/M\leq10$ & $35$ & $20-60$ & $0.50-1.83$ & $2024$ \\
    $>10$ & $35-81$ & $T_{\rm{core}}\pm30$ & $0.33-1.99$ & $3063$ \\
    \hline\\[2pt]
    \end{tabular}
\end{table}

\subsubsection{Mass sensitivity and completeness}
\label{Msens-compl}

The trend in Fig. \ref{Mcore_bias} shows that our mass sensitivity (i.e., the minimum detectable fragment mass) varies by more than $1$ order of magnitude due to heliocentric distance variations across our sample. While in the nearest targets ($d\simeq2$ kpc) we are sensitive to masses down to $\sim10^{-2}\,\rm{M_{\odot}}$, in the farthest ones ($d\geq6$ kpc) we are actually able to reveal only $M_{\mathrm{core}}\gtrsim0.1\,\mathrm{M_{\odot}}$. 
The dynamic range of recoverable masses is then higher for closer sources (about $4$ orders of magnitude in mass) than for farther ones ($\sim3$ orders of magnitude in mass). An estimate of the mass corresponding to the $1\sigma_\mathrm{rms}$ flux level can be derived by converting the average flux sensitivity of ALMAGAL maps of $0.05$ mJy/beam (see Sect. \ref{contmapprop}) with Eq. \ref{Mcore_eq} (using a median temperature of $35$ K, assuming this flux as the emission of a point-like source). This leads to a mass range of $\sim2\times10^{-3}-4\times10^{-2}\,\mathrm{M_{\odot}}$ across the covered range of distances ($\sim2-8$ kpc). 
Dedicated estimates of the $1\sigma_\mathrm{rms}$ mass for each field ($M_{1\sigma}$) can be achieved by employing the rms noise of the map $\sigma_\mathrm{rms}$ together with the proper distance and temperature of the clump. 
In these terms, we obtain a range of ratios between the minimum core mass ($M_{\mathrm{core}}^\mathrm{min}$) detected within each clump and its $M_{1\sigma}$ of $\sim5-50$, whereas the ratios between the maximum core mass for each clump and $M_{1\sigma}$ are within $\sim8-1000$. 
It must be noted that the ratios $M_{\mathrm{core}}^{\rm{min}}/M_{1\sigma}$ do not exactly correspond to the S/N of the core detection measured in the photometry stage (see Sect. \ref{cutex}), due to the local primary beam correction and background subtraction operations computed by \textit{CuTEx}. 

We derived an average mass completeness limit for the entire ALMAGAL sample from the flux limits estimated from experiments with synthetic sources (see Sect. \ref{FcomplPhotacc}). 
Using the median temperature of $35$ K and an average distance of $4.5$ kpc, the corresponding mass completeness limits for the three groups of targets (classified based on their rms noise, see Sect. \ref{contmapprop}) are $0.14$ (Group 1), $0.23$ (Group 2), and $0.41$ (Group 3) $\mathrm{M_{\odot}}$, sorted by increasing map rms. Hereafter, being Group 2 by far the most statistically representative subsample, the value of $0.23$ $\mathrm{M_{\odot}}$ obtained for it will be used as the reference mass completeness limit, derived from the corresponding flux completeness value of $0.94$ mJy. 
However, whereas integrated fluxes are essentially constant across the whole sample (as shown in Fig. \ref{Fint_plots}, right panel) and thus flux completeness limits can be considered largely independent of clump distance, the trend between estimated masses and distance (Fig. \ref{Mcore_bias}) suggests to make specific mass completeness limit computations for near and far sample targets, separately, to account for the different sensitivity available at different distances within our sample. 
Using the most statistically relevant flux completeness value of $0.94$ mJy (Group 2) and average distances of $3.5$ and $6$ kpc for near and far samples respectively, we obtain corresponding values of $0.13$ and $0.37$ $\mathrm{M_{\odot}}$ in terms of mass completeness limit. These values are within a factor $2$ from the average value of $0.23$ $\mathrm{M_{\odot}}$ adopted for the whole sample. Finally, we note that the effect of sensitivity in defining the range of detectable masses is substantially removed if we consider only masses above the average completeness limit.

\subsection{Mass-size relation, mass surface density and molecular volume density}
\label{MR_surfd_n_core}

\begin{figure*}[t!]
    \includegraphics[width=\columnwidth,valign=t]{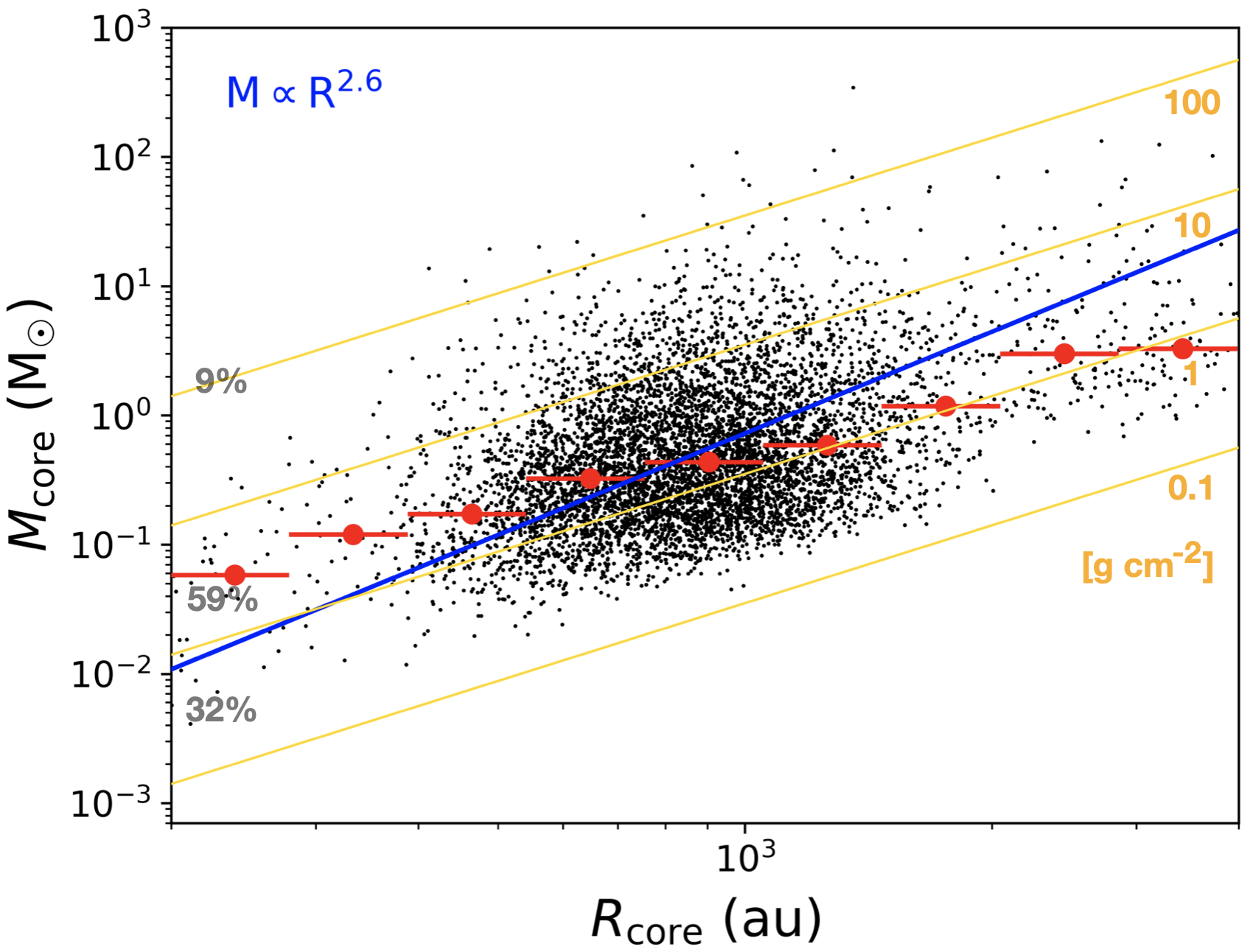}
    \includegraphics[width=1.02\columnwidth,valign=t]{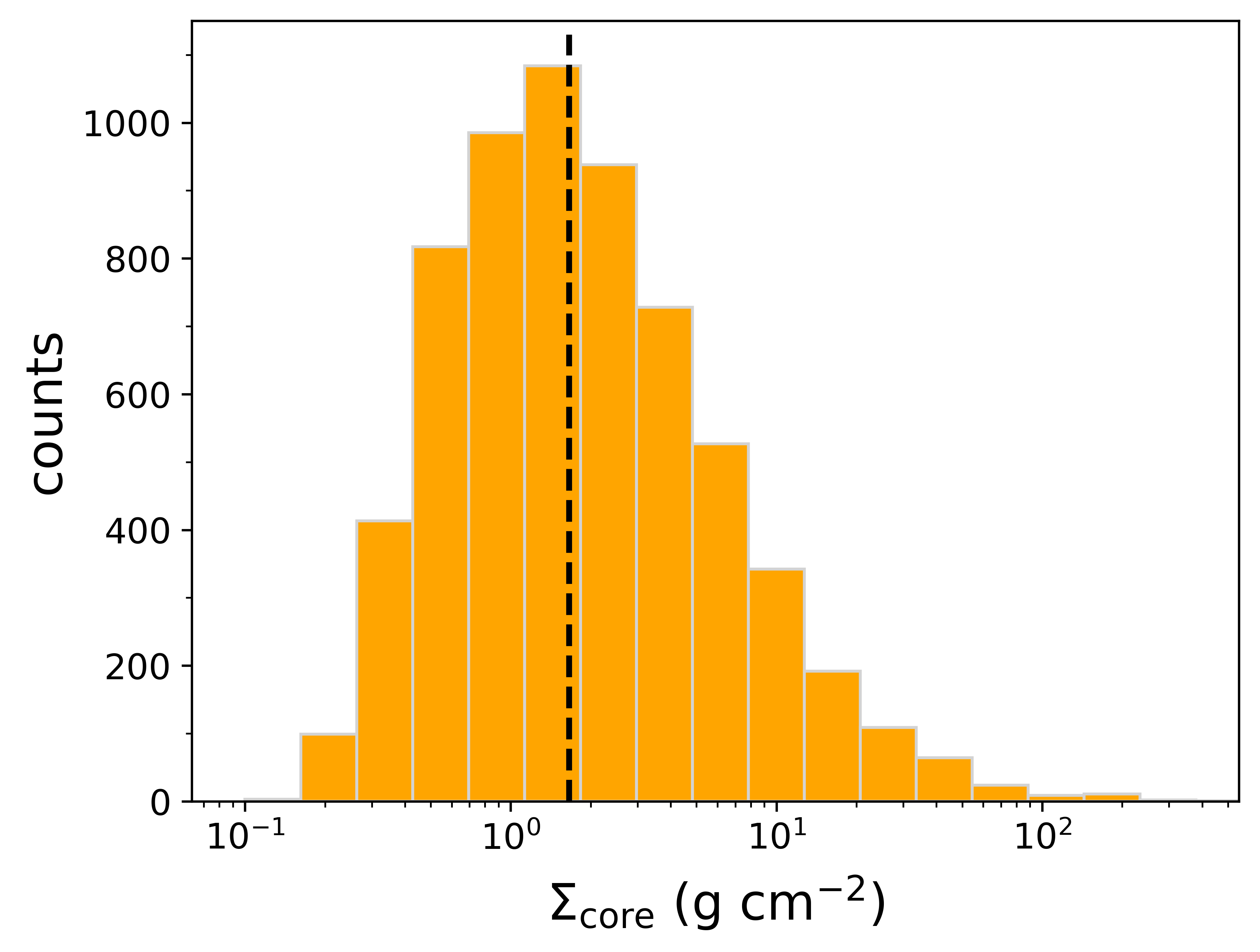}
    
    \caption{Mass-radius relation and mass surface density distribution for the core sample. 
    \textit{Left panel}: Core masses as a function of estimated radius. Solid gold lines represent different values of core surface density: from bottom to top, $0.1$, $1$, $10$, and $100$ g $\mathrm{cm^{-2}}$, respectively. Percentages of data points within the corresponding bins of surface density are given in grey. Red points indicate the median $M_\mathrm{core}$ values in given bins (red bars) of $R_\mathrm{core}$. The solid blue line is the best fit to the data points, corresponding to the power-law relation reported in the top-left corner. 
    \textit{Right panel}: Distribution of the mass surface density of the cores. The vertical dashed black line marks the median value of $1.7$ g $\mathrm{cm^{-2}}$.}
    \label{Mcore_Rcore-surfd_plots}
\end{figure*}

Figure \ref{Mcore_Rcore-surfd_plots} (left panel) shows the estimated core masses as a function of the core radius (see Sect. \ref{Dcore}). Red points mark the median $M_\mathrm{core}$ values within given bins of $R_\mathrm{core}$; they display a rather good correlation between the two quantities. Moreover, the best fit to the data points (solid blue line) corresponds to a power-law relation $M\propto R^{\alpha}$ with $\alpha\simeq2.6$ (namely $\alpha=2.612\pm0.001$), which is higher compared with the ones ($\sim1.6-2.5$) proposed in multiple literature studies for molecular clouds, clumps, and cores (e.g., \citealt{Larson81,Kauffmann+10b,Tan+14,Lombardi+10,Miville+17,Traficante+18b,Urquhart+18,Ballesteros+19}). Starting from these two quantities, we computed the average mass surface density of the cores assuming a spherical shape as
\begin{equation}
\label{surfd_eq}
\Sigma_{\rm{core}} = \frac{M_{\rm{core}}}{\pi\,R^{2}_{\rm{core}}}\,.
\end{equation}
The resulting distribution of core surface densities is shown in the right panel of Fig. \ref{Mcore_Rcore-surfd_plots}. Surface density $\Sigma_{\rm{core}}$ ranges from $\sim0.15$ to $\sim500$ g $\rm{cm^{-2}}$, with a mean value of $4.5$ and a median of $1.7$ g $\rm{cm^{-2}}$. This range is wider (especially in the upper end) than what was found in other similar studies, such as \citet{Palau+13} ($\sim0.7-10$ g $\rm{cm^{-2}}$), \citet{Sanhueza+19} ($\sim0.8-5$ g $\rm{cm^{-2}}$), \citet{Morii+23} ($\sim0.05-10$ g $\rm{cm^{-2}}$), and \citet{Traficante+23} ($\sim0.1-30$ g $\rm{cm^{-2}}$). 
It must be noted, however, that comparisons with those studies are not straightforward since, differently from the present work, they featured lower resolution observations of much more limited samples, and surface densities were in some cases computed using beam-deconvolved core sizes. 
Moreover, computed surface densities are affected by the uncertainties on temperatures used to derive the core masses. 
In more detail, $\sim32\%$ of our cores have surface densities $<1$ g $\rm{cm^{-2}}$, $\sim59\%$ are within $1$ and $10$ g $\rm{cm^{-2}}$, while only $\sim9\%$ of them exceed $10$ g $\rm{cm^{-2}}$. Therefore, $\sim68\%$ of the cores exceed $1$ g $\rm{cm^{-2}}$, which was suggested by \citet{Krumholz+08} as the minimum threshold above which coherent collapse is allowed to form high-mass stars (although the influence of magnetic fields may overcome this, \citealt{Tan+13b}). This percentage is considerably higher than the one found by \citet{Sanhueza+19} and \citet{Morii+23} ($\sim10\%$). This is probably due to the fact that given our observational setup and target sample we are able to probe more dense and compact substructures on smaller scales down to $\sim800$ au, in a broader variety of clumps in terms of evolutionary stage, including also more evolved and condensed protoclusters. Different levels of surface density are also displayed as a reference as solid gold lines in the left panel of Fig. \ref{Mcore_Rcore-surfd_plots}, from bottom to top: $0.1$, $1$, $10$, and $100$ g $\rm{cm^{-2}}$, respectively. 

The gas volume density of the cores was also estimated assuming a spherical shape as (e.g., \citealt{Zhang+18}):
\begin{equation}
\label{vold_eq}
n({\rm{H_2}}) = \frac{M_{\rm{core}}}{\bar{m}\,(4\pi R^{\,3}_{\rm{core}}/3)}
,\end{equation}
where $\bar{m}$ is the mean
molecular mass, for which we assume $2.8\,\rm{m_{H}}$ (\citealt{Kauffmann+08}), with $\rm{m_{H}}$ the hydrogen mass. We found a range for $n(\rm{H_2})$ of $\sim10^6-10^9$ $\rm{cm^{-3}}$ (with a median of $3\times10^7$ $\rm{cm^{-3}}$). Most of the cores ($\sim64\%$) lie between $\sim10^7$ and $10^8$ $\rm{cm^{-3}}$. 
Our estimates are roughly consistent with what found by \citet{Beuther+18}, while they are on average higher than in other recent works, which typically range from $\sim10^5$ to at most $\sim10^8$ $\rm{cm^{-3}}$ (e.g., \citealt{Sanhueza+19,Zhang+21,Pouteau+22,Morii+23,Xu+24}). 
This is likely due to the higher resolution achieved, and confirms that we are indeed revealing dense, compact substructures that represent the intermediate entities between clumps and (proto)stars in the star formation framework. The estimated mass surface densities and molecular volume densities of the cores are reported in Cols. 7-8 of Table 7, respectively.

\subsection{Estimated physical parameters of the cores}
\label{core_phys_props_sect}

Table 7 summarizes the main physical properties of the extracted population of cores. The table can be accessed at the CDS in machine-readable form. The content of the table is the following. 
Column 1) hosting clump name; 
Col. 2) ordinal core identifier number within the clump; 
Col. 3) core physical diameter as derived from the circularized FWHM of the fitted source ellipse ($D_{\rm{core}}$); 
Col. 4) clump $L/M$ evolutionary indicator, used to assign core temperatures (see text for explanation); 
Col. 5) core temperature used for mass estimation ($T_{\rm{core}}$); 
Col. 6) core mass ($M_{\rm{core}}$); 
Col. 7) core average mass surface density ($\Sigma_{\rm{core}}$);  
Col. 8) core average volume density ($n(\rm{H_2})$). 

The non-deconvolved core physical diameter has been derived from the circularized FWHM of the fitted source ellipse (see Sect. \ref{Dcore}). Core temperatures have been assigned to the cores according to the reported evolutionary stage of the hosting clump as described in Sect. \ref{Tcore_model}. The core masses and related surface and volume densities (Sects. \ref{Mcore} and \ref{MR_surfd_n_core}) are also reported.

\section{CMF variation with evolution}
\label{CMF_evol}

    In this section, the distribution of estimated core masses (Sect. \ref{Mcore}) is evaluated as a function of the clump evolutionary indicator $L/M$. 
    The CMF is shown and discussed in terms of inverse cumulative number density of objects. Such representation has the advantage, with respect to the differential form, of being independent from the adopted binning, while preserving the possibility to reproduce and compare the slope of the high-mass tail (see, e.g., \citealt{Testi+98,Olmi+13,Fiorellino+21,Pezzuto+23}). 
    To visualize and quantify the impact on the distribution of computed core masses of mass uncertainties due to the adopted model of core temperatures (Sect. \ref{Tcore_model}), which represent the largest source of uncertainty (see Sect. \ref{Mcore_err}), we performed a series of $1000$ Monte Carlo (MC) simulations of masses of the whole population of extracted cores (see, e.g., \citealt{Sadaghiani+20,Traficante+23}). We assigned to each core a random temperature within the range $\Delta T$ reported in Table \ref{Tcore_tab} and described in Sect. \ref{Mcore_err}. 
    The mass distributions shown in Sect. \ref{CMF_evol_trend} will then have the corresponding uncertainties associated with temperatures reported as colored bands drawn by the $1000$ simulated CMFs, lying in the background of the observed profile. \\

    \subsection{Core sample selection}
    \label{core_sample_sel}
    
    For the present analysis, we filtered the core catalog in order to enhance the robustness of the results. 
    First, we include here only cores with masses above the estimated average mass completeness limit for our catalog ($0.23\,\mathrm{M_{\odot}}$, Sect. \ref{Msens-compl}). This allows us to avoid any sensitivity bias due to distance variations on the analysis of the shape of the CMF (see Fig. \ref{Mcore_bias}). We also underline that in our clump sample distance and $L/M$ are not correlated (see Elia et al. in prep.). 
    Moreover, we exclude cores that were positionally matched with strong radio continuum emission based on the analysis of the available ALMAGAL counterparts from the CORNISH (\citealt{Purcell+08,Hoare+12}) and CORNISH-South (\citealt{Irabor+23}) surveys at $5$ GHz (or $6$ cm). We identified $909$ matched cores ($\sim14\%$ of the catalog, flagged in Col. 16 of Table 3), mostly associated, as expected, with the more evolved clumps (i.e., $L/M>10\,\mathrm{L_{\odot}/M_{\odot}}$ group). Such clumps are in fact compatible with the potential presence of HII regions, in which free-free continuum emission can bring a significant contribution to the measured core fluxes (e.g., \citealt{Liu+22b,Motte+22,Pouteau+22,Louvet+24}). For further safety, we also exclude cores belonging to clumps for which no CORNISH of CORNISH-South map was available ($582$ cores, $\sim9\%$ of the catalog). This procedure was done to avoid possible mass overestimates for the cores resulting from flux contamination by free-free emission. The highest core masses are in fact the ones that influence the most the shape of the high-mass end of the CMF. Full details of this analysis can be found in Appendix \ref{Appff_analysis}. 
    Lastly, we exclude cores from clumps farther than $9$ kpc, i.e., distance outliers ($166$ cores, $<3\%$). 
    After this selection process, we still end up with a large sample of $3324$ cores ($\sim53\%$ of the catalog), almost equally distributed among the three evolutionary groups (from least to most evolved group, $991$, $1181$, and $1152$ cores, respectively). \\

    \subsection{Evolutionary trend of the CMF}
    \label{CMF_evol_trend}

\begin{figure}[t!]
    \includegraphics[width=1.0\columnwidth]{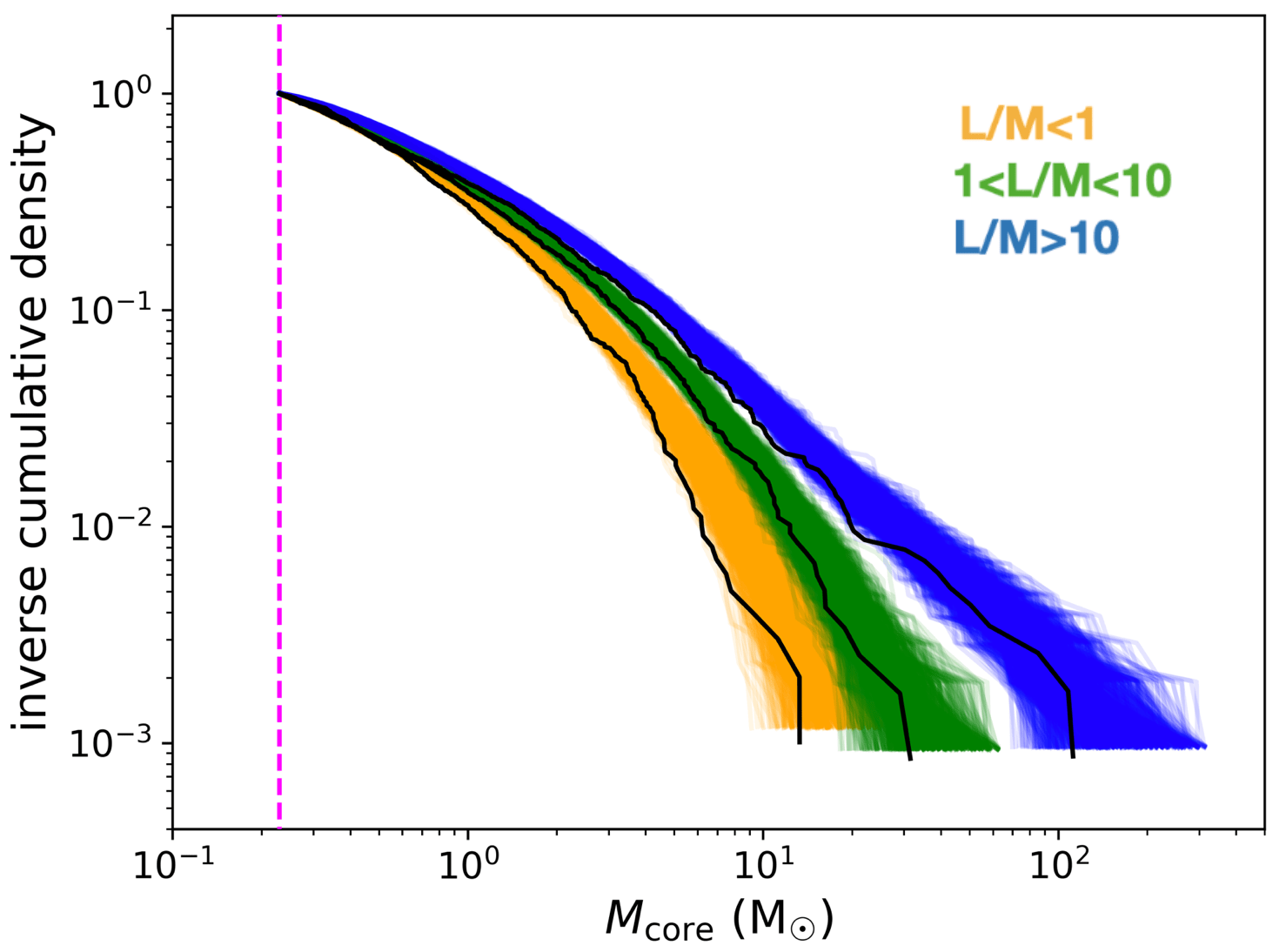}
    \caption{Observed CMFs (black lines, expressed as inverse cumulative number density distributions) for the three populations of ALMAGAL cores classified, as reported, according to the $L/M$ evolutionary indicator of the hosting clump, as explained in Sect. \ref{Tcore_model}. The three colored bands trace, for each population, the $1000$ MC realizations of CMFs computed by assigning to each core a random temperature within the ranges defined in Table \ref{Tcore_tab}. Less evolved sources (i.e., with $L/M\leq1\,\,\mathrm{L_{\odot}/M_{\odot}}$) are marked by golden lines, the intermediate ones ($1<L/M\leq10\,\,\mathrm{L_{\odot}/M_{\odot}}$) by green lines, and the more evolved sources ($L/M>10\,\mathrm{L_{\odot}/M_{\odot}}$) by blue lines. Note: we include only cores above the mass completeness value of $0.23\,\mathrm{M_{\odot}}$, marked by the vertical dashed magenta line (see text for explanation).}
    \label{CMF_evolution}
    \end{figure}

    Figure \ref{CMF_evolution} shows the distribution of estimated core masses within the three different subsamples of clumps selected based on their $L/M$ ratio, according to the evolutionary classification presented in Sect. \ref{Tcore_model}. 
    We remark that the evolutionary stage of the clump will not necessarily reflect the one of the individual embedded cores. Assigning the same temperature to all the cores within the same clump according to its properties, although representing a reasonably good approximation, inevitably ignores the distribution of temperatures that embedded cores actually have due to their different individual evolutionary stage and local physical conditions (e.g., \citealt{Hennebelle+20,Izumi+24}). 

The CMF profiles for the three different evolutionary groups appear as clearly shifted relative to each other, showing differences in shape and slope that go beyond the mass uncertainties due to temperature assumptions. Through the MC realizations, we overall cover a temperature range $10-111$ K (see Table \ref{Tcore_tab}). We note that if we assumed the same temperature for all the cores, as done by other works, the shift between the three CMF profiles would be even larger (Eq. \ref{Mcore_eq}). 
In more detail, the high-mass tail of the CMF appears to become flatter as evolution proceeds. In other words, more evolved clumps host more massive cores. We report core masses up to $\sim10\,\mathrm{M_{\odot}}$ in the $L/M\leq1$ group, up to $\sim30\,\mathrm{M_{\odot}}$ for $1<\,L/M\leq10$, while up to $\sim100\,\mathrm{M_{\odot}}$ for $L/M>10\,\mathrm{L_{\odot}/M_{\odot}}$. 
    
Our result is, in principle, consistent with the findings of some recent ALMA surveys, which however feature much more limited statistics and different samples. For example, \citet{Pouteau+23} (ALMA-IMF) analyzed the variation in the observed CMF across different subregions within W43-MM2\&MM3, which they broadly classified as corresponding to different evolutionary phases (from quiescent to post-burst). They found shallower high-mass end slopes with evolution, namely, more massive cores in more evolved regions. Nevertheless, it must be noted such study includes $205$ cores extracted within a single large region, resolved at a $\sim2500$ au resolution. Moreover, importantly, the inspected subregions are all evolved, so that only the more advanced part of the evolutionary sequence is traced. Analogous arguments can be made for similar results obtained in the W43-Main protocluster by \citet{Armante+24}. 
Furthermore, \citet{Liu+23} selected $19$ hub-filament systems from the samples of IRDCs of ASHES and ATOMS surveys and divided them into IR-dark and IR-bright sources based on the presence of IR emission in central hubs, to (non-uniformly) cover an evolutionary sequence spanning $\sim3$ orders of magnitude in $L/M$ ($\sim0.1-100\,\rm{L_{\odot}/M_{\odot}}$). They found larger masses of the most massive cores (MMCs) in more evolved regions at $\sim3000$ au average resolution. \citet{Morii+24} found higher mass dynamic range (i.e., higher MMC masses) in clumps with higher protostellar core fraction in ASHES. \citet{Xu+24} measured average core masses about $2-3$ times higher in ASSEMBLE evolved protoclusters with respect to ASHES earlier stage IRDCs (\citealt{Sanhueza+19,Morii+23}). 
Conversely, \citet{Anderson+21} ($\sim2500$ au resolution) found no clear indication of different behaviors at different evolutionary stages for the masses of the cores extracted from the $35$ ATLASGAL IRDCs selected from \citet{Csengeri+17}, which they subdivided into five evolutionary groups based on their mid-IR brightness. 

Before focusing on the physical interpretation of this result (Sect. \ref{CMF_evol_trend_interp}), we must take into account the fact that by constructing the overall CMF we are aggregating core populations extracted from a large and varied clump sample representing different physical conditions and environments. 
Such factors have an influence on core evolution, and thus on their mass distribution (e.g., \citealt{Lee+18,Hennebelle+24}).
Because of this, we do not focus on detailed computations of the CMF slopes (as done, e.g., in \citealt{Motte+18b,Sanhueza+19,Sadaghiani+20,Pouteau+22}), which require an in-depth study that will be the subject of a dedicated subsequent ALMAGAL work. Besides, we find our CMFs to be reasonably consistent with a power law (see, e.g., \citealt{Salpeter55,Kroupa01}) only starting from relatively high masses of about $3\,\mathrm{M_{\odot}}$, while more resembling a log-normal in the lower-mass range. 

Moreover, in addition to providing an excellent statistical relevance, the variety of our sample brings potential biases due to distance, resolution, and temperature effects. 
We already removed the influence of distance on mass sensitivity variation by cutting above the mass completeness limit. We also verified that the evolutionary trend of Fig. \ref{CMF_evolution} stands when considering near and far sample sources separately, as expected from the fact that distance and $L/M$ are not correlated in our targets (Sect. \ref{core_sample_sel}). We performed further analysis and tests to verify the reliability of the observed trend and reinforce our results, whose details are reported in Appendix \ref{AppCMFevoltests}. 
First, we evaluated the trend of Fig. \ref{CMF_evolution} for the subsamples of fields with spatial resolutions within a factor $2$ (namely between $1000$ and $2000$ au, including most of the ALMAGAL targets). 
Second, we further analyzed the effect of temperature on the CMF (besides what already done through MC realizations) by adopting an alternative model based on \citet{Sadaghiani+20}, in which the core temperature is proportional to its measured flux through a power law. 
Lastly, we considered a reduced version of the core mass formula Eq. \ref{Mcore_eq} (essentially removing the temperature term), evaluating the evolutionary trend for the $F_\mathrm{INT}d^2$ quantity. 
In all these cases (see Appendix \ref{AppCMFevoltests}), the trend of Fig. \ref{CMF_evolution} still holds, both in terms of shift and different slope among the three CMFs, confirming that none of these potential effects or biases are relevant in establishing the observed behavior. 

Therefore, given the preliminary selection of the core catalog to avoid flux contamination (Sect. \ref{core_sample_sel}), and based on the above described tests we performed to exclude potential underlying biases, we consider the result of Fig. \ref{CMF_evolution} to be definitely reliable and statistically robust.

\subsection{Physical interpretation of the observed trend}
\label{CMF_evol_trend_interp}

The presented result potentially has several relevant physical implications. 
Importantly, the observed trend of the CMF, which thanks to the ALMAGAL data we are able to inspect with unprecedentedly large statistics, suggests that the shape of the CMF is not constant throughout the star formation process (e.g., \citealt{Clark2007,Offner+14,Pouteau+22,Pouteau+23}). 
The CMF rather builds with evolution, being limited to lower masses in the early stages, and then gradually accumulating mass (and flattening) with time, forming the most massive structures ($>30\,\mathrm{M_{\odot}}$, Fig. \ref{CMF_evolution}) only at advanced stages. If we would assume a core-to-star formation efficiency of $\sim30\%$ (see, e.g., \citealt{Matzner+00,Alves+07,Elia+13,Elia+21}), this would imply that only clumps with $L/M>1\,\mathrm{L_{\odot}/M_{\odot}}$ could host cores already massive enough ($\sim25\,\mathrm{M_{\odot}}$) to be capable of forming high-mass stars ($M_{*}\geq8\,\mathrm{M_{\odot}}$). 
According to this interpretation, in practice, we could be observing cores of similar initial (low) mass at a different time across their lifetime, so that most evolved ones would have simply had more time to grow in mass. 
It has to be noted nonetheless that the one-core to one-star correspondence is not straightforward, since massive cores have been observed to further fragment and form multiple stars (\citealt{Cao+21,Li+24,Xu+24}). 

Interestingly, a compatible interpretation of the observed trend is a scenario in which more massive cores accrete mass at a higher rate than lower-mass ones, evolving over much shorter timescales when compared to the hosting clump, so that they soon dominate the global clump evolution (e.g., \citealt{Schmeja2004, Contreras+18,Redaelli+22}). Therefore, once an early stage clump (i.e., with low $L/M$) forms such massive fragment(s), it may accelerate its evolution, gradually (but rapidly) increasing its observed $L/M$ (e.g., \citealt{Sanhueza+19}). As a consequence, in this scenario, high core masses (and late clump evolutionary stages) are reached sooner in time, so that, observationally, it becomes relatively unlikely to reveal a clump with a low $L/M$ hosting massive fragments. 
Ultimately, the variation of the CMF with evolution implies that the stellar IMF is not a simple scaling of an observed CMF (see, e.g., \citealt{Motte+98,Andre+14,Cao+21,Hennebelle+24}). 

In more general terms, such result is in favor of a clump-fed scenario for massive star formation, in which cores gain their mass in a competitive way, with the more massive ones gradually accreting mass more efficiently from the surrounding intraclump medium thanks to their high gravitational attraction. 

Further results and discussions about these topics are presented in Sect. \ref{correlations}.

    \section{Core mass growth and fragmentation}
    \label{correlations}

    \begin{figure*}
    \includegraphics[width=0.52\textwidth,valign=t]{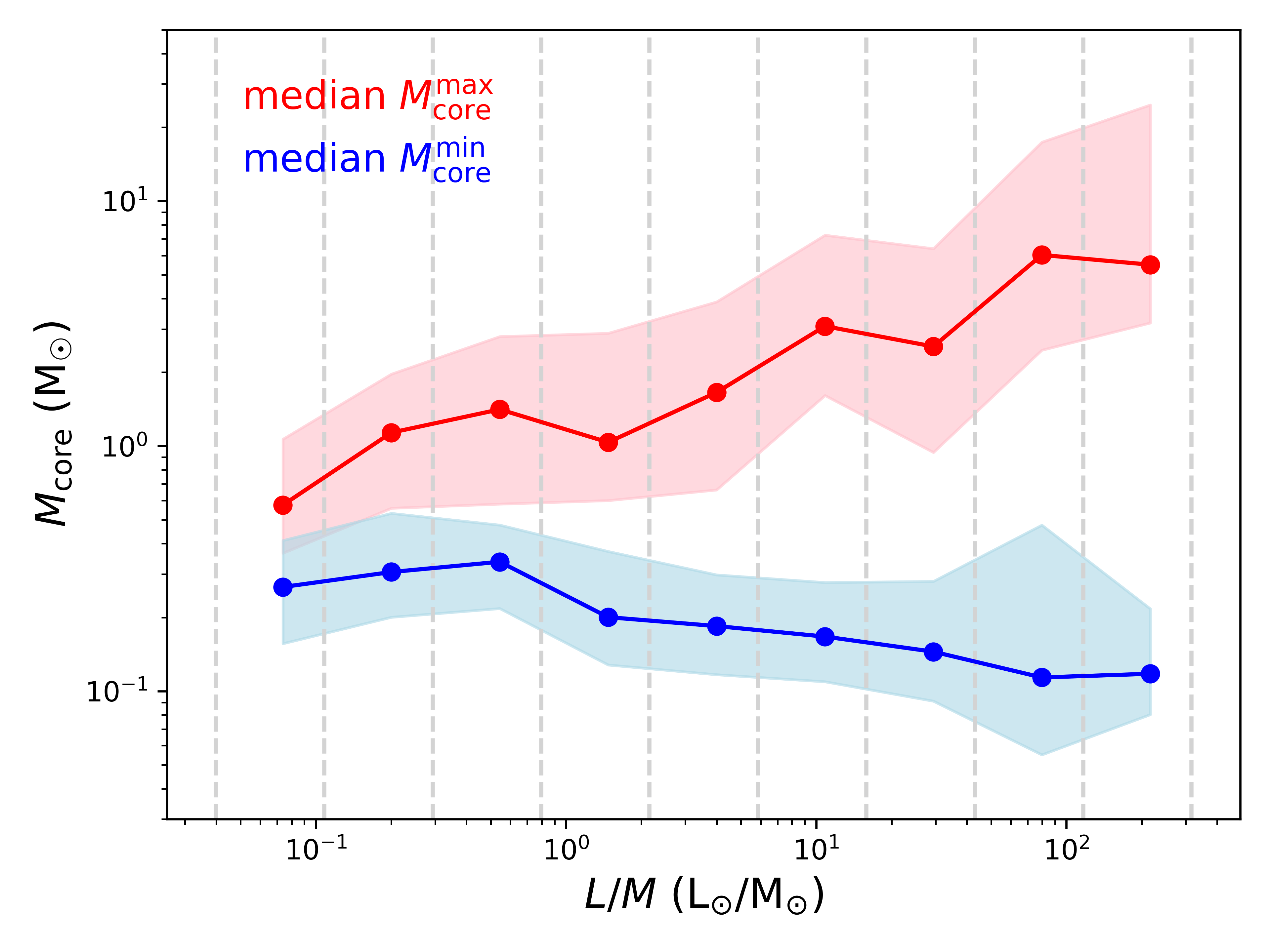}
    \includegraphics[width=0.51\textwidth,valign=t]{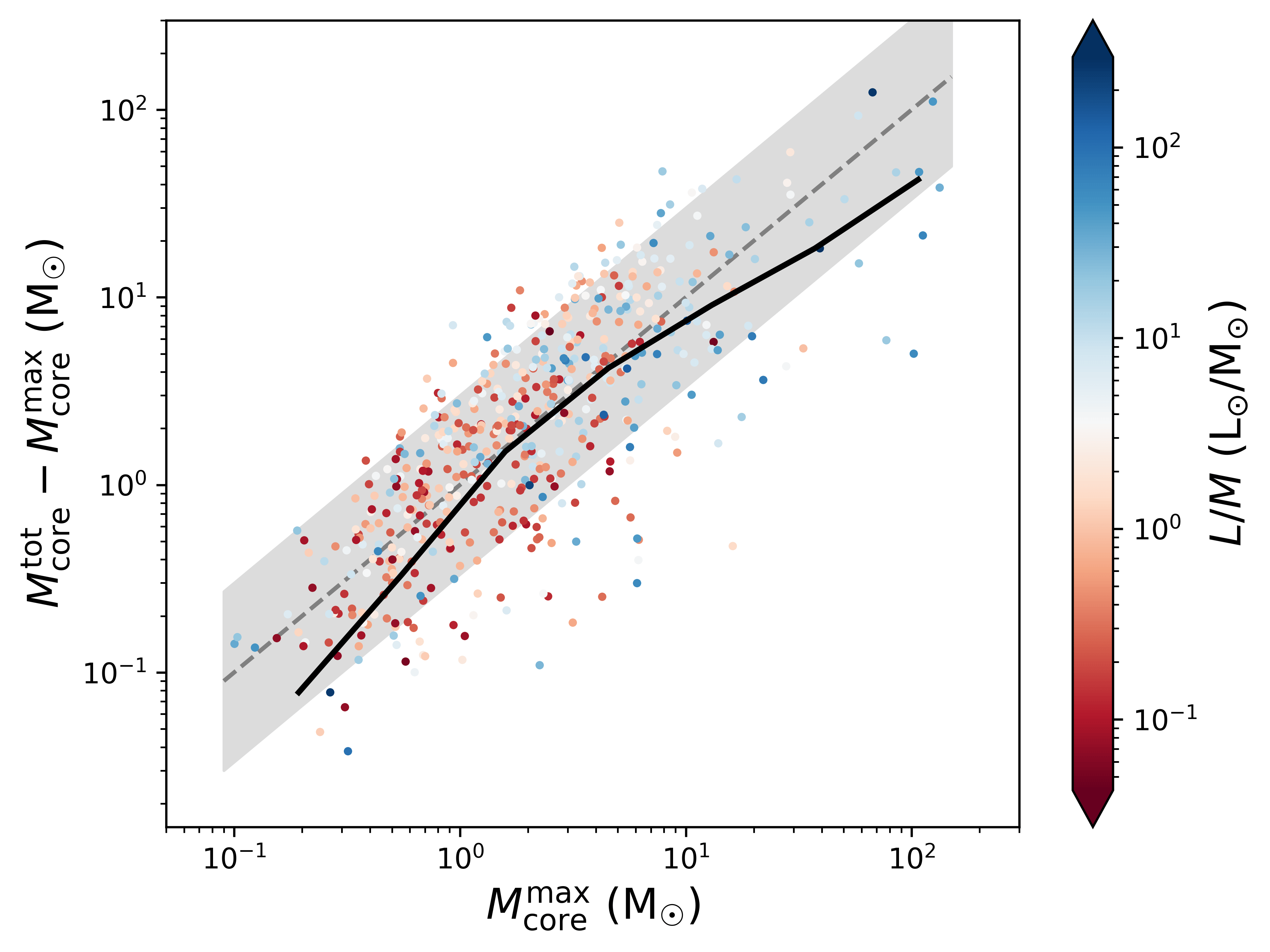}
    \caption{Evolution of core masses. \textit{Left panel}: Median values of the maximum ($M_{\rm{core}}^{\rm{max}}$, red) and the minimum ($M_{\rm{core}}^{\rm{min}}$, blue) core masses revealed within different bins of clump $L/M$ (delimited by vertical dashed grey lines). The respective colored shaded areas mark the 25th and 75th percentiles of the two distributions within the same $L/M$ bins. 
    \textit{Right panel}: Combined mass of the cores ($M_{\rm{core}}^{\rm{tot}}$) minus the mass of the MMC within each clump, as a function of the latter. Data points are color-coded by the clump $L/M$. The solid black line draws the trend of the median value of the quantity on the $y$-axis in given bins of $M_{\rm{core}}^{\rm{max}}$. The dashed grey line corresponds to the $M_{\rm{core}}^{\rm{max}}=50\%\,M_{\rm{core}}^{\rm{tot}}$ relation, while the shaded grey area marks (from top to bottom) the $25-75\%$ range.}
    \label{core_mass_growth}
    \end{figure*}

The result discussed in Sect. \ref{CMF_evol} suggests that compact cores grow in mass with time, so that massive cores ($>30\,\mathrm{M_{\odot}}$) are found only at more evolved stages. To investigate in more depth this scenario it is crucial to evaluate, in particular, whether it involves the entire population of cores embedded in a clump or just the more massive ones, and how this relates to (or affects) the fragmentation process in the region throughout the evolutionary sequence. 

To maximize the outcome of such analysis, we employ a slightly different core sample with respect to Sect. \ref{CMF_evol}. In detail, we now exclude all the cores embedded in clumps which reported at least one potentially free-free contaminated cores (see Appendix \ref{Appff_analysis}), i.e., also the ones not directly associated with strong local radio emission. This is done in order to not artificially alter the internal mass distribution of the interested clumps, especially when it comes to evaluate the most and least massive cores. On the other hand, we waive the completeness cut in mass to preserve the information on the lower-mass core population. 
Ultimately, the considered sample amounts to $3973$ cores ($63\%$ of the catalog). 

    \subsection{Core mass growth with evolution}
    \label{mcore_growth}
    
    In Fig. \ref{core_mass_growth} (left panel), we report the median value of the maximum and the minimum core masses ($M_{\rm{core}}^{\rm{max}}$ and $M_{\rm{core}}^{\rm{min}}$, respectively) revealed across the clumps within different bins of $L/M$. 
    The two quantities show a clearly different trend with the evolutionary tracer. The maximum core mass (i.e., the mass of the MMC) increases on average by a factor $\sim10$ over about $3$ orders of magnitude in $L/M$. The increasing trend essentially reflects the evidence of Fig. \ref{CMF_evolution} and confirms recent observational findings (e.g., \citealt{Liu+23,Pouteau+23,Morii+24,Xu+24}), but also agrees with numerical simulations produced by \citet{Bonnell+04}, assuming a subsequent competitive accretion model in the framework of clustered environment. 
    The minimum core mass, instead, remains roughly constant on average around $0.2\,\mathrm{M_{\odot}}$ (i.e, it slightly decreases within a factor of $\sim2$, between $\sim0.15-0.3\,\mathrm{M_{\odot}}$), roughly matching the mass completeness limit. As a consequence, the ratio between median $M_{\rm{core}}^{\rm{max}}$ and $M_{\rm{core}}^{\rm{min}}$ increases from $\sim2$ to $\sim40$ with evolution (e.g., \citealt{Morii+24}). 
    The roughly constant trend of $M_{\rm{core}}^{\rm{min}}$ may be interpreted in different ways: i) such objects might correspond to newly formed fragments that we reveal at all evolutionary stages before they start accreting mass in turn; or ii) we are tracing the "same" core population that forms at early stages and then stays unaltered throughout the cluster evolution. 
    In the first case, interestingly, these cores might then represent the low-mass seeds for massive star formation (e.g., \citealt{Bonnell+06,Traficante+23,Morii+24}). According to this scenario, core mass growth and clump fragmentation would proceed essentially in parallel through the star formation process (see below for further analysis on this), so that a population of relatively low-mass cores is present throughout. We note that this scenario does not necessarily imply that early formed cores will survive throughout evolution (as they could for example further fragment). 
    Nevertheless, the observed $M_{\rm{core}}^{\rm{min}}$ trend might also be the result of a combination of scenarios i) and ii), as it is argued later. 

Figure \ref{core_mass_growth} (left panel) displays only the most and least massive cores within the clumps. In Fig. \ref{core_mass_growth} (right panel), we report the total mass in cores within each clump minus the mass of the MMC, as a function of the latter. The color code represents the clump $L/M$. Although the quantities on the two axes inherently share a certain degree of correlation, the increasing trend on average between them, and the hint of a trend also with the evolutionary indicator, are particularly meaningful. 
They show that, on average, the whole core population within a clump (or at least the majority of it) undergoes mass growth during evolution (e.g., \citealt{Morii+24,Xu+24}). One would in fact expect a flatter (if not completely flat) trend in case where the MMC dominates over the companions to the point of monopolizing the mass reservoir available for accretion. In $\sim85\%$ of the clumps, the MMC hosts between $25\%$ and $75\%$ of the combined core mass, while in some extreme cases ($\sim2\%$ of the targets) it reaches $90\%$. 
The overall increasing trend appears to be present regardless of the maximum core mass value, thus explaining the shift observed among the three CMF profiles of Fig. \ref{CMF_evolution} throughout the mass range. Furthermore, we verified that the mass ratio between the MMC and its successive companions (in order of mass) increases with evolution, thus further suggesting a potential variation of the mass accretion rate of individual cores as a function of their mass, as discussed in Sect. \ref{CMF_evol}. However, detailed estimations of mass accretion rates cannot be performed from these data, and require further dedicated study. 
The same Fig. \ref{core_mass_growth} (right panel) trend between the mass of the MMC and the companions was predicted by the numerical realizations of \citet{Bonnell+04}, while the growth of the total core mass in the clump with evolution agrees with MHD and HD models simulations by \citet{Hennebelle+22}. We note that both works trace the formation and evolution of sinks, namely, (proto)stars. 

    \subsection{Relation with clump fragmentation}
    \label{mcore_growth_vs_frag}

    \begin{figure}[t!]
    \centering
    \includegraphics[width=1.0\columnwidth]{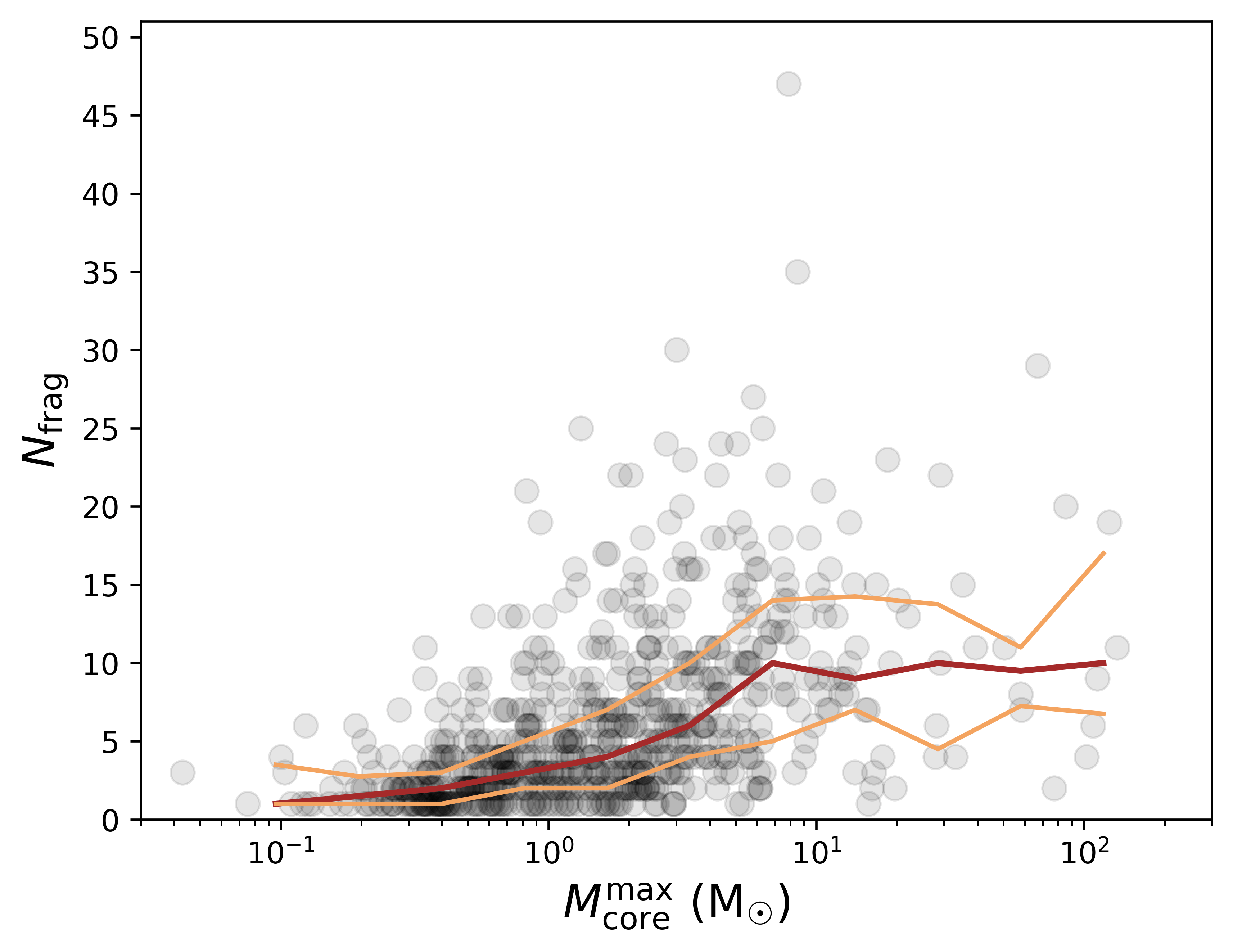}
    \caption{Number of detected compact fragments (cores) as a function of $M_{\rm{core}}^{\rm{max}}$ for each clump. Points are drawn in transparency to highlight the local density of the data. The solid brown line draws the trend of the median $N_\mathrm{frag}$ in given bins of $M_{\rm{core}}^{\rm{max}}$. The solid lighter brown lines correspond to the 25th (bottom) and 75th (top) percentiles of the $N_\mathrm{frag}$ distribution within the same bins.}
    \label{nfrag_mcoremax}
    \end{figure}
    
    To investigate whether such core mass growth is compatible with a continuous fragmentation occurring throughout evolution, as suggested by Fig. \ref{core_mass_growth} (left panel), we relate the number of detected fragments (cores) with $M_{\rm{core}}^{\rm{max}}$ (Fig. \ref{nfrag_mcoremax}). 
    An increasing trend of the median number of fragments (from $\sim2$ to $\sim12$) can be noticed between $\sim0.1$ and $\sim10\,\mathrm{M_{\odot}}$, so that, on average, more massive cores are found in more crowded clumps (e.g., \citealt{Bonnell+04}). This indicates that at least in such a mass range, core mass growth and further clump fragmentation can coexist. 
    Above $\sim10\,\mathrm{M_{\odot}}$, the trend instead stabilizes at around $10-12$ fragments. This might suggest that as soon as the most massive core(s) gets sufficiently massive, the fragmentation process within the clump is halted. Based on the previously discussed evidence from Figs. \ref{CMF_evolution} and \ref{core_mass_growth}, this most likely occurs at later evolutionary stages. Indeed, we verified that most ($\sim85\%$) of the MMCs above $\sim10\,\mathrm{M_{\odot}}$ belong to clumps with $L/M\sim10\,\rm{L_{\odot}/M_{\odot}}$ or higher (see also Elia et al. in prep.). 
    According to this interpretation, the roughly constant trend of $M_{\rm{core}}^{\rm{min}}$ in Fig. \ref{core_mass_growth} (left panel) could indeed result from a combination the above-presented scenarios i) and ii), i.e., with new fragments forming until the MMCs reach the $\sim10\,\mathrm{M_{\odot}}$ threshold, when the present low-mass population stabilizes in number and mass. 
    We also verified that the overall trend of $N_\mathrm{frag}$ is confirmed when considering (instead of their masses) the integrated fluxes of the cores, so that temperature- and distance-related effects can be ruled out. 
    The trend in Fig. \ref{nfrag_mcoremax} is compatible with the results of numerical simulations performed by \citet{Hennebelle+22}. 

However, the causality between the cores' mass growth and the simultaneous interruption of the fragmentation is not straightforward, as various mechanisms might be in action. For instance, the MMC could be able to directly inhibit further fragmentation in the clump by monopolizing the mass reservoir and/or through feedback, although this is actually not expected based on the trend in Fig. \ref{core_mass_growth} (right panel). On the other hand, such reservoir could be globally almost consumed at these late stages, so that there is not enough available material to form further fragments and/or to efficiently accrete them. 

Furthermore, some caveats must be put on interpreting the trend in Fig. \ref{nfrag_mcoremax}. First, the overall spread of the data points is rather wide, especially on the $x$-axis, so that in some cases we observe clumps with only a few fragments also at high $M_{\mathrm{core}}^\mathrm{max}$. Second, as it can be also noticed in Fig. \ref{core_mass_growth} (right panel), the relatively poor statistics available for masses above $\sim10\,\mathrm{M_{\odot}}$ makes the trend less robust over that range, so we refrain from making conclusive assertions in this respect. 

As a whole, the results discussed in Sects. \ref{CMF_evol} and \ref{correlations} depict a scenario in which the formation of high-mass stars is linked to the cluster formation, with cores initially forming with relatively low masses and then growing in mass with evolution while new fragments keep forming, at least until $M_{\mathrm{core}}^\mathrm{max}\simeq10\,\mathrm{M_{\odot}}$, where further fragmentation might be inhibited. 
Cores might gain their mass from the material reservoir of the surrounding intraclump medium, thus favoring a clump-fed mechanism in a dynamic, multi-scale framework (see, e.g., \citealt{Liu+23,Traficante+23,Morii+24}). 
Gravity could play a major role in regulating the process, facilitating both the assembly of more fragments and the mass accretion onto cores (see Elia et al. in prep. and, e.g., \citealt{Motte+98,Svoboda+19,Morii+23,Morii+24,Traficante+23}). 

Further exploration of the correlations among core- and clump-level properties (e.g., mass, surface density, evolutionary stage, age; see Elia et al. in prep.), as well as detailed study of the spatial distribution and mass segregation of the cores (see Schisano et al. in prep.), performed on the same ALMAGAL data, will be able to bring further, potentially crucial evidence in support of the scenario proposed in this work. 
    Moreover, crucial confirmations will come from comparing the observational findings with the outcome of numerical simulations performed under conditions fully comparable to the ALMAGAL data (Nucara et al. in prep.).

\section{Summary and conclusions}
\label{sum_concl}

In this paper, we present the results from an extensive physical analysis performed on the full sample of candidate high-mass star-forming clumps from the ALMAGAL survey. 
We investigated the fragmentation properties of a statistically relevant sample of $1013$ clumps, representing wide ranges of Galactic locations, distances, masses, and the full evolutionary path. They were observed at $1.38$ mm with ALMA to obtain a uniform high spatial resolution of $\sim1400$ au (median value), enough to resolve the typical scales of dense cores. A compact source detection and photometry procedure was performed on combined continuum images with an optimized version of the \textit{CuTEx} algorithm to obtain the ALMAGAL compact source catalog. We then characterized the fragmentation statistics and the photometric properties of the extracted core population, which were then used to estimate and analyze the main physical properties. 
The very large statistics obtained across the broad sample of clumps confers an unprecedented level of robustness and physical significance to our analysis. 

The main results of this work can be summarized as follows: 

   \begin{enumerate}
   
      \item The source extraction procedure revealed a population of $6348$ dense fragments within $844$ clumps ($83\%$ of the entire sample), with a mean number of detections per clump of $8$ (and median of $5$). A wide variety of clump fragmentation degrees is observed (from $1$ to $49$ fragments per clump), with $130$ clumps ($15\%$) reporting only $1$ detection and $58$ ($7\%$) with more than $20$ detected fragments. 
      Compared with other recent surveys, we find a larger fraction of clumps with a low number of cores ($28\%$ with only $1$ or $2$ fragments). \\
      
      \item The estimated core diameters are mostly ($\sim90\%$) within $\sim800-3000$ au, with a median value of $\sim1700$ au (slightly above the median resolution of $1400$ au). 
      Based on the size distribution, most of the detected compact sources are resolved, and prove to be slightly larger than the beam ($15\%$ median discrepancy), implying that the revealed fragments are compact and centrally peaked. \\
      
      \item The core masses were estimated from the measured integrated fluxes adopting for each core a dust temperature based on the evolutionary stage of the hosting clump as traced by the $L/M$ indicator (leading to $20-81$ K core temperatures). Thanks to the high spatial resolution and sensitivity of the ALMAGAL observations, we have been able to reveal compact fragments across about $5$ orders of magnitude in mass with large statistics, a span that has not been achieved by any other recent high-resolution survey. The measured core masses range $0.002-345\,\mathrm{M_\sun}$, with a median value of $0.4\,\mathrm{M_\sun}$. The average mass completeness limit for our sample is $\sim0.23\,\mathrm{M_\sun}$. We revealed $28$ cores with masses above $30\,\mathrm{M_\sun}$ ($0.05\%$ of the total). Moreover, the core masses show a rather good correlation with core radii, with a trend consistent with the widely proposed power-law relation $M\propto R^{\alpha}$ with $\alpha\simeq2.6$. The estimated surface density of the cores ranges within $\sim0.15-500$ g $\rm{cm^{-2}}$ (median of $1.7$ g $\rm{cm^{-2}}$). This range is wider (especially in the upper end) than that found in most of recent similar studies, likely due to our better spatial resolution allowing us to reveal denser, compact substructures. 
      This is also confirmed by the rather high estimated core volume densities of $\sim10^6-10^9$ $\rm{cm^{-3}}$ (median of $3\times10^7$ $\rm{cm^{-3}}$). \\
      
      \item We evaluated the variation in the observed CMF with evolution by comparing the distributions of three core subsamples selected according to the hosting clump $L/M$. 
      Interestingly, the three CMF profiles are clearly shifted between each other, and the high-mass tail gets flatter with evolution, going from the least ($L/M\leq1\,\mathrm{L_{\odot}/M_{\odot}}$) to the most evolved ($L/M>10\,\mathrm{L_{\odot}/M_{\odot}}$) sources. 
      In other terms, higher core masses are found at later evolutionary stages. In particular, cores above $30\,\mathrm{M_{\odot}}$ are found only at $L/M>10\,\mathrm{L_{\odot}/M_{\odot}}$. 
      We verified that such results stand when taking into account uncertainties and biases brought by distance, temperature assumptions, mass sensitivity, and resolution. 
      The observed trend suggests that the shape of the CMF is not constant throughout the star formation process, but rather builds (and flattens) with evolution, forming the most massive objects only at later stages. This result is also compatible with a scenario in which more massive cores accrete their mass at a higher rate than lower-mass ones, so that the higher masses are reached sooner in time. 
      Such a trend of the CMF with evolution is consistent with what was found in some recent literature studies; however, we have broadly expanded the analysis in terms of available statistics and range of physical parameters explored. \\

      \item We investigated the core mass growth as a function of evolution in detail. 
      The median value of the mass of the most massive core in the clumps clearly increases with evolution, by nearly $1$ order of magnitude over $3$ orders of magnitude in $L/M$. The other cores within the clump also grow in mass together with the most massive core (MMC), and the mass ratio between successive companion cores is found to increase with evolution, further suggesting different accretion rates depending on mass. However, the mass of the least massive core stays roughly constant throughout evolution ($\sim0.2\,\,\mathrm{M_{\odot}}$). 
      These trends suggest a scenario in which all cores grow in mass with evolution, while a population of lower-mass cores is present throughout, possibly corresponding to fragments newly formed at all stages. In this framework, such objects might represent low-mass seeds for high-mass star formation. \\

      \item We related the maximum core mass to the clump fragmentation statistics. 
      The number of detected compact fragments (cores) appears on average to grow together with the mass of the MMC, at least until (likely at later stages) the MMC reaches $\sim10\,\mathrm{M_{\odot}}$, when clump fragmentation is seen to be apparently halted (although limited statistics prevents conclusive statements in this respect). 
      Such a turning point might be directly due to the action of the MMC in the clump (e.g., feedback, monopolization of the mass reservoir) or related to more global effects, such as a gradual depletion of available reservoir in the clump environment that would subsequently reduce the efficiency for fragment formation and accretion. \\
      
   \end{enumerate}

\noindent Ultimately, our findings depict a likely clump-fed scenario for massive star formation, in which cores form as relatively low-mass seeds and then grow in mass with evolution, while further clump fragmentation proceeds in parallel (at least until late evolutionary stages). 
Cores might gain their mass from the intraclump medium in a dynamic, multi-scale framework, with more massive ones potentially accreting more efficiently than the others. 

Our results are in agreement with the outcome of various recent surveys, which we have been able to extend thanks to a much larger and representative sample. 
Future ALMAGAL studies, together with systematic comparisons with a new set of numerical simulations, will allow for additional and possibly crucial evidence to support the proposed scenario.

\begin{acknowledgements}
      We thank the anonymous referee for the warm reception of this paper and the insightful comments.\\
      AC and the Team at INAF-IAPS, RSK, P.\ Hennebelle, and LT, acknowledge financial support from the European Research Council via the ERC Synergy Grant ECOGAL (project ID 855130). 
      AS-M\ acknowledges support from the RyC2021-032892-I grant funded by MCIN/AEI/10.13039/501100011033 and by the European Union `Next GenerationEU’/PRTR, as well as the program Unidad de Excelencia María de Maeztu CEX2020-001058-M, and support from the PID2020-117710GB-I00 (MCI-AEI-FEDER, UE). 
      C.\ Battersby gratefully acknowledges funding from the National Science Foundation under Award Nos. 1816715, 2108938, 2206510, and CAREER 2145689, as well as from the National Aeronautics and Space Administration through the Astrophysics Data Analysis Program under Award No. 21-ADAP21-0179 and through the SOFIA archival research program under Award No.~09$\_$0540. 
      GAF gratefully acknowledges the Deutsche Forschungsgemeinschaft (DFG) for funding through SFB~1601 ``Habitats of massive stars across cosmic time’' (sub-project B2), the University of Cologne and its Global Faculty Programme. 
      P.\ Sanhueza was partially supported by a Grant-in-Aid for Scientific Research (KAKENHI Number JP22H01271 and JP23H01221) of JSPS. 
      RSK acknowledges financial support from the German Excellence Strategy via the Heidelberg Cluster of Excellence (EXC 2181 - 390900948) ``STRUCTURES'', and from the German Ministry for Economic Affairs and Climate Action in project ``MAINN'' (funding ID 50OO2206). RSK thanks for computing resources provided by the Ministry of Science, Research and the Arts (MWK) of the State of Baden-W\"{u}rttemberg through bwHPC and the German Science Foundation (DFG) through grants INST 35/1134-1 FUGG and 35/1597-1 FUGG, and also for data storage at SDS@hd funded through grants INST 35/1314-1 FUGG and INST 35/1503-1 FUGG. 
      Part of this research was carried out at the Jet Propulsion Laboratory, California Institute of Technology, under a contract with the National Aeronautics and Space Administration (80NM0018D0004). 
      SDC is supported by the National Science and Technology Council (NSTC) through grants 112-2112-M-001-066 and 111-2112-M-001-064. 
      TL acknowledges supports from the National Key R\&D Program of China (No. 2022YFA1603100); the National Natural Science Foundation of China (NSFC), through grants No. 12073061 and No. 12122307; the international partnership program of the Chinese Academy of Sciences, through grant No. 114231KYSB20200009; and the Shanghai Pujiang Program 20PJ1415500. 
      LB gratefully acknowledges support by the ANID BASAL project FB210003. 
      The National Radio Astronomy Observatory and Green Bank Observatory are facilities of the U.S. National Science Foundation operated under cooperative agreement by Associated Universities, Inc. 
\\
      This paper makes use of the following ALMA data: ADS/JAO.ALMA\#2019.1.00195.L. ALMA is a partnership of ESO (representing its member states), NSF (USA) and NINS (Japan), together with NRC (Canada), MOST and ASIAA (Taiwan), and KASI (Republic of Korea), in cooperation with the Republic of Chile. The Joint ALMA Observatory is operated by ESO, AUI/NRAO and NAOJ.
\end{acknowledgements}

%
%

\bibliographystyle{aa}
\bibliography{ref}    

\begin{thebibliography}{179}
\expandafter\ifx\csname natexlab\endcsname\relax\def\natexlab#1{#1}\fi

\bibitem[{{Adams}(2010)}]{Adams10}
{Adams}, F.~C. 2010, \araa, 48, 47

\bibitem[{{Alves} {et~al.}(2007){Alves}, {Lombardi}, \& {Lada}}]{Alves+07}
{Alves}, J., {Lombardi}, M., \& {Lada}, C.~J. 2007, \aap, 462, L17

\bibitem[{{Anderson} {et~al.}(2021){Anderson}, {Peretto}, {Ragan}, {Rigby}, {Avison}, {Duarte-Cabral}, {Fuller}, {Shirley}, {Traficante}, \& {Williams}}]{Anderson+21}
{Anderson}, M., {Peretto}, N., {Ragan}, S.~E., {et~al.} 2021, \mnras, 508, 2964

\bibitem[{{Andr{\'e}} {et~al.}(2014){Andr{\'e}}, {Di Francesco}, {Ward-Thompson}, {Inutsuka}, {Pudritz}, \& {Pineda}}]{Andre+14}
{Andr{\'e}}, P., {Di Francesco}, J., {Ward-Thompson}, D., {et~al.} 2014, in Protostars and Planets VI, ed. H.~{Beuther}, R.~S. {Klessen}, C.~P. {Dullemond}, \& T.~{Henning}, 27--51

\bibitem[{{Armante} {et~al.}(2024){Armante}, {Gusdorf}, {Louvet}, {Motte}, {Pouteau}, {Lesaffre}, {Galv{\'a}n-Madrid}, {Dell'Ova}, {Bonfand}, {Nony}, {Brouillet}, {Cunningham}, {Ginsburg}, {Men'shchikov}, {Bontemps}, {D{\'\i}az-Gonz{\'a}lez}, {Csengeri}, {Fern{\'a}ndez-L{\'o}pez}, {Gonz{\'a}lez}, {Herpin}, {Liu}, {Sanhueza}, {Stutz}, \& {Valeille-Manet}}]{Armante+24}
{Armante}, M., {Gusdorf}, A., {Louvet}, F., {et~al.} 2024, \aap, 686, A122

\bibitem[{{Avison} {et~al.}(2023){Avison}, {Fuller}, {Frimpong}, {Etoka}, {Hoare}, {Jones}, {Peretto}, {Traficante}, {van der Tak}, {Pineda}, {Beltr{\'a}n}, {Wyrowski}, {Thompson}, {Lumsden}, {Nagy}, {Hill}, {Viti}, {Fontani}, \& {Schilke}}]{Avison+23}
{Avison}, A., {Fuller}, G.~A., {Frimpong}, N.~A., {et~al.} 2023, \mnras, 526, 2278

\bibitem[{Ballesteros-Paredes {et~al.}(2019)Ballesteros-Paredes, Román-Zúñiga, Salomé, Zamora-Avilés, \& Jiménez-Donaire}]{Ballesteros+19}
Ballesteros-Paredes, J., Román-Zúñiga, C., Salomé, Q., Zamora-Avilés, M., \& Jiménez-Donaire, M.~J. 2019, Monthly Notices of the Royal Astronomical Society, 490, 2648

\bibitem[{Barnes {et~al.}(2021)Barnes, Henshaw, Fontani, Pineda, Cosentino, Tan, Caselli, Jiménez-Serra, Law, Avison, Bigiel, Feng, Kong, Longmore, Moser, Parker, Sánchez-Monge, \& Wang}]{Barnes+21}
Barnes, A.~T., Henshaw, J.~D., Fontani, F., {et~al.} 2021, Monthly Notices of the Royal Astronomical Society, 503, 4601

\bibitem[{{Battersby} {et~al.}(2010){Battersby}, {Bally}, {Jackson}, {Ginsburg}, {Shirley}, {Schlingman}, \& {Glenn}}]{Battersby+10}
{Battersby}, C., {Bally}, J., {Jackson}, J.~M., {et~al.} 2010, \apj, 721, 222

\bibitem[{{Battersby} {et~al.}(2017){Battersby}, {Bally}, \& {Svoboda}}]{Battersby+17}
{Battersby}, C., {Bally}, J., \& {Svoboda}, B. 2017, \apj, 835, 263

\bibitem[{{Beltr{\'a}n} {et~al.}(2011){Beltr{\'a}n}, {Cesaroni}, {Neri}, \& {Codella}}]{Beltran+11}
{Beltr{\'a}n}, M.~T., {Cesaroni}, R., {Neri}, R., \& {Codella}, C. 2011, \aap, 525, A151

\bibitem[{{Beuther} {et~al.}(2018){Beuther}, {Mottram, J. C.}, {Ahmadi, A.}, {Bosco, F.}, {Linz, H.}, {Henning, Th.}, {Klaassen, P.}, {Winters, J. M.}, {Maud, L. T.}, {Kuiper, R.}, {Semenov, D.}, {Gieser, C.}, {Peters, T.}, {Urquhart, J. S.}, {Pudritz, R.}, {Ragan, S. E.}, {Feng, S.}, {Keto, E.}, {Leurini, S.}, {Cesaroni, R.}, {Beltran, M.}, {Palau, A.}, {S\'anchez-Monge, \'A.}, {Galvan-Madrid, R.}, {Zhang, Q.}, {Schilke, P.}, {Wyrowski, F.}, {Johnston, K. G.}, {Longmore, S. N.}, {Lumsden, S.}, {Hoare, M.}, {Menten, K. M.}, \& {Csengeri, T.}}]{Beuther+18}
{Beuther}, {Mottram, J. C.}, {Ahmadi, A.}, {et~al.} 2018, A\&A, 617, A100

\bibitem[{{Beuther}(2007)}]{Beuther07}
{Beuther}, H. 2007, in IAU Symposium, Vol. 237, Triggered Star Formation in a Turbulent ISM, ed. B.~G. {Elmegreen} \& J.~{Palous}, 148--154

\bibitem[{{Beuther} {et~al.}(2007){Beuther}, {Churchwell}, {McKee}, \& {Tan}}]{Beuther+07}
{Beuther}, H., {Churchwell}, E.~B., {McKee}, C.~F., \& {Tan}, J.~C. 2007, in Protostars and Planets V, ed. B.~{Reipurth}, D.~{Jewitt}, \& K.~{Keil}, 165

\bibitem[{{Beuther} {et~al.}(2005){Beuther}, {Zhang}, {Sridharan}, \& {Chen}}]{Beuther+05}
{Beuther}, H., {Zhang}, Q., {Sridharan}, T.~K., \& {Chen}, Y. 2005, \apj, 628, 800

\bibitem[{{Bolatto} {et~al.}(2013){Bolatto}, {Warren}, {Leroy}, {Walter}, {Veilleux}, {Ostriker}, {Ott}, {Zwaan}, {Fisher}, {Weiss}, {Rosolowsky}, \& {Hodge}}]{Bolatto+13}
{Bolatto}, A.~D., {Warren}, S.~R., {Leroy}, A.~K., {et~al.} 2013, \nat, 499, 450

\bibitem[{{Bonnell} \& {Bate}(2006)}]{Bonnell+06}
{Bonnell}, I.~A. \& {Bate}, M.~R. 2006, \mnras, 370, 488

\bibitem[{{Bonnell} {et~al.}(2001){Bonnell}, {Bate}, {Clarke}, \& {Pringle}}]{Bonnell+01}
{Bonnell}, I.~A., {Bate}, M.~R., {Clarke}, C.~J., \& {Pringle}, J.~E. 2001, \mnras, 323, 785

\bibitem[{{Bonnell} {et~al.}(2007){Bonnell}, {Larson}, \& {Zinnecker}}]{Bonnell+07}
{Bonnell}, I.~A., {Larson}, R.~B., \& {Zinnecker}, H. 2007, in Protostars and Planets V, ed. B.~{Reipurth}, D.~{Jewitt}, \& K.~{Keil}, 149

\bibitem[{{Bonnell} {et~al.}(2004){Bonnell}, {Vine}, \& {Bate}}]{Bonnell+04}
{Bonnell}, I.~A., {Vine}, S.~G., \& {Bate}, M.~R. 2004, \mnras, 349, 735

\bibitem[{{Brouillet} {et~al.}(2022){Brouillet}, {Despois, D.}, {Molet, J.}, {Nony, T.}, {Motte, F.}, {Gusdorf, A.}, {Louvet, F.}, {Bontemps, S.}, {Herpin, F.}, {Bonfand, M.}, {Csengeri, T.}, {Ginsburg, A.}, {Cunningham, N.}, {Galv\'an-Madrid, R.}, {Maud, L.}, {Busquet, G.}, {Bronfman, L.}, {Fern\'andez-L\'opez, M.}, {Jeff, D. L.}, {Lefloch, B.}, {Pouteau, Y.}, {Sanhueza, P.}, {Stutz, A. M.}, \& {Valeille-Manet, M.}}]{Brouillet+22}
{Brouillet}, N., {Despois, D.}, {Molet, J.}, {et~al.} 2022, A\&A, 665, A140

\bibitem[{{Butler} \& {Tan}(2012)}]{Butler+12}
{Butler}, M.~J. \& {Tan}, J.~C. 2012, \apj, 754, 5

\bibitem[{{Cao} {et~al.}(2021){Cao}, {Qiu}, {Zhang}, {Wang}, \& {Xiao}}]{Cao+21}
{Cao}, Y., {Qiu}, K., {Zhang}, Q., {Wang}, Y., \& {Xiao}, Y. 2021, \apjl, 918, L4

\bibitem[{{Carpenter}(2000)}]{Carpenter00}
{Carpenter}, J.~M. 2000, \aj, 120, 3139

\bibitem[{{CASA Team} {et~al.}(2022){CASA Team}, {Bean}, {Bhatnagar}, {Castro}, {Donovan Meyer}, \& {et al.}}]{CASATeam22}
{CASA Team}, {Bean}, B., {Bhatnagar}, S., {et~al.} 2022, Publications of the Astronomical Society of the Pacific, 134, 114501

\bibitem[{{Caselli}(2005)}]{Caselli05}
{Caselli}, P. 2005, in Astrophysics and Space Science Library, Vol. 324, Astrophysics and Space Science Library, ed. M.~S.~N. {Kumar}, M.~{Tafalla}, \& P.~{Caselli}, 47

\bibitem[{Cesaroni {et~al.}(2019)Cesaroni, Beuther, Ahmadi, Beltr{\'a}n, Csengeri, Galv{\'a}n-Madrid, Gieser, Henning, Johnston, Klaassen, Kuiper, Leurini, Linz, Longmore, Lumsden, Maud, Moscadelli, Mottram, Palau, Peters, Pudritz, S{\'a}nchez-Monge, Schilke, Semenov, Suri, Urquhart, Winters, Zhang, \& Zinnecker}]{Cesaroni+19}
Cesaroni, R., Beuther, H., Ahmadi, A., {et~al.} 2019, {Astronomy and Astrophysics - A\&A}, 627, A68

\bibitem[{{Cesaroni} {et~al.}(2007){Cesaroni}, {Galli}, {Lodato}, {Walmsley}, \& {Zhang}}]{Cesaroni+07}
{Cesaroni}, R., {Galli}, D., {Lodato}, G., {Walmsley}, C.~M., \& {Zhang}, Q. 2007, in Protostars and Planets V, ed. B.~{Reipurth}, D.~{Jewitt}, \& K.~{Keil}, 197

\bibitem[{{Chenu} {et~al.}(2016){Chenu}, {Navarrini}, {Bortolotti}, {Butin}, {Fontana}, {Mahieu}, {Maier}, {Mattiocco}, {Serres}, {Berton}, {Garnier}, {Moutote}, {Parioleau}, {Pissard}, \& {Reverdy}}]{Chenu+16}
{Chenu}, J.-Y., {Navarrini}, A., {Bortolotti}, Y., {et~al.} 2016, IEEE Transactions on Terahertz Science and Technology, 6, 223

\bibitem[{{Clark} {et~al.}(2007){Clark}, {Klessen}, \& {Bonnell}}]{Clark2007}
{Clark}, P.~C., {Klessen}, R.~S., \& {Bonnell}, I.~A. 2007, \mnras, 379, 57

\bibitem[{{Coletta} {et~al.}(2020){Coletta}, {Fontani}, {Rivilla}, {Mininni}, {Colzi}, {S{\'a}nchez-Monge}, \& {Beltr{\'a}n}}]{Coletta+20}
{Coletta}, A., {Fontani}, F., {Rivilla}, V.~M., {et~al.} 2020, \aap, 641, A54

\bibitem[{{Commer{\c{c}}on} {et~al.}(2011){Commer{\c{c}}on}, {Hennebelle}, \& {Henning}}]{Commercon+11}
{Commer{\c{c}}on}, B., {Hennebelle}, P., \& {Henning}, T. 2011, \apjl, 742, L9

\bibitem[{{Contreras} {et~al.}(2018){Contreras}, {Sanhueza}, {Jackson}, {Guzm{\'a}n}, {Longmore}, {Garay}, {Zhang}, {Nguyễn-Lu'o'ng}, {Tatematsu}, {Nakamura}, {Sakai}, {Ohashi}, {Liu}, {Saito}, {Gomez}, {Rathborne}, \& {Whitaker}}]{Contreras+18}
{Contreras}, Y., {Sanhueza}, P., {Jackson}, J.~M., {et~al.} 2018, \apj, 861, 14

\bibitem[{Cornwell(2008)}]{Cornwell08}
Cornwell, T.~J. 2008, IEEE Journal of Selected Topics in Signal Processing, 2, 793

\bibitem[{{Csengeri} {et~al.}(2017){Csengeri}, {Bontemps}, {Wyrowski}, {Motte}, {Menten}, {Beuther}, {Bronfman}, {Commer{\c{c}}on}, {Chapillon}, {Duarte-Cabral}, {Fuller}, {Henning}, {Leurini}, {Longmore}, {Palau}, {Peretto}, {Schuller}, {Tan}, {Testi}, {Traficante}, \& {Urquhart}}]{Csengeri+17}
{Csengeri}, T., {Bontemps}, S., {Wyrowski}, F., {et~al.} 2017, \aap, 600, L10

\bibitem[{{Draine}(2011)}]{Draine11}
{Draine}, B.~T. 2011, {Physics of the Interstellar and Intergalactic Medium}

\bibitem[{{Elia} {et~al.}(2021){Elia}, {Merello}, {Molinari}, {Schisano}, {Zavagno}, {Russeil}, {M{\`e}ge}, {Martin}, {Olmi}, {Pestalozzi}, {Plume}, {Ragan}, {Benedettini}, {Eden}, {Moore}, {Noriega-Crespo}, {Paladini}, {Palmeirim}, {Pezzuto}, {Pilbratt}, {Rygl}, {Schilke}, {Strafella}, {Tan}, {Traficante}, {Baldeschi}, {Bally}, {Giorgio}, {Fiorellino}, {Liu}, {Piazzo}, \& {Polychroni}}]{Elia+21}
{Elia}, D., {Merello}, M., {Molinari}, S., {et~al.} 2021, \mnras, 504, 2742

\bibitem[{{Elia} {et~al.}(2013){Elia}, {Molinari}, {Fukui}, {Schisano}, {Olmi}, {Veneziani}, {Hayakawa}, {Pestalozzi}, {Schneider}, {Benedettini}, {di Giorgio}, {Ikhenaode}, {Mizuno}, {Onishi}, {Pezzuto}, {Piazzo}, {Polychroni}, {Rygl}, {Yamamoto}, \& {Maruccia}}]{Elia+13}
{Elia}, D., {Molinari}, S., {Fukui}, Y., {et~al.} 2013, \apj, 772, 45

\bibitem[{{Elia} {et~al.}(2017){Elia}, {Molinari}, {Schisano}, {Pestalozzi}, {Pezzuto}, {Merello}, {Noriega-Crespo}, {Moore}, {Russeil}, {Mottram}, {Paladini}, {Strafella}, {Benedettini}, {Bernard}, {Di Giorgio}, {Eden}, {Fukui}, {Plume}, {Bally}, {Martin}, {Ragan}, {Jaffa}, {Motte}, {Olmi}, {Schneider}, {Testi}, {Wyrowski}, {Zavagno}, {Calzoletti}, {Faustini}, {Natoli}, {Palmeirim}, {Piacentini}, {Piazzo}, {Pilbratt}, {Polychroni}, {Baldeschi}, {Beltr{\'a}n}, {Billot}, {Cambr{\'e}sy}, {Cesaroni}, {Garc{\'\i}a-Lario}, {Hoare}, {Huang}, {Joncas}, {Liu}, {Maiolo}, {Marsh}, {Maruccia}, {M{\`e}ge}, {Peretto}, {Rygl}, {Schilke}, {Thompson}, {Traficante}, {Umana}, {Veneziani}, {Ward-Thompson}, {Whitworth}, {Arab}, {Bandieramonte}, {Becciani}, {Brescia}, {Buemi}, {Bufano}, {Butora}, {Cavuoti}, {Costa}, {Fiorellino}, {Hajnal}, {Hayakawa}, {Kacsuk}, {Leto}, {Li Causi}, {Marchili}, {Martinavarro-Armengol}, {Mercurio}, {Molinaro}, {Riccio}, {Sano}, {Sciacca}, {Tachihara}, {Torii}, {Trigilio}, {Vitello}, \&
  {Yamamoto}}]{Elia+17}
{Elia}, D., {Molinari}, S., {Schisano}, E., {et~al.} 2017, \mnras, 471, 100

\bibitem[{{Elia} {et~al.}(2010){Elia}, {Schisano}, {Molinari}, {Robitaille}, {Angl{\'e}s-Alc{\'a}zar}, {Bally}, {Battersby}, {Benedettini}, {Billot}, {Calzoletti}, {di Giorgio}, {Faustini}, {Li}, {Martin}, {Morgan}, {Motte}, {Mottram}, {Natoli}, {Olmi}, {Paladini}, {Piacentini}, {Pestalozzi}, {Pezzuto}, {Polychroni}, {Smith}, {Strafella}, {Stringfellow}, {Testi}, {Thompson}, {Traficante}, \& {Veneziani}}]{Elia+10}
{Elia}, D., {Schisano}, E., {Molinari}, S., {et~al.} 2010, \aap, 518, L97

\bibitem[{{Escoffier} {et~al.}(2007){Escoffier}, {Comoretto}, {Webber}, {Baudry}, {Broadwell}, {Greenberg}, {Treacy}, {Cais}, {Quertier}, {Camino}, {Bos}, \& {Gunst}}]{Escoffier+07}
{Escoffier}, R.~P., {Comoretto}, G., {Webber}, J.~C., {et~al.} 2007, \aap, 462, 801

\bibitem[{{Fiorellino} {et~al.}(2021){Fiorellino}, {Elia}, {Andr{\'e}}, {Men'shchikov}, {Pezzuto}, {Schisano}, {K{\"o}nyves}, {Arzoumanian}, {Benedettini}, {Ward-Thompson}, {Bracco}, {Di Francesco}, {Bontemps}, {Kirk}, {Motte}, \& {Molinari}}]{Fiorellino+21}
{Fiorellino}, E., {Elia}, D., {Andr{\'e}}, P., {et~al.} 2021, \mnras, 500, 4257

\bibitem[{{Fontani} {et~al.}(2018){Fontani}, {Commer{\c{c}}on}, {Giannetti}, {Beltr{\'a}n}, {S{\'a}nchez-Monge}, {Testi}, {Brand}, \& {Tan}}]{Fontani+18b}
{Fontani}, F., {Commer{\c{c}}on}, B., {Giannetti}, A., {et~al.} 2018, \aap, 615, A94

\bibitem[{{Garay} \& {Lizano}(1999)}]{Garay&Lizano99}
{Garay}, G. \& {Lizano}, S. 1999, \pasp, 111, 1049

\bibitem[{{Giannetti} {et~al.}(2017){Giannetti}, {Leurini}, {K{\"o}nig}, {Urquhart}, {Pillai}, {Brand}, {Kauffmann}, {Wyrowski}, \& {Menten}}]{Giannetti+17}
{Giannetti}, A., {Leurini}, S., {K{\"o}nig}, C., {et~al.} 2017, \aap, 606, L12

\bibitem[{{Gieser} {et~al.}(2023{\natexlab{a}}){Gieser}, {Beuther}, {Semenov}, {Ahmadi}, {Henning}, \& {Wells}}]{Gieser+23b}
{Gieser}, C., {Beuther}, H., {Semenov}, D., {et~al.} 2023{\natexlab{a}}, in ALMA at 10 years: Past, Present, and Future, 65

\bibitem[{{Gieser} {et~al.}(2023{\natexlab{b}}){Gieser}, {Beuther}, {Semenov}, {Ahmadi}, {Henning}, \& {Wells}}]{Gieser+23}
{Gieser}, C., {Beuther}, H., {Semenov}, D., {et~al.} 2023{\natexlab{b}}, \aap, 674, A160

\bibitem[{{Girichidis} {et~al.}(2016){Girichidis}, {Naab}, {Walch}, {Hanasz}, {Mac Low}, {Ostriker}, {Gatto}, {Peters}, {W{\"u}nsch}, {Glover}, {Klessen}, {Clark}, \& {Baczynski}}]{Girichidis2016}
{Girichidis}, P., {Naab}, T., {Walch}, S., {et~al.} 2016, \apjl, 816, L19

\bibitem[{{Gounelle} \& {Meynet}(2012)}]{Gounelle+12}
{Gounelle}, M. \& {Meynet}, G. 2012, \aap, 545, A4

\bibitem[{{Grudi{\'c}} {et~al.}(2022){Grudi{\'c}}, {Guszejnov}, {Offner}, {Rosen}, {Raju}, {Faucher-Gigu{\`e}re}, \& {Hopkins}}]{Grudic+22}
{Grudi{\'c}}, M.~Y., {Guszejnov}, D., {Offner}, S. S.~R., {et~al.} 2022, \mnras, 512, 216

\bibitem[{{Hennebelle} {et~al.}(2011){Hennebelle}, {Commer{\c{c}}on}, {Joos}, {Klessen}, {Krumholz}, {Tan}, \& {Teyssier}}]{Hennebelle+11}
{Hennebelle}, P., {Commer{\c{c}}on}, B., {Joos}, M., {et~al.} 2011, \aap, 528, A72

\bibitem[{{Hennebelle} {et~al.}(2020){Hennebelle}, {Commer{\c{c}}on}, {Lee}, \& {Chabrier}}]{Hennebelle+20}
{Hennebelle}, P., {Commer{\c{c}}on}, B., {Lee}, Y.-N., \& {Chabrier}, G. 2020, \apj, 904, 194

\bibitem[{{Hennebelle} \& {Grudi{\'c}}(2024)}]{Hennebelle+24}
{Hennebelle}, P. \& {Grudi{\'c}}, M.~Y. 2024, \araa, 62, 63

\bibitem[{{Hennebelle} {et~al.}(2022){Hennebelle}, {Lebreuilly}, {Colman}, {Elia}, {Fuller}, {Leurini}, {Nony}, {Schisano}, {Soler}, {Traficante}, {Klessen}, {Molinari}, \& {Testi}}]{Hennebelle+22}
{Hennebelle}, P., {Lebreuilly}, U., {Colman}, T., {et~al.} 2022, \aap, 668, A147

\bibitem[{{Ho} {et~al.}(2004){Ho}, {Moran}, \& {Lo}}]{Ho+04}
{Ho}, P. T.~P., {Moran}, J.~M., \& {Lo}, K.~Y. 2004, \apjl, 616, L1

\bibitem[{Hoare {et~al.}(2005)Hoare, Lumsden, Oudmaijer, Urquhart, Busfield, Sheret, Clarke, Moore, Allsopp, Burton, Purcell, Jiang, \& Wang}]{Hoare+05}
Hoare, M.~G., Lumsden, S.~L., Oudmaijer, R.~D., {et~al.} 2005, in , 370--375

\bibitem[{{Hoare} {et~al.}(2012){Hoare}, {Purcell}, {Churchwell}, {Diamond}, {Cotton}, {Chandler}, {Smethurst}, {Kurtz}, {Mundy}, {Dougherty}, {Fender}, {Fuller}, {Jackson}, {Garrington}, {Gledhill}, {Goldsmith}, {Lumsden}, {Mart{\'\i}}, {Moore}, {Muxlow}, {Oudmaijer}, {Pandian}, {Paredes}, {Shepherd}, {Spencer}, {Thompson}, {Umana}, {Urquhart}, \& {Zijlstra}}]{Hoare+12}
{Hoare}, M.~G., {Purcell}, C.~R., {Churchwell}, E.~B., {et~al.} 2012, \pasp, 124, 939

\bibitem[{{H{\"o}gbom}(1974)}]{Hogbom74}
{H{\"o}gbom}, J.~A. 1974, \aaps, 15, 417

\bibitem[{{Hopkins} {et~al.}(2015){Hopkins}, {Whiting}, {Seymour}, {Chow}, {Norris}, {Bonavera}, {Breton}, {Carbone}, {Ferrari}, {Franzen}, {Garsden}, {Gonz{\'a}lez-Nuevo}, {Hales}, {Hancock}, {Heald}, {Herranz}, {Huynh}, {Jurek}, {L{\'o}pez-Caniego}, {Massardi}, {Mohan}, {Molinari}, {Orr{\`u}}, {Paladino}, {Pestalozzi}, {Pizzo}, {Rafferty}, {R{\"o}ttgering}, {Rudnick}, {Schisano}, {Shulevski}, {Swinbank}, {Taylor}, \& {van der Horst}}]{Hopkins+15}
{Hopkins}, A.~M., {Whiting}, M.~T., {Seymour}, N., {et~al.} 2015, \pasa, 32, e037

\bibitem[{{Iguchi} {et~al.}(2009){Iguchi}, {Morita}, {Sugimoto}, {Vilar{\'o}}, {Saito}, {Hasegawa}, {Kawabe}, {Tatematsu}, {Sakamoto}, {Kiuchi}, {Okumura}, {Kosugi}, {Inatani}, {Takakuwa}, {Iono}, {Kamazaki}, {Ogasawara}, \& {Ishiguro}}]{Iguchi+09}
{Iguchi}, S., {Morita}, K.-I., {Sugimoto}, M., {et~al.} 2009, \pasj, 61, 1

\bibitem[{{Irabor} {et~al.}(2023){Irabor}, {Hoare}, {Burton}, {Cotton}, {Diamond}, {Dougherty}, {Ellingsen}, {Fender}, {Fuller}, {Garrington}, {Goldsmith}, {Green}, {Gunn}, {Jackson}, {Kurtz}, {Lumsden}, {Marti}, {McDonald}, {Molinari}, {Moore}, {Mutale}, {Muxlow}, {O'Brien}, {Oudmaijer}, {Paladini}, {Pandian}, {Paredes}, {Richards}, {Sanchez-Monge}, {Spencer}, {Thompson}, {Umana}, {Urquhart}, {Wieringa}, \& {Zijlstra}}]{Irabor+23}
{Irabor}, T., {Hoare}, M.~G., {Burton}, M., {et~al.} 2023, \mnras, 520, 1073

\bibitem[{{Ishihara} {et~al.}(2024){Ishihara}, {Sanhueza}, {Nakamura}, {Saito}, {Chen}, {Li}, {Olguin}, {Taniguchi}, {Morii}, {Lu}, {Luo}, {Sakai}, \& {Zhang}}]{Ishihara+24}
{Ishihara}, K., {Sanhueza}, P., {Nakamura}, F., {et~al.} 2024, \apj, 974, 95

\bibitem[{{Izumi} {et~al.}(2024){Izumi}, {Sanhueza}, {Koch}, {Lu}, {Li}, {Sabatini}, {Olguin}, {Zhang}, {Nakamura}, {Tatematsu}, {Morii}, {Sakai}, \& {Tafoya}}]{Izumi+24}
{Izumi}, N., {Sanhueza}, P., {Koch}, P.~M., {et~al.} 2024, \apj, 963, 163

\bibitem[{{Kainulainen} {et~al.}(2017){Kainulainen}, {Stutz}, {Stanke}, {Abreu-Vicente}, {Beuther}, {Henning}, {Johnston}, \& {Megeath}}]{Kainulainen+17}
{Kainulainen}, J., {Stutz}, A.~M., {Stanke}, T., {et~al.} 2017, \aap, 600, A141

\bibitem[{{Kamazaki} {et~al.}(2012){Kamazaki}, {Okumura}, {Chikada}, {Okuda}, {Kurono}, {Iguchi}, {Mitsuishi}, {Murakami}, {Nishimura}, {Mita}, \& {Sano}}]{Kamazaki+12}
{Kamazaki}, T., {Okumura}, S.~K., {Chikada}, Y., {et~al.} 2012, \pasj, 64, 29

\bibitem[{{Kauffmann} {et~al.}(2008){Kauffmann}, {Bertoldi}, {Bourke}, {Evans}, \& {Lee}}]{Kauffmann+08}
{Kauffmann}, J., {Bertoldi}, F., {Bourke}, T.~L., {Evans}, N.~J., I., \& {Lee}, C.~W. 2008, \aap, 487, 993

\bibitem[{Kauffmann \& Pillai(2010)}]{Kauffmann+10}
Kauffmann, J. \& Pillai, T. 2010, The Astrophysical Journal Letters, 723, L7

\bibitem[{{Kauffmann} {et~al.}(2010){Kauffmann}, {Pillai}, {Shetty}, {Myers}, \& {Goodman}}]{Kauffmann+10b}
{Kauffmann}, J., {Pillai}, T., {Shetty}, R., {Myers}, P.~C., \& {Goodman}, A.~A. 2010, \apj, 712, 1137

\bibitem[{{Kerr} {et~al.}(2014){Kerr}, {Pan}, {Claude}, {Dindo}, {Lichtenberger}, {Effland}, \& {Lauria}}]{Kerr+14}
{Kerr}, A.~R., {Pan}, S.-K., {Claude}, S. M.~X., {et~al.} 2014, IEEE Transactions on Terahertz Science and Technology, 4, 201

\bibitem[{{Klessen} \& {Burkert}(2000)}]{Klessen+00}
{Klessen}, R.~S. \& {Burkert}, A. 2000, \apjs, 128, 287

\bibitem[{{Klessen} \& {Burkert}(2001)}]{Klessen2001}
{Klessen}, R.~S. \& {Burkert}, A. 2001, \apj, 549, 386

\bibitem[{{Klessen} \& {Glover}(2016)}]{Klessen+16}
{Klessen}, R.~S. \& {Glover}, S. C.~O. 2016, Saas-Fee Advanced Course, 43, 85

\bibitem[{{Kroupa}(2001)}]{Kroupa01}
{Kroupa}, P. 2001, \mnras, 322, 231

\bibitem[{{Krumholz} {et~al.}(2014){Krumholz}, {Bate}, {Arce}, {Dale}, {Gutermuth}, {Klein}, {Li}, {Nakamura}, \& {Zhang}}]{Krumholz+14}
{Krumholz}, M.~R., {Bate}, M.~R., {Arce}, H.~G., {et~al.} 2014, in Protostars and Planets VI, ed. H.~{Beuther}, R.~S. {Klessen}, C.~P. {Dullemond}, \& T.~{Henning}, 243

\bibitem[{{Krumholz} {et~al.}(2009){Krumholz}, {Klein}, {McKee}, {Offner}, \& {Cunningham}}]{Krumholz+09}
{Krumholz}, M.~R., {Klein}, R.~I., {McKee}, C.~F., {Offner}, S. S.~R., \& {Cunningham}, A.~J. 2009, Science, 323, 754

\bibitem[{{Krumholz} \& {McKee}(2008)}]{Krumholz+08}
{Krumholz}, M.~R. \& {McKee}, C.~F. 2008, \nat, 451, 1082

\bibitem[{{Krumholz} {et~al.}(2005){Krumholz}, {McKee}, \& {Klein}}]{Krumholz+05}
{Krumholz}, M.~R., {McKee}, C.~F., \& {Klein}, R.~I. 2005, \nat, 438, 332

\bibitem[{{Kurtz} {et~al.}(2000){Kurtz}, {Cesaroni}, {Churchwell}, {Hofner}, \& {Walmsley}}]{Kurtz+00}
{Kurtz}, S., {Cesaroni}, R., {Churchwell}, E., {Hofner}, P., \& {Walmsley}, C.~M. 2000, in Protostars and Planets IV, ed. V.~{Mannings}, A.~P. {Boss}, \& S.~S. {Russell}, 299--326

\bibitem[{{Lada} \& {Lada}(2003)}]{Lada&Lada03}
{Lada}, C.~J. \& {Lada}, E.~A. 2003, \araa, 41, 57

\bibitem[{Larson(1981)}]{Larson81}
Larson, R.~B. 1981, Monthly Notices of the Royal Astronomical Society, 194, 809

\bibitem[{{Lee} \& {Hennebelle}(2018)}]{Lee+18}
{Lee}, Y.-N. \& {Hennebelle}, P. 2018, \aap, 611, A88

\bibitem[{{Li} {et~al.}(2024){Li}, {Sanhueza}, {Beuther}, {Chen}, {Kuiper}, {Olguin}, {Pudritz}, {Stephens}, {Zhang}, {Nakamura}, {Lu}, {Kuruwita}, {Sakai}, {Henning}, {Taniguchi}, \& {Li}}]{Li+24}
{Li}, S., {Sanhueza}, P., {Beuther}, H., {et~al.} 2024, Nature Astronomy, 8, 472

\bibitem[{{Li} {et~al.}(2022){Li}, {Sanhueza}, {Lu}, {Lee}, {Zhang}, {Bovino}, {Sabatini}, {Liu}, {Kim}, {Morii}, {Tafoya}, {Tatematsu}, {Sakai}, {Wang}, {Li}, {Silva}, {Izumi}, \& {Allingham}}]{Li+22}
{Li}, S., {Sanhueza}, P., {Lu}, X., {et~al.} 2022, \apj, 939, 102

\bibitem[{{Liu} {et~al.}(2022{\natexlab{a}}){Liu}, {Liu}, {Evans}, {Wang}, {Garay}, {Qin}, {Li}, {Stutz}, {Goldsmith}, {Liu}, {Tej}, {Zhang}, {Juvela}, {Li}, {Wang}, {Bronfman}, {Ren}, {Wu}, {Kim}, {Lee}, {Tatematsu}, {Cunningham}, {Liu}, {Wu}, {Hirota}, {Lee}, {Li}, {Kang}, {Mardones}, {Ristorcelli}, {Zhang}, {Luo}, {Toth}, {Yi}, {Yun}, {Peng}, {Li}, {Zhu}, {Shen}, {Baug}, {Dewangan}, {Chakali}, {Liu}, {Xu}, {Wang}, {Zhang}, {Li}, {Zhang}, {Zhou}, {Tang}, {Xue}, {Issac}, {Soam}, \& {{\'A}lvarez-Guti{\'e}rrez}}]{Liu+22b}
{Liu}, H.-L., {Liu}, T., {Evans}, Neal~J., I., {et~al.} 2022{\natexlab{a}}, \mnras, 511, 501

\bibitem[{{Liu} {et~al.}(2022{\natexlab{b}}){Liu}, {Tej}, {Liu}, {Issac}, {Saha}, {Goldsmith}, {Wang}, {Zhang}, {Qin}, {Wang}, {Li}, {Soam}, {Dewangan}, {Lee}, {Li}, {Liu}, {Zhang}, {Ren}, {Juvela}, {Bronfman}, {Wu}, {Tatematsu}, {Chen}, {Li}, {Stutz}, {Zhang}, {Viktor Toth}, {Luo}, {Xu}, {Li}, {Liu}, {Zhou}, {Zhang}, {Tang}, {Zhang}, {Baug}, {Mannfors}, {Chakali}, \& {Dutta}}]{Liu+22a}
{Liu}, H.-L., {Tej}, A., {Liu}, T., {et~al.} 2022{\natexlab{b}}, \mnras, 510, 5009

\bibitem[{{Liu} {et~al.}(2023){Liu}, {Tej}, {Liu}, {Sanhueza}, {Qin}, {He}, {Goldsmith}, {Garay}, {Pan}, {Morii}, {Li}, {Stutz}, {Tatematsu}, {Xu}, {Bronfman}, {Saha}, {Issac}, {Baug}, {Toth}, {Dewangan}, {Wang}, {Zhou}, {Lee}, {Yang}, {Luo}, {Shen}, {Zhang}, {Wu}, {Ren}, {Liu}, {Soam}, {Zhang}, \& {Luo}}]{Liu+23}
{Liu}, H.-L., {Tej}, A., {Liu}, T., {et~al.} 2023, \mnras, 522, 3719

\bibitem[{{Liu} {et~al.}(2020){Liu}, {Evans}, {Kim}, {Goldsmith}, {Liu}, {Zhang}, {Tatematsu}, {Wang}, {Juvela}, {Bronfman}, {Cunningham}, {Garay}, {Hirota}, {Lee}, {Kang}, {Li}, {Li}, {Mardones}, {Qin}, {Ristorcelli}, {Tej}, {Toth}, {Wu}, {Wu}, {Yi}, {Yun}, {Liu}, {Peng}, {Li}, {Li}, {Lee}, {Shen}, {Baug}, {Wang}, {Zhang}, {Issac}, {Zhu}, {Luo}, {Soam}, {Liu}, {Xu}, {Wang}, {Zhang}, {Ren}, \& {Zhang}}]{Liu+20}
{Liu}, T., {Evans}, N.~J., {Kim}, K.-T., {et~al.} 2020, \mnras, 496, 2790

\bibitem[{{Lombardi} {et~al.}(2010){Lombardi}, {Alves}, \& {Lada}}]{Lombardi+10}
{Lombardi}, M., {Alves}, J., \& {Lada}, C.~J. 2010, \aap, 519, L7

\bibitem[{{Louvet} {et~al.}(2021){Louvet}, {Hennebelle}, {Men'shchikov}, {Didelon}, {Ntormousi}, \& {Motte}}]{Louvet+21}
{Louvet}, F., {Hennebelle}, P., {Men'shchikov}, A., {et~al.} 2021, \aap, 653, A157

\bibitem[{{Louvet} {et~al.}(2024){Louvet}, {Sanhueza}, {Stutz}, {Men'shchikov}, {Motte}, {Galv{\'a}n-Madrid}, {Bontemps}, {Pouteau}, {Ginsburg}, {Csengeri}, {Di Francesco}, {Dell'Ova}, {Gonz{\'a}lez}, {Didelon}, {Braine}, {Cunningham}, {Thomasson}, {Lesaffre}, {Hennebelle}, {Bonfand}, {Gusdorf}, {{\'A}lvarez-Guti{\'e}rrez}, {Nony}, {Busquet}, {Olguin}, {Bronfman}, {Salinas}, {Fernandez-Lopez}, {Moraux}, {Liu}, {Lu}, {Huei-Ru}, {Towner}, {Valeille-Manet}, {Brouillet}, {Herpin}, {Lefloch}, {Baug}, {Maud}, {L{\'o}pez-Sepulcre}, \& {Svoboda}}]{Louvet+24}
{Louvet}, F., {Sanhueza}, P., {Stutz}, A., {et~al.} 2024, \aap, 690, A33

\bibitem[{{Lu} {et~al.}(2020){Lu}, {Cheng}, {Ginsburg}, {Longmore}, {Kruijssen}, {Battersby}, {Zhang}, \& {Walker}}]{Lu+20}
{Lu}, X., {Cheng}, Y., {Ginsburg}, A., {et~al.} 2020, \apjl, 894, L14

\bibitem[{{Lu} {et~al.}(2018){Lu}, {Zhang}, {Liu}, {Sanhueza}, {Tatematsu}, {Feng}, {Smith}, {Myers}, {Sridharan}, \& {Gu}}]{Lu+18}
{Lu}, X., {Zhang}, Q., {Liu}, H.~B., {et~al.} 2018, \apj, 855, 9

\bibitem[{{Lu} {et~al.}(2014){Lu}, {Zhang}, {Liu}, {Wang}, \& {Gu}}]{Lu+14}
{Lu}, X., {Zhang}, Q., {Liu}, H.~B., {Wang}, J., \& {Gu}, Q. 2014, \apj, 790, 84

\bibitem[{{Lumsden} {et~al.}(2013){Lumsden}, {Hoare}, {Urquhart}, {Oudmaijer}, {Davies}, {Mottram}, {Cooper}, \& {Moore}}]{Lumsden+13}
{Lumsden}, S.~L., {Hoare}, M.~G., {Urquhart}, J.~S., {et~al.} 2013, \apjs, 208, 11

\bibitem[{{Mai} {et~al.}(2024){Mai}, {Liu}, {Liu}, {Zhu}, {Garay}, {Goldsmith}, {Juvela}, {Liu}, {Mannfors}, {Tej}, {Sanhueza}, {Li}, {Xu}, {Semadeni}, {Jiao}, {Peng}, {Baug}, {Yang}, {Dewangan}, {Bronfman}, {G{\'o}mez}, {Palau}, {Lee}, {Qin}, {Tatematsu}, {Chibueze}, {Yang}, {Lu}, {Luo}, {Gu}, {Issac}, {Zhang}, {Li}, {Zhang}, \& {T{\'o}th}}]{Mai+24}
{Mai}, X., {Liu}, T., {Liu}, X., {et~al.} 2024, \apjl, 961, L35

\bibitem[{{Mathis} {et~al.}(1977){Mathis}, {Rumpl}, \& {Nordsieck}}]{Mathis+77}
{Mathis}, J.~S., {Rumpl}, W., \& {Nordsieck}, K.~H. 1977, \apj, 217, 425

\bibitem[{{Matzner} \& {McKee}(2000)}]{Matzner+00}
{Matzner}, C.~D. \& {McKee}, C.~F. 2000, \apj, 545, 364

\bibitem[{{McKee} \& {Tan}(2003)}]{Mckee+03}
{McKee}, C.~F. \& {Tan}, J.~C. 2003, \apj, 585, 850

\bibitem[{{McMullin} {et~al.}(2007){McMullin}, {Waters}, {Schiebel}, {Young}, \& {Golap}}]{McMullin+07}
{McMullin}, J.~P., {Waters}, B., {Schiebel}, D., {Young}, W., \& {Golap}, K. 2007, in Astronomical Society of the Pacific Conference Series, Vol. 376, Astronomical Data Analysis Software and Systems XVI, ed. R.~A. {Shaw}, F.~{Hill}, \& D.~J. {Bell}, 127

\bibitem[{{Men'shchikov}(2021)}]{Men'shchikov+21}
{Men'shchikov}, A. 2021, \aap, 649, A89

\bibitem[{{Mininni} {et~al.}(2024){Mininni}, {Molinari}, {Soler}, {S{\'a}nchez-Monge}, {Coletta}, \& {et al.}}]{Mininni+25}
{Mininni}, C., {Molinari}, S., {Soler}, J., {et~al.} 2024, \aap, submitted

\bibitem[{{Miville-Desch{\^e}nes} {et~al.}(2017){Miville-Desch{\^e}nes}, {Murray}, \& {Lee}}]{Miville+17}
{Miville-Desch{\^e}nes}, M.-A., {Murray}, N., \& {Lee}, E.~J. 2017, \apj, 834, 57

\bibitem[{{Molinari} {et~al.}(2019){Molinari}, {Baldeschi}, {Robitaille}, {Morales}, {Schisano}, {Traficante}, {Merello}, {Molinaro}, {Vitello}, {Sciacca}, \& {Liu}}]{Molinari+19}
{Molinari}, S., {Baldeschi}, A., {Robitaille}, T.~P., {et~al.} 2019, \mnras, 486, 4508

\bibitem[{{Molinari} {et~al.}(2016{\natexlab{a}}){Molinari}, {Merello}, {Elia}, {Cesaroni}, {Testi}, \& {Robitaille}}]{Molinari+16}
{Molinari}, S., {Merello}, M., {Elia}, D., {et~al.} 2016{\natexlab{a}}, \apjl, 826, L8

\bibitem[{{Molinari} {et~al.}(2008){Molinari}, {Pezzuto}, {Cesaroni}, {Brand}, {Faustini}, \& {Testi}}]{Molinari+08}
{Molinari}, S., {Pezzuto}, S., {Cesaroni}, R., {et~al.} 2008, \aap, 481, 345

\bibitem[{{Molinari} {et~al.}(2024){Molinari}, {Schilke}, {Battersby}, {Ho}, {Sanchez-Monge}, \& {et al.}}]{Molinari+25}
{Molinari}, S., {Schilke}, P., {Battersby}, C., {et~al.} 2024, \aap, submitted

\bibitem[{{Molinari} {et~al.}(2016{\natexlab{b}}){Molinari}, {Schisano}, {Elia}, {Pestalozzi}, {Traficante}, {Pezzuto}, {Swinyard}, {Noriega-Crespo}, {Bally}, {Moore}, {Plume}, {Zavagno}, {di Giorgio}, {Liu}, {Pilbratt}, {Mottram}, {Russeil}, {Piazzo}, {Veneziani}, {Benedettini}, {Calzoletti}, {Faustini}, {Natoli}, {Piacentini}, {Merello}, {Palmese}, {Del Grande}, {Polychroni}, {Rygl}, {Polenta}, {Barlow}, {Bernard}, {Martin}, {Testi}, {Ali}, {Andr{\'e}}, {Beltr{\'a}n}, {Billot}, {Carey}, {Cesaroni}, {Compi{\`e}gne}, {Eden}, {Fukui}, {Garcia-Lario}, {Hoare}, {Huang}, {Joncas}, {Lim}, {Lord}, {Martinavarro-Armengol}, {Motte}, {Paladini}, {Paradis}, {Peretto}, {Robitaille}, {Schilke}, {Schneider}, {Schulz}, {Sibthorpe}, {Strafella}, {Thompson}, {Umana}, {Ward-Thompson}, \& {Wyrowski}}]{Molinari+16b}
{Molinari}, S., {Schisano}, E., {Elia}, D., {et~al.} 2016{\natexlab{b}}, \aap, 591, A149

\bibitem[{{Molinari} {et~al.}(2011){Molinari}, {Schisano}, {Faustini}, {Pestalozzi}, {di Giorgio}, \& {Liu}}]{Molinari+11}
{Molinari}, S., {Schisano}, E., {Faustini}, F., {et~al.} 2011, \aap, 530, A133

\bibitem[{{Molinari} {et~al.}(2017){Molinari}, {Schisano}, {Faustini}, {Pestalozzi}, {di Giorgio}, \& {Liu}}]{Molinari+17}
{Molinari}, S., {Schisano}, E., {Faustini}, F., {et~al.} 2017, {CUTEX: CUrvature Thresholding EXtractor}

\bibitem[{{Molinari} {et~al.}(2010){Molinari}, {Swinyard}, {Bally}, {Barlow}, {Bernard}, {Martin}, {Moore}, {Noriega-Crespo}, {Plume}, {Testi}, {Zavagno}, {Abergel}, {Ali}, {Andr{\'e}}, {Baluteau}, {Benedettini}, {Bern{\'e}}, {Billot}, {Blommaert}, {Bontemps}, {Boulanger}, {Brand}, {Brunt}, {Burton}, {Campeggio}, {Carey}, {Caselli}, {Cesaroni}, {Cernicharo}, {Chakrabarti}, {Chrysostomou}, {Codella}, {Cohen}, {Compiegne}, {Davis}, {de Bernardis}, {de Gasperis}, {Di Francesco}, {di Giorgio}, {Elia}, {Faustini}, {Fischera}, {Fukui}, {Fuller}, {Ganga}, {Garcia-Lario}, {Giard}, {Giardino}, {Glenn}, {Goldsmith}, {Griffin}, {Hoare}, {Huang}, {Jiang}, {Joblin}, {Joncas}, {Juvela}, {Kirk}, {Lagache}, {Li}, {Lim}, {Lord}, {Lucas}, {Maiolo}, {Marengo}, {Marshall}, {Masi}, {Massi}, {Matsuura}, {Meny}, {Minier}, {Miville-Desch{\^e}nes}, {Montier}, {Motte}, {M{\"u}ller}, {Natoli}, {Neves}, {Olmi}, {Paladini}, {Paradis}, {Pestalozzi}, {Pezzuto}, {Piacentini}, {Pomar{\`e}s}, {Popescu}, {Reach}, {Richer}, {Ristorcelli},
  {Roy}, {Royer}, {Russeil}, {Saraceno}, {Sauvage}, {Schilke}, {Schneider-Bontemps}, {Schuller}, {Schultz}, {Shepherd}, {Sibthorpe}, {Smith}, {Smith}, {Spinoglio}, {Stamatellos}, {Strafella}, {Stringfellow}, {Sturm}, {Taylor}, {Thompson}, {Tuffs}, {Umana}, {Valenziano}, {Vavrek}, {Viti}, {Waelkens}, {Ward-Thompson}, {White}, {Wyrowski}, {Yorke}, \& {Zhang}}]{Molinari+10}
{Molinari}, S., {Swinyard}, B., {Bally}, J., {et~al.} 2010, Publications of the Astronomical Society of the Pacific, 122, 314

\bibitem[{{Morii} {et~al.}(2023){Morii}, {Sanhueza}, {Nakamura}, {Zhang}, {Sabatini}, {Beuther}, {Lu}, {Li}, {Garay}, {Jackson}, {Olguin}, {Tafoya}, {Tatematsu}, {Izumi}, {Sakai}, \& {Silva}}]{Morii+23}
{Morii}, K., {Sanhueza}, P., {Nakamura}, F., {et~al.} 2023, \apj, 950, 148

\bibitem[{{Morii} {et~al.}(2024){Morii}, {Sanhueza}, {Zhang}, {Nakamura}, {Li}, {Sabatini}, {Olguin}, {Beuther}, {Tafoya}, {Izumi}, {Tatematsu}, \& {Sakai}}]{Morii+24}
{Morii}, K., {Sanhueza}, P., {Zhang}, Q., {et~al.} 2024, \apj, 966, 171

\bibitem[{{Motte} {et~al.}(1998){Motte}, {Andre}, \& {Neri}}]{Motte+98}
{Motte}, F., {Andre}, P., \& {Neri}, R. 1998, \aap, 336, 150

\bibitem[{{Motte} {et~al.}(2022){Motte}, {Bontemps}, {Csengeri}, {Pouteau}, {Louvet}, {Stutz}, {Cunningham}, {L{\'o}pez-Sepulcre}, {Brouillet}, {Galv{\'a}n-Madrid}, {Ginsburg}, {Maud}, {Men'shchikov}, {Nakamura}, {Nony}, {Sanhueza}, {{\'A}lvarez-Guti{\'e}rrez}, {Armante}, {Baug}, {Bonfand}, {Busquet}, {Chapillon}, {D{\'\i}az-Gonz{\'a}lez}, {Fern{\'a}ndez-L{\'o}pez}, {Guzm{\'a}n}, {Herpin}, {Liu}, {Olguin}, {Towner}, {Bally}, {Battersby}, {Braine}, {Bronfman}, {Chen}, {Dell'Ova}, {Di Francesco}, {Gonz{\'a}lez}, {Gusdorf}, {Hennebelle}, {Izumi}, {Joncour}, {Lee}, {Lefloch}, {Lesaffre}, {Lu}, {Menten}, {Mignon-Risse}, {Molet}, {Moraux}, {Mundy}, {Nguyen Luong}, {Reyes}, {Reyes Reyes}, {Robitaille}, {Rosolowsky}, {Sandoval-Garrido}, {Schuller}, {Svoboda}, {Tatematsu}, {Thomasson}, {Walker}, {Wu}, {Whitworth}, \& {Wyrowski}}]{Motte+22}
{Motte}, F., {Bontemps}, S., {Csengeri}, T., {et~al.} 2022, \aap, 662, A8

\bibitem[{{Motte} {et~al.}(2018){Motte}, {Bontemps}, \& {Louvet}}]{Motte+18}
{Motte}, F., {Bontemps}, S., \& {Louvet}, F. 2018, \araa, 56, 41

\bibitem[{Motte {et~al.}(2018)Motte, Nony, Louvet, Marsh, Bontemps, Whitworth, Men’shchikov, Nguyễn~Lương, Csengeri, Maury, \& et~al.}]{Motte+18b}
Motte, F., Nony, T., Louvet, F., {et~al.} 2018, Nature Astronomy, 2, 478–482

\bibitem[{{Nony} {et~al.}(2023){Nony}, {Galv{\'a}n-Madrid}, {Motte}, {Pouteau}, {Cunningham}, {Louvet}, {Stutz}, {Lefloch}, {Bontemps}, {Brouillet}, {Ginsburg}, {Joncour}, {Herpin}, {Sanhueza}, {Csengeri}, {Towner}, {Bonfand}, {Fern{\'a}ndez-L{\'o}pez}, {Baug}, {Bronfman}, {Busquet}, {Di Francesco}, {Gusdorf}, {Lu}, {Olguin}, {Valeille-Manet}, \& {Whitworth}}]{Nony+23}
{Nony}, T., {Galv{\'a}n-Madrid}, R., {Motte}, F., {et~al.} 2023, \aap, 674, A75

\bibitem[{{Offner} {et~al.}(2014){Offner}, {Clark}, {Hennebelle}, {Bastian}, {Bate}, {Hopkins}, {Moraux}, \& {Whitworth}}]{Offner+14}
{Offner}, S.~S.~R., {Clark}, P.~C., {Hennebelle}, P., {et~al.} 2014, in Protostars and Planets VI, ed. H.~{Beuther}, R.~S. {Klessen}, C.~P. {Dullemond}, \& T.~{Henning}, 53--75

\bibitem[{{Olguin} {et~al.}(2021){Olguin}, {Sanhueza}, {Guzm{\'a}n}, {Lu}, {Saigo}, {Zhang}, {Silva}, {Chen}, {Li}, {Ohashi}, {Nakamura}, {Sakai}, \& {Wu}}]{Olguin+21}
{Olguin}, F.~A., {Sanhueza}, P., {Guzm{\'a}n}, A.~E., {et~al.} 2021, \apj, 909, 199

\bibitem[{{Olmi} {et~al.}(2013){Olmi}, {Angl{\'e}s-Alc{\'a}zar}, {Elia}, {Molinari}, {Montier}, {Pestalozzi}, {Pezzuto}, {Polychroni}, {Ristorcelli}, {Rodon}, {Schisano}, {Smith}, {Testi}, \& {Thompson}}]{Olmi+13}
{Olmi}, L., {Angl{\'e}s-Alc{\'a}zar}, D., {Elia}, D., {et~al.} 2013, \aap, 551, A111

\bibitem[{{Ossenkopf} \& {Henning}(1994)}]{Ossenkopf&Henning94}
{Ossenkopf}, V. \& {Henning}, T. 1994, \aap, 291, 943

\bibitem[{{Padoan} {et~al.}(2020){Padoan}, {Pan}, {Juvela}, {Haugb{\o}lle}, \& {Nordlund}}]{Padoan+20}
{Padoan}, P., {Pan}, L., {Juvela}, M., {Haugb{\o}lle}, T., \& {Nordlund}, {\r{A}}. 2020, \apj, 900, 82

\bibitem[{{Palau} {et~al.}(2014){Palau}, {Estalella}, {Girart}, {Fuente}, {Fontani}, {Commer{\c{c}}on}, {Busquet}, {Bontemps}, {S{\'a}nchez-Monge}, {Zapata}, {Zhang}, {Hennebelle}, \& {di Francesco}}]{Palau+14}
{Palau}, A., {Estalella}, R., {Girart}, J.~M., {et~al.} 2014, \apj, 785, 42

\bibitem[{{Palau} {et~al.}(2013){Palau}, {Fuente}, {Girart}, {Estalella}, {Ho}, {S{\'a}nchez-Monge}, {Fontani}, {Busquet}, {Commer{\c{c}}on}, {Hennebelle}, {Boissier}, {Zhang}, {Cesaroni}, \& {Zapata}}]{Palau+13}
{Palau}, A., {Fuente}, A., {Girart}, J.~M., {et~al.} 2013, \apj, 762, 120

\bibitem[{{Palau} {et~al.}(2021){Palau}, {Zhang}, {Girart}, {Liu}, {Rao}, {Koch}, {Estalella}, {Chen}, {Liu}, {Qiu}, {Li}, {Zapata}, {Bontemps}, {Ho}, {Beuther}, {Ching}, {Shinnaga}, \& {Ahmadi}}]{Palau+21}
{Palau}, A., {Zhang}, Q., {Girart}, J.~M., {et~al.} 2021, \apj, 912, 159

\bibitem[{{Peretto} {et~al.}(2013){Peretto}, {Fuller}, {Duarte-Cabral}, {Avison}, {Hennebelle}, {Pineda}, {Andr{\'e}}, {Bontemps}, {Motte}, {Schneider}, \& {Molinari}}]{Peretto+13}
{Peretto}, N., {Fuller}, G.~A., {Duarte-Cabral}, A., {et~al.} 2013, \aap, 555, A112

\bibitem[{{Peters} {et~al.}(2010){Peters}, {Banerjee}, {Klessen}, {Mac Low}, {Galv{\'a}n-Madrid}, \& {Keto}}]{Peters2010}
{Peters}, T., {Banerjee}, R., {Klessen}, R.~S., {et~al.} 2010, \apj, 711, 1017

\bibitem[{Pezzuto {et~al.}(2023)Pezzuto, Coletta, Klessen, Schisano, Benedettini, Elia, Molinari, Soler, \& Traficante}]{Pezzuto+23}
Pezzuto, S., Coletta, A., Klessen, R.~S., {et~al.} 2023, Monthly Notices of the Royal Astronomical Society, 525, 4744

\bibitem[{{Pfalzner} {et~al.}(2015){Pfalzner}, {Davies}, {Gounelle}, {Johansen}, {M{\"u}nker}, {Lacerda}, {Portegies Zwart}, {Testi}, {Trieloff}, \& {Veras}}]{Pfalzner+15}
{Pfalzner}, S., {Davies}, M.~B., {Gounelle}, M., {et~al.} 2015, \physscr, 90, 068001

\bibitem[{{Pouteau} {et~al.}(2022){Pouteau}, {Motte}, {Nony}, {Galv{\'a}n-Madrid}, {Men'shchikov}, {Bontemps}, {Robitaille}, {Louvet}, {Ginsburg}, {Herpin}, {L{\'o}pez-Sepulcre}, {Dell'Ova}, {Gusdorf}, {Sanhueza}, {Stutz}, {Brouillet}, {Thomasson}, {Armante}, {Baug}, {Bonfand}, {Busquet}, {Csengeri}, {Cunningham}, {Fern{\'a}ndez-L{\'o}pez}, {Liu}, {Olguin}, {Towner}, {Bally}, {Braine}, {Bronfman}, {Joncour}, {Gonz{\'a}lez}, {Hennebelle}, {Lu}, {Menten}, {Moraux}, {Tatematsu}, {Walker}, \& {Whitworth}}]{Pouteau+22}
{Pouteau}, Y., {Motte}, F., {Nony}, T., {et~al.} 2022, \aap, 664, A26

\bibitem[{{Pouteau} {et~al.}(2023){Pouteau}, {Motte}, {Nony}, {Gonz{\'a}lez}, {Joncour}, {Robitaille}, {Busquet}, {Galv{\'a}n-Madrid}, {Gusdorf}, {Hennebelle}, {Ginsburg}, {Csengeri}, {Sanhueza}, {Dell'Ova}, {Stutz}, {Towner}, {Cunningham}, {Louvet}, {Men'shchikov}, {Fern{\'a}ndez-L{\'o}pez}, {Schneider}, {Armante}, {Bally}, {Baug}, {Bonfand}, {Bontemps}, {Bronfman}, {Brouillet}, {D{\'\i}az-Gonz{\'a}lez}, {Herpin}, {Lefloch}, {Liu}, {Lu}, {Nakamura}, {Luong}, {Olguin}, {Tatematsu}, \& {Valeille-Manet}}]{Pouteau+23}
{Pouteau}, Y., {Motte}, F., {Nony}, T., {et~al.} 2023, \aap, 674, A76

\bibitem[{{Purcell} {et~al.}(2008){Purcell}, {Hoare}, \& {Diamond}}]{Purcell+08}
{Purcell}, C.~R., {Hoare}, M.~G., \& {Diamond}, P. 2008, in Astronomical Society of the Pacific Conference Series, Vol. 387, Massive Star Formation: Observations Confront Theory, ed. H.~{Beuther}, H.~{Linz}, \& T.~{Henning}, 389

\bibitem[{{Rathjen} {et~al.}(2021){Rathjen}, {Naab}, {Girichidis}, {Walch}, {W{\"u}nsch}, {Dinnbier}, {Seifried}, {Klessen}, \& {Glover}}]{Rathjen+21}
{Rathjen}, T.-E., {Naab}, T., {Girichidis}, P., {et~al.} 2021, \mnras, 504, 1039

\bibitem[{{Redaelli} {et~al.}(2022){Redaelli}, {Bovino}, {Sanhueza}, {Morii}, {Sabatini}, {Caselli}, {Giannetti}, \& {Li}}]{Redaelli+22}
{Redaelli}, E., {Bovino}, S., {Sanhueza}, P., {et~al.} 2022, \apj, 936, 169

\bibitem[{{Rosen} \& {Krumholz}(2020)}]{Rosen+20}
{Rosen}, A.~L. \& {Krumholz}, M.~R. 2020, \aj, 160, 78

\bibitem[{{Rosolowsky} {et~al.}(2008){Rosolowsky}, {Pineda}, {Kauffmann}, \& {Goodman}}]{Rosolowsky+08}
{Rosolowsky}, E.~W., {Pineda}, J.~E., {Kauffmann}, J., \& {Goodman}, A.~A. 2008, \apj, 679, 1338

\bibitem[{{Sadaghiani} {et~al.}(2020){Sadaghiani}, {S{\'a}nchez-Monge}, {Schilke}, {Liu}, {Clarke}, {Zhang}, {Girart}, {Seifried}, {Aghababaei}, {Li}, {Ju{\'a}rez}, \& {Tang}}]{Sadaghiani+20}
{Sadaghiani}, M., {S{\'a}nchez-Monge}, {\'A}., {Schilke}, P., {et~al.} 2020, \aap, 635, A2

\bibitem[{{Salpeter}(1955)}]{Salpeter55}
{Salpeter}, E.~E. 1955, \apj, 121, 161

\bibitem[{{S{\'a}nchez-Monge} {et~al.}(2025){S{\'a}nchez-Monge}, {Brogan}, {Hunter}, {Ahmadi}, {Avison}, \& {et al.}}]{Sanchez+25}
{S{\'a}nchez-Monge}, {\'A}., {Brogan}, C., {Hunter}, T., {et~al.} 2025, \aap, accepted

\bibitem[{S{\'a}nchez-Monge {et~al.}(2013)S{\'a}nchez-Monge, Palau, Fontani, Busquet, Juárez, Estalella, Tan, Sepúlveda, Ho, Zhang, \& Kurtz}]{Sanchez-Monge+13b}
S{\'a}nchez-Monge, {\'A}., Palau, A., Fontani, F., {et~al.} 2013, Monthly Notices of the Royal Astronomical Society, 432, 3288

\bibitem[{Sanhueza {et~al.}(2019)Sanhueza, Contreras, Wu, Jackson, Guzm{\'{a}}n, Zhang, Li, Lu, Silva, Izumi, Liu, Miura, Tatematsu, Sakai, Beuther, Garay, Ohashi, Saito, Nakamura, Saigo, Veena, Nguyen-Luong, \& Tafoya}]{Sanhueza+19}
Sanhueza, P., Contreras, Y., Wu, B., {et~al.} 2019, The Astrophysical Journal, 886, 102

\bibitem[{{Sanhueza} {et~al.}(2012){Sanhueza}, {Jackson}, {Foster}, {Garay}, {Silva}, \& {Finn}}]{Sanhueza+12}
{Sanhueza}, P., {Jackson}, J.~M., {Foster}, J.~B., {et~al.} 2012, \apj, 756, 60

\bibitem[{Sanhueza {et~al.}(2017)Sanhueza, Jackson, Zhang, Guzm{\'{a}}n, Lu, Stephens, Wang, \& Tatematsu}]{Sanhueza+17}
Sanhueza, P., Jackson, J.~M., Zhang, Q., {et~al.} 2017, The Astrophysical Journal, 841, 97

\bibitem[{{Schmeja} \& {Klessen}(2004)}]{Schmeja2004}
{Schmeja}, S. \& {Klessen}, R.~S. 2004, \aap, 419, 405

\bibitem[{{Silva} {et~al.}(2017){Silva}, {Zhang}, {Sanhueza}, {Lu}, {Beltran}, {Fallscheer}, {Beuther}, {Sridharan}, \& {Cesaroni}}]{Silva+17}
{Silva}, A., {Zhang}, Q., {Sanhueza}, P., {et~al.} 2017, \apj, 847, 87

\bibitem[{{Smith} {et~al.}(2014){Smith}, {Glover}, \& {Klessen}}]{Smith2014}
{Smith}, R.~J., {Glover}, S. C.~O., \& {Klessen}, R.~S. 2014, \mnras, 445, 2900

\bibitem[{{Smith} {et~al.}(2016){Smith}, {Glover}, {Klessen}, \& {Fuller}}]{Smith2016}
{Smith}, R.~J., {Glover}, S. C.~O., {Klessen}, R.~S., \& {Fuller}, G.~A. 2016, \mnras, 455, 3640

\bibitem[{{Smith} {et~al.}(2009){Smith}, {Longmore}, \& {Bonnell}}]{Smith+09}
{Smith}, R.~J., {Longmore}, S., \& {Bonnell}, I. 2009, \mnras, 400, 1775

\bibitem[{{Stahler} \& {Palla}(2004)}]{Stahler&Palla04}
{Stahler}, S.~W. \& {Palla}, F. 2004, {The Formation of Stars}

\bibitem[{{Stahler} {et~al.}(2000){Stahler}, {Palla}, \& {Ho}}]{Stahler+00}
{Stahler}, S.~W., {Palla}, F., \& {Ho}, P.~T.~P. 2000, in Protostars and Planets IV, ed. V.~{Mannings}, A.~P. {Boss}, \& S.~S. {Russell}, 327--352

\bibitem[{Svoboda {et~al.}(2019)Svoboda, Shirley, Traficante, Battersby, Fuller, Zhang, Beuther, Peretto, Brogan, \& Hunter}]{Svoboda+19}
Svoboda, B.~E., Shirley, Y.~L., Traficante, A., {et~al.} 2019, The Astrophysical Journal, 886, 36

\bibitem[{{Tan} {et~al.}(2014){Tan}, {Beltr{\'a}n}, {Caselli}, {Fontani}, {Fuente}, {Krumholz}, {McKee}, \& {Stolte}}]{Tan+14}
{Tan}, J.~C., {Beltr{\'a}n}, M.~T., {Caselli}, P., {et~al.} 2014, in Protostars and Planets VI, 149

\bibitem[{{Tan} {et~al.}(2013){Tan}, {Kong}, {Butler}, {Caselli}, \& {Fontani}}]{Tan+13b}
{Tan}, J.~C., {Kong}, S., {Butler}, M.~J., {Caselli}, P., \& {Fontani}, F. 2013, \apj, 779, 96

\bibitem[{{Tan} \& {McKee}(2003)}]{Tan+03}
{Tan}, J.~C. \& {McKee}, C.~F. 2003, arXiv e-prints, astro

\bibitem[{{Tang} {et~al.}(2019){Tang}, {Koch}, {Peretto}, {Novak}, {Duarte-Cabral}, {Chapman}, {Hsieh}, \& {Yen}}]{Tang+19}
{Tang}, Y.-W., {Koch}, P.~M., {Peretto}, N., {et~al.} 2019, \apj, 878, 10

\bibitem[{{Taniguchi} {et~al.}(2023){Taniguchi}, {Sanhueza}, {Olguin}, {Gorai}, {Das}, {Nakamura}, {Saito}, {Zhang}, {Lu}, {Li}, \& {Chen}}]{Taniguchi+23}
{Taniguchi}, K., {Sanhueza}, P., {Olguin}, F.~A., {et~al.} 2023, \apj, 950, 57

\bibitem[{{Testi} {et~al.}(1998){Testi}, {Felli}, {Persi}, \& {Roth}}]{Testi+98}
{Testi}, L., {Felli}, M., {Persi}, P., \& {Roth}, M. 1998, \aaps, 129, 495

\bibitem[{{Traficante} {et~al.}(2018){Traficante}, {Duarte-Cabral}, {Elia}, {Fuller}, {Merello}, {Molinari}, {Peretto}, {Schisano}, \& {Di Giorgio}}]{Traficante+18b}
{Traficante}, A., {Duarte-Cabral}, A., {Elia}, D., {et~al.} 2018, \mnras, 477, 2220

\bibitem[{{Traficante} {et~al.}(2015){Traficante}, {Fuller}, {Pineda}, \& {Pezzuto}}]{Traficante+15}
{Traficante}, A., {Fuller}, G.~A., {Pineda}, J.~E., \& {Pezzuto}, S. 2015, \aap, 574, A119

\bibitem[{{Traficante} {et~al.}(2023){Traficante}, {Jones}, {Avison}, {Fuller}, {Benedettini}, {Elia}, {Molinari}, {Peretto}, {Pezzuto}, {Pillai}, {Rygl}, {Schisano}, \& {Smith}}]{Traficante+23}
{Traficante}, A., {Jones}, B.~M., {Avison}, A., {et~al.} 2023, \mnras, 520, 2306

\bibitem[{{Urquhart} {et~al.}(2007){Urquhart}, {Busfield}, {Hoare}, {Lumsden}, {Oudmaijer}, {Moore}, {Gibb}, {Purcell}, {Burton}, \& {Marechal}}]{Urquhart+07}
{Urquhart}, J.~S., {Busfield}, A.~L., {Hoare}, M.~G., {et~al.} 2007, \aap, 474, 891

\bibitem[{{Urquhart} {et~al.}(2018){Urquhart}, {K{\"o}nig}, {Giannetti}, {Leurini}, {Moore}, {Eden}, {Pillai}, {Thompson}, {Braiding}, {Burton}, {Csengeri}, {Dempsey}, {Figura}, {Froebrich}, {Menten}, {Schuller}, {Smith}, \& {Wyrowski}}]{Urquhart+18}
{Urquhart}, J.~S., {K{\"o}nig}, C., {Giannetti}, A., {et~al.} 2018, \mnras, 473, 1059

\bibitem[{{V{\'a}zquez-Semadeni} {et~al.}(2019){V{\'a}zquez-Semadeni}, {Palau}, {Ballesteros-Paredes}, {G{\'o}mez}, \& {Zamora-Avil{\'e}s}}]{Vazquez+19}
{V{\'a}zquez-Semadeni}, E., {Palau}, A., {Ballesteros-Paredes}, J., {G{\'o}mez}, G.~C., \& {Zamora-Avil{\'e}s}, M. 2019, \mnras, 490, 3061

\bibitem[{{Wang} {et~al.}(2014){Wang}, {Zhang}, {Testi}, {van der Tak}, {Wu}, {Zhang}, {Pillai}, {Wyrowski}, {Carey}, {Ragan}, \& {Henning}}]{Wang+14}
{Wang}, K., {Zhang}, Q., {Testi}, L., {et~al.} 2014, \mnras, 439, 3275

\bibitem[{{Wang} {et~al.}(2011){Wang}, {Zhang}, {Wu}, \& {Zhang}}]{Wang+11}
{Wang}, K., {Zhang}, Q., {Wu}, Y., \& {Zhang}, H. 2011, \apj, 735, 64

\bibitem[{{Wang} {et~al.}(2010){Wang}, {Li}, {Abel}, \& {Nakamura}}]{Wang+10}
{Wang}, P., {Li}, Z.-Y., {Abel}, T., \& {Nakamura}, F. 2010, \apj, 709, 27

\bibitem[{{Wells} {et~al.}(2024){Wells}, {Beuther}, {Molinari}, {Schilke}, {Battersby}, {Ho}, {S{\'a}nchez-Monge}, {Jones}, {Scheuck}, {Syed}, {Gieser}, {Kuiper}, {Elia}, {Coletta}, {Traficante}, {Wallace}, {Rigby}, {Klessen}, {Zhang}, {Walch}, {Beltr{\'a}n}, {Tang}, {Fuller}, {Lis}, {M{\"o}ller}, {van der Tak}, {Klaassen}, {Clarke}, {Moscadelli}, {Mininni}, {Zinnecker}, {Maruccia}, {Pezzuto}, {Benedettini}, {Soler}, {Brogan}, {Avison}, {Sanhueza}, {Schisano}, {Liu}, {Fontani}, {Rygl}, {Wyrowski}, {Bally}, {Walker}, {Ahmadi}, {Koch}, {Merello}, {Law}, \& {Testi}}]{Wells+24}
{Wells}, M.~R.~A., {Beuther}, H., {Molinari}, S., {et~al.} 2024, \aap, 690, A185

\bibitem[{{Wootten} \& {Thompson}(2009)}]{Wootten+09}
{Wootten}, A. \& {Thompson}, A.~R. 2009, IEEE Proceedings, 97, 1463

\bibitem[{{Xu} {et~al.}(2024){Xu}, {Wang}, {Liu}, {Tang}, {Evans}, {Palau}, {Morii}, {He}, {Sanhueza}, {Liu}, {Stutz}, {Zhang}, {Chen}, {Li}, {G{\'o}mez}, {V{\'a}zquez-Semadeni}, {Li}, {Mai}, {Lu}, {Liu}, {Chen}, {Li}, {Shi}, {Ren}, {Li}, {Garay}, {Bronfman}, {Dewangan}, {Juvela}, {Lee}, {Zhang}, {Yue}, {Wang}, {Ge}, {Jiao}, {Luo}, {Zhou}, {Tatematsu}, {Chibueze}, {Su}, {Sun}, {Ristorcelli}, \& {Toth}}]{Xu+24}
{Xu}, F., {Wang}, K., {Liu}, T., {et~al.} 2024, \apjs, 270, 9

\bibitem[{{Yamamoto}(2017)}]{Yamamoto17}
{Yamamoto}, S. 2017, {Introduction to Astrochemistry: Chemical Evolution from Interstellar Clouds to Star and Planet Formation}

\bibitem[{{Zhang} {et~al.}(2018){Zhang}, {Xu}, {Vasyunin}, {Semenov}, {Wang}, {Dib}, {Liu}, {Liu}, {Zhang}, {Liu}, {Wang}, {Li}, {Wu}, {Yuan}, {Li}, \& {Gao}}]{Zhang+18}
{Zhang}, G.-Y., {Xu}, J.-L., {Vasyunin}, A.~I., {et~al.} 2018, \aap, 620, A163

\bibitem[{{Zhang} {et~al.}(2002){Zhang}, {Hunter}, {Sridharan}, \& {Ho}}]{Zhang+02}
{Zhang}, Q., {Hunter}, T.~R., {Sridharan}, T.~K., \& {Ho}, P. T.~P. 2002, \apj, 566, 982

\bibitem[{{Zhang} {et~al.}(2014){Zhang}, {Qiu}, {Girart}, {Liu}, {Tang}, {Koch}, {Li}, {Keto}, {Ho}, {Rao}, {Lai}, {Ching}, {Frau}, {Chen}, {Li}, {Padovani}, {Bontemps}, {Csengeri}, \& {Ju{\'a}rez}}]{Zhang+14b}
{Zhang}, Q., {Qiu}, K., {Girart}, J.~M., {et~al.} 2014, \apj, 792, 116

\bibitem[{{Zhang} \& {Wang}(2011)}]{Zhang+11}
{Zhang}, Q. \& {Wang}, K. 2011, \apj, 733, 26

\bibitem[{{Zhang} {et~al.}(2015){Zhang}, {Wang}, {Lu}, \& {Jim{\'e}nez-Serra}}]{Zhang+15}
{Zhang}, Q., {Wang}, K., {Lu}, X., \& {Jim{\'e}nez-Serra}, I. 2015, \apj, 804, 141

\bibitem[{{Zhang} {et~al.}(2009){Zhang}, {Wang}, {Pillai}, \& {Rathborne}}]{Zhang+09}
{Zhang}, Q., {Wang}, Y., {Pillai}, T., \& {Rathborne}, J. 2009, \apj, 696, 268

\bibitem[{{Zhang} {et~al.}(2021){Zhang}, {Zavagno}, {L{\'o}pez-Sepulcre}, {Liu}, {Louvet}, {Figueira}, {Russeil}, {Wu}, {Yuan}, \& {Pillai}}]{Zhang+21}
{Zhang}, S., {Zavagno}, A., {L{\'o}pez-Sepulcre}, A., {et~al.} 2021, \aap, 646, A25

\bibitem[{{Zinnecker}(1982)}]{Zinnecker82}
{Zinnecker}, H. 1982, Annals of the New York Academy of Sciences, 395, 226

\bibitem[{{Zinnecker} \& {Yorke}(2007)}]{Zinn&York07}
{Zinnecker}, H. \& {Yorke}, H.~W. 2007, \araa, 45, 481

\end{thebibliography}


\begin{appendix}

\twocolumn

\section{Source extraction tests using different algorithms}
\label{AppSE_codes_tests}

\noindent
Source extraction runs were performed with the \textit{astrodendo} (\citealt{Rosolowsky+08}) and \textit{CuTEx} (\citealt{Molinari+11}) codes on ALMAGAL continuum maps, to characterize and compare their performance in terms of detecting compact sources and measuring their fluxes. 

Because of this test and the following flux completeness analysis (see Appendix \ref{AppFcomplPhotacc}), an \textit{IDL} routine was developed to generate sets of "hybrid" continuum images of real ALMAGAL fields with synthetic compact sources injected. Synthetic sources are ideal tools to characterize the performance of a source extraction algorithm in a fully controlled and tunable environment (see, e.g., \citealt{Molinari+11,Hopkins+15,Molinari+16b}). Instead of producing fully simulated fields with artificial noise and emission patterns, we chose to use as a basis continuum images of ALMAGAL fields with different properties, on which we injected populations of synthetic sources at different flux levels, similarly to the approach recently used, for example, by \citet{Pouteau+22} and \citet{Nony+23}. This was made to test the algorithms in conditions as similar as possible to the real maps. 

To this end, a sample of $60$ ALMAGAL fields was selected to properly represent the different features observed across the whole target sample: from relatively empty fields mainly characterized by diffuse emission to more complex ones showing a certain degree of structuring, to even crowded fields with multiple apparent compact bright objects. 
Synthetic sources are assumed to have a 2D elliptical Gaussian spatial brightness profile, and locations and position angles are randomly generated with uniform probability distribution within defined regions of the field. FHWMs are randomly generated allowing variation up to $\pm40\%$ of the map beam size. A dedicated control is applied within the code on the potential blending among simulated sources, although we checked that, given the large statistics provided by the number of different map realizations and synthetic sources, and the broad region considered for source placement, source blending cases turn out to be extremely rare across our sample. In each simulation, a population of synthetic sources with a defined flux is generated. Different flux levels are also explored, so that multiple realizations for each ALMAGAL field are created. 

For the source extraction test, $10$ ALMAGAL fields were selected, in each of which $10$ synthetic sources were injected at three different flux levels, thus generating a set of $30$ maps with a total of $300$ added simulated sources. Injected flux levels were $0.2$, $0.5$, and $0.8$, in terms of fraction of the peak flux of the native map. 
To optimize their performance and provide a fairer comparison, a series of tests and tuning employing different combinations of source extraction parameters was preliminarily performed for both the \textit{astrodendro} and \textit{CuTEx} codes. 
They were then executed on all maps realizations, and their outputs were compared with the truth tables from the source simulation routine, adopting as metric the number of synthetic sources detected and the accuracy in measuring their integrated flux and size. The positional matching between extracted and simulated sources was performed using a $2$ pixel radius as a threshold (nearly one-third of the beam size of 7M+TM2+TM1 maps). We point out that, differently from what done by \textit{CuTEx} (which is based on the position of the local peak flux, see Sect. \ref{cutex}), \textit{astrodendro} defines as source position the mean positions of the structure (leaf) along the \textit{x} and \textit{y} directions, since it does not assume any specific morphology for the sources to detect. This could make the positional matching with perfectly elliptical simulated sources less accurate in some cases, especially when the leaf is highly irregular in shape.

\begin{figure*}[ht!]
    \centering
        \includegraphics[width=1\textwidth]{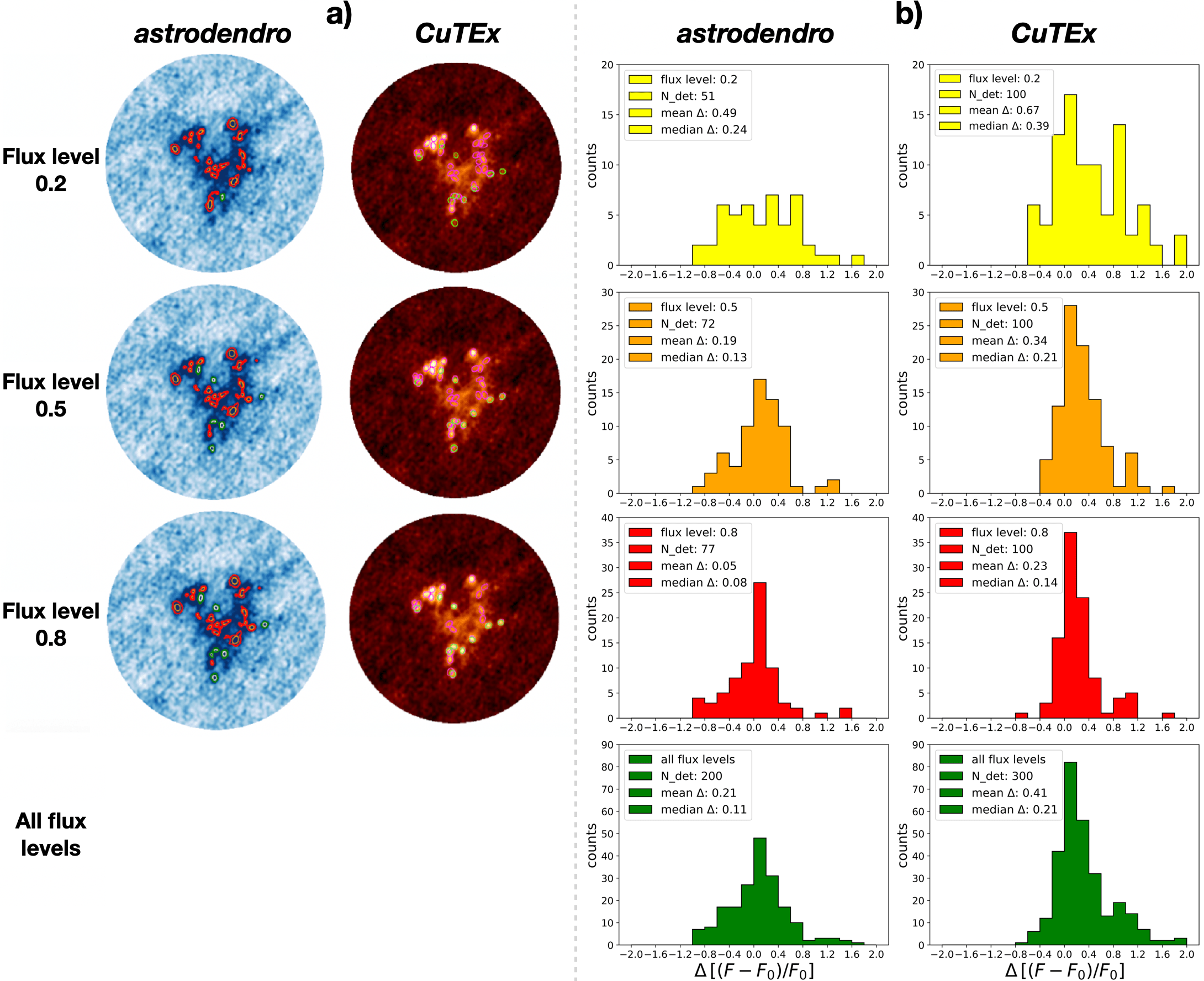}
 
        \caption{
    Comparison of the outcome of the source extraction tests performed with \textit{astrodendro} and \textit{CuTEx}. 
    \textit{a)} Comparison of the detections obtained with \textit{astrodendro} (left column panels) and \textit{CuTEx} (right column panels) on the continuum map of the ALMAGAL field AG304.5567+0.3273 added with $10$ synthetic sources with three different flux levels (see text for details). The three flux levels reported to the side refer to fractions of the native map peak flux. Green contours/ellipses mark the detected sources that were matched with the truth tables from the simulation code (i.e., the recovered synthetic sources), while red/magenta contours/ellipses indicate detected but unmatched sources. 
    \textit{b)} Comparison of the outcome of the photometry procedure performed with the implemented version of \textit{astrodendro} (left column panels) and \textit{CuTEx} (right column panels) on the set of simulated maps (see text for details). In detail, histograms of the relative discrepancies (referred to as $\Delta$) between measured integrated fluxes ($F$) and generated ones ($F_0$) are shown for the matched synthetic sources detected at each flux level ($100$ injected sources in total, first three rows) and for all flux levels combined ($300$ injected sources in total, last row). Boxes within individual panels report the corresponding number of detections and the mean and median values of the relative discrepancy in the flux.}
    \label{ad-cutex_comp}
\end{figure*}

Figure \ref{ad-cutex_comp} (panel \textit{a}) compares, for the three different generated flux levels, the detection maps obtained with \textit{astrodendro} (left column panels) and \textit{CuTEx} (right column panels) for a relatively complex field. Predictably, the number of synthetic sources recovered increases as they become brighter. However, \textit{CuTEx} is able to detect all $10$ sources already at the faintest source flux level, whereas \textit{astrodendro} fails. The reason behind this is that \textit{CuTEx} is able to deal with the presence of a strong background pattern by filtering it, while \textit{astrodendro}, which works on plain thresholding of intensity image, can pick up a source only if it lies above the adopted threshold. 

The two codes were then compared in terms of accuracy of the integrated flux measurement. \textit{Astrodendro} does not include any background emission subtraction by default (see Sect. \ref{SE_intro}). This approach could still provide fairly accurate flux estimates in case of  isolated compact sources in relatively simple fields. However, it is worth noting that in our continuum images of star-forming regions we observe that compact substructures often appear to reside within or on top of diffuse emission zones, structured background, and/or filamentary structures (see, e.g., Fig. \ref{cutex_outcome}). 
As a consequence, based on \textit{CuTEx} estimates, the contribution of the background emission to the total source flux is not negligible in our maps, being on average of about $\sim25\%$, but varying from $\sim10\%$ to up to $\sim90\%$ in the most complex, extreme cases. To enhance its capability to recover correct fluxes and for a more significant comparison with \textit{CuTEx}, an additional post-processing module was implemented in the \textit{astrodendro} software, able to obtain a reliable estimate of the local background emission as the median intensity across a ring of pixels around each source, which was then subtracted from the total leaf flux. 
The compared outcome of the photometry procedure performed with \textit{CuTEx} and this implemented version of \textit{astrodendro} on the above described set of simulated maps is reported in Fig. \ref{ad-cutex_comp} (panel \textit{b}). 
The implemented version of \textit{astrodendro} achieves a rather good accuracy in recovering the integrated fluxes of the detected synthetic sources overall ($\sim10\%$ better than that achieved by \textit{CuTEx}, in terms of median value of the relative discrepancy $\Delta$), mostly due to the introduced background estimation. However, \textit{astrodendro} records a relatively poor performance in terms of number of synthetic sources detected, going from the $\sim50\%$ for faintest sources to a $\sim75\%$ at most for brighter ones (for a combined $\sim66\%$ over the whole population). This suggests that the detection approach applied by \textit{astrodendro} is likely not the most suited to efficiently reveal compact sources over a wide range of fluxes across our continuum maps sample. On the other hand, \textit{CuTEx} performs perfectly in the source detection stage, where it records $100\%$ of matched sources regardless of the flux level considered, while maintaining a rather high accuracy in measuring source fluxes ($\lesssim20\%$ overall; see also Appendix \ref{AppFcomplPhotacc}). 

In addition, the accuracy in recovering correct source sizes was also compared between the two codes, with \textit{CuTEx} reporting excellent results overall: measured FWHMs were consistent with simulated ones within $10-15\%$ at most. While this is not surprising given the analogous assumption of elliptical Gaussian sources (see Sect. \ref{cutex}), it is worth noting that the sources are simulated over the local interferometric extended emission, which presents sufficient fluctuations to influence the observed source profile. Since \textit{astrodendro} considers for the source size the ellipse best matching the leaf mask, in such cases even small distortions in the mask can substantially change the estimated size. 

Ultimately, by compensating for a slightly worse flux recovery accuracy with a significantly higher source detection rate and size recovery accuracy, \textit{CuTEx} reported the best overall performance in this series of compact source extraction tests, confirming to be more suitable for the aims of this work given the properties of the ALMAGAL continuum maps. 

\subsection{Aperture photometry tests with \textit{hyper}}
\label{AppSE_codes_tests_hyper}

\noindent Additionally, we performed source extraction runs with \textit{hyper} (\citealt{Traficante+15}) on the same set of simulated maps presented above, with the aim of testing the aperture photometry approach and its outcome compared to \textit{CuTEx}. 

The \textit{hyper} code provides reliable compact source photometry through a combination of multi-Gaussian fitting and aperture
photometry. Local background subtraction and source deblending are performed to account for variable backgrounds and crowded regions. Full details on its operation and application can be found in \citet{Traficante+15} and \citet{Traficante+23}. For this work, we tuned the parameter setup to efficiently perform based on the properties of our images. 

\begin{figure}[ht!]
    \centering
        \includegraphics[width=0.4\textwidth]{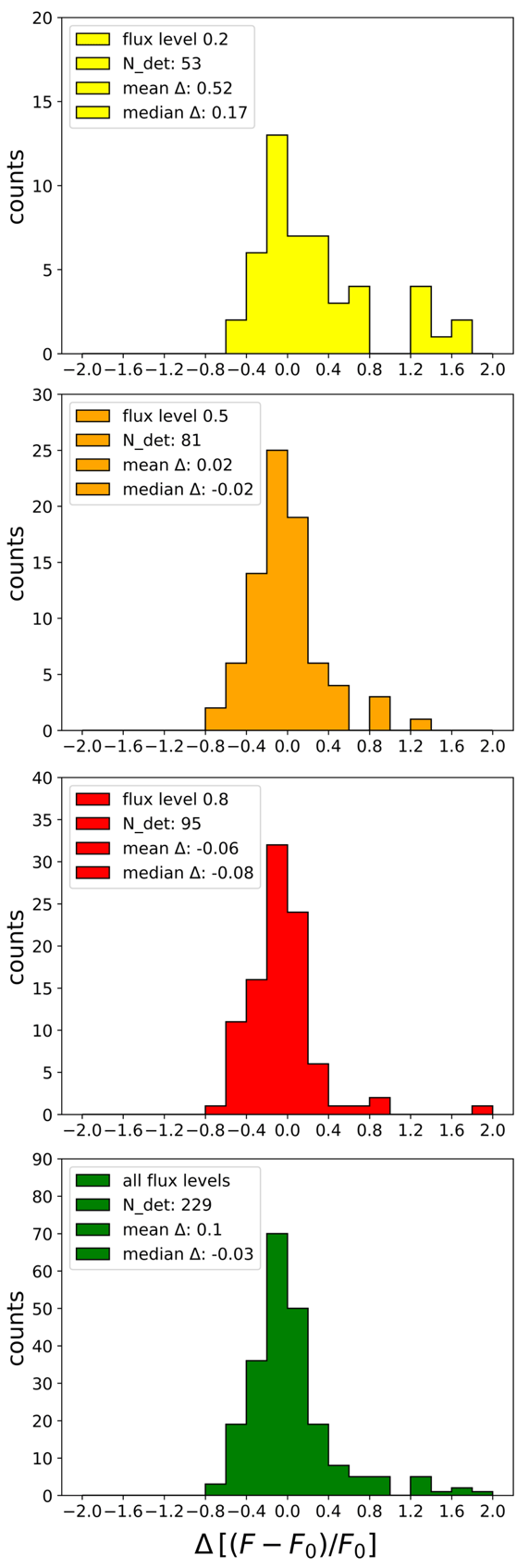}
 
        \caption{
    Same as panel \textit{b)} of Fig. \ref{ad-cutex_comp}, but for the source extraction tests performed with the \textit{hyper} code.}
    \label{hyper-sims_comp}
\end{figure}

\begin{figure}[ht!]
    \centering
        \includegraphics[width=0.4\textwidth]{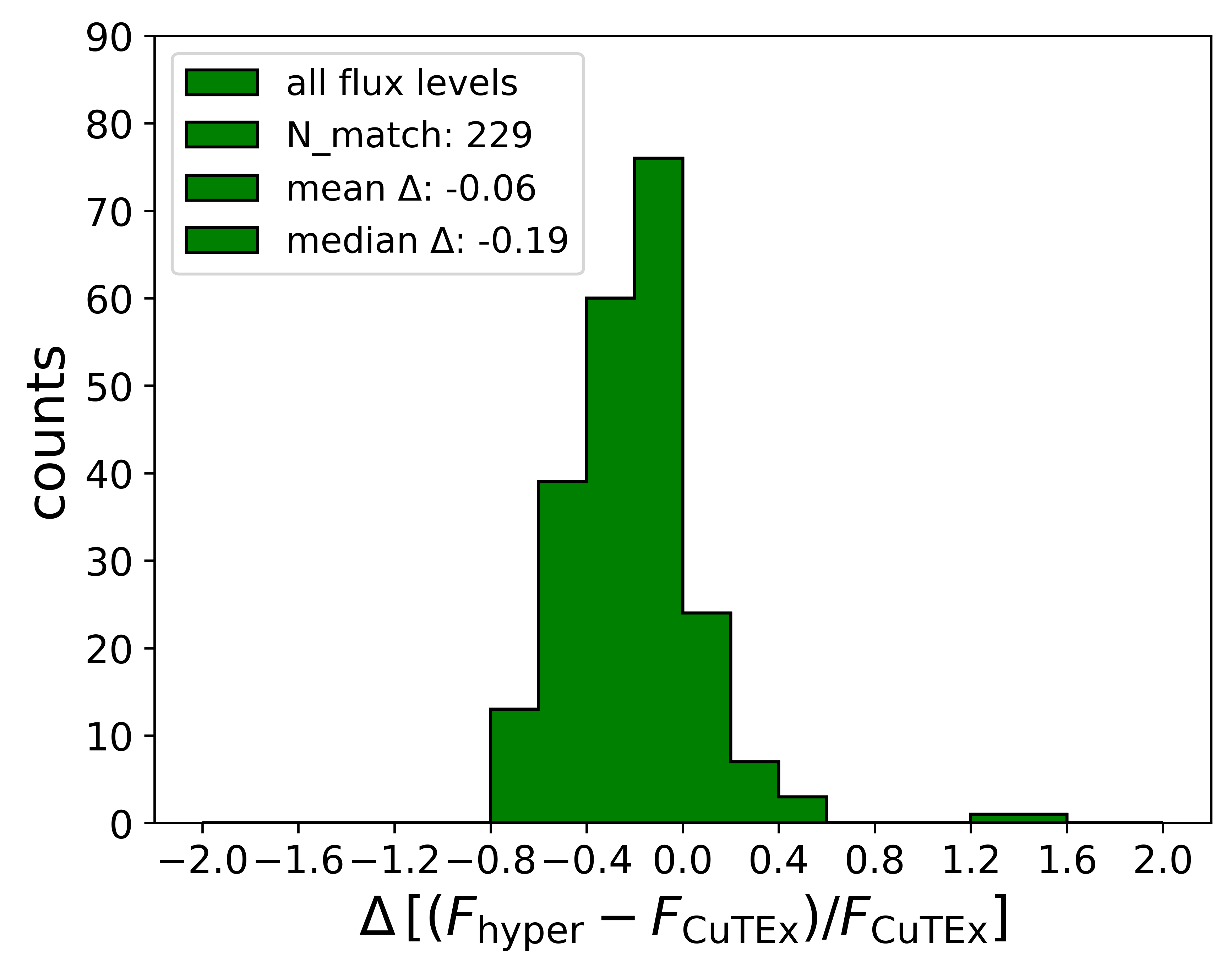}
 
        \caption{
    Histogram of the relative discrepancies (referred to as $\Delta$) between the integrated fluxes measured by \textit{hyper} and \textit{CuTEx}, for the matched synthetic sources combining all flux levels.}
    \label{hyper-cutex_comp}
\end{figure}

Figure \ref{hyper-sims_comp} shows the results obtained with \textit{hyper} on the simulated maps in terms of compact source detection and flux estimation accuracy. 
The detection rate ranges from $\sim50\%$ (lower flux level) to $\sim95\%$ (highest flux level), for a combined $\sim75\%$ overall. The integrated flux recovery is rather accurate, being within $10\%$ on average over the whole population. 
A direct comparison between \textit{hyper} and \textit{CuTEx} is shown in Fig. \ref{hyper-cutex_comp}, in terms of relative discrepancy of the estimated fluxes for the detected sources in common. The fluxes are generally lower for \textit{hyper}, but overall compatible within at most $20\%$ on average, a value which is consistent with the flux uncertainty we estimate for \textit{CuTEx} (see Appendix \ref{AppFcomplPhotacc}).

%

\clearpage

\section{Full description and outcome of the flux completeness and photometric accuracy analysis}
\label{AppFcomplPhotacc}

    With the aim of estimating the flux completeness levels for our extracted compact source catalog, and contextually determining the accuracy of the photometric algorithm applied, we tested the code on $60$ fields from the ALMAGAL target sample, which we divided into three groups of $20$ each representing different map noise properties (see Table \ref{fl_compl_stats}). 
    Based on the rms noise distribution shown in Fig. \ref{map_RMS_hist} (see Sect. \ref{FcomplPhotacc}), fields were selected within the ranges $\sigma_{\rm{rms}}<0.06$ mJy/beam (lower noise, Group 1), $0.06\leq\sigma_{\rm{rms}}\leq0.1$ mJy/beam (medium noise including the peak of the distribution, Group 2), and $\sigma_{\rm{rms}}>0.1$ mJy/beam (higher noise, Group 3). Moreover, through a visual inspection of the maps, we made sure i) to select fields showing only few possible evident compact sources at most, in order to avoid frequent cases of blending with the incoming simulated ones, and ii) that maps with various emission patterns (e.g., localized, filamentary, extended, etc.) were adequately represented, in order to increase the reliability of the completeness estimation on the selected sample with respect to the whole target sample. 

For each field, using the same method presented in Appendix \ref{AppSE_codes_tests}, we generated a set of simulated images by injecting $20$ synthetic elliptical 2D Gaussians with constant integrated flux, but with a distribution of angular sizes. 
    The integrated flux levels to be injected were initially chosen in order to properly sample the measured flux distribution reported across our compact source catalog (Fig. \ref{Fint_plots}), taken as reference to reproduce in the simulations the same properties of the observed core population for best results in code testing. Then, additional flux levels were added to the test runs for each group of fields to be able to more accurately sample the specific range where the $90\%$ detection threshold defining the flux completeness level lied. 
    Integrated flux levels ($F_{\rm{INT}}^{\,\rm{inj}}$, Table \ref{fl_compl_stats}) were generated in the range $0.2-2$ mJy (11 levels) for Group 1 fields, $0.5-2.5$ mJy (12 levels) for Group 2, and $0.7-2.4$ mJy (15 levels) for Group 3. Considering all combinations, a total of about $4000$ synthetic sources were generated for each group of fields, ensuring high statistical significance to the analysis. 
    Source positions were randomly generated within the inner region of the map (up to a primary beam value of $0.6$), and their size (i.e., FWHMs of the ellipse) was left free to randomly vary within $40\%$ from the beam size. A limited control is applied to avoid full source overlapping. However, given the number of synthetic sources injected in each simulation, blending is not affecting the displayed results. In the case of real observations, there is a limited number of cases where sources are close enough that the partial blending affects their flux estimation. 

We ran our \textit{CuTEx} compact source extraction code on all the generated maps with the same parameters configuration used for the catalog extraction, and compared the photometry output with the "truth tables" of simulated sources adopting a matching radius of $2$ pixels (about one-third of the map beam size). 
We then analyzed the detection statistics as a function of $F_{\rm{INT}}^{\,\rm{inj}}$, in order to determine the flux completeness level for each of the three groups (see plots of Fig. \ref{fl_compl_plots}). 
Determined completeness fluxes ($F_{\rm{INT}}^{\,\rm{compl}}$, Table \ref{fl_compl_stats}) were $0.58$ mJy for Group 1, $0.94$ mJy for Group 2, and $2.05$ mJy for Group 3, respectively. 
    
\begin{figure*}
        \includegraphics[width=0.68\columnwidth]{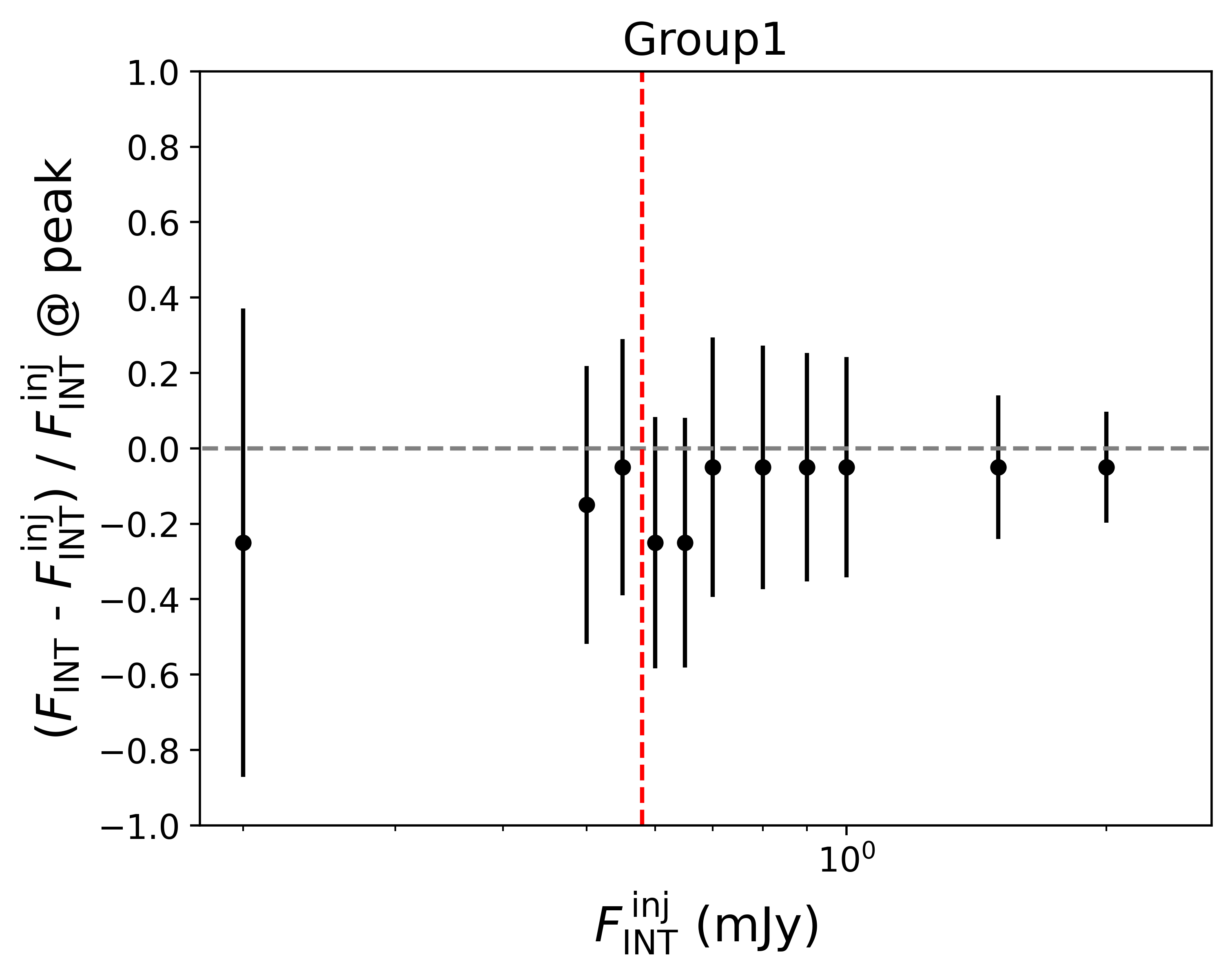}
        \includegraphics[width=0.68\columnwidth]{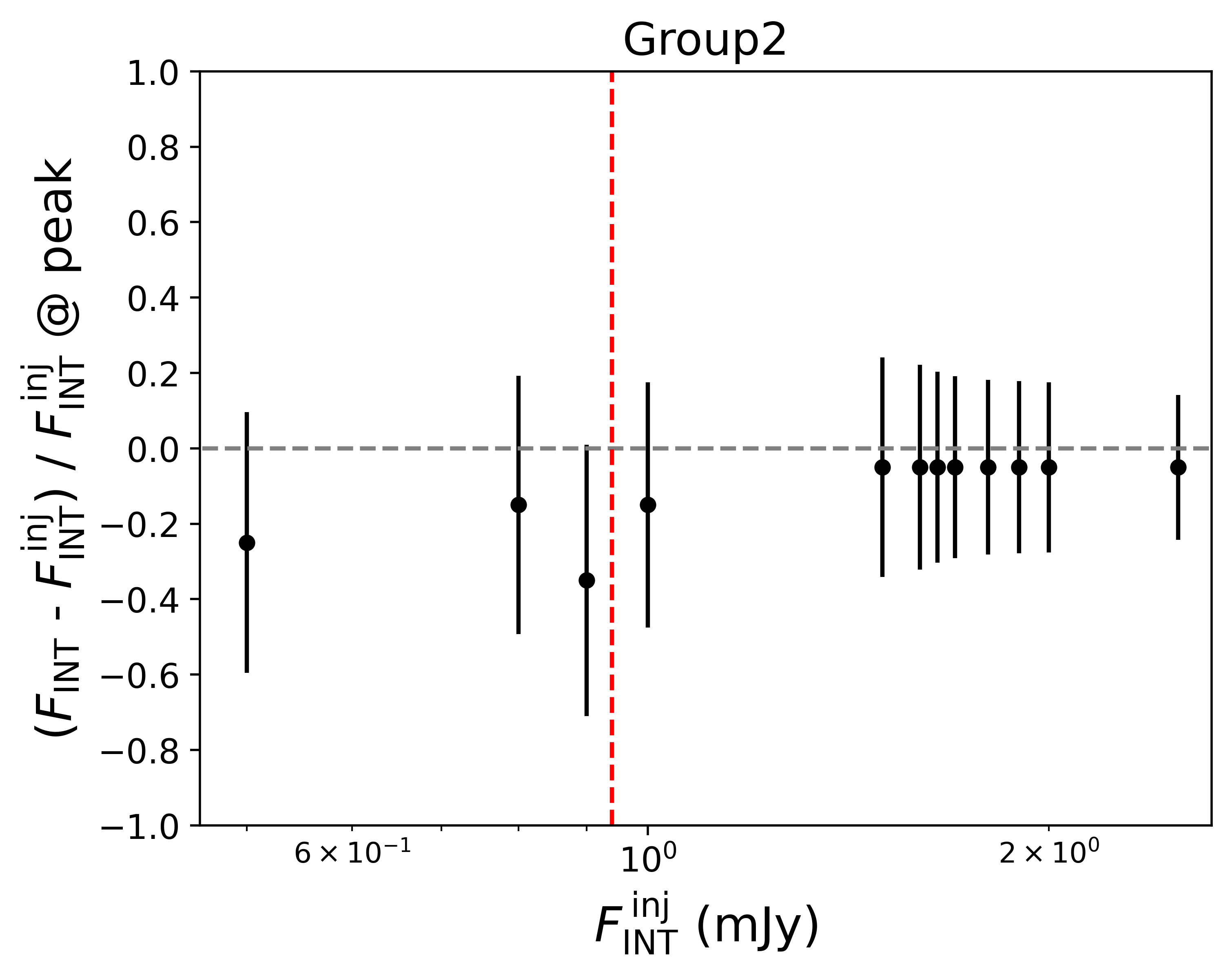}
        \includegraphics[width=0.68\columnwidth]{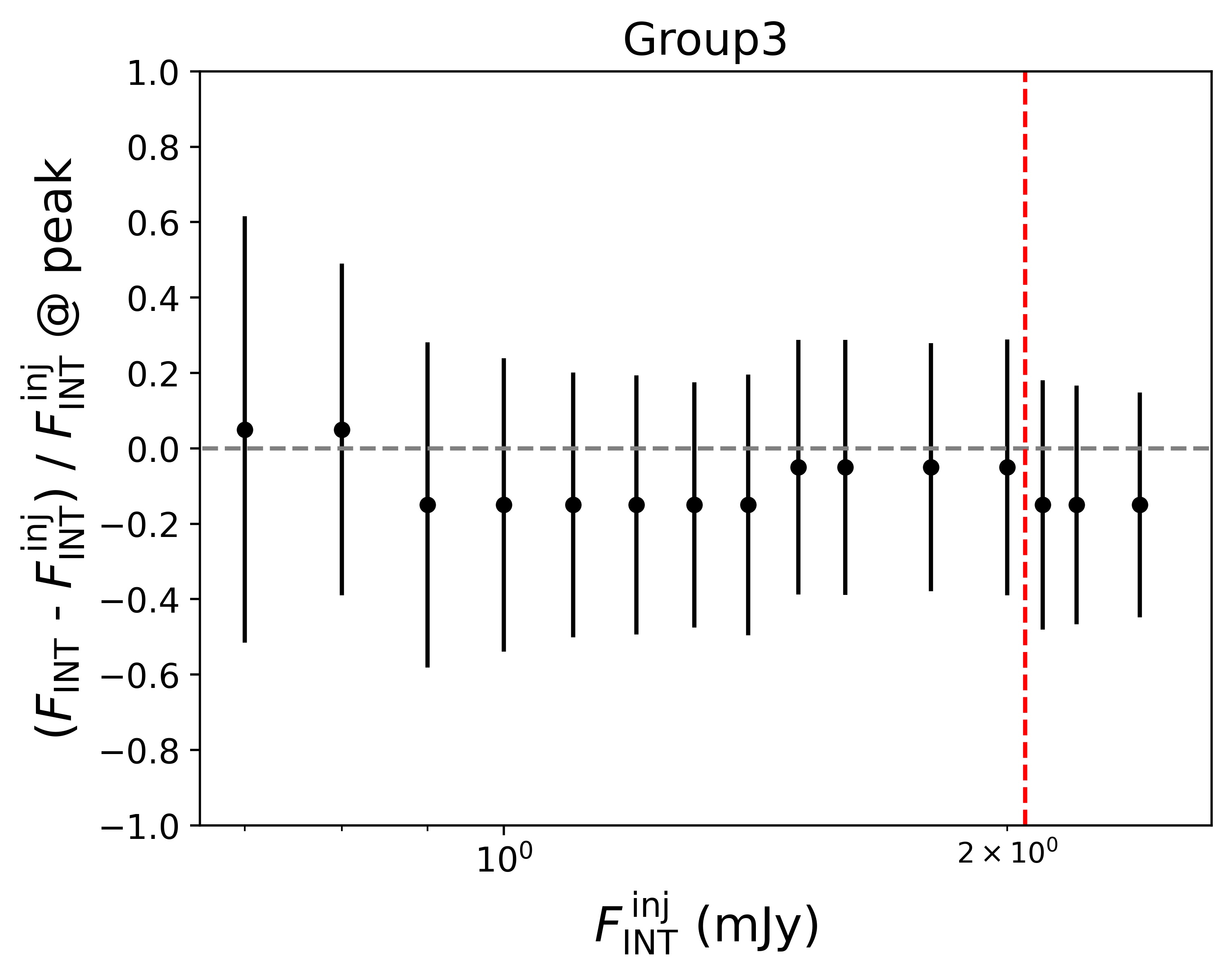}
        \caption{Photometric accuracy analysis: source integrated fluxes. Peak value (the mode) of the distribution of the relative discrepancy between measured ($F_{\rm{INT}}$) and simulated ($F_{\rm{INT}}^{\,\rm{inj}}$) integrated fluxes, as a function of the injected flux level $F_{\rm{INT}}^{\,\rm{inj}}$ (Group 1 left panel, Group 2 middle panel, Group 3 right panel). In all panels, the vertical error bars mark the interquartile range of the distribution within each level, while the vertical dashed red line reports the estimated flux completeness limit for the current group. The horizontal dashed grey line marks the equality between $F_{\rm{INT}}$ and $F_{\rm{INT}}^{\,\rm{inj}}$.} 
    \label{flux_acc_plots}

        \includegraphics[width=0.68\columnwidth]{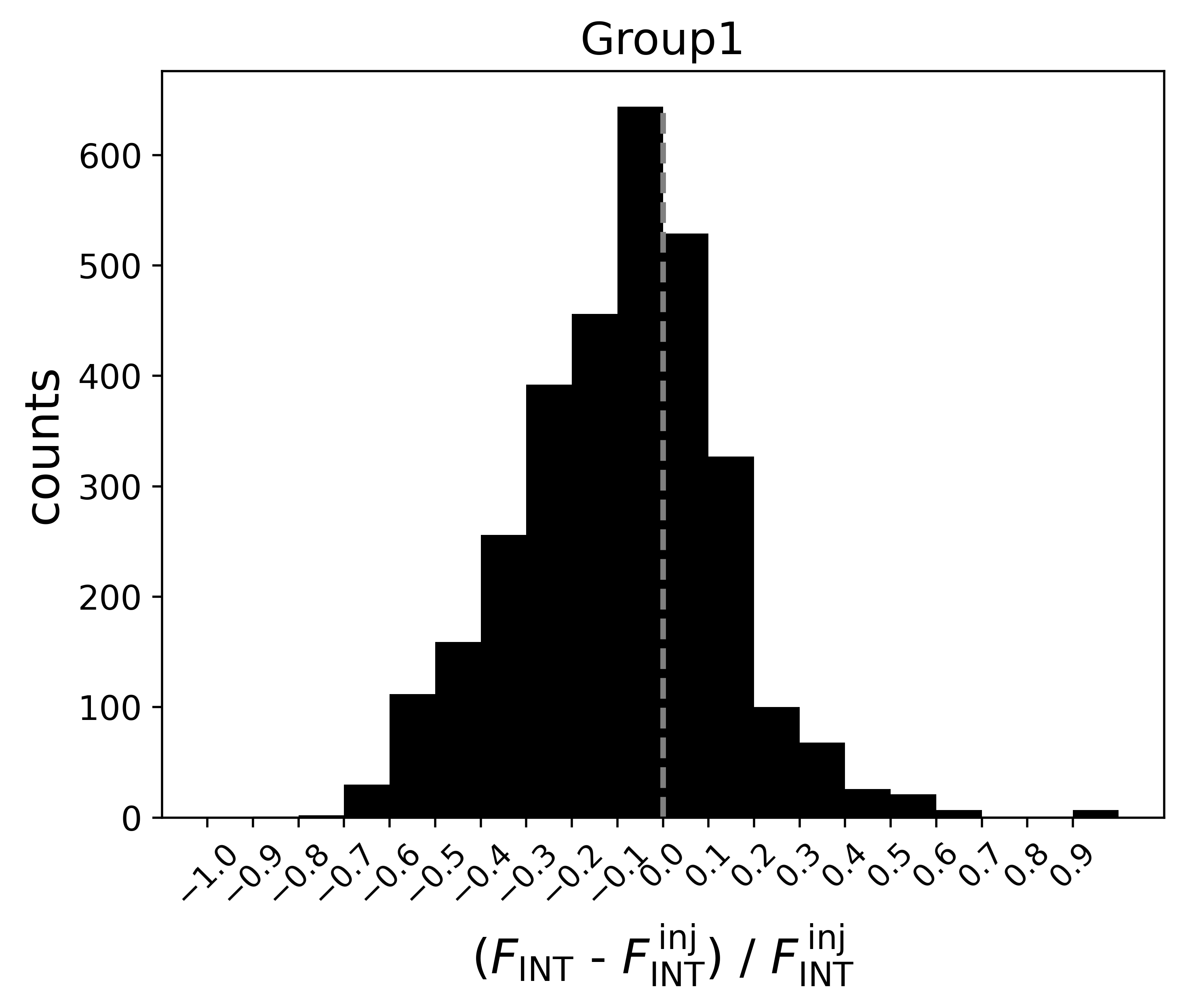}
        \includegraphics[width=0.68\columnwidth]{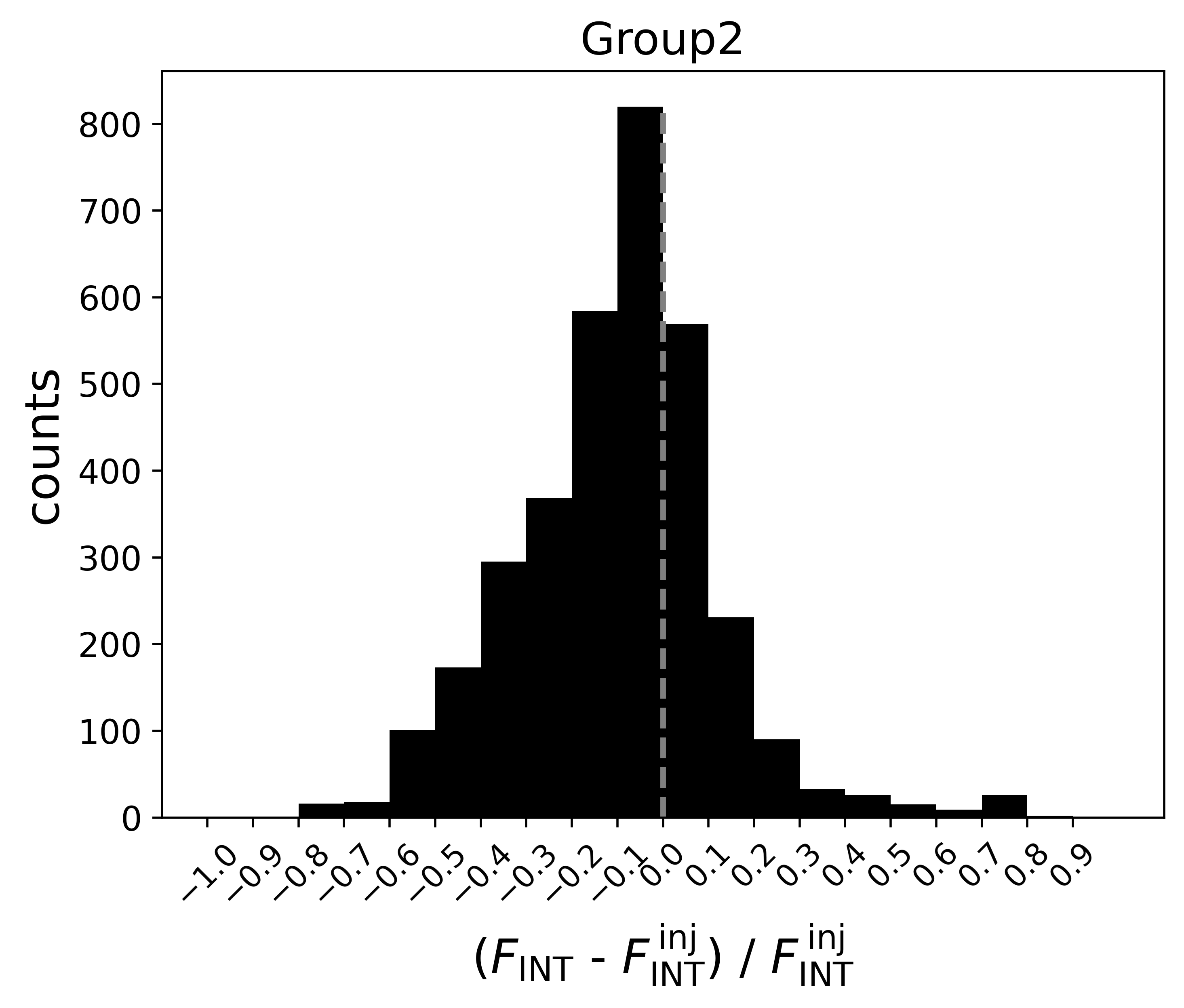}
        \includegraphics[width=0.68\columnwidth]{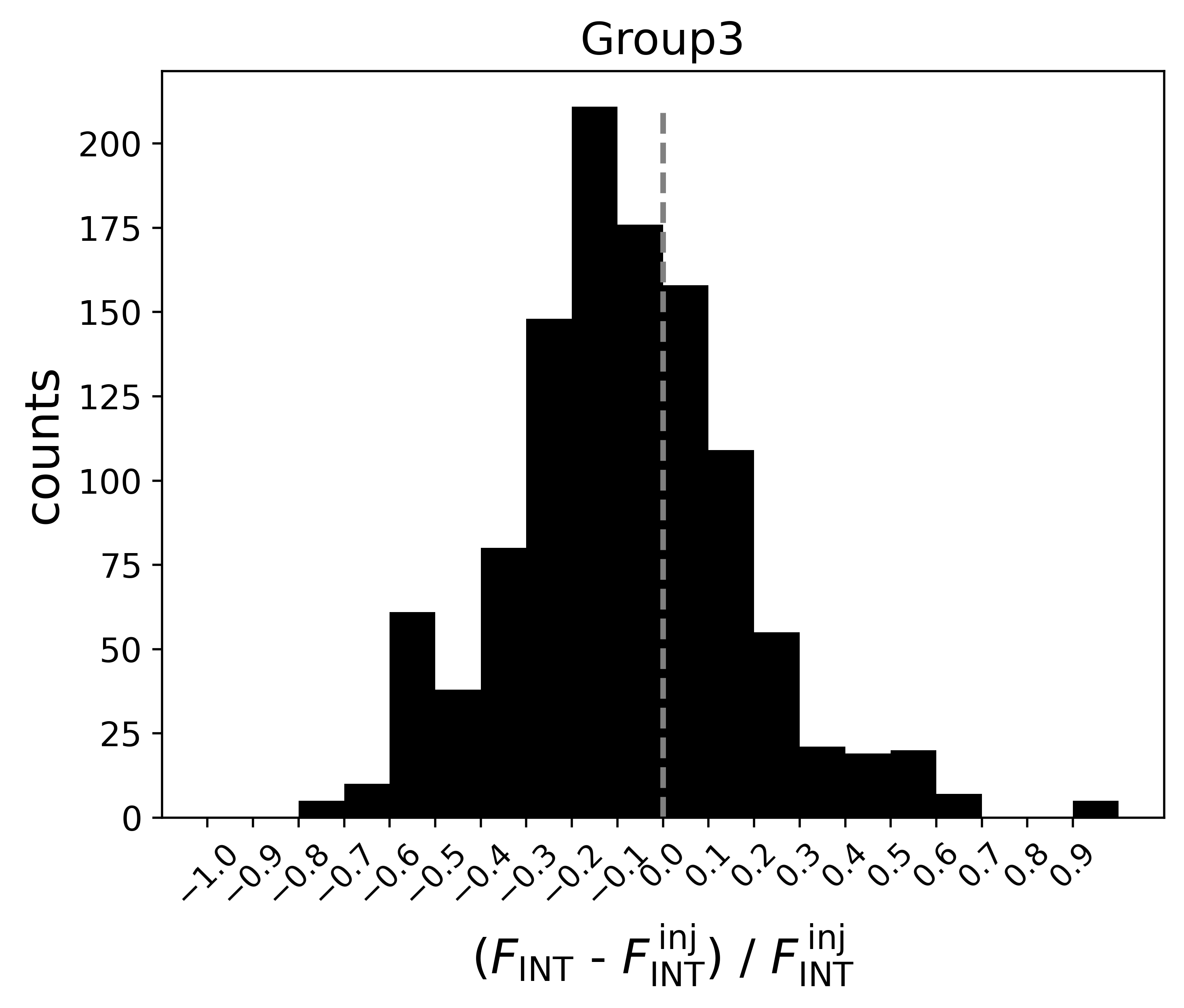}
        \caption{Photometric accuracy analysis: source integrated fluxes. Flux relative discrepancy distribution aggregating all flux levels above the completeness limit of the group (Group 1 left panel, Group 2 middle panel, Group 3 right panel). The vertical dashed grey line marks the equality between $F_{\rm{INT}}$ and $F_{\rm{INT}}^{\,\rm{inj}}$.}
    \label{flux_acc_hists}

        \includegraphics[width=0.68\columnwidth]{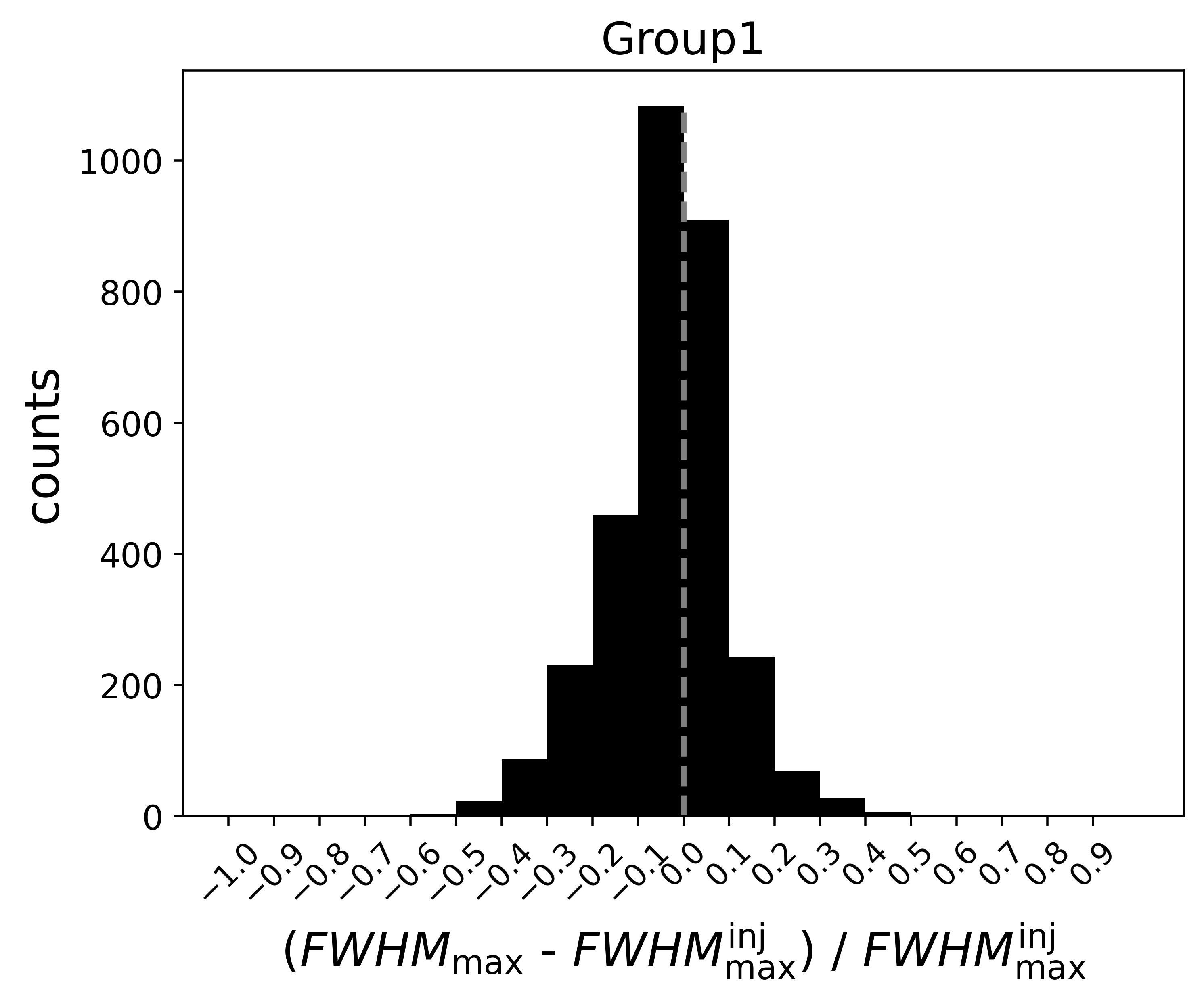}
        \includegraphics[width=0.68\columnwidth]{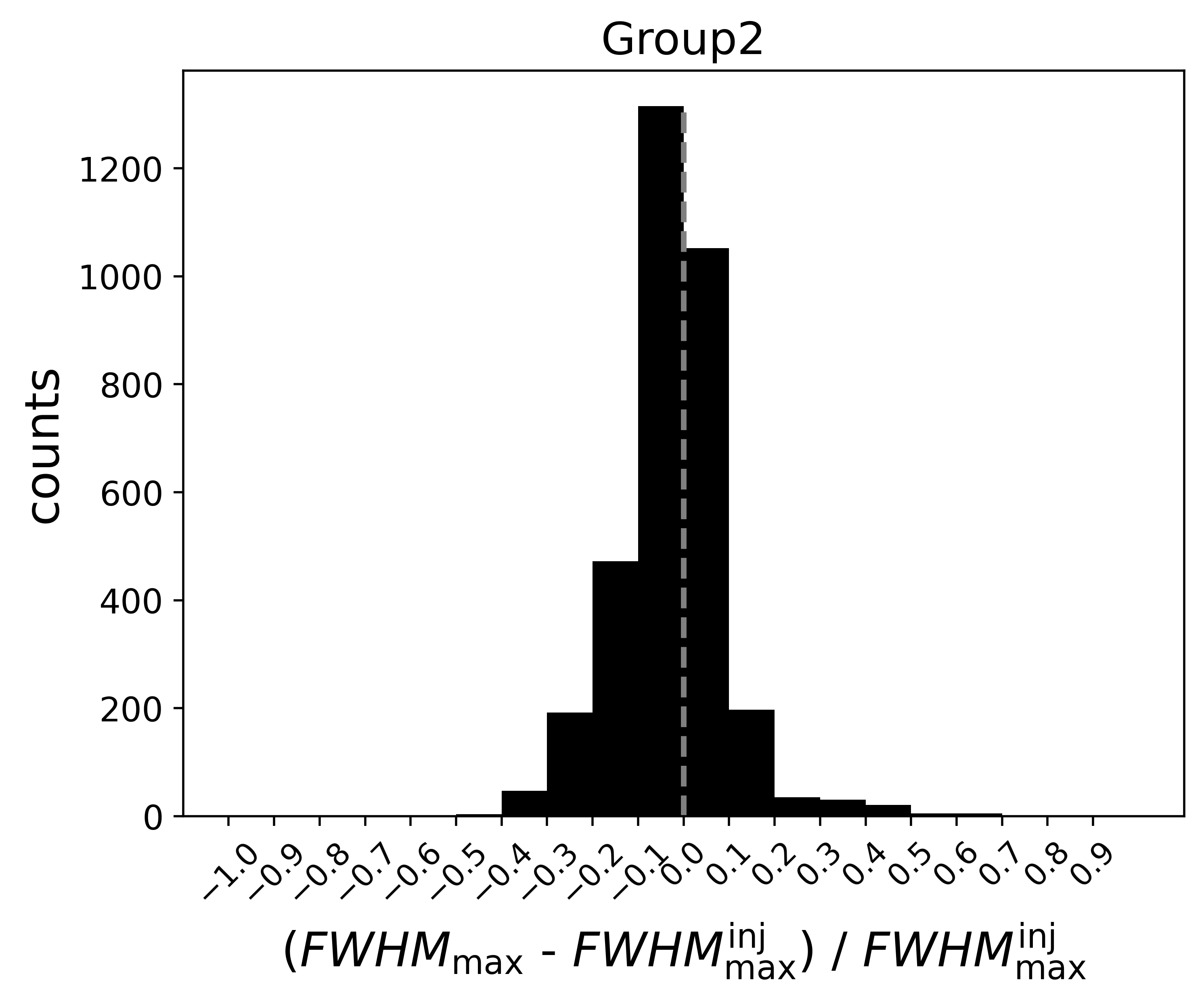}
        \includegraphics[width=0.67\columnwidth]{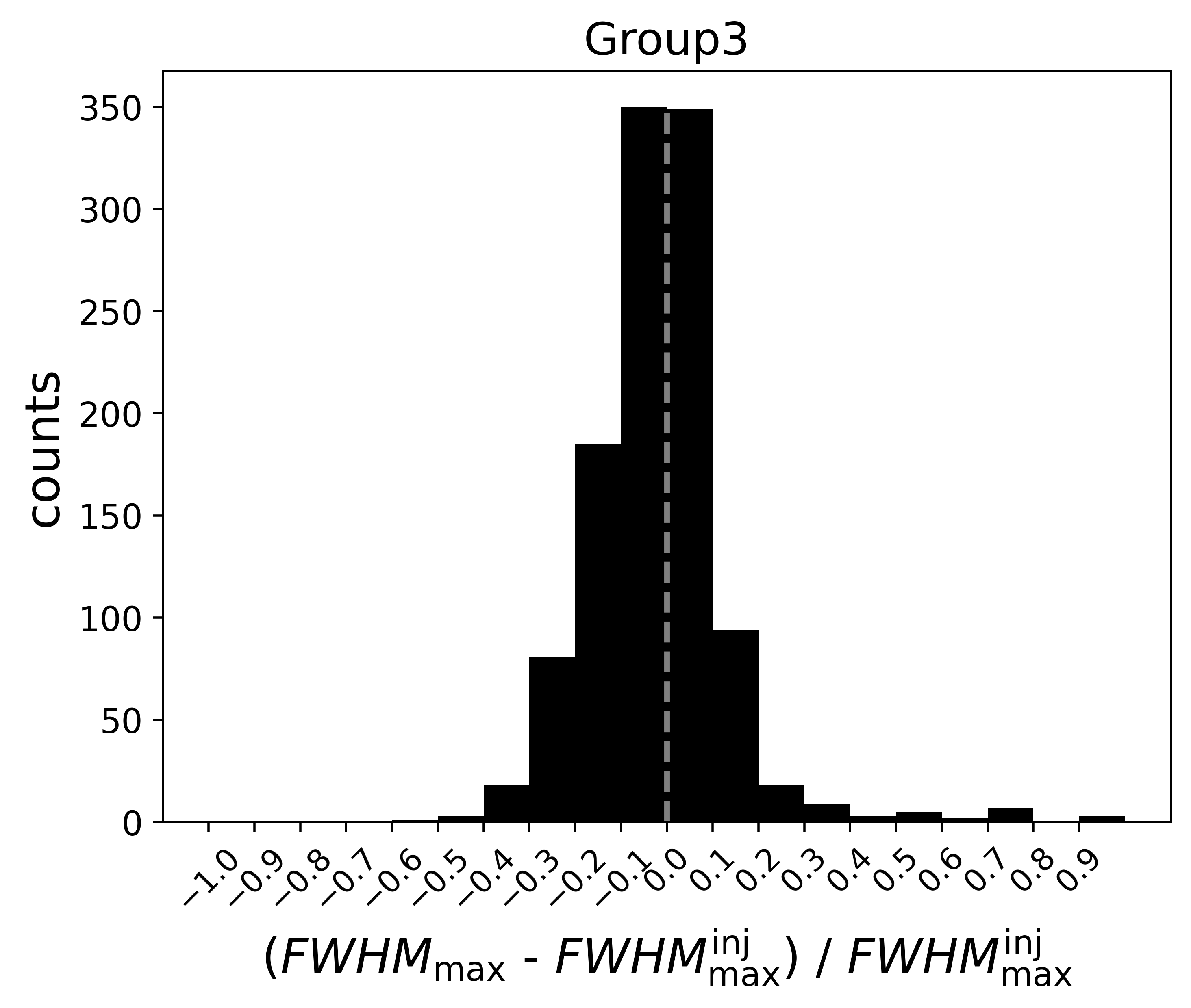}
        \includegraphics[width=0.68\columnwidth]{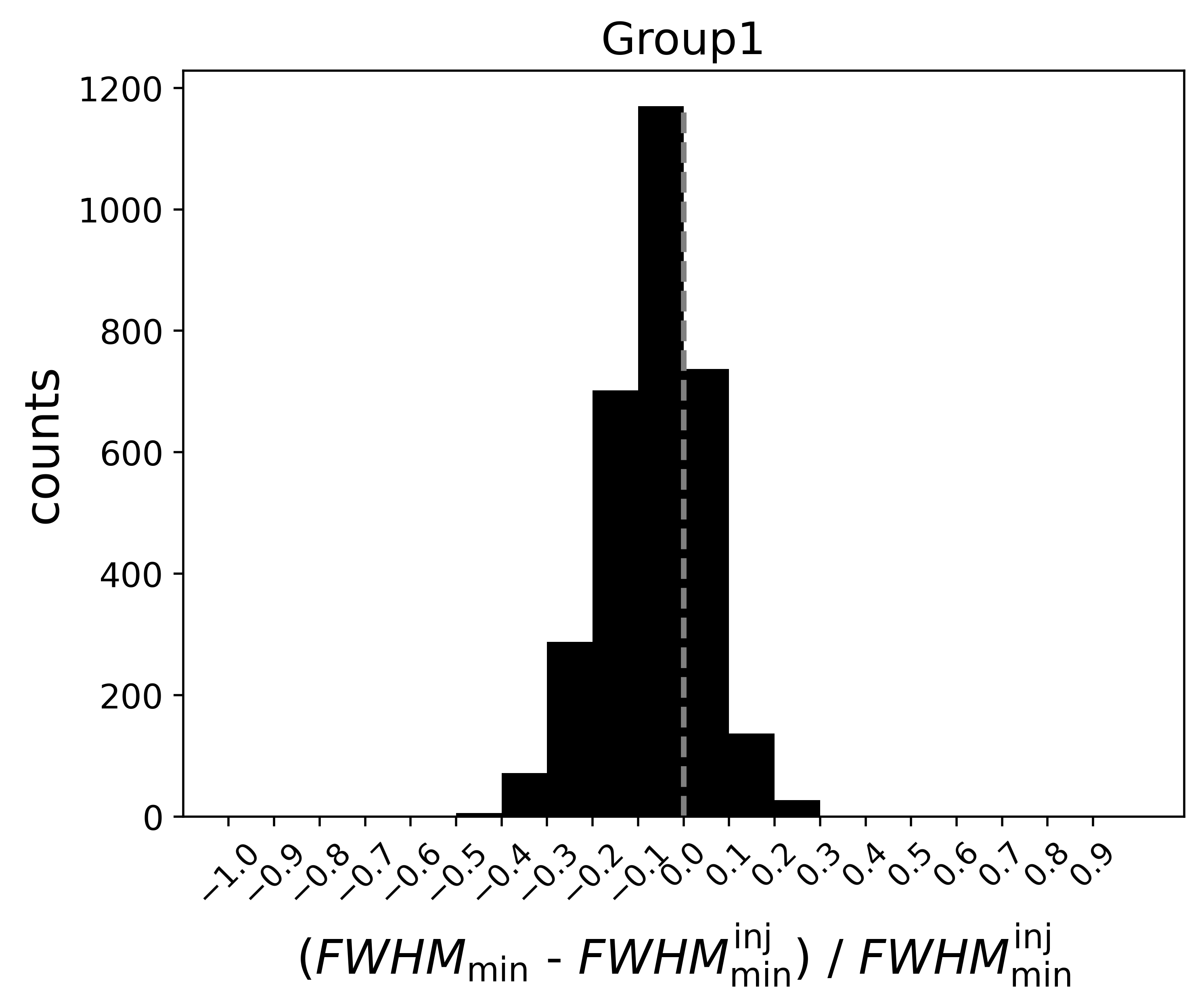}
        \includegraphics[width=0.68\columnwidth]{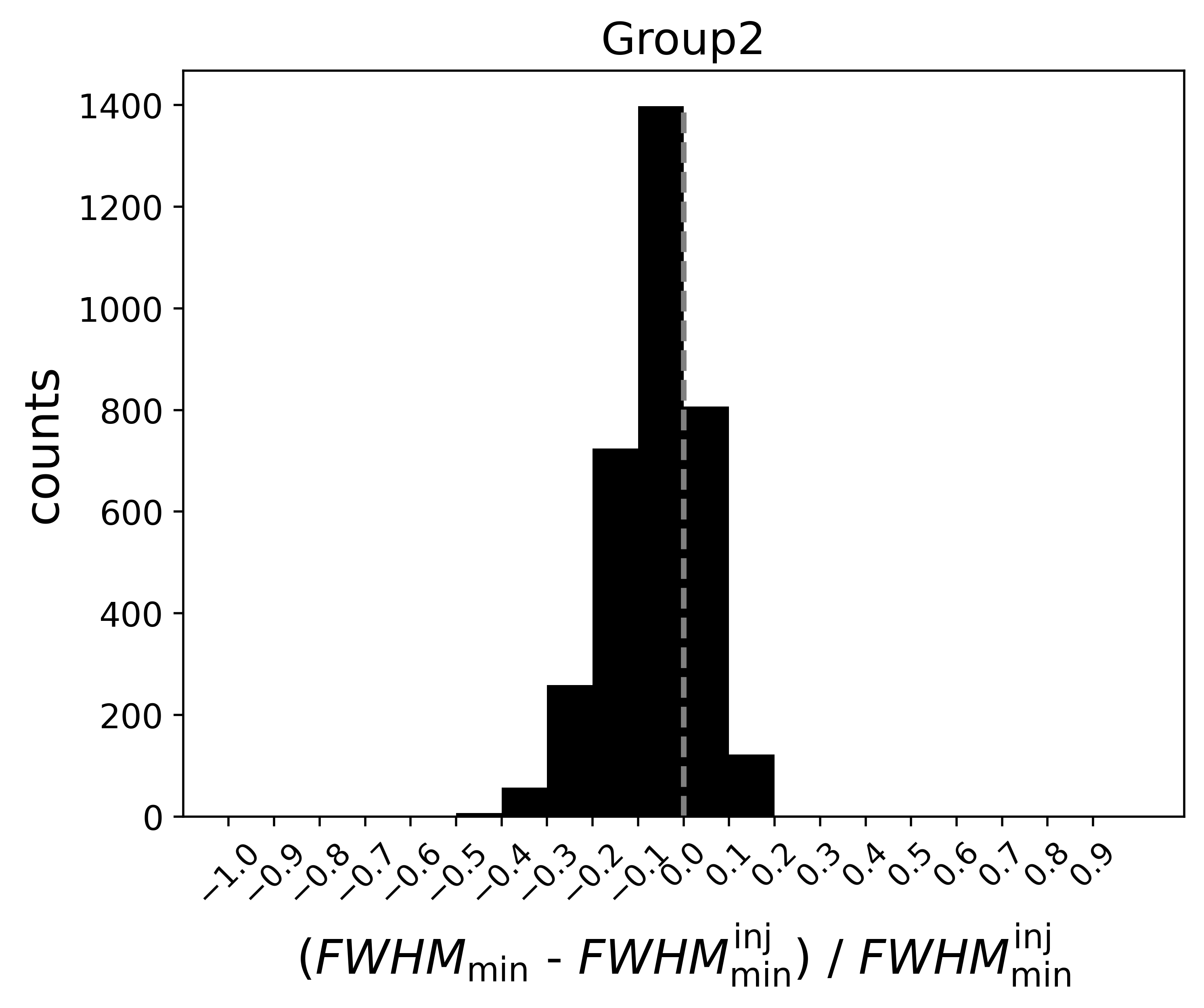}
        \includegraphics[width=0.67\columnwidth]{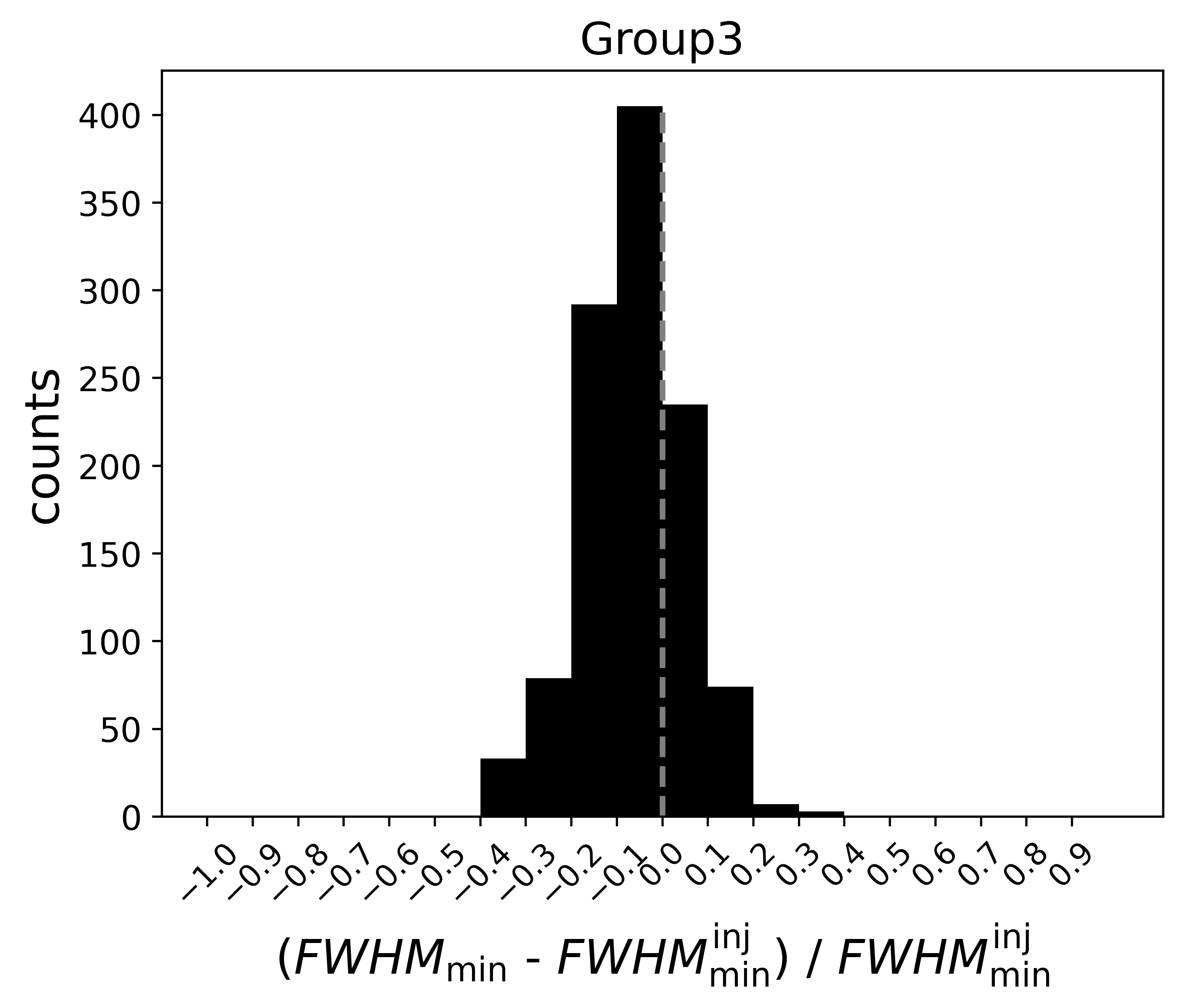}
        \caption{Photometric accuracy analysis: source angular sizes. Distribution of the relative discrepancy between measured ($FWHM$) and simulated ($FWHM^{\rm{inj}}$) sizes, in terms of major (top row) and minor (bottom row) axes of the detected source ellipse (Group 1 left panels, Group 2 middle panels, Group 3 right panels). As in Fig. \ref{flux_acc_hists}, only flux levels above the completeness limit are considered for each group. In all panels, the vertical dashed grey line marks the equality between $FWHM$ and $FWHM^{\rm{inj}}$.}
    \label{source_size_acc_hists}
    \end{figure*}

To characterize the photometric accuracy of our algorithm, we then went to compare, for the matched detections, the estimated integrated fluxes ($F_{\rm{INT}}$) with the generated ones, recording the relative discrepancy $(F_{\rm{INT}}-F_{\rm{INT}}^{\,\rm{inj}})/F_{\rm{INT}}^{\,\rm{inj}}$, for each of the simulated flux levels. 
Figure \ref{flux_acc_plots} shows the peak value (mode) of the flux discrepancy distribution as a function of the injected flux level, for the three groups of fields. 
Figure \ref{flux_acc_hists} instead reports the discrepancy distributions for the three groups aggregating all flux levels above the corresponding completeness limit. 
Recovered integrated fluxes are overall consistent with the simulations, presenting average values correct within $\sim20\%$ for all groups across nearly all the flux levels. 

Similarly, Figure \ref{source_size_acc_hists} shows the distributions of source angular size discrepancies between the generated and the recovered ones ($FWHM^{\rm{inj}}$) expressed in terms of major and minor axes of the source ellipse, still considering all flux levels above completeness. Again, recovered sizes are correct within $\sim20\%$ for all groups. \\

\pagebreak

\section{Further details of the adopted core temperature model}
\label{AppTcoremodel}

\begin{figure}
    \centering
    \includegraphics[width=\columnwidth]{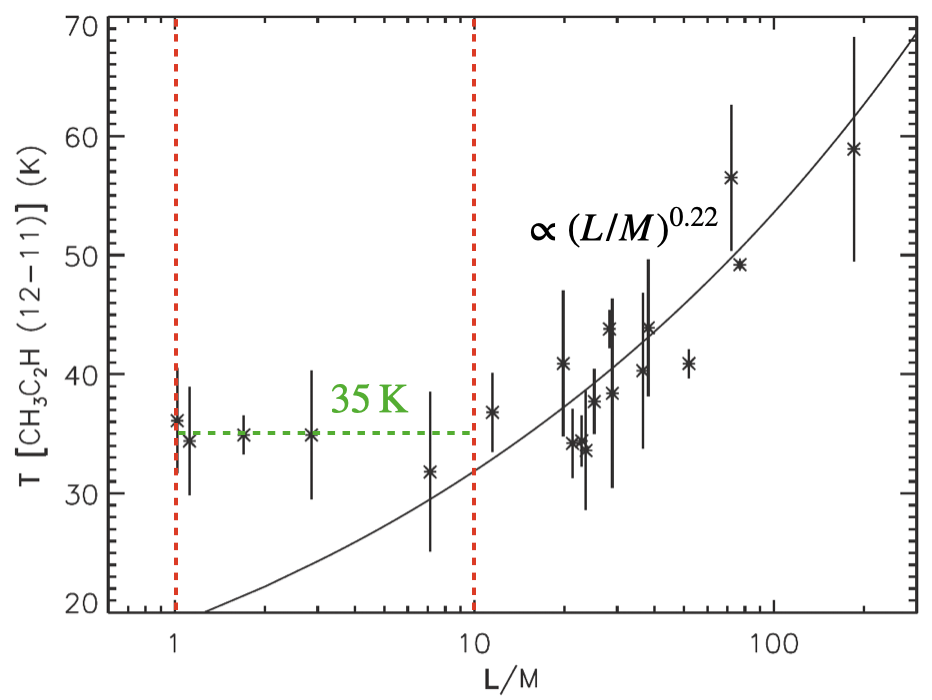}
    \includegraphics[width=\columnwidth]{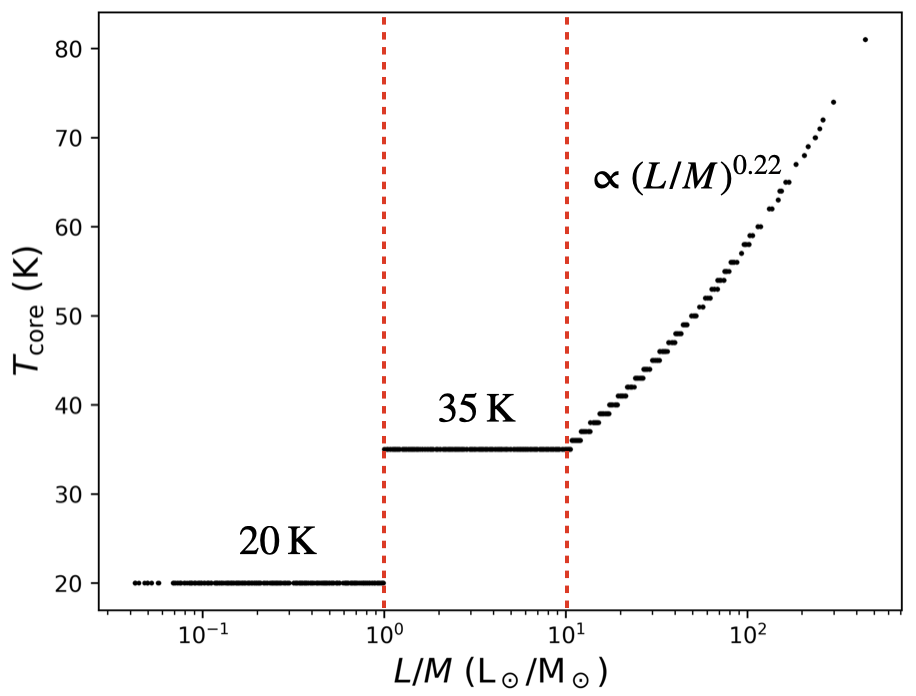}
    
    \caption{Details of the temperature model adopted for the cores. 
    \textit{Top panel}: Calibration of the clump $L/M$ with the rotational temperature of $\mathrm{CH_3C_2H}$, edited from \citet{Molinari+16}. The vertical dashed red lines mark different ranges of $L/M$ as done in our work. The horizontal dashed green line mark the $35$ K temperature value fitting the point within the $1-10\,\mathrm{L_{\odot}/M_{\odot}}$ range, which has been adopted in this work for the correspondent evolutionary group. The full black line represents the power law fitting the points at $L/M>10\,\mathrm{L_{\odot}/M_{\odot}}$ (see \citealt{Molinari+16}), also adopted in this work. \textit{Bottom panel}: Overall distribution of the assumed core temperatures as a function of the clump $L/M$, resulting from the model used in this work. The vertical dashed red lines serve as in the top panel.}
    \label{Tcore_model_plots}
\end{figure}

The top panel of Fig. \ref{Tcore_model_plots} shows the calibration of the clump $L/M$ with the rotational temperature of the high-density tracer $\mathrm{CH_3C_2H}$ performed by \citet{Molinari+16}, on which we based our core temperature model (see Sect. \ref{Tcore_model}). Their study covers the range $\sim1-200\,\mathrm{L_{\odot}/M_{\odot}}$ in $L/M$, so we took $35$ K as the temperature value fitting the points within $1-10\,\mathrm{L_{\odot}/M_{\odot}}$ for the corresponding evolutionary group, and applied the power-law relation $T_{\rm{core}}\,\mathrm{(K)}=21.1\cdot (L/M){^{0.22}}$, representing the curve that fits the points at $L/M>10\,\mathrm{L_{\odot}/M_{\odot}}$, for the more evolved sources. For the $L/M\leq1\,\mathrm{L_{\odot}/M_{\odot}}$ range, not covered by the \citet{Molinari+16} model, we assumed a typical temperature of $20$ K (see Sect. \ref{Tcore_model}). 

The distribution of core temperature resulting from our assumptions is shown in the bottom panel of Fig. \ref{Tcore_model_plots}, and ranges between $20$ and $81$ K. 


\clearpage

\section{Analysis of potential free-free emission contamination on ALMAGAL cores}
\label{Appff_analysis}

Potential flux contamination from free-free continuum emission on detected ALMAGAL cores is to some extent expected in the ALMAGAL sample, given the very wide range of evolutionary stages that also includes evolved regions ($L/M$ up to $500\,\mathrm{L_{\odot}/M_{\odot}}$). 
Such clumps might in fact host HII regions, which are typically characterized by strong continuum emission at radio frequencies. This might bring significant contribution to the integrated fluxes of the sources that we measure on our maps (see, e.g., \citealt{Liu+22b,Pouteau+22}), ultimately leading to potentially overestimated core masses. 

\begin{figure}[h!]
    \centering
    
    \vspace{-1mm}
    \includegraphics[width=0.4\textwidth]{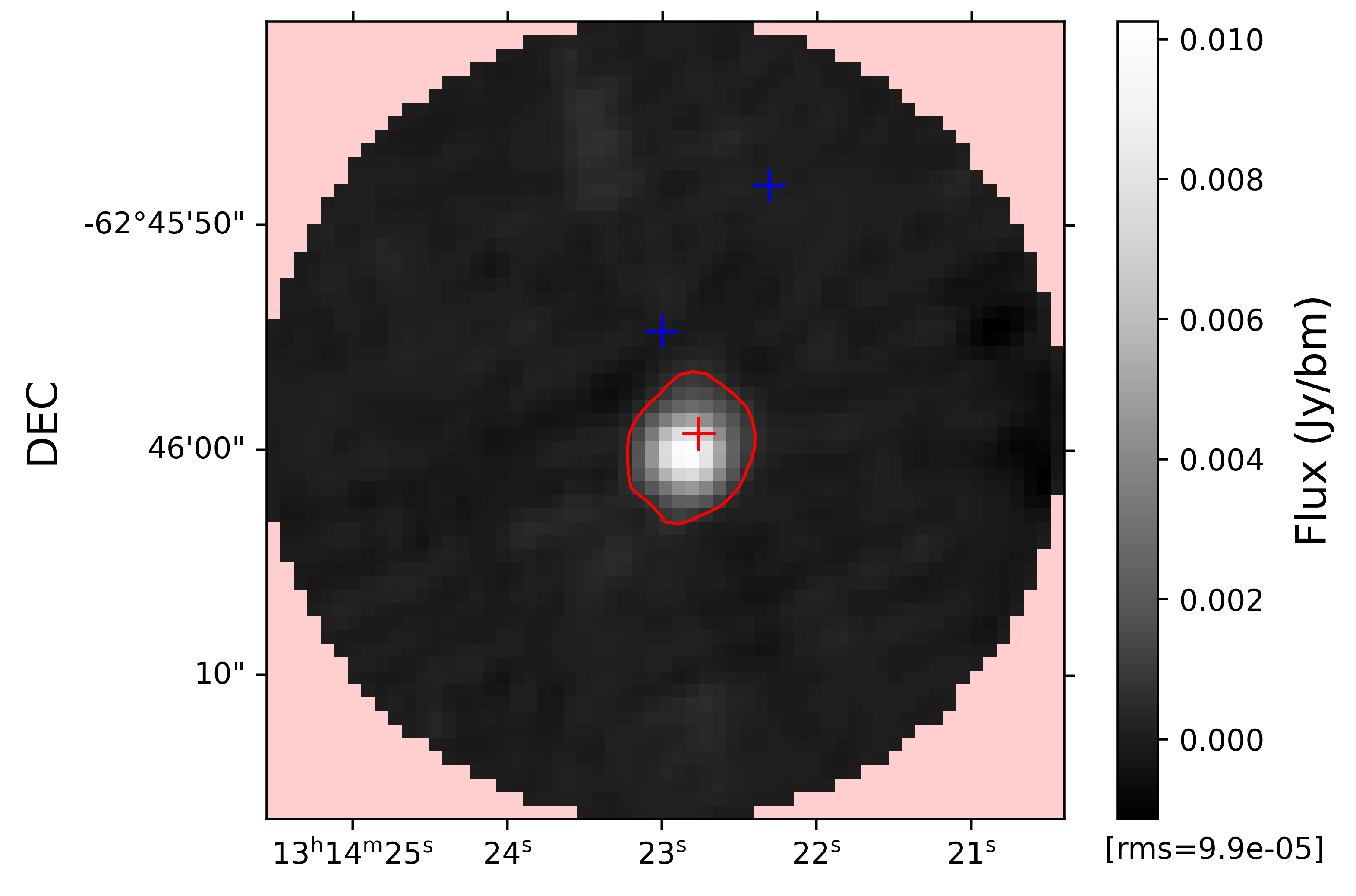}
    \vspace{-1mm}
    \includegraphics[width=0.4\textwidth]{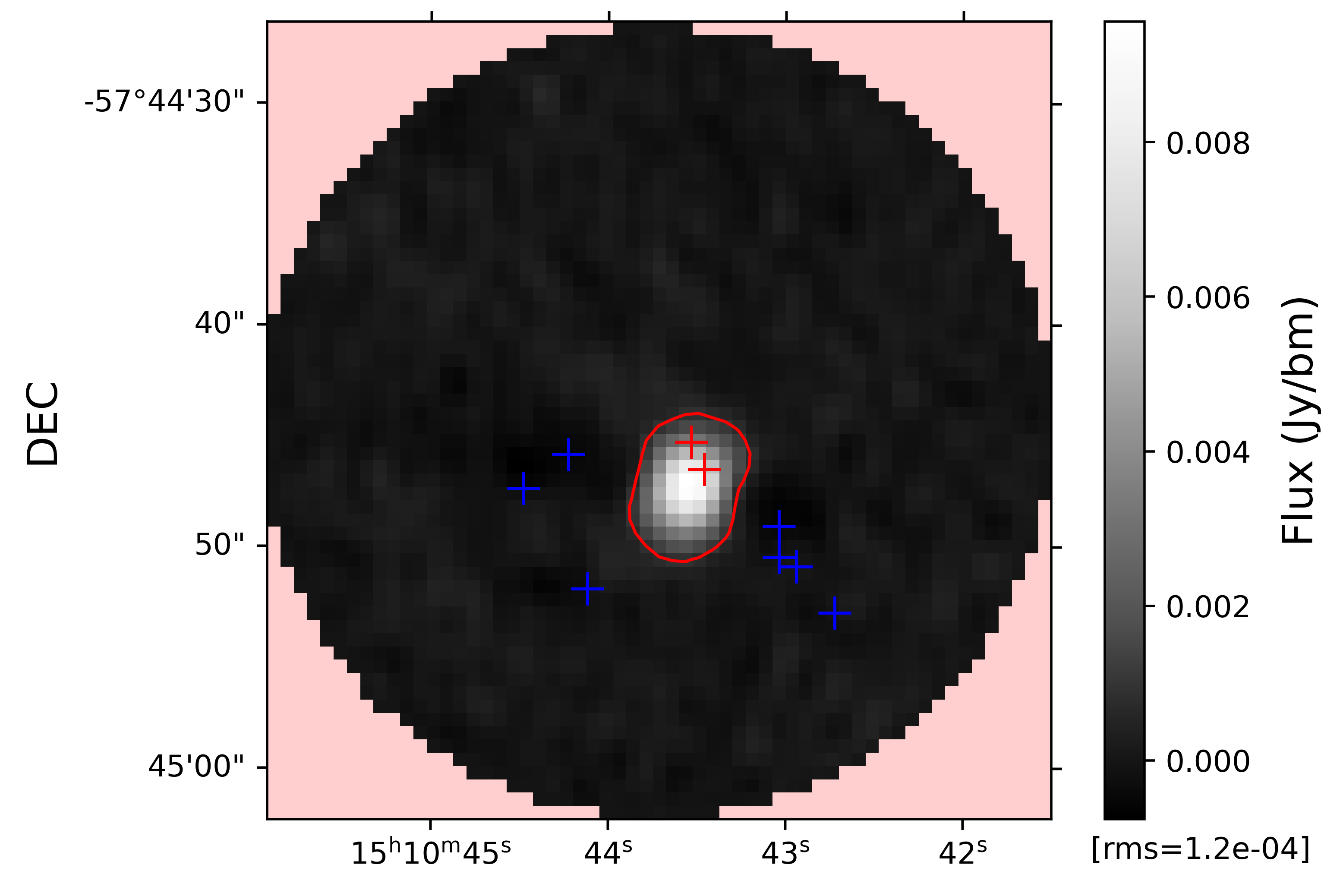}
    \vspace{-1mm}
    \includegraphics[width=0.4\textwidth]{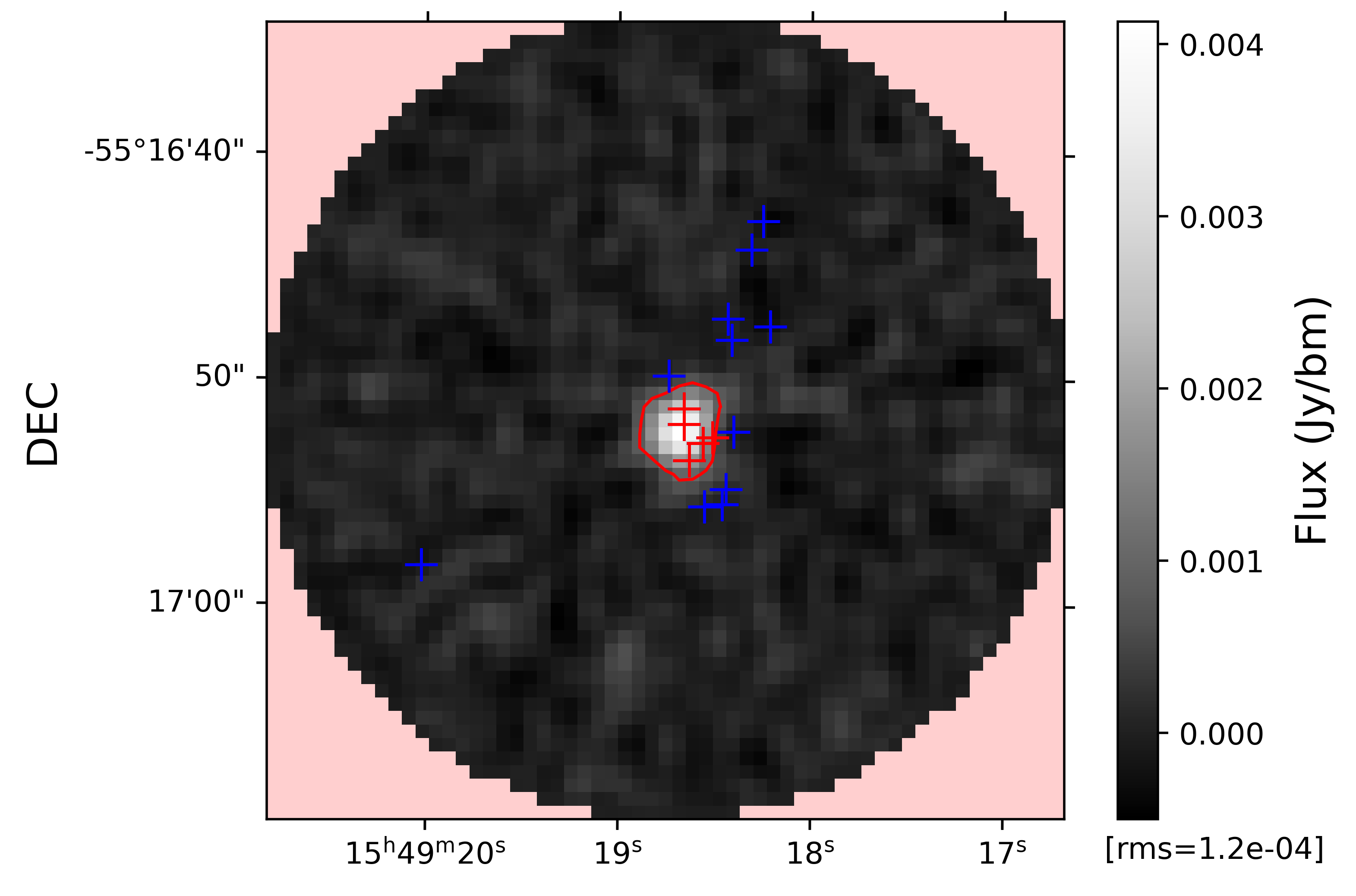}
    \vspace{-1mm}
    \includegraphics[width=0.4\textwidth]{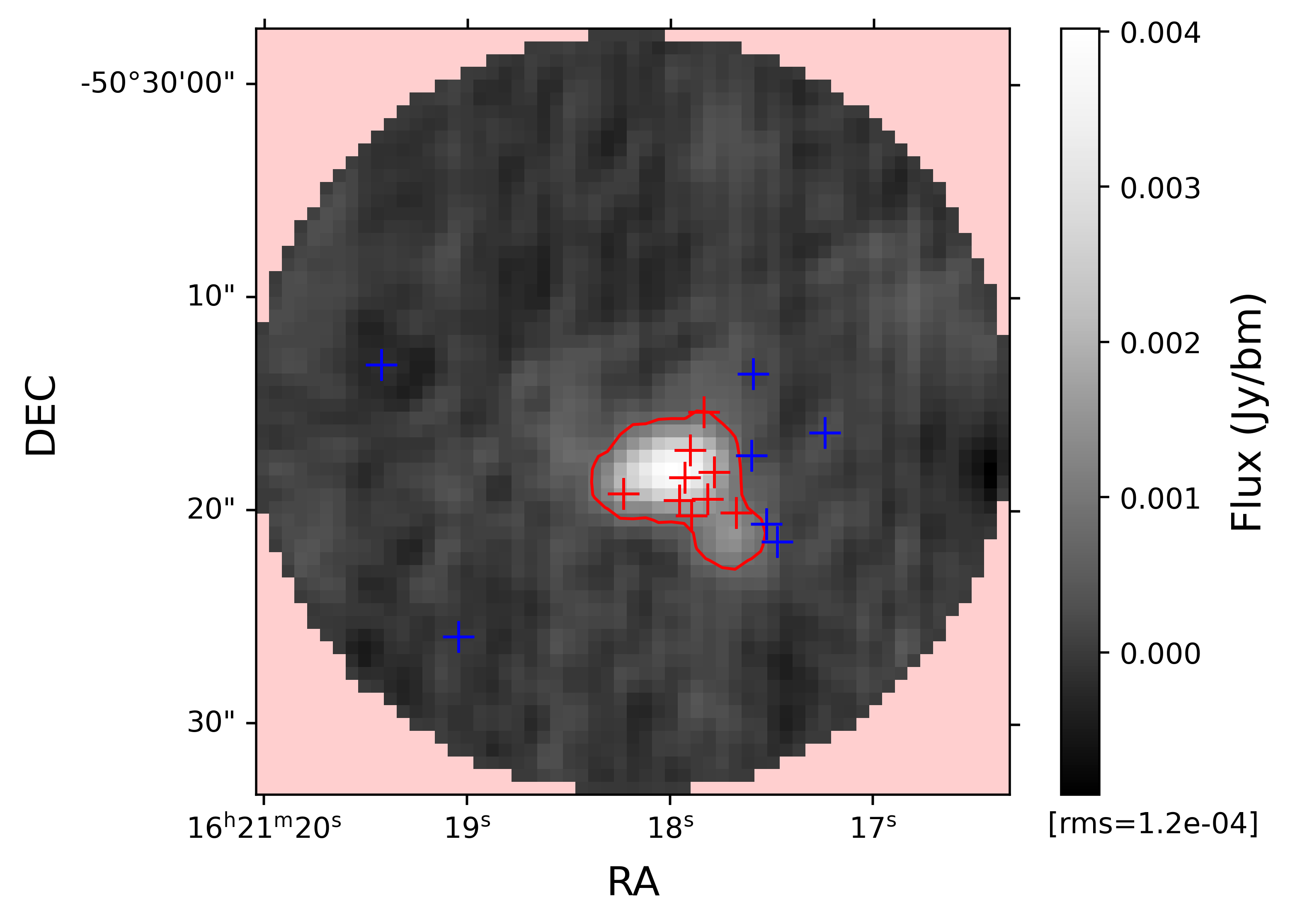}
    
    \caption{Cutouts of CORNISH/CORNISH-South radio continuum emission maps ($5$ GHz frequency, $\sim1''$ angular resolution) corresponding to the ALMAGAL fields (from top to bottom): AG305.5535-0.0114 (first panel), AG320.6753+0.2459 (second panel), AG326.4486-0.7484 (third panel), and AG333.2217-0.4020 (fourth panel). Red contour levels indicate $7\,\sigma_{\rm{rms}}^{\rm{rad}}$ emission regions. Red crosses mark the locations of the ALMAGAL cores that fell within such radio emission contours and are thus considered as potentially contaminated, whereas blue crosses mark the unmatched ALMAGAL cores.}
    \label{ff_analysis_example}
\end{figure}

To evaluate this occurrence, we searched for counterparts of ALMAGAL fields in the CORNISH (COordinated Radio and Infrared Survey for High-mass star formation, \citealt{Purcell+08,Hoare+12}) and CORNISH-South (\citealt{Irabor+23}) surveys, providing radio continuum maps at $5$ GHz (or $6$ cm) with $\sim1''$ angular resolution, obtained with the VLA. We found $894$ correspondences ($88\%$ of the ALMAGAL target sample), for which we produced cutouts matching the ALMAGAL FOV ($\sim35''$). 
We computed the rms noise of the radio continuum maps ($\sigma_{\rm{rms}}^{\rm{rad}}$) through a sigma clipping procedure. In some regions, such estimation is not straightforward, since the noise can vary due to the presence of artifacts. In those cases, we visually inspected the maps to verify that the automatic estimation was reliable. We then generated flux contour levels at $3$, $5$, and $7$ times the $\sigma_{\rm{rms}}^{\rm{rad}}$. 
Based on visual inspection, the latter ones proved to be the most realistic, i.e., to best represent the observed emission pattern on the maps, whereas lower levels included regions that did not seem to be really significant. Moreover, a stricter flux threshold leads to more reliable results in the source matching, reducing the impact of potential spurious correlations induced by the difficult estimation of the map noise. 
The clumps showing the presence of strong radio emission (i.e., above the $7\,\sigma_{\rm{rms}}^{\rm{rad}}$ threshold) were $192$ ($21\%$ of the inspected maps) and are flagged in Col. $17$ of Table 3. 
The coordinates of the ALMAGAL cores (Table 3) were matched with the $7\,\sigma_{\rm{rms}}^{\rm{rad}}$ radio contours allowing a tolerance of $1$ radio continuum map pixel. 
Figure \ref{ff_analysis_example} shows some examples of the analysis conducted. 
We noted the cores that fell within such contours, which we consider to be potentially significantly contaminated in their flux by radio continuum emission. 
The clumps containing at least $1$ matched core were $110$ ($11\%$ of the ALMAGAL sample), combining for a total of $909$ matched cores ($14\%$ of the ALMAGAL core catalog). Considering also the other unmatched cores pertaining to such clumps, we get a total of $1793$ cores ($28\%$). We underline that, as actually expected due to the nature of the regions, $\sim90\%$ of matched cores belong to the more evolved clumps (i.e., with $L/M>10\,\mathrm{L_{\odot}/M_{\odot}}$). To avoid contamination on the results coming from flux and mass overestimations, those cores were excluded from the physical analysis of Sects. \ref{CMF_evol} and \ref{correlations}. Furthermore, we also excluded cores detected within clumps for which no CORNISH or CORNISH-South counterpart was available for the above described analysis ($582$ cores or $9\%$ of the catalog, coming from $119$ clumps). We remark that this represents a rather conservative choice since, based on the above reported numbers, only a small fraction of such cores will actually be potentially contaminated. More generally, we emphasize the use of the term "potentially" throughout, since it is still possible that some of the positionally matched cores might actually not be significantly contaminated by the presence of local free-free emission, as well as some not matched nearby cores might be. 

Overall, this analysis led to the identification of $1491$ potentially contaminated cores ($\sim23\%$ of the catalog).

\newpage
\clearpage
\twocolumn

\section{Further analysis on the effect of resolution and temperature on the observed trend of the CMF with evolution}
\label{AppCMFevoltests}

In this appendix we report details of the further tests and analysis that were conducted to verify and reinforce the result of the evolution of the observed ALMAGAL CMF, discussed in Sect. \ref{CMF_evol}. 
In particular, we inspect the potential influence of differences in resolution and of the temperature assumptions on determining the evolutionary trend shown in Fig. \ref{CMF_evolution}. 

\begin{figure}[h!]
    \centering
    \includegraphics[width=0.96\columnwidth]{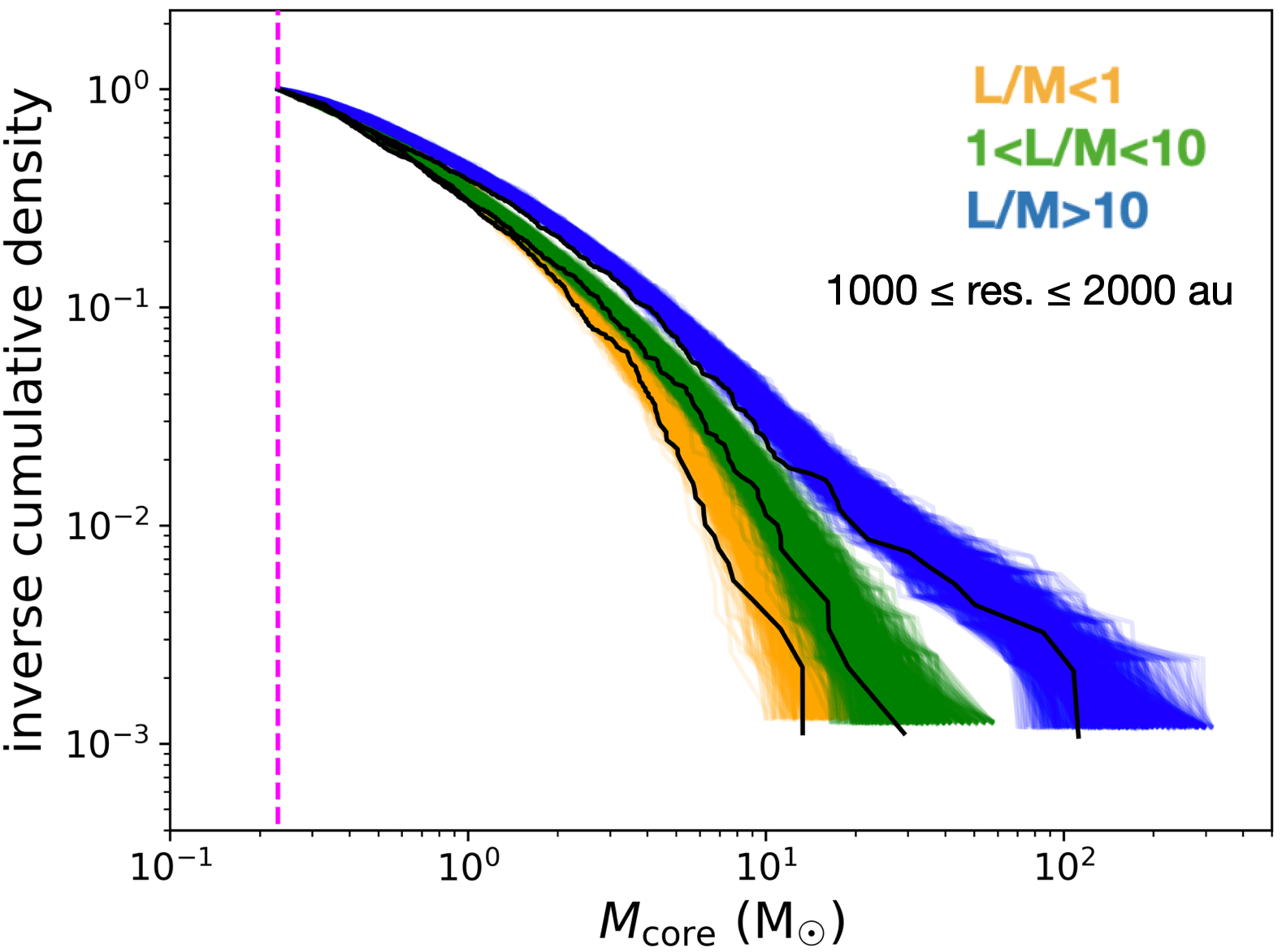}
    \includegraphics[width=0.96\columnwidth]{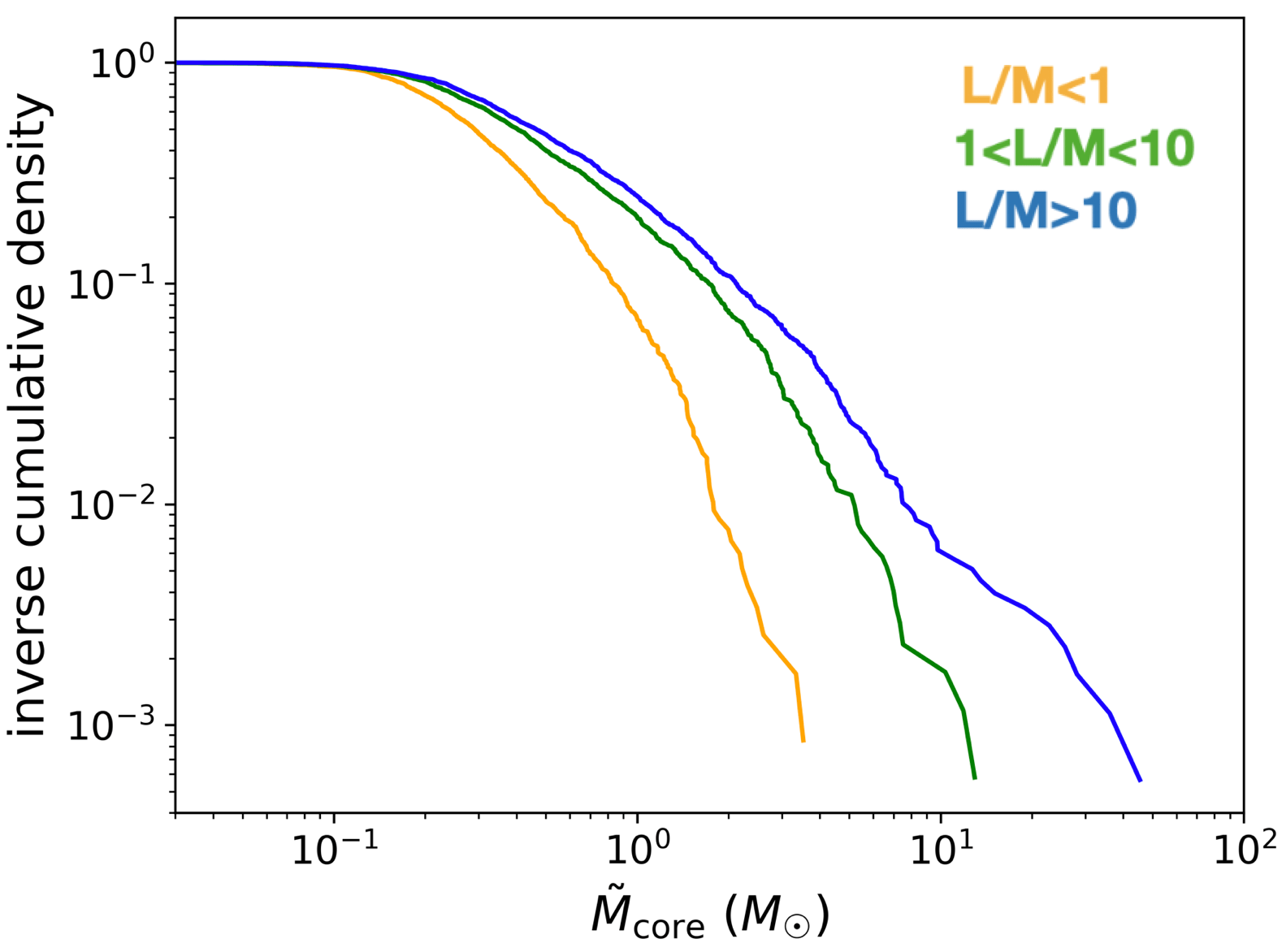}
    \includegraphics[width=0.96\columnwidth]{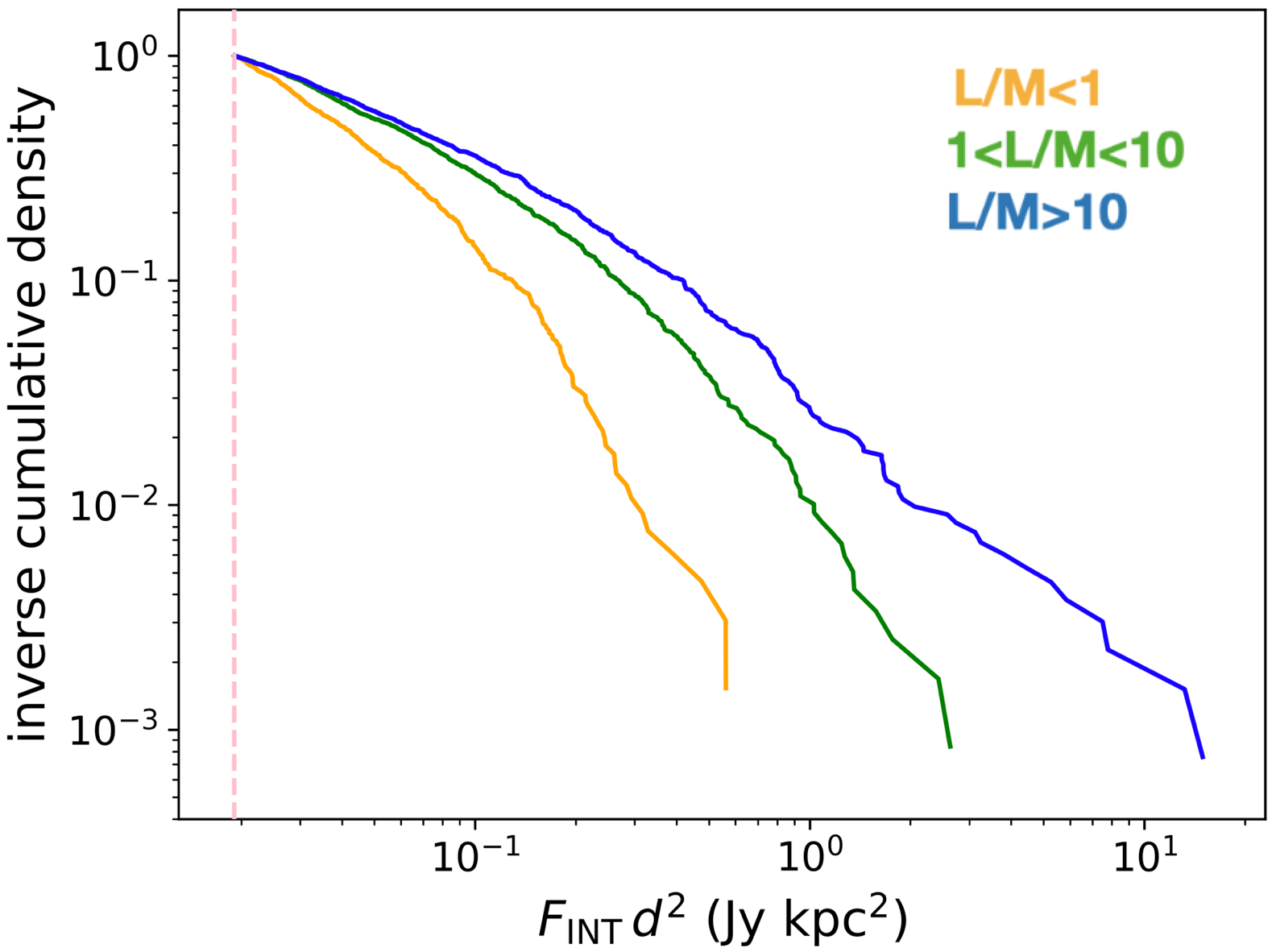}
    
    \caption{Analysis conducted to test the result of Fig. \ref{CMF_evolution}. \textit{Top panel}: Same as Fig. \ref{CMF_evolution}, but for a subsample of cores selected based on the spatial resolution of the corresponding continuum map (within $1000-2000$ au). \textit{Middle panel}: CMF for the three evolutionary groups with core masses obtained using the temperature model from \citet{Sadaghiani+20} (see text for explanation). \textit{Bottom panel}: Distribution of the $F_\mathrm{INT}d^2$ quantity for the three evolutionary groups, computed to obtained an analog of the CMF but without the influence of the core temperature assumption. The vertical dashed pink line marks the minimum threshold we imposed for the plotted quantity, calculated by converting our mass completeness limit.}
    \label{CMF_evolution_tests}
\end{figure}

The spatial resolution of the observations can have a significant effect on the measured masses of the detected objects and thus on the shape of the CMF (see, e.g., \citealt{Sadaghiani+20,Louvet+21}). 
Although the distribution of the linear sizes of the beam is rather uniform across ALMAGAL maps (right panel of Fig. \ref{beam_circ_props_plots}), the resolution can still vary in some cases by a factor up to $\sim3$ with respect to the median value of $\sim1400$ au. This might influence the shape of the CMF, especially in the higher-mass end where the tail can be driven by relatively few objects (e.g., unresolved resolution outliers). In principle, this should not be related to the evolutionary stage of the observed clump. We nevertheless evaluated the evolutionary trend of Fig. \ref{CMF_evolution} for a more homogeneous subsample of cores observed with resolution between $1000$ and $2000$ au (thus within a factor $2$ of discrepancy at most). Such range actually includes most of the ALMAGAL targets (see Fig. \ref{beam_size_dist_plot}). The outcome of this is shown in Fig. \ref{CMF_evolution_tests} (top panel). 
As for Fig. \ref{CMF_evolution}, we consider only cores above the $0.23\,\mathrm{M_{\odot}}$ mass completeness limit. The evolutionary trend holds, with the more evolved profile actually appearing here to be more shifted with respect to the other two groups. We then consider the result to be not significantly affected by potential resolution effects across our sample. \\
To further evaluate the impact of temperature, we first applied an alternative model taken from \citet{Sadaghiani+20}, for which the core temperature is assumed to be proportional to its measured flux through a power law, so that
    \begin{align}
    \label{Tcore_sad_eq}
    \tilde{T}_{\rm{core}}=T_{\rm{core}}^{\rm{min}}\cdot\Big(\frac{F_\mathrm{PEAK}}{F_\mathrm{PEAK}^{\rm{min}}}\Big)^\gamma\;,
    \end{align}
where $T_{\rm{core}}^{\rm{min}}$ is the minimum assumed core temperature ($20$ K in our case), $F_\mathrm{PEAK}$ the measured core peak flux (Table 3), and $F_\mathrm{PEAK}^{\rm{min}}$ its minimum value. A power law exponent $\gamma=0.238$ is used from \citet{Sadaghiani+20} (found to reproduce the Salpeter IMF slope of $-1.35$, see \citealt{Salpeter55}). 
In this way, we are waiving our assumption of a single temperature for all cores within the same clump. 
We obtain a temperature range of $20-137$ K, and a range of corresponding masses (computed with the same Eq. \ref{Mcore_eq}) $\tilde{M}_{\mathrm{core}}\simeq0.02-73\,\mathrm{M_{\odot}}$. We evaluated the evolutionary trend, which is shown in Fig. \ref{CMF_evolution_tests} (middle panel). Again, the shift among the three CMF profiles is clearly present. 

As a further, conclusive test, we analyzed the behavior of the $F_\mathrm{INT}d^2$ quantity, thus basically considering a reduced version of the core mass expression (Eq. \ref{Mcore_eq}) in which the temperature term is removed. The result is shown in the bottom panel of Fig. \ref{CMF_evolution_tests}. The three profiles corresponding to the evolutionary groups are still clearly separated. The evolutionary trend is thus already well visible from the measured fluxes. Fluxes are of course physically related to the temperature of the sources, but the assumptions made to convert them into masses have no role in creating the evolutionary trend of the CMF. 

Based on these tests, we can definitely rule out that the evolutionary trend that we observe for the overall ALMAGAL CMF (Fig. \ref{CMF_evolution}) is determined by the adopted temperature model. Ultimately, the result is then robust against both resolution and temperature effects.

\end{appendix}


\end{document}